\documentclass[10pt]{article}
\usepackage[margin=0.6in]{geometry}
\usepackage{hyperref}
\usepackage{multicol}
\usepackage{blindtext}
\usepackage{amsmath}
\usepackage{amssymb}
\usepackage{mathrsfs}
\usepackage{dsfont}
\usepackage{framed}

\newcommand\unmarkedfootnote[1]{%
  \begingroup
  \renewcommand\thefootnote{}\footnote{#1}%
  \addtocounter{footnote}{-1}%
  \endgroup
}

\usepackage{algorithm}
\usepackage{algpseudocode}
\usepackage{authblk}
\usepackage{graphicx}
\usepackage[table]{xcolor}
\usepackage{subcaption}
\usepackage{wrapfig}
\usepackage{float}
\usepackage{mdframed}
\usepackage{multirow}
\mdfdefinestyle{mddefault}{
rightline=true,innerleftmargin=4pt,innerrightmargin=4pt,innertopmargin=4pt,innerbottommargin=2pt
frametitlerule=false,backgroundcolor=black!5}

\usepackage{framed,color}
\definecolor{shadecolor}{rgb}{0.85,0.85,0.85}
\usepackage{ulem}
\usepackage{url}
\usepackage{cleveref}
\usepackage{hyperref,xcolor}
\hypersetup{
colorlinks,
linkcolor={red!50!black},
citecolor={blue!50!black},
urlcolor={blue!80!black}
}
\usepackage{xr}
\makeatletter
\newcommand*{\addFileDependency}[1]{
\typeout{(#1)}
\@addtofilelist{#1}
\IfFileExists{#1}{}{\typeout{No file #1.}}
}\makeatother

\usepackage{diagbox}
\usepackage{caption}
\usepackage{subcaption}

\RequirePackage{tikz} 
\newcommand{\orcidicon}{\includegraphics[width=0.32cm]{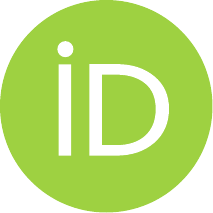}}

\foreach \x in {A, ..., Z}{%
\expandafter\xdef\csname orcid\x\endcsname{\noexpand\href{https://orcid.org/\csname orcidauthor\x\endcsname}{\noexpand\orcidicon}}
}

\title{Disc Game Dynamics: \\ A Latent Space Perspective on Selection and Learning in Games}
\author{Pablo Lechon-Alonso $^1$, Andrew Dennehy $^1$, Ruizheng Bai $^2$, Nicolas Sanchez $^3$, \\ Derek K.~Wise $^4$  \orcidB{} , David Sewell $^4$, David Rosenbluth $^4$, and Alexander Strang $^{3+}$ \orcidA{} }

\affil{\footnotesize $^{1}$ University of Chicago \quad
$^{2}$ Texas A\&M \quad
$^{3}$ University of California, Berkeley \quad $^{4}$ Lockheed Martin.\\
$^{+}$ Correspondence: alexstrang@berkeley.edu}

\date{\vspace{-0.6em} \today}

\date{\today}

\begin{document}

\maketitle

\tableofcontents

\vfill
\unmarkedfootnote{PIRA CET2025010209}

\pagebreak 

\section{Abstract}

Evolutionary game theory studies populations that change in response to an underlying game. Often, the functional form relating outcome to player attributes or strategy is complex, preventing mathematical progress. In this work, we axiomatically derive a latent space representation for pairwise, symmetric, zero-sum games by seeking a coordinate space in which the optimal training direction for an agent responding to an opponent depends only on their opponent's coordinates. The associated embedding represents the original game as a linear combination of copies of a simple game, the disc game, in a new coordinate space. In this article, we show that disc-game embedding is useful for studying learning dynamics. We demonstrate that a series of classical evolutionary processes simplify to constrained oscillator equations in the latent space. In particular, the continuous replicator equation reduces to a Hamiltonian system of coupled oscillators that exhibit Poincar\'e recurrence. This reduction allows exact, finite-dimensional closure when the underlying game is finite-rank, and optimal approximation otherwise. It also establishes an exact equivalence between the continuous replicator equation and adaptive dynamics in the transformed coordinates. By identifying a minimal rank representation, the disc game embedding offers numerical methods that could decouple the cost of simulation from the number of attributes used to define agents. These results generalize to metapopulation models that mix inhomogeneously, and to any time-differentiable dynamic where the rate of growth of a type, relative to its expected payout, is a nonnegative function of its frequency. We recommend disc-game embedding as an organizing paradigm for learning and selection in response to symmetric two-player zero-sum games.

\section{Introduction}

\subsection{Paradigms in Evolutionary Game Theory} \label{sec: paradigms}

Evolutionary game theory studies the dynamics of populations adapting in response to an underlying game \cite{smith1982evolution}. Evolutionary game theory has a rich history in biology. Example applications include the evolution of sex-ratios \cite{hardy2002sex,trivers1983evolution}, ritual fighting \cite{houston1999models,smith1982evolution}, mate choice \cite{iwasa1995continual}, sibling rivalry \cite{mock1997evolution}, social habitat selection and dispersal \cite{doebeli1997evolution}, cooperative behavior \cite{pfeiffer2001cooperation,velicer2003evolution}, honest and dishonest signalling, parasite-host interactions \cite{abrams2001modelling}, and infection virulence \cite{anderson1991infectious}. For a survey see \cite{nowak2004evolutionary}. Evolutionary game theory is also important in learning theory, where it presents the generic framework for studying multi-agent optimization, mini-max optimization, and adversarial learning \cite{bailey2019multi,balduzzi2018mechanics,goodfellow2014generative,immorlica2011dueling,schuurmans2016deep}.

Before proceeding, we emphasize two contrasts that position our narrative: evolutionary processes need not behave as noisy optimizers, and, co-evolutionary dynamics require dynamic, not static, solution concepts.  

First, evolution is not, generically, optimization. In the folk narrative, random perturbation and selection combine to drive populations upwards on a fitness landscape \cite{frauenfelder1997p,wright1932roles}. If taken literally, this model presupposes the existence of a fitness function, i.e.~the stochastic process specifying evolution over possible types can be shown to, in some sense, optimize a function that assigns ``fitness" to types. No such function need exist \cite{Kerr,sinervo1996rock}. Moreover, even if such a function exists, it may not provide a useful dynamical summary \cite{cairns2022strong,de2018utility}, as when it is highly rough \cite{de2014empirical,kouyos2012exploring,pitt2010rapid,rowe2010analysis,szendro2013quantitative}, quickly varying due to environmental perturbations \cite{trubenova2019surfing}, or when the dynamics are dominated by noise, as in neutral theory \cite{chan2002perspectives,munoz2021constructive,ohta1992nearly,schuster2006prediction,stoltzfus1999possibility,wagner2008neutralism}. 

If taken figuratively, the optimization perspective should, at least, provide generalizable intuition. This point is also circumspect. Evolution is an emergent feature of ecological dynamics, which are almost universally frequency-dependent. When equated to per-capita growth rate, the fitness of a type often depends on the current frequency of other types. Then, even if the population adapts to increase the frequency of fit types, it does not optimize any fixed objective, since the objective changes in tandem with the population. To quote Nowak and Sigmund \cite{nowak2004evolutionary}, ``The landscape metaphor... neglects one-half the evolutionary mechanism." Namely, co-evolution.

Frequency-dependent selection can produce evolutionary dynamics that are incompatible with any optimization process \cite{bailey2019multi}. Optimization implies constant improvement, and directed motion towards, or convergence to, local optima \cite{nowak2004evolutionary}. More subtly, instantaneous step-wise improvement ensures cumulative progress. In contrast, when frequency-dependent, evolutionary dynamics need not admit any notion of improvement, need not admit any notion of direction, and need not converge, since they may circulate periodically, recur, or mix chaotically \cite{akin1984evolutionary,boone2019darwin,flokas2020no,hofbauer1996hamiltonian,legacci2024geometric,mertikopoulos2018cycles,sato2002chaos}. 

So, in the generic co-evolutionary setting, optimization, at best, provides an instantaneous intuition via a constantly shifting, population-dependent, fitness ``seascape" \cite{merrell1994adaptive,mustonen2009fitness}. Even proponents of the seascape model admit that it is difficult to analyze and is hard to communicate as a minimal mental model \cite{cairns2022strong}. 

Collected, these observations beg a critical question; \textit{what is the utility of an optimization metaphor whose objective is ill-defined or constantly changing in response to the system's state}? 

In response, we will seek alternate paradigms. What paradigm should complement optimization?  

To scope alternatives, let's step back to a more fundamental problem frame. Population dynamics are often parameterized by a fixed set of rules governing interactions between individuals and types. Formally, these rules fix, or are fixed by, an underlying game \cite{smith1982evolution}. Accordingly, game theory provides a candidate frame \cite{nowak2004evolutionary}. 

Formally, a game is a function that accepts the strategies selected by a group of players, and returns a utility to each based on the collection of chosen strategies. Thus, games model multi-agent interactions. For ecological, or evolutionary, realizations, exchange strategy for type, and utility for growth rate. Importantly, only narrow classes of games, e.g. potential games \cite{monderer1996potential}, admit an exact analogy to optimization. Instead, generic games exhibit a mixture of potential and harmonic components \cite{candogan2011flows,legacci2024geometric}. If the harmonic component is non-zero, then the game does not admit an exact reduction to optimization. Generic evolutionary dynamics in response to harmonic games are poorly understood \cite{legacci2024geometric}.

While game theory is a classically static theory, evolutionary game theory is \textit{dynamic}. Classical game theory studies equilibria as solution concepts for games. While an equilibria is commonly understood as a fixed point of some dynamic \cite{holt2004nash}, the dynamic in traditional game theory is usually implicit, if not unspecified \cite{mailath1998people}. Equilibria are typically justified by rationality criteria, most famously, that a set of strategies cannot be stable if any agent could improve their utility by adjusting their strategy unilaterally \cite{nash1950equilibrium}. That said, while stationarity conditions are often considered necessary for rationality (c.f.~\cite{mailath1998people}), they are not sufficient to show that equilibria can be, will be, or frequently are discovered or approached by evolving populations. Nor can they identify which equilibrium a population will select.

Evolutionary game theory aims to solve the equilibrium selection problem by posing specific evolutionary dynamics \cite{mailath1998people}. Equilibria are often fixed points for these dynamics. Indeed, all asymptotically stable rest points of a monotone dynamic are Nash equilibria \cite{mailath1998people}. However, not all equilibria are asymptotically stable or attracting. Additional criteria are almost always needed \cite{fudenberg1998learning}. For example, inefficient equilibria are often unstable and invasible under simple dynamics \cite{mailath1998people}. Since generic equilibria are not attracting for all game categories or all reasonable dynamics \cite{newton2018evolutionary}, evolutionary game theory has produced a refined taxonomy of equilibrium classes that satisfy increasingly stringent conditions. In some cases, as for evolutionary-stable strategies, simple non-invasibility conditions suffice, and ensure global attraction for a wide class of games and dynamics \cite{hofbauer2003folk,smith1982evolution}. In other settings, stronger conditions are required (c.f.~\cite{Apaloo, dieckmann1996dynamical,eshel1983evolutionary,leimar2005multidimensional,oechssler2001evolutionary}). For extensions to stochastic processes see \cite{cabrales2000stochastic,foster1990stochastic,fudenberg1998learning,imhof2005long,kandori1993learning}. 

The diversity of solution concepts reflects a key impossibility result; there is no uncoupled learning dynamic that converges to a Nash equilibrium in all games from all initial conditions \cite{hart2003uncoupled,hart2006stochastic}. For example games whose Nash equilibria are not approached by any uncoupled dynamic, or by any adjustment dynamic, see \cite{hart2003uncoupled,hofbauer1998evolutionary}. In retrospect, it is not surprising that reasonable learning dynamics, which typically react myopically to limited game information with boundedly rational updates, do not have the power to discover generic equilibria. Nash equilibria are hard to compute \cite{chen2009settling,daskalakis2009complexity,mehta2014constant}, to approximate \cite{rubinstein2015inapproximability,rubinstein2017settling,skopalik2008inapproximability}, and to communicate \cite{babichenko2017communication,hart2010long}. Moreover, even when players can communicate, Nash equilibria need not be self-enforcing \cite{aumann1990nash,clark2001nash}. Experimental evidence for Nash play is similarly qualified \cite{chiappori2002testing,holt2004nash}, even in foundational studies on canonical examples \cite{flood1958some,raiffa1992game}. For laboratory studies, see \cite{brown1990testing,levitt2010what,mccabe2000experimental,mookherjee1994learning,o1987nonmetric,ochs1995games,palacios2008experientia,rapoport1992mixed}. For observational studies in real-world games, see \cite{azar2011soccer,arrieta2020professionals,dohmen2018further,gauriot2016nash,gilbert1994wagering,kovash2009professionals,palacios2003professionals}. For example controversies, see \cite{o1987nonmetric} versus \cite{brown1990testing}, or \cite{palacios2003professionals,palacios2008experientia} versus \cite{kovash2009professionals,levitt2010what}. 

Convergence time is also problematic. Even if an equilibrium is attracting, convergence may be too slow to plausibly restrict observed populations near it. This concern mirrors a broader doubt that dynamical systems theories in ecology have focused too narrowly on linearized dynamics near steady-states (c.f.~\cite{may2019stability}), and have not fully addressed the possibility that real systems are better characterized by slowly decaying transients \cite{hastings2001transient,hastings2004transients,hastings2018transient}.

These concerns raise another critical question; \textit{are equilibria sufficient solution concepts for evolutionary games?} 


At strongest, all generic no-regret learning dynamics converge to equilibria in the sense that the time-averaged population converges to a subset of the game's coarse correlated equilibria \cite{greenwald2003general,hart2000simple,hannan1957approximation}. In general two-player games, the coarse correlated equilibria may be exclusively supported on strictly dominated strategies \cite{viossat2013no}, so fail basic rationality criteria \cite{flokas2020no}. However, in constant-sum two-player games, all coarse correlated equilibria are Nash equilibria \cite{macqueen2023proof}. So, generic no-regret learners may discover Nash equilibria. Importantly, this convergence result only holds for the time-averaged population. So, hindsight rationality only guarantees that long trajectories are rational when taken as a whole. The population at any particular time is only justified by the adaption process that produced it, not by any static notion independent of the trajectory.\footnote{Whether convergence of the time average is sufficient or insufficient depends on the application. In artificial settings, the simulator can time-average post hoc. In real settings, time averaging can be performed if a long trajectory is recorded, but longitudinal data is expensive to collect, and does not characterize the population at any fixed time. If the population can be subdivided into parts that evolve independently, then a time average may be realized by an average over separate populations. This, however, either requires an ergodic evolutionary dynamic, or, in the absence of ergodicity, specially chosen initial conditions. The latter explanation begs the question: what process would initialize the populations appropriately?} 

At weakest, while equilibria necessarily exist, and can be computed using linear programming, in two-player zero-sum games \cite{von2007theory}, they are not attracting under the most basic evolutionary game dynamics unless they correspond to pure strategies that dominate all other strategies \cite{akin1984evolutionary,boone2019darwin,hofbauer1996hamiltonian}. Instead, even in the two-player zero-sum setting, basic evolutionary dynamics, including a broad class of no-regret dynamics, are recurrent \cite{bailey2019multi,boone2019darwin,flokas2020no,legacci2024geometric,mertikopoulos2018cycles}. Recurrent dynamics orbit persistently, thus fail to converge. For instance, the replicator dynamic, which is the most widely used model in evolutionary game theory, either retains a constant KL divergence away from a fully-mixed equilibria, or diverges from that equilibria when simulated in discrete time \cite{akin1984evolutionary,boone2019darwin}. Recurrent dynamics return arbitrarily close to any initial condition arbitrarily often, so do not satisfy any notion of directed learning. While these results only apply when a fully mixed equilibrium exists, if no such equilibrium exists, then at least one strategy is dominated, and will be eliminated by the dynamic \cite{akin1980domination}. Thus, in two-player zero-sum games, the canonical no-regret learning dynamic only succeeds in iteratively eliminating dominated strategies, before orbits recurrently among the surviving strategies \cite{akin1984evolutionary}.

Both critical questions interrogate widely used, if often criticized (c.f.~\cite{gould1979spandrels}), paradigms. Both paradigms remain attractive mental models because they provide clear solution concepts, simplify complex dynamics, and promise guiding intuition by analogy to familiar processes. Both have directed large bodies of successful work.\footnote{Examples include robust linearized analyses in certain ecological systems \cite{abbott2020non,macarthur1967limiting,nicholson1935balance,rosenzweig1971paradox}, empirical efforts to measure fitness landscapes with applications to drug sequencing, the evolution of antibiotic resistance, and HIV treatment \cite{goulart2013designing,hall2002predicting,kouyos2012exploring,otwinowski2013genotype,palmer2013understanding,pitt2010rapid,rowe2010analysis,szendro2013quantitative}, and observational demonstration of equilibrium play in real games \cite{azar2011soccer,chiappori2002testing,gauriot2016nash,palacios2003professionals}.} A compelling alternative needs to offer a similar suite of advantages \cite{efron1986isn,kuhn1997structure}.

We argue, in the restricted setting of two-player, symmetric, constant-sum games, for an alternative that addresses each of the shortcomings of the standard paradigms while encompassing them as special cases and critical points. Our alternative is either exact or allows arbitrarily accurate approximation in essentially all constant-sum games. It allows precise mathematical conversion between learning dynamics and physical processes. These conversions allow visualization, and supply clear dynamical intuition which, unlike optimization, correctly captures the \textit{generic} behavior of evolution subject to zero-sum games. Importantly, the paradigm produces \textit{global, dynamical}, rather than local, static, intuition. The paradigm admits solutions that converge and that orbit persistently, and provides a clean geometric criteria separating the convergent and recurrent cases.

Advocating for a paradigm requires detailed exposition. In this pursuit, we adopt Polya's imperative regarding vague analogies: ``clarify them" \cite{gentner1993shift}. We aim to show that our paradigm can be derived from simple aims regarding the choice of coordinates used to represent the game, that the resulting representation is general, exact, unique, and interpretable. Finally, we show that it is \textit{useful} through an in-depth case-study. En route we recover and generalize a representation first proposed in \cite{bailey2019multi,balduzzi2019open,balduzzi2018mechanics,balduzzi2018re}, then use it to extend classical results constraining the behavior of the replicator dynamic subject to zero-sum games to continuous trait spaces \cite{akin1984evolutionary,hofbauer1996hamiltonian,boone2019darwin}. In addition, we demonstrate an exact equivalence between the replicator dynamic and adaptive dynamics. We show that our key dynamical conclusions generalize to metapopulations, and to any dynamic where the rate of growth in the density of a type, relative to its expected payout, is a nonnegative function of its density. These results strongly recommend the paradigm.

\subsection{Game Formalism}

\begin{figure}[t]
    \centering
    \includegraphics[trim = 60 50 60 50, clip, width = \textwidth]{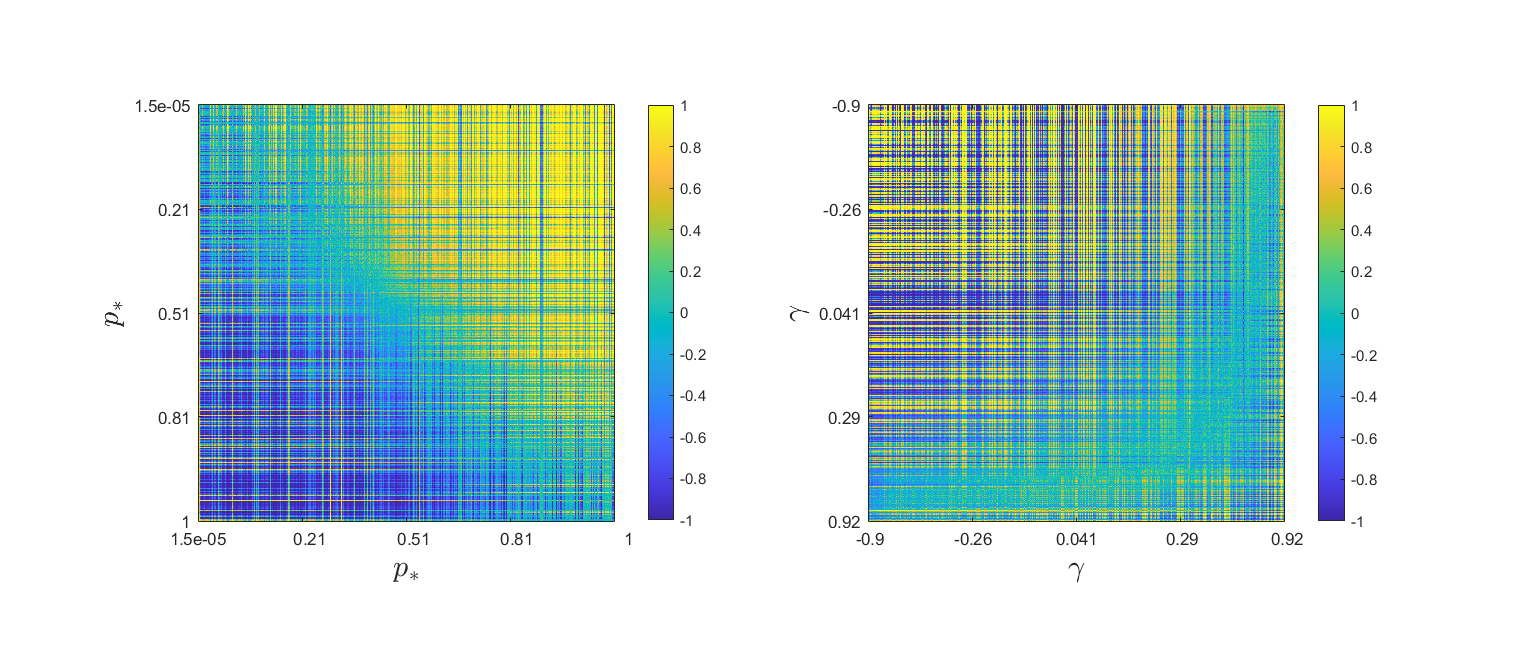}
    \caption{Payout matrices for an population implementation of the iterated prisoner's dilemma (See Appendix Section \ref{app: example game}). The color of the $i,j$ entry represents $f(x(i),x(j))$ for a population of 800 randomly drawn agents with attributes $\{x(i) \}_{i=1}^{800}$. The matrices represent the same data, but with the population ordered according to different attributes. Note that, while performance could be roughly approximated as a simple function of the attributes $p_*$ and $\gamma$, in both cases an exact description is difficult. For example, the left panel shows that agents with large $p_*$ (rows near the bottom), tend to lose to agents with a small $p_*$ (columns near the left. However, this game is not transitively ordered by $p_*$ alone, as there exist upsets against this prediction. The upper left half is mostly, but not entirely yellow, and the bottom half is mostly, but not entirely, blue. The upsets correspond to particular choices of the pair $(p_*,\gamma)$ that reverse the general trend in $p_*$. Even in this two-parameter case, it is not immediately apparent how to extract a simple strategic description from this data, despite the apparent structure. }
    \label{fig: IPD Payout Matrices}
\end{figure}

This paper pursues a canonical coordinate system for learning and selection dynamics driven by a pairwise, symmetric, zero-sum, functional form game. A functional form game, $(\Omega, f)$, is defined by a space of attributes $\Omega$, where $x \in \Omega$ is collection of attributes used to characterize agents, and $f: \Omega \times \Omega \rightarrow \mathbb{R}$ is a payout function where $f(x,x')$ returns the payout to an agent with attributes $x$ when paired against an opponent with attributes $x'$. If the game is symmetric, then agents do not take on distinguishable roles in the game, so share the same payout function. 
If the game is symmetric and zero-sum, then the payout function $f$ must satisfy $f(x,x') + f(x',x) = 0$, so $f$ must be skew-symmetric; $f(x,x') = - f(x',x)$. 

The attributes $x$ are any list of agent characteristics that determine payout \cite{strang2022network,strang2022principal}. Attributes are not limited to physical characteristics, play statistics, decision probabilities, etc. Instead, attributes differ by context. Since we will seek a procedure for standardizing agent description, the generality of this initial description is a strength of the subsequent analysis.
Some examples follow:
\begin{enumerate}
\item In biology, attributes may include phenotypic, morphological, or behavioral characteristics. Common examples include body size, aggressiveness, and testosterone levels \cite{chang2012aggression,yan2024winner}. As specific examples, \cite{Beacham, Haley, Stuart} study the relationship between attributes and competitive success in pumpkinseed sunfish, elephant seals, and Cape dwarf chameleon. For sunfish, relevant attributes include body size and experience \cite{Beacham}. For elephant seals, relevant attributes include body mass, length, age, and time of beach arrival \cite{Haley}. For chameleons, relevant attributes include body mass, morphology (torso, tail, and jaw length, head width, casque size), and the size of a signaling patch \cite{Stuart}. Additional examples are studied in \cite{Jackson_b,Kerr,Sapolsky,Solberg,vieira2013winners}. 

\item In sports analytics, attributes may include physical characteristics and play statistics. Example predictive performance models that attempt to relate play statistics to performance are reviewed in \cite{chazan2020sports,kapadia2022sport,Kovalchik,sarlis2020sports,Soto,vistro2019cricket,wilkens2021sports}. 

\item In artificial settings, attributes are typically parameters. For example, in traditional game theory, mixed strategies are parameterized by decision probabilities. A collection of decision probabilities, or policy, may be the output of a more complicated model, as is usually the case in artificial intelligence, where neural network or decision tree architectures may be employed to produce a policy, or to predict value, based on game state (c.f.~\cite{brown2019superhuman,schrittwieser2020mastering,silver2017mastering,silver2018general,tesauro1990neurogammon,moravvcik2017deepstack}). In this setting, the explicit parameters of the policy and value networks are a valid, if exhaustive, collection of attributes. These attributes often carry little to no meaning when isolated from the full list of parameters, which is typically too long to interpret without tools that can reorganize and compress the agent description. At a higher level the hyperparameters used for tuning training may be used as attributes. 
\end{enumerate}

\subsection{Objectives}
Suppose you can observe a population evolving in response to a functional form game. The game is defined by a payout function $f$ whose specific form depends on the attributes used to describe the agents. You can choose what attributes to measure. Different attributes may be easier or harder to estimate, and may be more or less related to competitive success. Often, it is not clear a priori \textit{what} variables are relevant, that is, which variables are sufficient to predict advantage. Often, it is also unclear which \textit{combination} of these variables provides the simplest coordinate space to represent competition.

For example, is body mass, experience, or priority more important among elephant seals \cite{Haley}?  
What combination of play statistics are most predictive of performance in sports \cite{Soto}? What changes in the weights of a policy network correspond to interpretable changes in strategy? 

Both problems matter. Identifying a minimal, sufficient set of attributes is important since measurements require effort. Organizing those attributes to simplify the mapping from attributes to outcome is important when extracting a strategic interpretation or predicting selection for specific traits. This organization problem is closely related to feature engineering in artificial intelligence \cite{bengio2013representation,franccois2019combined,zhang2020learning,zheng2018feature} and credit assignment in reinforcement learning \cite{arumugam2021information,ferret2019self,minsky1961steps,sutton1984temporal,sutton1988learning,van2021expected}. 
Identifying a latent space that organizes parameters is a longstanding area of research in both fields \cite{agogino2004unifying,pignatelli2023survey,sutton1984temporal}. Organizing attributes by importance can also guide attribute selection, so can help solve the first problem.

What, then, should you measure and in what combination? We will attempt to answer the second question. 

Suppose that you can measure all relevant attributes, but do not know, a priori, how they determine competitive outcome. For example, consider an artificial setting where agents compete in a defined game via a simulator, and act according to a policy that involves many parameters. We will distinguish extrinsically motivated coordinates, those chosen based on considerations separate from the payout, $f$, such as ease of measurement or convenience of parameterization, from intrinsically motivated coordinates, those chosen based on $f$ alone. Intrinsically motivated coordinates should be selected to simplify the functional form of $f$. For consistency, these coordinates should only depend on $f$ and the frequency of agent types, not the initial attributes used for description. 

Identifying an intrinsically meaningful coordinate system is a classical problem in model reduction and exploratory data analysis. For example, principal component analysis, multi-dimensional scaling, diffusion mapping, sparse dictionary learning, and variational auto-encoding all seek a mapping into a latent space that simplifies a data set by encoding some of its structure in an embedded geometry \cite{abdi2010principal,coifman2006diffusion,hall1970r,kingma2019introduction,koren2003spectral,kruskal1978multidimensional,pearson1901liii,torgerson1952multidimensional}. That geometry is typically equipped with a canonical interpretation that gives meaning to the coordinates, and provides a frame for deriving a specific representation from an arbitrary initial coordinate system. The striking success of transformer models suggests that complex functional relationships can be effectively learned from sample data by first embedding entirely arbitrary token representations into such latent spaces \cite{karvonen2024emergent,li2022emergent,toshniwal2022chess}. 

We seek a meaningful latent space by mapping from an arbitrary, sufficient trait space into a new trait space, selected to simplify the relationship between agent description and agent performance. 
We pursue a coordinate change that:
\begin{enumerate}
    \item simplifies $f$ so that the selective response of a population is intuitive, analytically tractable, and/or efficiently computable when working in the latent space, and
    \item uses only as many variables as needed, and allows optimal approximation upon truncation. 
\end{enumerate}


To select a canonical coordinate system, we introduce a constraint on the desired representation. The constraint is chosen to plausibly accomplish the first goal for a range of selection dynamics. Section \ref{sec: desiderata} specifies the constraint. Section \ref{sec: canonical coordinates} explores a special case when the constraint is satisfied, then shows that, if satisfied, the chosen constraint fixes the desired representation. Namely, if the constraint is satisfied, then the game may be expressed as a combination of simple games after changing coordinates. These are disc games. Disc games encode strategic relations in a simultaneously expressive and intuitive geometry \cite{balduzzi2018re,balduzzi2019open,strang2022principal}. In Section \ref{sec: disc games} we list sample relations between strategy and geometry to show that the coordinates in a disc game are interpretable. These relations recommend the disc game as a natural, desired representation. Section \ref{sec: embedding} shows that such a representation exists generically, and is unique in the sense that it produces a consistent representation that is invariant to changes in agent description. We specify the method needed to perform the coordinate change, and show that it allows optimal model reduction, thus satisfies our second objective. 


To show that the new coordinates are actually useful for particular dynamics we focus on generalizations of the continuous replicator equation. Other sample dynamics are discussed in the appendix. In Section \ref{sec: latent space dynamics} we show that the coordinate change can be used to reduce the continuous replicator dynamic to a Hamiltonian system of coupled oscillators that is equivalent to adaptive dynamics in the new coordinates. These results show that the intuition suggested by the representation is accurate. In Section \ref{sec: numerics} we show that the coordinate change also allows faster computation by reducing the dimension used for agent description. Collectively, we argue that disc game dynamics offer a compelling complement to optimization as a paradigm for evolution in response to pairwise, symmetric, zero-sum games.

\pagebreak

\section{Constructing a Latent Space} \label{sec: latent space}

\subsection{Desiderata} \label{sec: desiderata}
Let $\Omega$ represent a set of possible agent attribute where $x \in \Omega$ represents a particular collection of attributes. Unless stated otherwise, we will assume that $\Omega$ may be represented as a subset of a finite-dimensional vector space.  
We assume that $\Omega$ is sufficient to predict payout so there exists a function $f:\Omega \times \Omega \rightarrow \mathbb{R}$ where $f(x,x')$ returns the competitive advantage (payout) an agent of type $x$ receives when interacting with an opponent $x'$. We assume that $f$ is deterministic. If the game is stochastic, or there are relevant agent attributes outside $\Omega$ that can be sensibly averaged given the attributes $x \in \Omega$, then $f$ is defined as an expected payout. We restrict our attention to symmetric, zero-sum games, so assume that $f$ is skew-symmetric: $f(x,x') = -f(x',x)$ for all pairs $(x,x') \in \Omega \times \Omega$.

Consider an agent with attributes $x$, paired against an opponent with attributes $x'$. How should the agent $x$ adapt, to best respond to $x'$? To focus our construction, we will start by assuming that $\Omega$ is a Euclidean vector space, and $f$ is differentiable. We will relax both constraints later. 

At simplest, the first agent might adapt their attributes to maximize $f(x,x')$ over all possible choices of $x$. This would define a best-response dynamic \cite{fudenberg1998learning}. If maximization is impossible, then the agent might simply aim to change their attributes so as to increase $f(x,x')$. If the agent is restricted to adapt continuously, or via small steps, then the agent should adapt along a direction which maximizes $f(\cdot,x')$ over all possible choices near $x$. If $f$ is sufficiently regular, then, in the limit of small steps, the first agent should adapt along the direction of fastest ascent about $x,x'$. That is, along the gradient of $f$ with respect to the first input, evaluated at $x$: $\nabla_w f(w,x')|_{w = x}$. Let $v_f(x,x')$ denote the optimal local training direction:
\begin{equation} \label{eqn: optimal local training}
    v_f(x,x') = \nabla_{w} f(w,x')|_{w = x}
\end{equation}

The optimal local training vector plays a fundamental role in many selection dynamics including self-play, ficitious self-play, simultaneous gradient ascent, and adaptive dynamics \cite{fudenberg1998learning,nowak2004evolutionary}. See Appendix Section \ref{app: dynamics} for details. These dynamics need not assume rational agents, nor rational designers, as the optimal local training vector fixes the local linear approximation to $g$ about $y$ and $y'$. Consequently, models that implicitly enforce a form of myopic gradient ascent may be parameterized by $v$ without requiring explicit calculation of gradients \cite{abrams2001modelling}. In Section \ref{sec: replicator} we show that $v$ remains relevant for an essential ecological dynamic that does not require local perturbations to agents' types. 

The optimal training vector $v_f$ is defined on the product space consisting of pairs of agent attributes. Selection is easier to study if the vector field is constant in one of its arguments. Since $v_f$ is defined as the optimal local training vector for $x$ when paired with $x'$ we will consider $x$ the student and $x'$ their opponent, who acts as a teacher in the sense of a sparring partner. Since we can't guarantee that $v_f$ is constant in one of its arguments, we will seek a procedure for changing coordinates such that the vector field in the new coordinates, $v_g$, is constant in one of its arguments. 

Let $T:\Omega \rightarrow \Psi$ denote a one-to-one transformation mapping from $\Omega$ to $\Psi$. If $T$ is diffeomorphic, then $y = T(x)$ is the new representation of an agent with attributes $x$ after a change of coordinates. Let $g(y,y') = f(T^{-1}(y),T^{-1}(y'))$, where $T^{-1}$ denotes the inverse transformation over $\Psi$. Then $g(y,y') = f(x,x')$ when $y = T(x)$ and $y' = T(x')$. We will constrain $T$ by enforcing simplifying criteria on $v_g$, which will, in turn, enforce simplifying criteria on $g$. Since we seek constraints on $v_g$ we will, from here on out, suppress the subscript, and let $v$ denote $v_g = \nabla_w g(w,y)|_{w = y}$. 


\begin{snugshade}
\noindent \textbf{Desiderata: } Identify a transform $T$ such that:
\vspace{-0.05 in}
\begin{enumerate}
    \item[D1.] $v(y,y')$ depends only on the student's attributes, $y$, and is independent of the opponent.
    \vspace{-0.05 in}
    \item[D2.] $v(y,y')$ depends only on the opponent's attributes, $y'$, and is independent of the student.
\end{enumerate}
\end{snugshade}

Surprisingly, desideratum (D1) is only achievable for an extremely limited subset of games, while desideratum (D2) is achievable without placing any substantive restrictions on $f$. We will show that desideratum (D2) can be accomplished in general (see Secion \ref{sec: embedding}) and uniquely specifies $T$ up to linear transformation and a choice of measure on $\Omega$.

Desideratum (D1) implies that the optimal training direction for $y$ depends on $y$ alone and is independent of their opponent. Then selection and learning reduce to optimization of an opponent-independent fitness or rating. Not all payout functions can be reduced to a fitness comparison. Those that can are perfectly transitive \cite{cebra2023similarity}:

\begin{snugshade}
\noindent \textbf{Definition:} A skew-symmetric function $f:\Omega \times \Omega \rightarrow \mathbb{R}$ is \textbf{perfectly transitive} if there exists a rating function $r:\Omega \rightarrow \mathbb{R}$ such that $f(x,x') = r(x) - r(x')$ for all pairs $(x,x') \in \Omega \times \Omega$. 
 \end{snugshade}

 Perfectly transitive payout functions induce well-ordered hierarchies and do not admit any cyclic (intransitive) competitive relationships. Generic payouts need not be perfectly transitive, as they may admit cycles \cite{strang2022network}. 

If $f$ is perfectly transitive, then $g(y,y') = r(T^{-1}(y)) - r(T^{-1}(y'))$, so is also perfectly transitive. 
If $g(y,y') = r(y) - r(y')$ for some $r$, then $f(x,x') = g(T(x),T(x')) = r(T(x)) - r(T(x'))$, so is also perfectly transitive. If desideratum (D1) is satisfied, then the fundamental theorem of calculus (FTC) implies that $g$ must be perfectly transitive, so $f$ must also be perfectly transitive.

 \begin{snugshade}
     \noindent \textbf{Lemma 1: [Opponent-Independent Learning Implies Transitivity]} If there exists a transformation $T$ such that $g$ satisfies desideratum (D1) and $\Psi$ is connected\footnote{Can we generalize to drop this constraint?}, then $f$ is perfectly transitive.
 \end{snugshade}

\noindent \textbf{Proof Outline:} Set $r(y) = \int_{w = y_0}^y v(w,z) dy$ for any choice of $y_0$ and $z$ in $\Psi$. Then, by the FTC, $g(y,y') = r(y) - r(y')$. See Appendix Section \ref{app: lemma 1} for details. $\square$
\vspace{0.1 in}

Therefore, desideratum (D1) cannot be achieved in general. 
Consider desideratum (D2) instead.

\begin{snugshade}
    \noindent \textbf{Lemma 2: [Student-Independent Learning Implies Bi-Affine]} The vector field $v(y,y')$ does not depend on $y$ if and only if $g(y,y')$ is a bi-affine function of $y$ and $y'$ of the form:
    \begin{equation} \label{eqn: bilinear form}
        g(y,y') = \hat{y}^{\intercal} \hat{G} \hat{y}' =  \left[y^{\intercal},1 \right] \left[\begin{array}{cc} G & a \\ -a^{\intercal} & 0  \end{array}\right] \left[ \begin{array}{c} y' \\ 1 \end{array} \right] = y^{\intercal} G y' + a^{\intercal}(y - y') 
    \end{equation}
    where $G$ is skew-symmetric.
\end{snugshade}

\noindent \textbf{Proof Outline:} If $g$ is bi-affine, then $v$ is student-independent since, for each fixed $y'$, $g(y,y')$ is an affine function of $y$. If $v(y,y')$ does not depend on $y$, then, for fixed $y'$, $g(y,y')$ must be an affine function of $y$. By skew-symmetry, if $g$ is an affine function of $y$ for fixed $y'$, it is also an affine function of $y'$ for fixed $y$. Thus, $g$ is bi-affine and admits the general form \eqref{eqn: bilinear form}. Expanding to $\mathbb{R}^d$, by replacing $y$ with $[y,1]$, expresses the bi-affine function as a bilinear function in $\mathbb{R}^{d+1}$, restricted to a $\mathbb{R}^d$ dimensional affine subspace. See Appendix Section \ref{app: lemma 2} for details. $\square$
\vspace{0.1 in}

The appended coordinate fixed to 1 in Equation \eqref{eqn: bilinear form} contributes a perfectly transitive component with rating function $r(y) = a^{\intercal} y$. We will, from now on, expand the coordinates $y$ to include the appended $1$ whenever $a \neq 0$, so that bi-affine functions may be expressed as restrictions of bilinear functions. So, with some abuse of notation, we discard the hats, and use $y$ to mean $\hat{y}$ and $G$ to mean $\hat{G}$. This reduction is worthwhile since all subsequent statements hold with an appended transitive term if expressed explicitly, and we will subsequently focus on a class of functional form games that can be expressed in a bilinear form, restricted to a subset of a larger Euclidean space.

Not all bilinear payouts are perfectly transitive. For example, the simple bilinear game $g(y,y') = y_1 y'_2 - y_2 y'_1$ allows cyclic advantage relationships if $\Psi$ includes the origin in its convex hull. Thus, while desideratum (D1) requires perfect transitivity, desideratum (D2) does not. desideratum (D2) is more general. It retains some coupling between student and opponent, so learning and selection in games that satisfy desideratum (D2) remain a multi-agent dynamics.

Bilinearity is attractive for two reasons. First, if bilinear, then the optimal training vector $v(y,y') = v(y')$ completely parameterizes performance for all agents paired against the opponent $y'$. In particular $v(y') = G y'$ and $g(y,y') = y \cdot v(y')$. In this sense, the local linearization of payout $g$ defined by the gradient $v$ extends globally when $g$ is bilinear. Third, many selective dynamics update agents or populations based on either expected best responses or expected payouts when sampling an opponent $y'$ from a distribution $\pi$. Expectations are linear operations, so:
\begin{equation}
    \bar{v}_{\pi} = \mathbb{E}_{y' \sim \pi}[v(y,y')] = \mathbb{E}_{y' \sim \pi}[v(y')] = \mathbb{E}_{y' \sim \pi}[G y'] = G \mathbb{E}_{y' \sim \pi}[y'] = v(\bar{y}_{\pi}).
\end{equation}

Therefore, when desideratum (D2) holds, the averaged optimal training vector can always be reduced to the optimal training vector evaluated at an average opponent. Then, by the second property listed above, the expected payout to a student $y$ against a pool of opponents $y' \sim \pi$ equals the payout of the student $y$ against the average opponent $\bar{y}_{\pi}$:
\begin{equation}
    \mathbb{E}_{y' \sim \pi}[g(y,y')] = \mathbb{E}_{y' \sim \pi} [y \cdot v(y')] = y \cdot \bar{v}_{\pi} = y^{\intercal} G \bar{y}_{\pi}.
\end{equation}

These properties can dramatically simplify selection dynamics that depend on weighted averages against the current and past population. They also explain why the pursuit of desideratum (D2) will simplify dynamics that do not depend on local perturbations (c.f.~Section \ref{sec: latent space dynamics}). Once bilinear, local linearization predicts payouts globally and averages over a population reduce to the local best response at an average agent.

\subsection{Bilinear Games and a Canonical Coordinate System} \label{sec: canonical coordinates}
To start, we consider a special case where desideratum (D2) can be accomplished trivially; if $f$ is bilinear in $x$ and $x'$, then any invertible affine transformation produces bi-affine $g$ that can be expressed as a bilinear function restricted to an affine subspace. 

Bilinear payouts occur naturally in normal form games with finite strategy sets. For example, consider a simultaneous, symmetric, zero-sum, normal form game. Then, both players may select a pure strategy from a finite set of available strategies, or, more generally, may pick a strategy at random from a distribution over possible strategies. In the latter case, the agents use a mixed strategy. A mixed strategy assigns a distribution $p$ over the possible strategies. Then, given $s$ possible strategies, the agents may be parameterized by selecting $d = s-1$ probabilities which act as as traits.

Let $f_{ij}$ represent the payout received by the first agent if they use pure strategy $i$ when the second agent uses pure strategy $j$. Let $F$ denote the matrix storing all the payouts. When symmetric and zero-sum, $F$ is skew-symmetric. Let $p$ represent the mixed strategy for the first agent, and $p'$ the mixed strategy for the second agent. Then:
\begin{equation} \label{eqn: normal form f}
    f(p,p') = \mathbb{E}_{I \sim p, J \sim p'}[f_{IJ}] =  p^{\intercal} F p'.
\end{equation}

Equation \eqref{eqn: normal form f} is bilinear. So, all symmetric, zero-sum, normal-form games satisfy desideratum (D2) before transformation. Nevertheless, it may still be possible to select an affine coordinate transformation that simplifies the normal form game by replacing $F$ with $G$, where $G$ is sparser than $F$. 

In particular, since $F$ is real, skew-symmetric, it has even rank, $r$, and admits a real Schur-form \cite{youla1961normal, zumino1962normal}:
\begin{equation} \label{eqn: real Schur form}
    F = Q W Q^{\intercal}, \quad W = \left[\begin{array}{ccc} \omega_1 R & 0 & \hdots \\ 0 & \omega_2 R & \hdots \\ \vdots & \vdots & \ddots \end{array}\right], \quad R = \left[ \begin{array}{cc} 0 & 1 \\ -1 & 0 \end{array} \right]\end{equation}
where $Q$ is $d \times r$ with orthonormal columns $\{q_j\}_{j=1}^r$, $W$ is $r \times r$ block-diagonal with $2 \times 2$ diagonal blocks of the form $\omega_j R$, and $\{\omega_j\}_{j=1}^{r/2}$ are strictly positive and decreasing. Notice that multiplication by $R$ performs a ninety-degree rotation so the quadratic form $v^{\intercal} R w$ evaluates a cross-product: $v^{\intercal} R w = v_1 w_2 - v_2 w_1 =  v \times w$. 

The components of real Schur form can be recovered from the eigenvalue decomposition of $F$. Set $\omega_{k} = |\lambda_{2k-1}|$, and $[q_{2k-1},q_{2k}] \propto [\text{Real}(v_{2k-1}),\text{Imag}(v_{2k-1})]$ where $\{(\lambda_j,v_j)\}_{j}$ are the eigenvalues and vectors of $F$ listed so that the eigenvalues decrease in magnitude, and so that the eigenvalue with positive imaginary part appears first in every pair. 

Let $T(x) = M x$. Then, since $f$ is bilinear, and $T$ is linear, $g$ is bilinear. Next, let $M = D_{\omega}^{1/2} Q^{\intercal}$ where $D_{\omega} = \text{diag}(\omega_1,\omega_1,\omega_2,
\omega_2,\hdots)$. If $r < d$, then append $d - r$ rows to the bottom of $M$, chosen to span the component of $\mathbb{R}^d$ perpendicular to the range of $F$. Then:
\begin{equation} \label{eqn: U matrix}
    g(y,y') = T^{-1}(y)^{\intercal} F T^{-1}(y') =  y^{\intercal} U y' \text{ where } \quad U = \left[\begin{array}{ccc} R & 0 & \hdots \\ 0 & R & \hdots \\ \vdots & \vdots & \ddots \end{array}\right] .
\end{equation}

In this coordinate system, the normal form game decouples blockwise. In fact:
\begin{equation} \label{eqn: cross product decomp}
    g(y,y') = y^{\intercal} U y' = \sum_{k=1}^{r/2} [y_{2k-1},y_{2k}]^{\intercal} R \left[\begin{array}{c} y'_{2k-1} \\ y'_{2k} \end{array} \right] = \sum_{k=1}^{r/2} y_{2k-1} y'_{2k} - y_{2k} y'_{2k-1} = \sum_{k=1}^{r/2} y^{(k)} \times {y'}^{(k)}
\end{equation}
where $\times$ denotes the cross product and $y^{(k)} = [y_{2k-1},y_{2k}]$ represents the $k^{th}$ consecutive pair of coordinates \cite{balduzzi2018re,balduzzi2019open,strang2022principal}. The cross-product defines a canonical functional form game.

\begin{snugshade}
    \noindent \textbf{Definition:} A functional form game $(f,\Omega)$ with $\Omega \subseteq \mathbb{R}^2$ and $f(x,x') = \text{disc}(x,x') = x \times x'$ is a \textbf{disc-game} \cite{balduzzi2019open}. 
\end{snugshade}

Equation \eqref{eqn: cross product decomp} states that any symmetric, zero-sum normal form game can be represented with a sum of disc-games, in the coordinate system produced by applying $T(x) = D_{\omega}^{1/2} Q^{\intercal} x$:
\begin{equation} \label{eqn: disc-game decomposition}
    g(y,y') = \sum_{k} \text{disc}(y^{(k)},{y'}^{(k)}).
\end{equation}
Notice that $Q^{\intercal} x$ is the orthonormal change of coordinates associated with the projection onto the eigenvectors of $F$. We will use this observation to generalize from the spectral decomposition of a skew-symmetric matrix to the spectral decomposition of a skew-symmetric function $f$.

Equation \eqref{eqn: disc-game decomposition} is an example of a disc-game decomposition \cite{balduzzi2018re}. It represents a game as a linear combination of disc-games, each acting on separate pairs of coordinates. Disc game decomposition has been proposed as a tool for expanding player evaluation \cite{balduzzi2018re}, designing population level learning algorithms which seek diverse populations \cite{balduzzi2019open}, and as an exploratory tool that visualizes strategic trade-offs \cite{strang2022principal}. In the next section, we explain how geometry encodes strategy in a disc game. This geometry recommends the disc game as an interpretable representation of otherwise complicated games. 

\pagebreak

\subsection{Disc Games: Interpretation and Geometry} \label{sec: disc games}


\begin{figure}[t]
    \centering
    \includegraphics[trim = 50 40 50 40, clip, scale = 0.37]{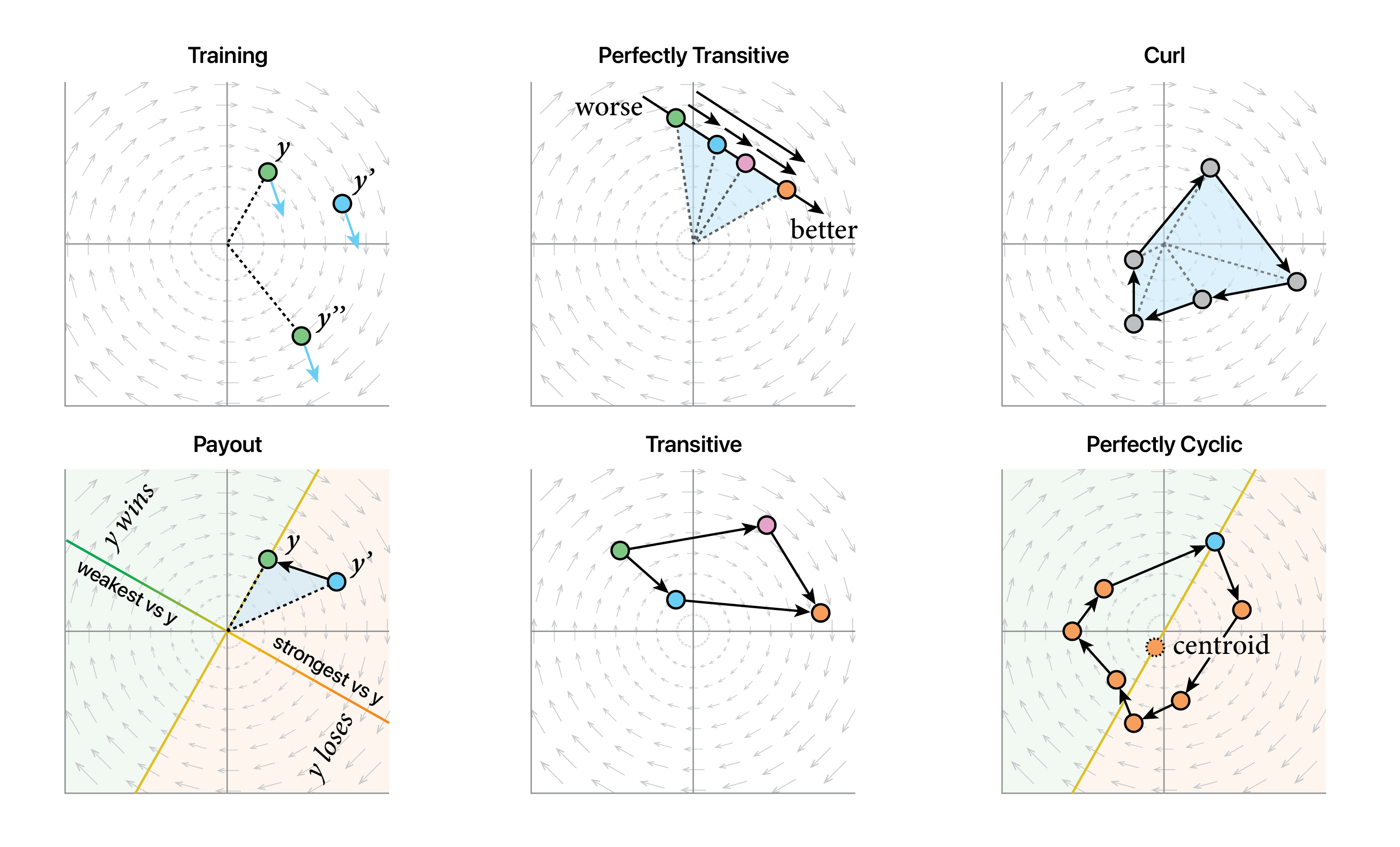}
    \caption{Disc game geometries. In all panels, the two coordinates correspond to $y_1$ and $y_2$, and the circulating grey vector field represents the optimal training response to an agent at each possible location in the disc-game. This is the optimal self-play vector field, $v(y,y)$. It represents the direction in which agents should move when training against themselves.  }
    \label{fig: Disc Game Geometries}
\end{figure}

Disc-game decompositions are useful since each individual disc-game is simple, yet can encode rich strategic information in its geometry \cite{balduzzi2019open,strang2022principal}. Since the disc-game determines payout via a cross-product, strategy is intimately related to agents' coordinates $y(x)$.  Figure \ref{fig: Disc Game Geometries} illustrates the geometric features associated with training, payout, transitivity, intransitivity, and cyclic competition. The geometric features are explained in Box 1.

\begin{wrapfigure}{l}{0.67\textwidth}
\vspace{-1.2\baselineskip} 
\begin{mdframed}[style=mddefault]
\begin{small}
\begin{centering}{\textbf{Box 1.} \it Disc Game Geometry\\}\end{centering}

\begin{enumerate}
    \item \textbf{Training:} optimal training vectors depend only on the opponent. Opponent, $y'$: small blue circle. Optimal training vector, $v(y')$: small black arrow pointing parallel to the background vector field. Students, $y$ and $y''$: green circles. Students should both move in the direction $v(y')$ to optimize their payout against $y'$: light blue arrows. 
    \item \textbf{Payout:} advantage circulates clockwise. Agents counterclockwise from $y$ lose to $y$, while agents clockwise from $y$ beat $y$. Advantage is maximized at $90^{\circ}$ off-sets. Agents lying along any line passing through the origin are equally matched. The payout $\text{disc}(y,y')$ equals twice the signed area of the blue triangle $[0,0] \rightarrow y' \rightarrow y \rightarrow [0,0]$. Equivalently, $\text{disc}(y,y')$ equals the path integral over the line from $y'$ to $y$ against the optimal training vector field. 
    \item \textbf{Perfectly Transitive:} Competition among a set of agents is perfectly transitive \cite{strang2022network} if and only if the agents are colinear. Ratings equal signed distance along the line. Rating is represented by color: green (weak) to orange (strong). 
    \item \textbf{Transitive:} Agents may be assigned a partial order, so that $g(y,y') \geq 0$ if $y \succeq y'$. A population is transitive if and only if the convex hull of the agents (outlined with black arrows) does not contain the origin (see Section \ref{sec: analysis}).
    \item \textbf{Curl:} The sum of the payout between pairs of agents, moving around a cycle of agents: arrows pointing between agents \cite{strang2022network}. In a disc game, the curl equals twice the signed area of the region enclosed by the loop: blue polygon. 
    \item \textbf{Perfectly Cyclic:} Every agent receives zero payout, on average, against a random opponent \cite{strang2022network}. A population is perfectly cyclic if and only if the centroid, after removing any agent, lies along the line connecting the origin to that agent. 
    \item \textbf{Origin:} The origin marks a competitor who ties against all opponents.
\end{enumerate}

\end{small}
\end{mdframed}
\vspace{-1\baselineskip}
\end{wrapfigure}

By encoding strategic information in geometry, disc-game decompositions imbue their coordinates $y_1, y_2$ with strategic meaning. For example, in Kuhn poker, the axes associated with a disc-game decomposition represent interpretable betting behaviors such as honesty and skepticism \cite{strang2022principal}.

The geometric relations listed in Box 1 suggest alternative axioms. For example, advantage in a disc-game is perfectly transitive along any line. Thus, training advantages are additive along lines; the running sum of the payout between a sequence of agents along a line segment adds to the payout between the agents at the endpoints. 
This observation suggests an alternate desideratum; pursue coordinates such that payout compounds predictably along training paths. Alternatively, one could pursue coordinates where the sum of performance around any loop (the curl) equals its embedded area. Each approach points back to disc-game embedding (See Appendix Section \ref{app: vector field}).


\subsection{Disc Game Embedding} \label{sec: embedding}

The existence of a disc-game decomposition for all symmetric, zero-sum, normal form games, suggests an additional desideratum for bilinear games:
\begin{snugshade}
    \noindent \textbf{Desideratum:} Identify a transform $T$ such that:
    \vspace{-0.05 in}
    \begin{enumerate}
        \item[3.] $g(y,y')$ admits a disc-game decomposition of the form \eqref{eqn: disc-game decomposition}.
    \end{enumerate}
\end{snugshade}

\begin{figure}[t]
    \centering
    \includegraphics[scale = 0.14]{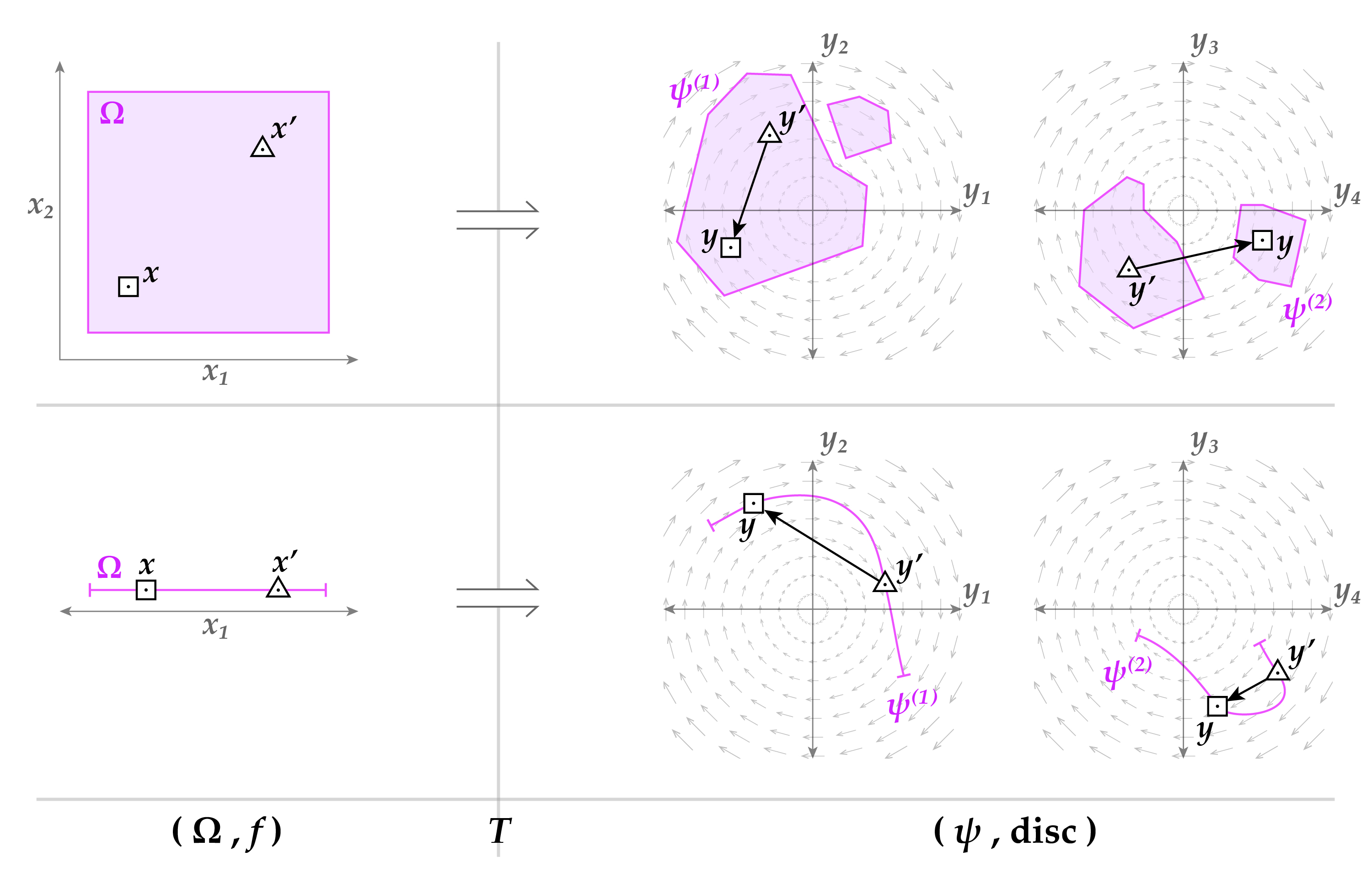}
    \caption{Disc game embedding for a two-dimensional trait space (top row) and a one-dimensional trait space (bottom row). The left-hand column shows the original trait spaces, $\Omega$ (magenta region), and a pair of agents in each, with attribute vectors $x$ and $x'$ (marked with a square and a triangle). Competitive advantage is evaluated using $f$ in the original space. In the original attribute space, $f$ may be arbitrarily complicated. The arrow represents the transformation $T$ which maps $\Omega$ to $\Psi$. The right-hand column represents the disc game embedding. The coordinates, $y$ are broken into consecutive pairs, each composing a separate game. The image of the original attribute space under the embedding, $\Psi$, is shown in magenta. The image of $x$ and $x'$, $y = T(x)$ and $y' = T(x')$ are shown with matching markers. In the embedding space, the competitive advantage between $y$ and $y'$ is simple, and is determined by a disc game, or, equivalently, cross-product between the embedding coordinates. The value of the cross-product 
    equals the value of the line integral from $y'$ to $y$ against the optimal local training vector field, $v$ (shown in grey in each disc game). Note: although we motivated the disc game embedding assuming smooth $T$, disc game embeddings may use nondifferentiable and discontinuous $T$. The regularity of $T$ is constrained by the regularity of $f$ (see Section \ref{sec: regularity}).  }
    \label{fig: Embedding Visualization}
\end{figure}

Notice that, if desideratum (D2) is possible, then $g$ is bilinear, so desideratum (3) is also possible. Thus, any $f$ that admits a transform satisfying desideratum (D2), also admits a coordinate transformation such that:
\begin{equation} \label{eqn: generic disc-game embedding}
    f(x,x') = g(y,y') = \sum_{k} \text{disc}(y^{(k)},y'^{(k)}), \text{ where } y = T(x). 
\end{equation}

Equation \eqref{eqn: generic disc-game embedding} is a disc-game embedding. The desired transformation is represented schematically in Figure \ref{fig: Embedding Visualization}. The left-hand column of Figure \ref{fig: Embedding Visualization} illustrate simple two-dimensional (top-row), and one-dimensional (middle-row), attribute spaces. These are intentionally drawn as simple regions since the attribute spaces are usually defined by the user. For example, the possible parameter values defining a value or policy network would usually be a set of real-valued weights and biases. While the attribute space is simple, its relation to performance is opaque, and, for any example with more than two attributes cannot be visualized directly since $f$ is a function of four or more variables. The disc-game embedding exchanges a simple attribute space $\Omega$ with a, possibly complicated, performance function, for a, possibly complicated, embedded attribute space $\Psi$, and a simple, indeed canonical, performance function: $\text{disc}(y,y')$. In essence, any complexity in the original performance function $f$ is absorbed into the transform $T$, and the image of the original attribute space under the embedding, $\Psi$.

\begin{figure}[t]
    \centering
    \includegraphics[trim = 140 90 100 80, clip, width = \textwidth]{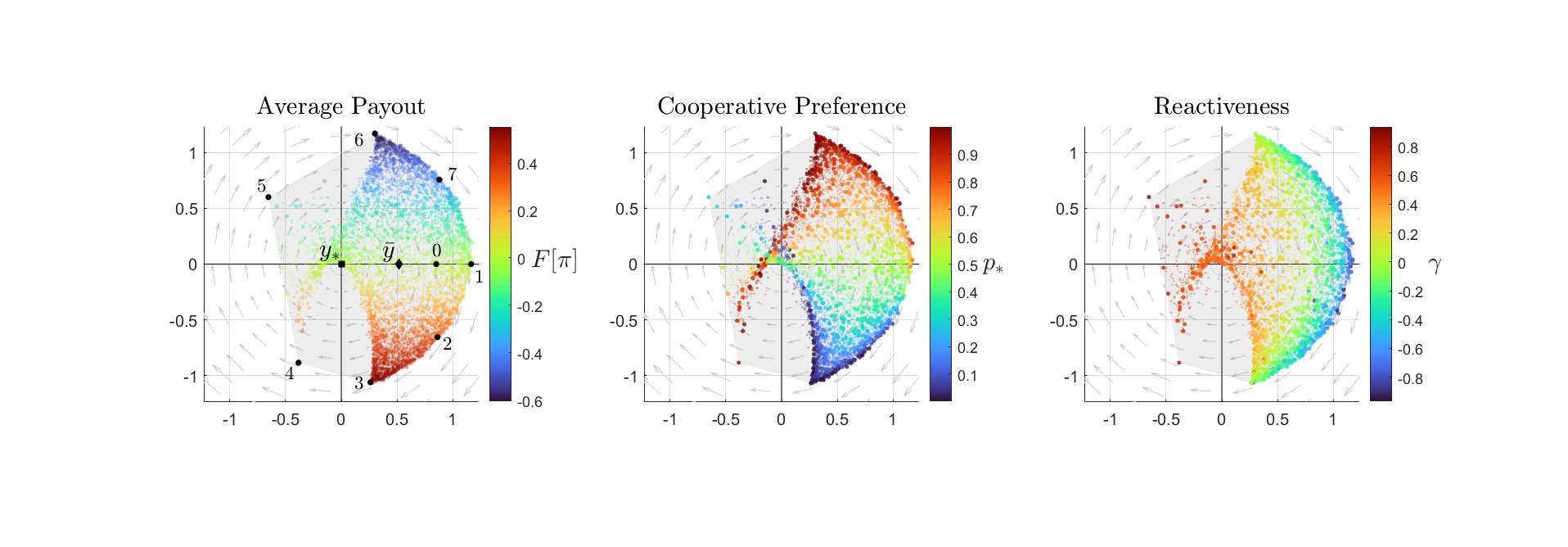}
    \caption{The disc game representation of the IPD payout matrices presented in Figure \ref{fig: IPD Payout Matrices}. For the full game specification, see Appendix Seection \ref{app: example game}. All three panels plot the coordinates, $[y^{(1)}_1(x),y^{(1)}_2(x)]$ for each agent of type $x$ in the sampled population. The first 800 agents are bolded. Small scatter points correspond to an interpolation of the embedded 800 agents using 3,000 agents. The interpolation is added to smooth the visual trends, and to highlight the apparently simple underlying functional relation mapping from $x$ to $y$. Here $T$ is only solved for pointwise. Only the first disc game is shown since it accounts for 90\% of the total variance in performance. \textbf{Left:} Agents are colored according to their average performance, in the first disc game, against the 800 agent population. Since the disc game is bilinear, the average payout is determined by a cross-product against the centroid in the disc game space, marked with a black diamond and denoted $\bar{y}$. Agents with embedded attributes $y_*$, marked with a black square, performs neutrally in the disc game. They correspond to a Nash Equilibrium policy as their disc-game payout against any opponent is zero, and their opponent's payout is also zero. The remaining marked agents correspond to agents with distinct behaviors. For an interprative key, see Box 1. \textbf{Middle:} Agents are colored according to their innate preference for cooperation, $p_*$. The value of $p_*$ sets the probability with which they cooperate before interacting with an opponent, and is the policy they would return to in absence of information about their opponent. Notice that phase around the origin is a smooth function of $p_*$, with two fan-shaped lobes. Within each lobe, agents with smaller $p_*$ posses an advantage over agents with a larger $p_*$. That is, within each lobe, it is better to be distrustful. \textbf{Right:} Agents are colored by an attribute, $\gamma$, which controls how they react to their opponent's actions. Agents with large positive $\gamma$ imitate their opponent, agents with $\gamma$ near zero mostly don't respond to their opponent, and agents with large negative $\gamma$ play the opposite action their opponent last played. Again, note that position in the disc game is a smooth function of the attribute $\gamma$. In particular, the horizontal coordinate $y_1$ is approximately monotonic in $\gamma$. The two fans correspond to $\gamma < 0.4$ (right fan), and $\gamma > 0.4$ (left fan).   }
    \label{fig: IPD Disc Game Embedding}
\end{figure}

\begin{wrapfigure}{l}{0.67\textwidth}
\vspace{-1.2\baselineskip} 
\begin{mdframed}[style=mddefault]
\begin{small}
\begin{centering}{\textbf{Box 2.} \it Example Disc Game Interpretation\\}\end{centering}
Figure \ref{fig: IPD Disc Game Embedding} illustrates the first pair of disc-game coordinates produced by embedding the payout matrices illustrated in Figure \ref{fig: IPD Payout Matrices}. Position in the disc game space is related to an agent's policies by smooth functions (see the middle and left most panel). 8 embedded policies are numbered in the left most panel. These form an advantage cycle in which $1 \succ 7 \succ 6 \succ \hdots \succ 2 \succ 1$.

Each agent is an IPD agent. The numbered policies are interpreted below:

\begin{enumerate}
    \item[0:]  plays a uniform random strategy (coin toss) and does not respond to their opponent (stoic/stubborn),
    \item[1:] prefers a coin toss, but responds in opposition to their opponent by doing whatever the opponent didn't do (deceptive)
    \item[2:] usually defects (distrusting) and is deceptive
    \item[3:] defects deterministically (distrusting, stoic),
    \item[4:] prefers to cooperate (trusting) and attempts to enforce a social-contract by imitating their opponent (law-enforcing). 4 is the famous ``tit-for-tat" agent \cite{axelrod1981evolution}.
    \item[5:] is distrusting and law-enforcing,
    \item[6:] deterministically cooperates (trusting, stoic),
    \item[7:] usually cooperates (trusting) and is deceptive.
\end{enumerate}

\end{small}
\end{mdframed}
\vspace{-1\baselineskip}
\end{wrapfigure}

Figure \ref{fig: IPD Disc Game Embedding} illustrates a point-wise disc-game embedding constructed using the Schur decomposition, as described in Section \ref{sec: canonical coordinates} and in \cite{strang2022principal}. The input matrix $F$ used to create the embeddings is shown in Figure \ref{fig: IPD Payout Matrices}, and the underlying game is described in Appendix Section \ref{app: example game}. Box 2 offers a simple interpretation based on the relations between strategy and embedded geometry presented in Box 1.

 It is unclear whether such an embedding is possible in general. Can we identify a disc-game embedding for all $f$? If not, what properties must $f$ satisfy to admit a disc game embedding? Are those properties satisfied by most games of interest?  Alternately, if a game cannot be embedded, can it be approximated to sufficient accuracy via a disc-game embedding?

We will show that essentially all payout functions $f$ admit a disc-game embedding, a function $f$ is disc-game embeddable if and only if desideratum (D2) is achievable, and, all transforms $T$ satisfying desideratum (D2) are affine transformations of a disc-game embedding. 

\subsubsection{Existence and Approximation}
\begin{snugshade}
    \noindent \textbf{Theorem 1: [Essentially All Games are Disc Game Embeddable]} If there exists a finite, positive measure $\nu$ that is absolutely continuous with respect to the Lebesgue measure on $\Omega$ such that $\|f\|_{\nu \times \nu}  < \infty$ where
    $\|f\|_{\nu \times \nu}^2 =  \iint_{x,x' \in \Omega\times\Omega} f(x,x')^2 d\nu(x) d\nu(x')$
    then $f$ admits a disc-game embedding which can be constructed using the real Schur form of the integral operator:
    \begin{equation} \label{eqn: integral operator}
        F_{\nu}[h](x) = \int_{y \in \Omega} f(x,y) h(y) d\nu(y). 
    \end{equation}
\end{snugshade}

\noindent \textbf{Remark:} The measure $\nu$ is not prescribed by the game. Rather, it is chosen by the user. Technically, the measure is needed to define an inner product on the space of functions used for constructing the coordinate transformation. This inner product induces notions of orthogonality, orthogonal projection, and optimal approximation with respect to an associated norm. For example, we will see that, after choosing $\nu$, the embedding coordinates $\{y_j(X)\}_{j}$ are uncorrelated if $X$ is drawn from $\nu$, and have variances associated with the importance of each embedding coordinate in the overall reconstruction of $f$. Formally, the measure is used to define a Hilbert space, within which functions on the product space, such as $f$ and $g$, may be treated like matrices. This analogy follows from the fact that matrices are scalar-valued functions of pairs of index-valued inputs, whereas $f$ and $g$ are scalar-valued functions of pairs of generic inputs.

The measure weights $\Omega$. Regions where the measure places a large mass are important, and regions where the measure places a small mass are not important. Since the measure of the full space does not matter, we will assume that the measure is normalized, and can be treated as a probability measure when convenient. Then, the measure could be chosen to preferentially focus on certain trait combinations of interest, to reflect the distribution of a particular population, to reflect the empirical distribution of some observed set of types, or to represent a generating distribution producing observed agents. Note that the uniform measure is, itself, a finite measure in compact spaces, so replacing with $d\nu(y)$ with $dy$ corresponds to a more restricted theory. 

Even though the embedding depends on the measure, we will see that some crucial characteristics, namely, the number of embedding coordinates required, are independent of the choice of measure. Moreover, the introduction of a measure will prove useful since it fixes the embedding in terms of a reference population, allowing consistency across different initial choices of $\Omega$. The latter property is essential. If the embedding depends on the initial choice of $\Omega$, then it is extrinsic. $\square$
\vspace{0.1 in}

\noindent \textbf{Proof:} The proof is a special application of the Hilbert-Schmidt theorem \cite{royden2010real} to skew-symmetric linear operators. The operator $F_{\nu}[\cdot]$ is skew-symmetric, so $i F_{\nu}[\cdot]$ is Hermitian and thus is self-adjoint. If $\|f\|_{\nu}$ is finite, then the operator is also compact so is a Hilbert operator \cite{bogachev2020real}. 

Hilbert operators obey the Hilbert-Schmidt theorem, namely, they admit an expansion onto a sequence of eigenpairs, $\{(\lambda_j,\phi_j)\}_{j=1}^{\text{rank}(F)}$ \cite{royden2010real}. When real, skew-symmetric, the eigenvalues $\lambda_j$ are purely imaginary, come in complex conjugate pairs, and, by convention, are listed in non-increasing order, with positive imaginary eigenvalues listed first. The eigenfunctions $\phi_j:\Omega \rightarrow \mathbb{C}$ also come in complex conjugate pairs and are orthonormal under the inner product induced by $\nu$, $\langle u,w \rangle_{\nu} = \int_{x \in \Omega} \bar{u}(x) w(x) d\nu(x)$. The eigenfunctions and eigenvalues satisfy the equation:
\begin{equation}
    F_{\nu}[\phi_j](x) = \lambda_j \phi_j(x)
\end{equation}
and depend on the choice of $\nu$. 

The rank of $F$ is the total number of eigenpairs such that $\lambda_j \neq 0$. Unlike the specific eigenvalues and eigenfunctions, the rank is independent of $\nu$ provided $\nu$ is continuous and is supported everywhere on $\Omega$. Thus, the rank is uniquely defined by the payout $f$ and the support of the reference measure. For details, see Lemma 5. Then, we can write $\text{rank}(f)$ unambiguously. If the rank is finite, then $f$ is degenerate.

The Hilbert-Schmidt theorem expresses $F_{\nu}$ in terms of its eigenfunctions:
\begin{equation}
    F_{\nu}[h](x) = \sum_{j=1}^{\text{rank}(f)} \lambda_j \phi_j(x) \langle \phi_j, h \rangle_{\nu}
\end{equation}
for all functions $h$ that are square integrable with respect to $\nu$.

By selecting appropriate test functions $h$, the payout function also admits an expansion onto the eigenfunctions:
\begin{equation}
    f(x,x') = \sum_{j=1}^{\text{rank}(f)} \lambda_j \phi_j(x) \bar{\phi}_j(x')
\end{equation}
for almost all $(x,x')$ with respect to the product measure $\nu \times \nu$. 

Since the eigenfunctions and eigenvalues form complex conjugate pairs, group the sum in consecutive pairs. Then, setting:
\begin{equation}
    y^{(k)}(x) = \sqrt{|\lambda_{2k-1}|}\left[\text{Real}(\phi_{2k-1}(x)), \text{Imag}(\phi_{2k-1}(x))\right]
\end{equation}
yields:
\begin{equation} \label{eqn: eigen disc-game}
    f(x,x') = g(T(x),T(x')) = \sum_{j=1}^{\text{rank}(f)} \text{disc}(T(x),T(x')) \text{ where } T(x) = [y^{(1)}(x),y^{(2)}(x),\hdots,y^{(\text{rank}(f)/2)}(x)].
\end{equation}
Thus, if $\|f\|_{\nu \times \nu} < \infty$ , $f$ admits a disc-game embedding. $\square$

\vspace{0.03 in}
\noindent \textbf{Remark:} The decomposition is unique up to arbitrary rotations of each pair of coordinates $y^{(k)} = [y^{(k)}_1,y^{(k)}_2]$ provided we require that each coordinate $y_j(X)$ has equal variance when the input $X$ is drawn from the measure $\nu$, and all of the eigenvalues are distinct. If some of the eigenvalues are identical, then the decomposition is unique up to arbitrary rotations among the set of coordinates corresponding to matching eigenvalues (c.f.~\cite{strang2022principal}). $\square$ 

\noindent \textbf{Remark:} Since the eigenfunctions of a Hermitian operator are orthonormal, the disc-game embedding functions $\{y_j\}_{j=1}^{\text{rank}(F)}$ are mutually orthogonal with respect to the inner product induced by $\nu$. That is, all distinct embedding coordinates $y_i(X)$, $y_j(x)$ are uncorrelated when $X$ is drawn from $\nu$. $\square$

\vspace{0.05 in}
\noindent \textbf{Remark:} 
The disc-game embedding introduced in Theorem 1 is an example of a spectral embedding. Spectral embeddings represent objects via the eigenvectors or eigenfunctions of an operator. Other popular examples include principal component analysis, multi-dimensional scaling, diffusion maps, and spectral graph embeddings \cite{abdi2010principal,coifman2006diffusion,hall1970r,koren2003spectral,kruskal1978multidimensional,pearson1901liii,torgerson1952multidimensional}. Like other spectral embeddings, disc-game embedding may be defined by minimizing a squared error between a reconstruction of some reference object, in our case, $f$, and its approximation, where the approximation is bilinear (a quadratic form), in some nonlinear transformation of the input coordinates. 

Here, the complexity of the original object is absorbed into the nonlinear coordinate transformation, while its essence is distilled by the quadratic form. Often, the quadratic form is simply an inner product. For example, large-language models and natural language processing frequently use token-embeddings where a comparison is reduced to an inner product after some change of coordinates. Similar spectral approaches are popular in methods that leverage reproducing kernel Hilbert spaces \cite{berlinet2011reproducing,paulsen2016introduction}, where either the Hilbert-Schmidt theorem \cite{royden2010real}, or Mercer's theorem \cite{williams2006gaussian}, are used to decompose a kernel into its eigenfunctions, the eigenfunctions are used for embedding, and the kernel is replaced with an inner product in the embedding space \cite{williams2006gaussian}. The disc-game embedding used here simply replaces the inner product with a cross-product. For an in-depth discussion of the significance of this exchange, see \cite{strang2022principal}. $\square$

\vspace{0.05 in}
Many important examples of functional form game satisfy the requirements of Theorem 1. For example, Theorem 1 applies to all bounded functions $f$ defined over compact $\Omega$. This case applies to all extensive form games with finite payouts with traits equal to action probabilities. It requires only that the variance in payout, when sampling $X$ and $X'$ independently and identically from $\nu$ is finite for some continuous $\nu$ supported everywhere on $\Omega$. 


Importantly, the finite norm requirement introduced in Theorem 1 can be dropped when the rank of $f$ is finite. In particular, all degenerate $f$ are also disc-game embeddable. 

\begin{snugshade}
    \noindent \textbf{Definition:} A function $f:\Omega \times \Omega \rightarrow \mathbb{R}$ is \textbf{separable} if it can be expressed as a finite sum of \textbf{product} functions:
    \begin{equation} \label{eqn: linear combination of separable functions}
        f(x,x') = \sum_{j=1}^r g_j(x) h_j(x') \text{ where } r < \infty.
    \end{equation}
    
\end{snugshade}

Separable skew-symmetric functions are degenerate by construction. They may be expressed:
\begin{equation} \label{eqn: degenerate f}
    f(x,x') = b(x)^{\intercal} F b(x')
\end{equation}
where $F \in \mathbb{R}^{m \times m}$ is skew-symmetric, $b(x)^{\intercal} = [b_1(x),b_2(x),\hdots, b_m(x)]$ is a vector-valued function whose entries $\{b_j\}_{j=1}^m$ each map from $\Omega$ to $\mathbb{R}$, and $m$ is finite. The collection of functions $\mathcal{B} = \{b_j\}_{j=1}^m$ is a function basis. 
If the basis functions are linearly independent, then the rank of the linear operator $F_{\nu}$ coincides with the rank of the coefficient matrix $F$. 

\begin{snugshade}
\noindent  \textbf{Theorem 2: [Separable Functions are Disc Game Embeddable]} If there exists a function basis $\mathcal{B}$ containing finitely many functions $\{b_j\}_{j=1}^m$, $b_j:\Omega \rightarrow \mathbb{R}$, such that $f(x,x') = b(x)^{\intercal} F b(x')$ for some skew-symmetric $F \in \mathbb{R}^{m \times m}$, then $f$ is disc-game embeddable using $r = \text{rank}(F)$ coordinates. 
\end{snugshade}

\noindent \textbf{Proof:} Given a payout function of the form \eqref{eqn: degenerate f}, the real Schur form of $F$ offers a disc-game decomposition. If $F = Q D_{\omega}^{1/2} U D_{\omega}^{1/2} Q^{\intercal}$, then setting $T(x) = D_{\omega}^{1/2} Q^{\intercal} b(x)$ gives:
\begin{equation}
    f(x,x') = b(x)^{\intercal} F b(x') = T(x)^{\intercal} U T(x') = \sum_{k=1}^{\text{rank}(F)} \text{disc}(y^{(k)},y'^{(k)}) \text{ where } y^{(k)} = [T_{2k-1}(x),T_{2k}(x)].
\end{equation}
Thus, any separable function is disc-game embeddable. $\square$

Since $\mathcal{B}$ and $m$ were arbitrary, the class of separable payouts of the form \eqref{eqn: degenerate f} is quite large. It includes all skew-symmetric polynomials, as well as all piecewise polynomials with finitely many pieces. Since the payout function of any finite extensive form game is a polynomial in the decision probabilities assigned to each branch in the game tree, the payout for all finite extensive form games are disc-game embeddable. Indeed, any finite combination of any classical function bases, whether polynomial, piecewise polynomial, Fourier, wavelet, or kernel are disc-game embeddable.

Theorem 1 and Theorem 2 are related. When the linear operator $F_{\nu}$ has finite rank, then the Hilbert-Schmidt theorem guarantees that $f$ can be expressed as a separable function, where each basis function is an eigenfunction of the operator. Thus, when $F_{\nu}$ is degenerate, $f$ is disc-game embeddable using the eigenfunctions as a function basis. If all of the eigenfunctions themselves can be expressed as a linear combination of other basis functions $\mathcal{B}$, then $f$ is also a finite linear combination of product functions built from the basis $\mathcal{B}$.

Conversely, when $\|f\|_{\nu}$ is finite, then Theorem 2 can be extended to recover Theorem 1 by considering the closure of the set of separable functions as the order $m$ increases. This is, in essence, an approximation argument. If $f$ cannot be expressed as a finite sum of product functions, but can be approximated to arbitrary accuracy by sufficiently many product functions, then $f$ is in the closure of the set of separable functions. For example, consider analytic functions, which may be expressed locally as power series, thus approximated to arbitrary accuracy by finite linear combinations of monomials of increasing degree. 

The closure of a set of separable functions depends on the form of convergence enforced. The form of convergence requires a norm on functions used to measure the error in an approximation. Enter $\nu$. 

Let $\nu$ denote a non-negative measure supported on $\Omega$. Let $\langle \cdot, \cdot \rangle_{\nu}$ be the inner product with respect to $\nu$, and $\| \cdot\|_{\nu}$ the induced two-norm. Similarly, let $\nu \times \nu$ be the product measure on $\Omega \times \Omega$, $\langle \cdot, \cdot \rangle_{\nu \times \nu}$ the associated inner product, and $\|\cdot\|_{\nu\times \nu}$ the associated norm.  Then, let $\mathcal{B}^{(m)}$ be a collection of $m$ basis functions that are orthogonal with respect to $\langle \cdot, \cdot \rangle_{\nu}$, and normalized with respect to $\|\cdot\|_{\nu}$. This basis can be extended to a product basis by letting $\mathcal{B} \times \mathcal{B}$ be the set of all product functions of the form $b_i(x) b_j(x')$. Then the product basis is $\langle \cdot, \cdot\rangle_{\nu \times \nu}$ orthogonal and all of product basis functions are $\|\cdot\|_{\nu \times \nu}$ normalized. Let $\text{Proj}_{\nu \times \nu,\mathcal{B} \times \mathcal{B}}$ be the projection operator such that:
\begin{equation}
    \text{Proj}_{\nu \times \nu,\mathcal{B} \times \mathcal{B}}[f] = \sum_{i,j=1}^m F_{ij} b_i(x) b_j(x')  = \text{argmin}_{\hat{f} \in \text{span}(\mathcal{B} \times \mathcal{B})}\{\|f - \hat{f}\|_{\nu \times \nu}\}.
\end{equation}

When the basis functions are orthonormal:
\begin{equation} \label{eqn: coefficient matrix}
F_{ij} = \langle b_i b_j, f \rangle_{\nu\times\nu} = \iint_{x,x' \in \Omega \times \Omega'} b_i(x) f(x,x') b_j(x') d\nu(x) d\nu(x').
\end{equation}

Otherwise, the coefficient matrix $F$ can be recovered by either orthonormalizing the basis via a Gram-Schmidt procedure, or by inverting the Gram matrix storing all the inner products of the basis functions. Notice that these equations only make sense when working in the space of functions whose $\| \cdot\|_{\nu}$ and $\| \cdot\|_{\nu \times \nu}$ norms are finite. That is, within the Hilbert space associated with the inner product defined by the product measure \cite{royden2010real}. 

Since any separable payout function is disc-game embeddable, $\hat{f}^{(m)} = \text{Proj}_{\nu \times \nu, \mathcal{B} \times \mathcal{B}}[f]$ is disc-game embeddable. So, if $\hat{f}^{(m)}$ converges to $f$ in $\|\cdot\|_{\nu \times \nu}$, then there exists a sequence of disc-game embeddings that converge to $f$, and there exists a disc-game embedding that recovers $f$ to arbitrarily small error. 
Importantly, the disc-game embedding produced by embedding $\hat{f}^{(m)}$ provides the closest approximation to $f$ in $\| \cdot \|_{\nu \times \nu}$ among all transformations $T \in \text{span}(\mathcal{B}^{(m)})$. 

Theorem 1 and Theorem 2 are unified by searching for a basis $\mathcal{B}$ such that $\hat{f}^{(r)}$ is the closest approximation to $f$ in $\| \cdot \|_{\nu \times \nu}^2$ among all separable functions of rank $r$. In particular, by extending the Eckart-Mirsky-Young theorem \cite{eckart1936approximation,mirsky1960symmetric,strang2019linear}, the best rank $r < \text{rank}(F)$ embedding is recovered by truncating spectrum of $F_{\nu}$:
\begin{snugshade}
\noindent \textbf{Lemma 3: [Optimal Rank-$2r$ Approximation]} The rank $2r \leq \text{rank}(F)$ approximation:
    \begin{equation} \label{eqn: truncated disc game expansion}
        \hat{f}^{(2r)}(x,x') = \sum_{k=1}^{r} \text{disc}(y^{(k)},y'^{(k)})
    \end{equation}
    defined by projection onto the eigenbasis $\mathcal{B} = \{\text{Real}(\phi_1),\text{Imag}(\phi_1),\hdots\,\text{Real}(\phi_{2r-1}),\text{Imag}(\phi_{2r-1})\}$ is equivalent to truncating the exact disc-game embedding, \eqref{eqn: eigen disc-game}, after $r$ terms, and minimizes the approximation error $\|\hat{f}^{(2r)} - f\|_{\nu \times \nu}$ among all rank $2r$ separable functions (all combinations of $2r$ linearly independent, product functions). 
\end{snugshade}

As long as $\|f\|_{\nu \times \nu} < \infty$, the sequence of approximations converges, and the error in the approximation is:
\begin{equation}
    \|\hat{f}^{(m)} - f\|_{\nu \times \nu}^2 = 2\sum_{k > r} |\lambda_{2k}|^2. 
\end{equation}
Thus, the sequence of eigenvalues of $F_{\nu}$ control the rate of convergence of all separable approximations to $f$ by setting an achievable lower bound on the $\| \cdot\|_{\nu \times \nu}$ error over all rank $2r$ functions. 

Together, Theorems 1 and 2 establish that essentially all games of interest are either exactly disc-game embeddable, or may be approximated to arbitrary accuracy with a disc-game embedding. To approximate, either start by identifying the eigenfunctions under a particular choice of $\nu$, then truncate, or, choose a complete function basis (e.g. polynomials), project onto the function basis, then embed the projection.

More strongly still:
\begin{snugshade}
\noindent  \textbf{Theorem 3: [Disc Game Embeddings Are Sufficient]} Any coordinate transformation satisfying desideratum (D2) using finitely many coordinates is related to a disc-game embedding by a linear transformation.
\end{snugshade}

\noindent \textbf{Proof Outline:} Any coordinate change satisfying desideratum (D2) with finitely many coordinates produces bilinear $g$, and all finite-dimensional bilinear $g$ are separable, thus disc-game embeddable. See Appendix Section \ref{app: theorem 3} for details $\square$
\vspace{0.1 in}

It follows, that any coordinate transformation $T$ that satisfies desideratum (D2) and uses finitely many coordinates, may be reduced to disc-game embedding by a linear mapping. Since the disc-game embedding separates the coordinates into non-interacting pairs, there is little reason to adopt a transform satisfying desideratum (D2) that is not a disc-game embedding.  

\subsubsection{Consistency and Uniqueness}
To isolate an intrinsically meaningful coordinates from any set of sufficient coordinates, the coordinate system produced cannot depend on the initial coordinate system chosen. Suppose that $w = M(x)$ for some invertible mapping $M$. Then an embedding is intrinsic if and only if the coordinates assigned to any agent after the embedding are the same no matter whether the agent was initially described using $x$ or $w$. 

Let $M: \Omega \rightarrow \Phi$ denote a one-to-one map and $w = M(x)$ the change of coordinates associated with $M$. Let $h(w,w') = f(M^{-1}(w),M^{-1}(w'))$. Let $\nu$ denote a measure over $\Omega$, and $\mu$ denote the associated measure over $\Phi$ where $\mu(A) = \nu(M^{-1}(A))$ for any set $A$. 

\begin{snugshade}
\noindent    \textbf{Definition: [Embedding Equivalency]} Two disc-game embeddings are equivalent if the coordinates assigned to all agents are equal up to a rotation within each set of coordinates sharing an eigenvalue.
\end{snugshade}

\begin{snugshade}
\noindent    \textbf{Lemma 4: [Composition Consistency]} Given $M:\Omega \rightarrow \Phi$, the disc-game embedding constructed using $F_{\nu}$ is equivalent to the disc-game embedding constructed using $H_{\mu}$ in the sense that $\{y_j(x)\} \sim \{z_j(M(x))\}$ if $y$ and $z$ are the disc-game embeddings for $(f,\Omega,\nu)$ and $(h,\Phi,\mu)$ respectively, and $\mu(A) = \nu(M^{-1}(A))$ for any $A \subseteq \Phi$.
\end{snugshade}

\noindent \textbf{Remark:} Lemma 4 shows that, by introducing a measure over $\Omega$, it is possible define an embedding procedure that is entirely intrinsic, that is, whose output coordinates are independent of the choice of initial attributes up to invertible transformations. In other words, the embedding of the payout is unique up to the choice of measure.

Note that this consistency is impossible in the absence of the measure, since it is the consistency of the measures across representations that ensures a consistent geometry in the Hilbert space used to construct the embedding. More direct methods based on the expansion of a separable function do not satisfy Lemma 4, since they depend on the specific functional form of $f$, and thus, on the particular choice of $\Omega$. 

In this sense, Theorem 1 and Lemma 3 provide transformations into an intrinsic coordinate system. Lemma 4 shows that the embedding produced by Theorem 1 or Lemma 3 depends only on payout $f$ and the measure $\nu$, not the initial representation selected. This is natural when $\nu$ models a population, or distribution generating samples. The embedding is a function of \textit{who }makes up a population and \textit{how} they interact, not the attributes used to \textit{describe} the members of the population. $\square$ 

Since the embedding is defined with respect to a user-chosen reference measure, $\nu$, it is essential to distinguish which features of the embedding depend on $\nu$ and which do not. In particular:
\begin{snugshade}
\noindent \textbf{Lemma 5: [Range and Rank Invariance]} If $\nu$ and $\mu$ are each continuous reference measures, both supported on the same set $S \subseteq \Omega$, then $\text{range}(F_{\nu}) = \text{range}(F_{\mu})$ so $\text{rank}(F_{\nu}) = \text{rank}(F_{\mu})$ and the embeddings are related by invertible linear transformation.  
\end{snugshade}

\noindent \textbf{Proof:} Suppose that $h(x) \in \text{range}(F_{\nu})$. Then there exists a function $g$ such that $h(x) = \int_{x' \in S} f(x,x') g(x') d\nu(x')$. Since $\nu$ and $\mu$ are both supported on $S$, $h(x) = \int_{x' \in S} f(x,x') \left(g(x') \tfrac{d\nu}{d\mu}(x') \right) d\mu(x')$, where $\tfrac{d\nu}{d\mu}(x')$ is the Radon-Nikodym derivative of $\nu$ with respect to $\mu$. The derivative is finite since $\mu$ and $\nu$ are both supported on $S$, and are continuous, so the derivative equals a ratio of finite, non-zero densities. Then $h(x) \in \text{range}(F_{\mu})$ since $h(x) = F_{\mu}[g(\cdot) \tfrac{d\nu}{d \mu}(\cdot)](x)$.

The rank of $F_{\nu}$ is the dimension of its range. Since the range of $F_{\nu}$ equals the range of $F_{\mu}$, their ranks also match. The eigenfunctions under either measure form orthonormal bases for this function space, so must be related by an invertible linear transformation. $\square$

Lemma 5 establishes that both the range and rank of the operator $F_{\nu}$ depend only on $f$ and $S = \text{support}(\nu)$, and are otherwise invariant to changes in $\nu$ that preserve the support set $S$. This establishes, that the rank is well defined by $f$ and $S$ alone. We will use this fact when study learning dynamics to show that changing the reference measure while preserving the support set $S$ produces topologically equivalent dynamics. This will allow more flexible analysis, since the reference measure can be adjusted strategically.

\subsubsection{Regularity} \label{sec: regularity}

Recognizing the disc game embedding functions as eigenfunctions of the operator $F_{\nu}$ ensures that the embeddings inherit regularity properties of $f$. For instance, if $f$ is Lipschitz continuous, then, since the embeddings are scaled versions of the eigenfunctions of $f$, and since the eigenfunctions of $f$ live inside the range of $F_{\nu}$, all of the embedding functions will be Lipschitz continuous with Lipschitz constants determined by the eigenvalues of $F_{\nu}$ and the Lipschitz constant of $f$. Similar statements hold for differentiability. It also follows that, if $f$ is restricted to a particular function class, then its embeddings may be as well. For example, if $f$ is a polynomial, as for finite extensive form games, then its range contains polynomials of matching order, so the embeddings must be polynomials of matching order. Symmetries of $f$ also constrain the behavior of its eigenfunctions. For example, if $f$ is stationary in the sense that it is translationally invariant, $f(x,x') = s(x - x')$ for some $s:\Delta \Omega \rightarrow \mathbb{R}$, then the eigenfunctions of $f$ may be restricted to Fourier modes \cite{strang2022principal}. The last example arises naturally for certain auctions and bidding games. Observations of this kind can be used to guess a relevant function basis when estimating the embeddings from data, or, to ensure that the embeddings are not exceedingly bizarre functions of the initial traits. Conceptually, regularity results of this kind ensure that the embeddings are only as exotic as $f$, so only as exotic as need be given the original agent descriptions. 

One regularity statement is relevant for the following analysis. Namely, the embeddings are continuous functions of $f$ in the sense that, if two distinct agents $x$ and $x'$, share similar outcomes for all opponents $x''$, then they will share similar embeddings $y$ and $y'$. In short, agents who play together, stay together once embedded.

\begin{snugshade}
\noindent \textbf{Lemma 6: [Outcome Continuity]} If $|f(x,x'') - f(x',x'')| \leq \delta$ for almost all $x''$ with respect to $\nu$, or $\|f(x,\cdot) - f(x',\cdot)\|_{\nu} = \mathbb{E}_{X'' \sim \nu}[(f(x,x'') - f(x',x''))^2] \leq \delta^2$, then $\|y^{(k)}(x) - y^{(k)}(x')\|_2 \leq |\lambda_{2k-1}|^{-1/2} \delta$.
\end{snugshade}

For the proof, see Appendix section \ref{app: lemma 6}. 

This is a desirable property for an intrinsic set of coordinates. It ensures that agents who perform similarly are embedded near each other. Lipschitz continuity and differentiability follow if $f$ is Lipschitz or differentiable since any smoothness statement relating perturbations in traits to perturbations in outcome can be extended, via Lemma 6, to perturbations in embedding. It also ensures that agents who cannot be distinguished by their payout alone are assigned the same coordinates once embedded (see Lemma 11). This reduction by outcome equivalency will be useful when analyzing selection dynamics, since agents who are outcome equivalent are subject to the same selective pressures in canonical evolutionary dynamics.

\subsection{Disc Game Embeddings as a Candidate Latent Space for Learning}

Since disc-games satisfy desideratum (D2), allow easy manipulation of averages via bilinearity, decouple blockwise, represent strategy with intuitive geometry, and can be defined uniquely up to the choice of a reference measure, we propose the disc-game coordinates as a candidate latent space for learning and selection dynamics. For an alternate derivation based on accumulated training advantages, see Appendix Section \ref{app: vector field}. By adopting a canonical coordinate transformation, we, in essence, adopt a canonical game. Since essentially all symmetric, zero-sum games admit a disc-game embedding, up to a change of coordinates, essentially all symmetric, zero-sum games \textit{are} combinations of disc-games. 

In this sense, the disc-game payout replaces other canonical payout models, such as differences in fitness \cite{balduzzi2018re}. Unlike fitness landscapes, which only describe transitive competition, the disc-game payout is generic. Moreover, unlike fitness ``seascapes" \cite{merrell1994adaptive,mustonen2009fitness}, or frequency-dependent fitness landscapes \cite{nowak2004evolutionary}, which are generic, but change as a population changes, the disc-game payout is fixed. Instead, individual games differ only through the set of admissable agents, $\Psi$. In turn, fixing the payout function will simplify dynamical analysis, since, up to a change of coordinates, all selection dynamics respond to the same payout. Based on its generality and interpretability, we contend that selection in response to a disc-game payout should replace optimization of a fitness landscape as the canonical mental model for evolution in response to a skew-symmetric payout.

A successful mental model should guide intuition and ease communication through analogy or extension of familiar ideas. This process is often aided by visualization, especially when the analogy builds on real-world processes. To work as a successful analogy the model should accurately map the structural relations between components of the unfamiliar and the familiar system \cite{gentner1993shift}. 

Does optimization satisfy these criteria? It is familiar, visualizable, even concrete when related to ``hill-climbing." Yet, the intuition it provides is inaccurate, even misleading. Optimization suggests directed progress towards optima. As discussed in Section \ref{sec: paradigms}, and shown in Section \ref{sec: analysis}, this behavior is not generic, indeed is atypical, in simple settings. 

Do disc-games offer a compelling alternative? They are strictly more general (see Section \ref{sec: embedding}), visualizable, and interpretable (see Section \ref{sec: disc games}). What intuition do they provide?

Disc games suggest recurrent motion, even periodic circulation, unless barred by the boundaries of $\Psi$. The basic physical processes suggested by disc-games are all oscillatory - planetary motion, pendulum motion - etc. In the next section, we demonstrate that this intuition is not only qualitatively correct, it is mathematically precise.

\pagebreak
\section{Learning in the Latent Space} \label{sec: latent space dynamics}



To demonstrate the utility of the disc-game embedding, we turn to selection and learning dynamics. For classical reviews, see \cite{friedman1991evolutionary,fudenberg1998learning,hofbauer1998evolutionary,weibull1997evolutionary}. We focus on the replicator dynamic \cite{taylor1978evolutionary} as a case study. Sample results for other dynamics are presented in Box 3. These dynamics are detailed in Appendix Section \ref{app: dynamics}, and solved in Appendix Section \ref{app: explicit solutions}.

\begin{wrapfigure}{l}{0.69\textwidth}
\vspace{-0.6\baselineskip} 
\begin{mdframed}[style=mddefault]
\begin{small}
\begin{centering}{\textbf{Box 3.} \it Explicit Solutions\\}\end{centering}

Explicit solutions are described below for three standard learning dynamics. These all set the rate of change in the attributes of an agent, or set of agents, proportional to the average optimal training vector for the agent in response to a reference population. 
In particular, all of the dynamics take the form:
\begin{equation}
    \frac{\mathrm d}{\mathrm d t} y(j,t) \propto \mathbb{E}_{y' \sim \pi(t)}[v(y,y')]
\end{equation}
where $j$ indexes a particular agent, and $\pi(t)$ is the reference population at time $t$. \\

These dynamics all admit explicit solutions when posed in the disc-game space, and when the set of agents are contained in the interior of $\Psi$. The following solutions all leverage the simplifying characteristics used to design the disc-game embedding. Namely, since disc-games satisfy desideratum (D2), the optimal response directions are student-independent, and linear in the opponent attributes. Since averages are linear operations, the average optimal response direction equals the optimal response direction at the reference population centroid. Then, since the disc-game is defined by a block-diagonal matrix, consecutive pairs of disc-game coordinates form separate blocks that evolve independently. Thus, the problem can be solved blockwise. That is, a single disc-game at a time. Accordingly, all solutions are presented for one pair of coordinates. Finally, since the disc-game is bilinear, and the optimal training vector field is a rotationally symmetric linear function, all three dynamics are highly symmetric linear systems differential equations which can be solved by finding a particular solution for one initial condition. For detailed solution techniques, see Appendix Section \ref{app: explicit solutions}.

\vspace{0.5\baselineskip}
\begin{tabular}{c l}
\multirow{3}{0.18\textwidth} {\includegraphics[trim = 35 40 35 40, clip, width=0.17\textwidth]{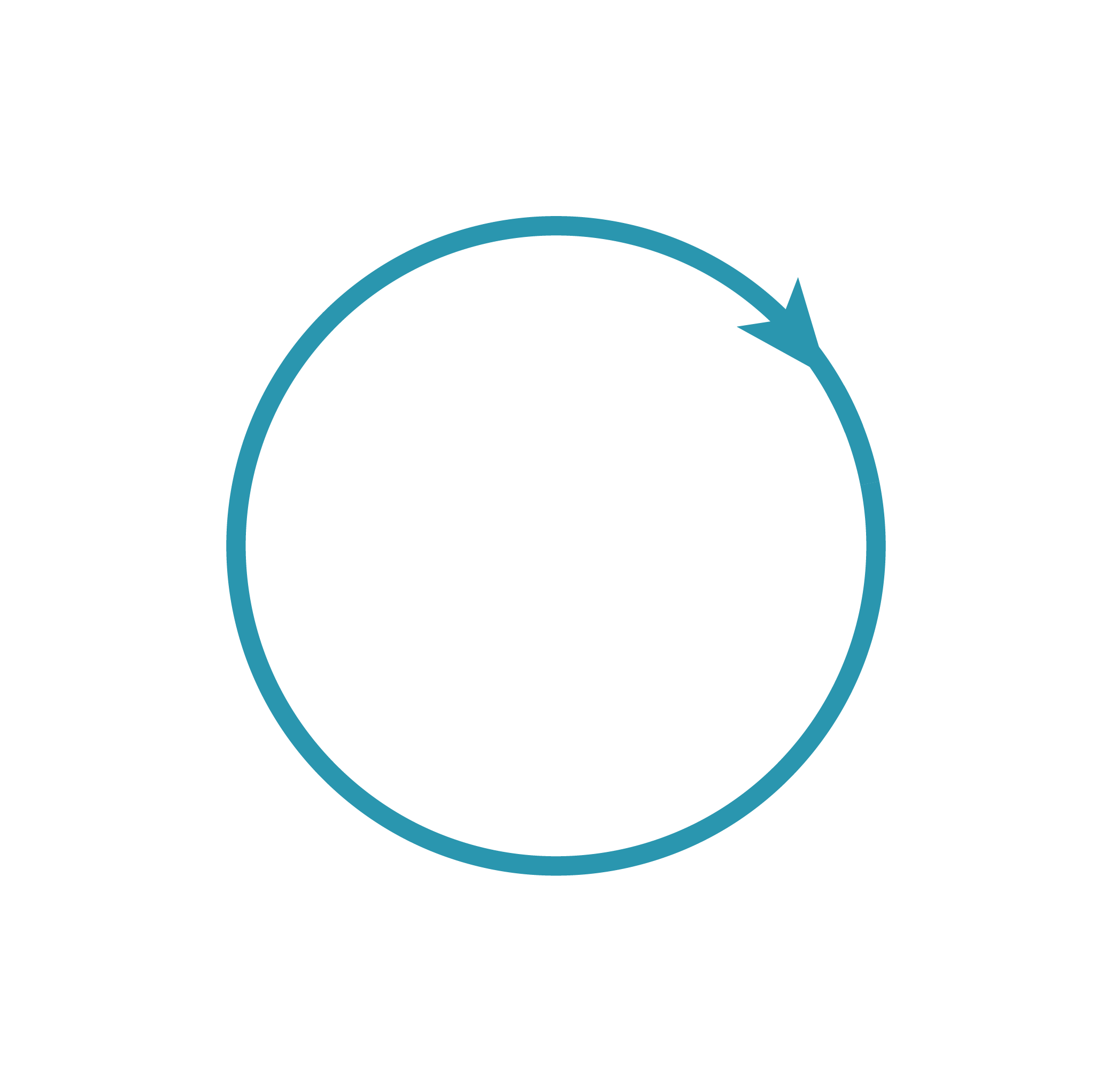}} & \textbf{Self-Play \cite{fudenberg1998learning}}\\[4pt]
 & {\it Reference Population:} The agent \\[4pt]
& {\it Solution:} Simple harmonic motion. Coordinate pairs orbit on \\[-2 pt] & circular paths at a fixed rate. Orbits move clockwise. \\[8 pt] \hline \\[-2 pt]
\multirow{3}{0.18\textwidth} {\includegraphics[trim = 35 40 35 40, clip, width=0.17\textwidth]{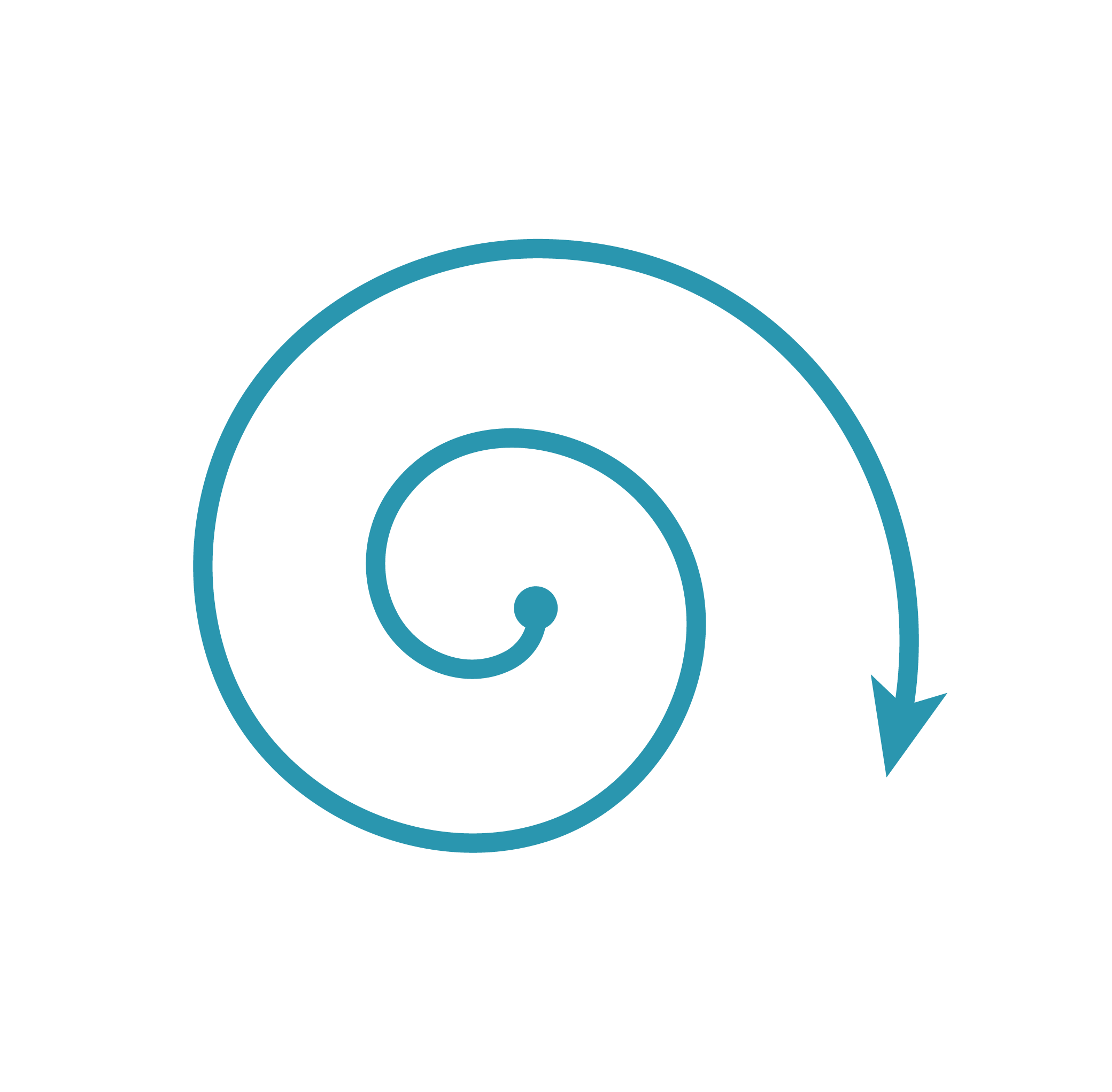}} & \textbf{Fictitious Self-Play \cite{fudenberg1998learning}}\\[4 pt]
 & {\it Reference Population:} All past agents.  \\[4 pt]
& {\it Solution:} Asymptotic to logarithmic spirals. The phase converges \\[-2pt] &to $-\sqrt{2 t}$, while the radius grows exponentially in the phase. Exact \\[-2pt] &solutions are Kelvin spirals. The figure (left) is log-scaled. \\[4 pt] \hline \\[-2 pt]
\multirow{3}{0.18\textwidth} {\includegraphics[trim = 32 40 30 40, clip, width=0.18\textwidth]{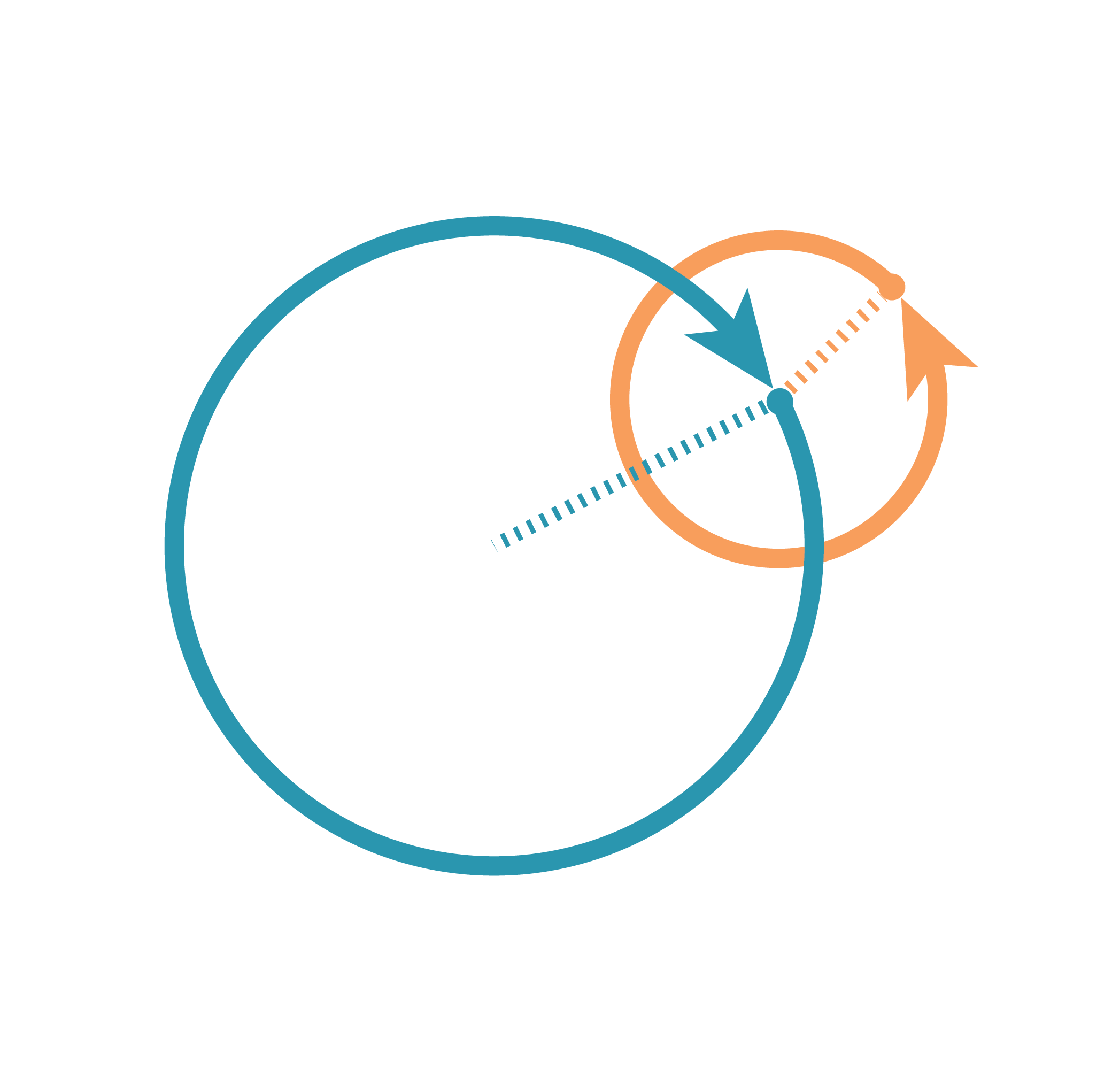}} &  \textbf{Simultaneous Gradient Ascent \cite{balduzzi2018mechanics}}\\[2pt]
 & {\it Reference Population:} All current agents. \\[2pt]
& {\it Solution:} Epicycles. The population centroid (blue) orbits as in  \\[-2 pt]& self-play. Each agent (orange) orbits backwards about the centroid \\[-2 pt] & at rate $1/n$, where $n$ is the number of agents.\\[4 pt]
\end{tabular}
\vspace{0.5\baselineskip}

\end{small}
\end{mdframed}
\vspace{-3 \baselineskip}
\end{wrapfigure}

All of the dynamics described in Box 3 are based on advective models in which agents change type according to the best-response gradient $v$, but without changing the overall population size. In this sense, these are explicitly learning dynamics. They consist of a fixed population of agents each moving to optimize their payout against an average over a population. In these cases, the optimal training vector field enters explicitly, so a coordinate system that simplifies $v$ simplifies the dynamics. 

In contrast, ecological population dynamics typically start from a birth-death process, in which individuals are produced and destroyed, but do not explicitly change types. Here, selection replaces training as the primary mechanism. Nevertheless, we will show that the disc-game embedding is a useful tool even when the dynamic of interest does not explicitly reference the training vector field.

\subsection{The Replicator Dynamic} \label{sec: replicator}

The replicator dynamic \cite{hofbauer1998evolutionary,taylor1978evolutionary} is the canonical ecological dynamic of evolutionary game theory. As the Lotka-Volterra equations \cite{lotka1920analytical,lotka1920undamped,volterra1927fluctuations} set baseline expectations in ecology, the replicator equation sets baseline expectations in evolutionary game theory. In fact, the two dynamics are equivalent up to an appropriate transformation \cite{hofbauer1981occurrence,hofbauer1998evolutionary,page2002unifying}. 

The replicator dynamic admits multiple reasonable derivations \cite{page2002unifying}. These include mean-field approximations, and infinite population limits, for imitation learning in finite-population models \cite{binmore,borgers1997learning}, for bandits \cite{schlag1998imitate}, and for selection in population genetics \cite{nowak2004evolutionary}, as a limited rationality model for inference \cite{karev2010replicator}, as a self-play dynamic in the space of mixed strategy distributions equipped with an appropriate metric \cite{legacci2024geometric,mertikopoulos2018riemannian}, or as the time evolution of mixed strategies under an exponential/multiplicative-weight learning rule \cite{auer1995gambling,littlestone1994weighted,vovk1990aggregating,flokas2020no} which is the gold-standard example of a no-regret learning rule since it is both easy to implement and satisfies a constant regret bound \cite{arora2012multiplicative,lattimore2020bandit,legacci2024geometric}. Indeed, the replicator dynamic satisfies a \textit{constant} cumulative regret upper-bound, with incremental regret converging to zero faster than $t^{-1}$ \cite{boone2019darwin}. See \cite{Cressman_a,nowak2004evolutionary} for reviews. 

The replicator dynamic sets the per-capita growth rate of a population of type $x$ proportional to the expected payout of type $x$ against an opponent drawn at random from the current population \cite{nowak2004evolutionary}. In particular, we focus on the continuous replicator equation \cite{cressman2004coevolution,cressman2005measure}, which extends the classical replicator dynamic to a measure-valued dynamic on continuous trait spaces \cite{cressman2004coevolution,cressman2004dynamic,Cressman_b,cressman2005measure,cressman2006stability}. We focus on the continuous setting since it is less well explored, results in the continuum often provide useful analytic limits for discrete spaces, and continuous trait spaces are natural in learning problems that involve a continuous range of possible parameter values or action probabilities. For analogous results in the discrete setting see \cite{akin1984evolutionary,hofbauer1996hamiltonian,boone2019darwin}.

Formally, a measure-valued dynamic uses a time-evolving measure $\Pi[\cdot](t)$, where $\Pi[S](t)$ is the number of individuals with traits in the set $S \subset \Omega$ at time $t$. The existence and uniqueness of measure-valued solutions to the continuous replicator equation is established in \cite{oechssler2001evolutionary,oechssler2002dynamic} provided $f$ is continuous.

If the population distribution is continuous, then the measure admits a density $\pi$, where $\pi(x,t)$ is the population density in trait space at type $x$ and time $t$. Let:
\begin{equation}
    P(t) = \Pi[\Omega](t) = \int_{x \in \Omega} \pi(x,t) dx
\end{equation}
denote the total population size. Note that $P(t)$ need not equal 1. That is $\pi$ need not be normalized.

Then, given skew-symmetric $f$, and a continuous measure, the replicator equation sets:
\begin{equation} \label{eqn: replicator equation}
    \partial_t \pi(x,t) = \frac{\rho(P(t),t)}{P(t)}\mathbb{E}_{X' \sim \pi(\cdot,t)}[f(x,X')] \pi(x,t)
\end{equation}
where $\rho(P,t) \in \mathbb{R}^+$ represents the per capita interaction rate between individuals given a total population of size $P$ at time $t$ in an underlying agent based model \cite{akin1984evolutionary}. For agent based derivations of the replicator equation, see \cite{binmore,borgers1997learning}. 

The standard replicator equation studies the change in the relative frequency of types implied by the dynamic \eqref{eqn: replicator equation}. Since we assume that $f$ is skew-symmetric, and ignore scalar factors, there is no need to distinguish between the frequency of types, and the distribution representing the population. When $f$ is skew-symmetric, the two dynamics differ by a constant factor equal to the total population size. Scaling the right-hand side of the differential equation by a constant is equivalent to scaling the units used to measure time, so the trajectories generated by the traditional replicator dynamic and the form given above are equivalent up to a change of time units. 

Notice that, since $f$ is skew-symmetric, the total population is necessarily conserved by the dynamic:
$$
\frac{\mathrm d}{\mathrm d t} P(t) \propto \int_{x \in \Omega} \mathbb{E}_{X' \sim \pi(\cdot,t)}[f(x,X')] \pi(x,t) dx = \mathbb{E}_{X,X' \sim \pi \times \pi}[f(X,X')] = 0
$$
where the latter inequality follows since $\mathbb{E}_{X,X' \sim \pi \times \pi}[f(X,X')] = -\mathbb{E}_{X,X' \sim \pi \times \pi}[f(X',X)]$ since $f$ is skew-symmetric, but $\mathbb{E}_{X,X' \sim \pi \times \pi}[f(X,X')] = \mathbb{E}_{X,X' \sim \pi \times \pi}[f(X',X)]$ since $X$ and $X'$ are drawn identically so are exchangeable. It follows that $P(t) = P(0) = P$ is constant. 

To standardize the analysis, we will make two assumptions, one without loss of generality.

First, we will restrict our attention to solutions that can be expressed with time-evolving densities. This is sufficient to explain transient finite-time dynamics and mixing, though may not be sufficient to explain all stability concerns for singular (monomorphic) steady-state distributions associated with equilibria. For stability considerations see \cite{cressman2004dynamic,cressman2006stability}. Note that not all static equilibria (e.g.~Nash equilibria) are stable under the replicator dynamic, nor do sufficient conditions (interior evolutionary stability \cite{smith1974theory}) developed in discrete trait spaces necessarily extend to continuous trait spaces unless the topology on measures is chosen appropriately \cite{oechssler2001evolutionary,oechssler2002dynamic}. Moreover, no-regret learning dynamics, which include the replicator dynamic, fail to converge to interior, fully-mixed equilibria in zero-sum normal form games with finitely many strategies. So, unless there exists a single dominating strategy, no-regret learning dynamics wander persistently \cite{akin1984evolutionary,bailey2019multi,hofbauer1996hamiltonian,mertikopoulos2018cycles,flokas2020no}. Since we are concerned with a dynamical analysis, rather than a limiting analysis, we leave further discussion of measure-theoretic complications to the literature. 

Second, we will also assume that  the interaction rate is linear in the total population size: $\rho(P(t),t) = P(t)$. This assumption is made without loss of generality. If $\rho(P,t)$ is not equal to $P$, then all solutions to the replicator equation are equal to solutions of the replicator equation with a linear reaction rate after transforming the time coordinate.

In particular, let $\tau(t)$ represent a monotonically increasing function of time. Then $\tau(t)$ is invertible, so we can exchange time coordinates freely. Write $\pi(\cdot,t) = \pi(\cdot,t(\tau))$. Then, 
\begin{equation}
    \begin{aligned}
    & \pi(x,t) = \tilde{\pi}(x,\tau(t)) \text{ where }  \begin{cases}
        & \partial_t \pi(x,t) = \frac{\rho(P(t),t)}{P(t)} \mathbb{E}_{X' \sim \pi(\cdot,t)}[f(x,X')] \pi(x,t) \\
        & \partial_{\tau} \tilde{\pi}(x,\tau) = \mathbb{E}_{X' \sim \tilde{\pi}(\cdot,\tau)}[f(x,X')] \tilde{\pi}(x,\tau) \\
        \end{cases} \\
        & \text{and: } \tau(t) = \int_{s = 0}^t  \frac{\rho(P(s),s)}{P(s)}, \quad t(\tau) = \int_{s = 0}^{\tau} \frac{P(s)}{\rho(P(s,\tau))}. 
    \end{aligned}
\end{equation}

Thus, all solutions to \eqref{eqn: replicator equation} can be recovered from the solution when $\rho(P,t) = P$ by changing the time coordinate. From now on, we will write:
\begin{equation} \label{eqn: replicator standardized}
    \partial_t \pi(x,t) \propto \mathbb{E}_{X'\sim \pi(\cdot,t)}[f(x,X')] \pi(x,t)
\end{equation}
with the understanding that the proportionality represents an unspecified interaction rate $\rho$, and will solve all systems using the convention that $\rho(P,t) = P$ so that the proportionality is an equality.

\subsection{Replicator Parameter Dynamics Via Disc Game-Embedding}

Here, we demonstrate that the disc-game embedding proposed in Section \ref{sec: latent space} is a useful coordinate change when studying parametric solutions to the replicator dynamic.

\subsubsection{Objectives and Special Cases} \label{sec: special cases}

A parametric solution to a density-valued dynamic is defined by fixing a parametric family of densities that is invariant under the dynamic, then solving for the parameter dynamics. Specifically, we write $\pi(x;\theta(t))$ for the density at traits $x$ given a parameter vector $\theta$ that evolves in time. We seek solutions for families of densities with finitely many degrees of freedom. Typically, we will seek $\theta(t) \in \mathbb{R}^r$ for some integer-valued $r < \infty$. When no such solution is available, we seek consistent approximation methods that converge in the limit as $r$ diverges. The latter problem is an example of a closure problem (e.g.~\cite{naghnaeian2017robust}). Exact closure is possible if the replicator dynamics can be reduced to the dynamics of a finite set of parameters without any approximation error for any initial distribution. 

The formal problem follows. Let $\Theta \subseteq \mathbb{R}^r$ be an admissable set of parameters. Let $\mathcal{R}(\Theta) = \text{range}_{\theta \in \Theta}(\pi(\cdot;\theta)) = \{\pi(\cdot;\theta) \mid \theta \in \Theta\}$ denote the set of all density functions that can be expressed by choosing parameters from $\Theta$. A set of densities is invariant under the replicator dynamic \eqref{eqn: replicator equation} if, given $\pi(\cdot,t) \in \mathcal{R}(\Theta)$ at some time $t$, then $\pi(\cdot,t') \in \mathcal{R}(\Theta)$ for all future times $t' \geq t$. We aim to find $(\Omega,f,\mathcal{R}(\Theta))$ that are invariant under the replicator dynamic. 

\begin{snugshade}
    \noindent \textbf{Problem:} Given a functional form game $(\Omega,f)$, does the replicator dynamic \eqref{eqn: replicator equation} admit an exact closure using $r < \infty$ degrees of freedom, and, if so, for what range of parametric densities $\mathcal{R}(\theta) = \{\pi(\cdot;\theta) \mid \theta \in \Theta\}$?
\end{snugshade}

We will use the disc-game embedding to answer both questions. To build intuition for the solution, and the ensuing parameter dynamics, we consider two special cases.

\begin{enumerate}
\item{\textbf{Perfectly Transitive Functions:}} Suppose that $f(x,x') = r(x) - r(x')$ for some rating function $r:\Omega \rightarrow \mathbb{R}$. Then, $f$ is a perfectly transitive function \cite{strang2022network}. When $f$ is perfectly transitive, the replicator Equation \eqref{eqn: replicator standardized} can be solved exactly for any initial distribution. In particular (see Appendix Section \ref{app: transitive solution}),
\begin{equation} \label{eqn: transitive soln}
    \pi(x,t) \propto e^{t P r(x)}\pi(x,0), \text{ where } P = P(0).
\end{equation}

Equation \eqref{eqn: transitive soln} is an example of a parametric solution with two degrees of freedom that evolve in time. Given $f(x,x') = r(x) - r(x')$ consider parametric distributions of the form $\pi(x,t) = \pi(x,0) \exp(\theta_0(t) + \theta_1(t) r(x))$. Then the solution \eqref{eqn: transitive soln} sets $\theta_1(t) = P t$, where $\theta_0(t)$ is implied by $\theta_1(t)$ via population conservation. More generally, we could write, $\pi(x,t) = e^{u(x,t)}$ where $u(x,t)$ is the log-density at traits $x$ and time $t$. Then, the transitive solution spans two-dimensional affine subspace of functions of the form $u(\cdot,t) \in u(\cdot,0) + \text{span}\{1,r(\cdot)\}$ where $\text{span}\{1,r(\cdot)\}$ is all possible linear combinations of the constant function and the rating function. 

The constant function and rating function correspond to the disc-game embedding functions when $f$ is perfectly transitive since they correspond to the real and imaginary parts of the eigenfunctions of the operator $F$. In fact, $f(x,y) = r(x) 1 - 1 r(y) = [r(x),1] \times [r(y),1]$ is already in the form of a disc-game decomposition with embedding functions $r(\cdot)$ and 1. We will show that the parameters $\theta_{2k-1}$ and $\theta_{2k}$ corresponding to paired embedding coordinates are always related by a Hamiltonian system of ODE's. 

The transitive solution \eqref{eqn: transitive soln} also illustrates a generic reduction technique that will be accomplished automatically by disc-game embedding. Notice that if $x$ and $x'$ satisfy $r(x) = r(x')$ then $\pi(x,t)/\pi(x',t) = \pi(x,0)/\pi(x',0)$ is constant. It follows that the conditional population distribution remains constant along every level set of the rating function $r$. This is an example of a reduction by outcome equivalency.

\begin{snugshade}
    \noindent \textbf{Definition: [Outcome Equivalency]} Trait vectors $x$ and $x'$ are \textbf{outcome equivalent}, $x \sim x'$, if $f(x,y) = f(x',y)$ for almost all $y \in \Omega$. 
\end{snugshade}

When $\pi(x,t)$ obeys the replicator dynamic, the population at outcome equivalent traits evolve in tandem. Indeed, for any $x \sim x'$, $\pi(x,t)/\pi(x',t) = \pi(x,0)/\pi(x',0)$ for all $t$. It follows that the conditional population distribution on each set of outcome equivalent traits remains constant under the replicator dynamic. Thus, the dynamic can be reduced by grouping outcome equivalent traits into a single representative trait for each equivalency class, then studying the replicator equation on the equivalency classes. 

We will show that all solutions generated using a disc-game embedding automatically reduce by equivalency class since all outcome equivalent trait vectors are mapped to the same set of embedded coordinates. To retain the information needed to separate distinct but outcome equivalent types, we can append additional coordinates to the disc-game embedding that do not contribute to the dynamics. In practice, this is never necessary since the conditional distributions within each equivalency class match the initial conditionals. Solutions can always be expressed by first solving for the time evolution of the marginal densities, then redistributing the marginal density of each class proportionally according to the initial conditional distribution over outcome equivalent types. 


\item{\textbf{Quadratic Functions:}} The transitive solution solves the replicator Equation \eqref{eqn: replicator equation} for all linear $f$ since any linear, skew-symmetric function is perfectly transitive \cite{cebra2023similarity}. Consider quadratic $f$ next.  

When $f$ is quadratic, the family of Gaussian distributions is invariant under the replicator dynamics \cite{cebra2024almost,cressman2004coevolution,Cressman_b}. Proofs are provided in \cite{cressman2004dynamic}. This result is weaker than the result presented for the linear case (transitive case), as it only applies if $\pi(x,0)$ is Gaussian, and provides no guarantee regarding the stability of the manifold of Gaussian distributions. For example, it does not guarantee that the family of Gaussian distributions is attracting in some neighborhood of the family. Using the disc-game embedding, we will show that the family of Gaussian distributions is neutrally stable (Lyapunov stable) when $f$ is quadratic, so is neither attracting or repelling. For now, we recapitulate the extant result.

Any Gaussian distribution is parameterized by its mean and covariance, so, we assume the parametric form:
\begin{equation}
    \pi(x;\theta) \propto \exp\left(- \frac{1}{2} (x - \bar{x})^{\intercal} \Sigma_x^{-1} (x - \bar{x}) \right)
\end{equation}
Then, $\theta$ consists of the independent entries of the centroid $\bar{x}$ and covariance $\Sigma_x$. Since the family of Gaussian distributions is invariant under the replicator dynamic when $f$ is quadratic, we can replace the density-valued dynamic with a finite dimensional system of ODE's relating the rate of change in $\bar{x}$ and $\Sigma_x$ to $f$.

To express this ODE, we need a canonical form for quadratic $f$. All skew-symmetric quadratic functions can be expressed:
\begin{equation} \label{eqn: quadratic f}
    f(x,x') = r(x) - r(x') + x^{\intercal} H_{xx'} x' \text{ where } r(x) = g^{\intercal} x + \frac{1}{2} x^{\intercal} H_{xx} x
\end{equation}
where $g$ is a $T \times 1$ vector corresponding to the gradient of $f(x,x')$ with respect to $x$ at $x = x' = 0$, and $H_{xx}$ and $H_{x'x'}$ are $T \times T$ matrices corresponding to the on-diagonal and off-diagonal blocks of the Hessian of $f$ evaluated at $x = x' = 0$. The Hessian blocks are symmetric and skew-symmetric respectively. Then $J = H_{xx} + H_{xx'}$ is the Jacobian of the self-play vector field $v(z) = \nabla_{x} f(x,x')|_{x = x' = z} = g + J z$ \cite{cebra2023similarity}.

\begin{snugshade}
    \noindent \textbf{Lemma 7: [Gaussian Invariance] \cite{cebra2024almost}} If $f$ is a quadratic function, then the manifold of multivariate normal distributions is invariant under the continuous replicator dynamic \eqref{eqn: replicator equation} so admits an exact closure in terms of its mean and covariance. Given \eqref{eqn: replicator standardized}, the mean and covariance satisfy the closed system of ODE's:
    \begin{equation} \label{eqn: moment dynamics}
    \begin{aligned}
    &\frac{\mathrm d}{\mathrm d t}\bar{x}(t) =  \Sigma_x(t) (g + J \bar{x}(t))\\
    &\frac{\mathrm d}{\mathrm d t} \Sigma(t) = \Sigma_x(t)^{\intercal} H_{xx} \Sigma_x(t) 
\end{aligned}
\end{equation}
\end{snugshade}

Equation \eqref{eqn: moment dynamics} is an example of a set of parameter dynamics. It replaces the continuous replicator equation, which is density-valued, so operates in an infinite dimensional space, with a closed system of ODE's in finitely many parameters. In this case, $r = T + T(T+1)/2$ degrees of freedom are needed. We will show, using the disc-game embedding, that $r$ corresponds with the $\text{rank}(F)$. 

The covariance evolves independently of the centroid, and can be solved explicitly:
\begin{equation} \label{eqn: covariance solution}
\Sigma_x(t) = \left(\Sigma_x(0) - H_{xx} t \right)^{-1} 
\end{equation}

If $H_{xx}$ is negative definite, as when near a local maximum in the rating function, then the solution is valid for all times, and the population distribution approaches a delta distribution at the maximizer of the rating function $g^{\intercal} x + \frac{1}{2} x^{\intercal} H_{xx} x$. If $H_{xx}$ has any positive eigenvalue, then the solution only exists for $t$ up until one of the eigenvalues of $\Sigma_x(0) - H_{xx} t$ crosses zero \cite{cebra2024almost}. Then, the population density escapes all bounded neighborhoods in finite time. This illustrates that, while working with densities is sufficient for transient dynamics, it does not explain the full measure-valued dynamic, even for finite time. 

Given, $\Sigma_x(t)$ satisfying \eqref{eqn: covariance solution}, the centroid obeys the equation:
\begin{equation} \label{eqn: adaptive dynamics}
    \frac{\mathrm d}{\mathrm d t} \bar{x}(t)  = \Sigma_x(t) (g + J \bar{x}(t)) = \Sigma_x(t) \nabla_{x} f(x,\bar{x}(t))|_{x = \bar{x}(t)}.
\end{equation}

Equation \eqref{eqn: adaptive dynamics} is an example of an \textit{adaptive dynamics} equation \cite{abrams2001modelling,mcgill2007evolutionary}, which is itself a generalization of self-play \cite{fudenberg1998learning} that incorporates scaling by a positive definite matrix (see Appendix \ref{app: dynamics}). In adaptive dynamics a (typically concentrated) population is approximated with a monomorphism at its centroid. The motion of the centroid is directed by the self-play vector field $v(z) = \nabla_x f(x,z)|_{x =z}$ \cite{nowak2004evolutionary}. In particular, $\frac{\mathrm d}{\mathrm d t} \bar{x}(t) = C(\bar{x}(t),t) v(\bar{x}(t))$ where $\bar{x}(t)$ is the population centroid, $v(z)$ is the self-play vector field, and $C(x,t)$ is a symmetric positive semi-definite matrix representing either $\Sigma_x$, variation introduced by mutation, or a metric on $\Omega$ \cite{abrams2001modelling,cressman2004coevolution,dieckmann1996dynamical,leimar2005multidimensional,mcgill2007evolutionary,meszena2002evolutionary,zeeman1981dynamics}.

So, as illustrated in \cite{cressman2004coevolution}, when $f$ is quadratic, $\pi(x,0)$ is Gaussian, and $\pi(x,t)$ obeys the replicator dynamic, then the centroid $\bar{x}(t) = \mathbb{E}_{X \sim \pi(\cdot,t)}[X]$ obeys adaptive dynamics. Unfortunately, this result does not hold for arbitrary initial distributions, nor does it hold for arbitrary $f$. We will show that, for any separable $f$, that is, any $f$ that admits a finite-dimensional disc-game embedding, the centroid of the population distribution in the disc-game coordinates obeys adaptive dynamics, no matter the initial distribution. More strongly, we will show that this adaptive dynamics equation fully specifies the original replicator dynamic, so is sufficient to recover the full distributional dynamic.

The quadratic-Gaussian case also illustrates a generic dichotomy in the dynamical behavior of replicator systems. If $H_{xx}$ is nonzero then the distribution either converges to a delta or escapes to infinity. In either case, the density function converges to a limiting function that is ill-defined or is only supported on a subset of $\Omega$. If $H_{xx}$ is zero, then $\Sigma_x(t) = \Sigma_x(0)$ is constant in time, and the centroid obeys a time-autonomous affine ODE. Moreover, there is either a unique solution to $g + H_{xx'} \bar{x} = 0$ corresponding to an interior, fully mixed Nash equilibrium, or there is not. Setting $\bar{x}$ equal to the equilibrium produces a steady-state distribution. The dynamics of $\bar{x}$ are linear when centered about the equilibrium, with all imaginary eigenvalues, so the equilibrium is a center. If no such equilibrium exists, then there is a component of $g$ in the nullspace of $H_{xx'}$, and $\bar{x}(t)$ will diverge to infinity along that direction. Thus, given quadratic $f$, and Gaussian $\pi_0$, $\pi(x,t)$ either escapes the space of well-defined densities, or orbits about a fully mixed Nash equilibrium. This dichotomy has been shown for normal form games on finite strategy space and for more general no-regret dynamics \cite{akin1980domination,boone2019darwin,flokas2020no,mertikopoulos2018cycles,piliouras2014optimization}. We will illustrate the same dichotomy for all symmetric, constant-sum, functional form games.  

\end{enumerate}


In the Section \ref{sec: closure} we will show that the embedding maps used for disc-game embedding provide a natural function space for answering the closure problem. Using that function space, we will solve the closure problem in complete generality. We will show that the replicator dynamic admits an exact finite-dimensional closure if and only if $f$ is separable, in which case solutions to the replicator equation require $r = \text{rank}(F)$ dynamic degrees of freedom. Moreover, we will show that the parameter dynamics are block Hamiltonian, so take the generic form of nonlinear coupled oscillators. This generalizes the special solutions developed for the transitive and quadratic cases. It also generalizes the observation that simultaneous gradient ascent is Hamiltonian given a zero-sum normal form game on a finite strategy space to gradient-free learning dynamics and to functional form games on continuous trait spaces \cite{balduzzi2018mechanics}. 


\subsubsection{Closure} \label{sec: closure}

To simplify the replicator dynamic, recall a standard result (c.f.~\cite{akin1984evolutionary}):
\begin{snugshade}
    \noindent \textbf{Lemma 8: [Support Preservation]} Let $S(t) = \{x \mid \pi(x,t) > 0\}$ denote the support of $\pi$ at time $t$. If $\pi(\cdot,t)$ obeys the continuous replicator dynamic, then $S(t) = S(0)$ for all $t$ on the interval containing 0 for which the continuous replicator equation admits unique, density-valued, solutions. 
\end{snugshade}

\noindent \textbf{Proof Outline:} If $t' > t$ then $S(t') \subseteq S(t)$, since, if $\pi(x,t) = 0$ then $\partial_t \pi(x,s) = 0$. That is, the support can never expand. The reverse statement is established by repeating the same argument but differentiating with respect to time running in reverse. Sufficient conditions for existence and uniqueness are provided in \cite{oechssler2001evolutionary,oechssler2002dynamic}. For instance, if $f$ is continuous and $S(0)$ is compact, or if $f$ is bounded, then the solutions to the continuous replicator equation exist and are unique for all finite time. $\square$ \vspace{0.1 in}

Since the support is preserved for all $t$ such that solutions to the continuous replicator equation remain density valued, we can, without loss of generality, restrict our attention to solutions on the domain $S(t) = S(0) = S$. In particular, dividing Equation \eqref{eqn: replicator standardized} by the population density at each $x \in S$ yields:
\begin{equation} \label{eqn: replicator log}
    \partial_t \log(\pi(x,t)) = \frac{\partial_t \pi(x,t)}{\pi(x,t)}  = \mathbb{E}_{x' \sim \pi(\cdot,t)}[f(x,x')] \text{ for all } x \in S.
\end{equation}

Let:
\begin{equation}
    u(x,t) = \log(\pi(x,t)) 
\end{equation}
denote the log population density at all $x \in S$. 

Since the logarithm is invertible, any parametric solution to the continuous replicator equation $\pi(x;\theta(t))$ can be expressed via a parametric sequence of log-densities $u(x;\theta(t))$. By the chain rule, the continuous replicator equation admits an exact solution in terms of a parametric form if and only if:
\begin{equation} \label{eqn: closure equation}
    \nabla_{\theta} u(\cdot;\theta(t)) \cdot \frac{\mathrm d}{\mathrm d t} \theta(t) = \mathbb{E}_{x' \sim \pi(\cdot;\theta(t))}[f(\cdot,x')] 
\end{equation}
for some vector $\frac{\mathrm d}{\mathrm d t} \theta(t)$ and where $\pi(\cdot;\theta) = \exp(u(\cdot;\theta))$. The closure equation, \eqref{eqn: closure equation}, must hold for all $x \in S$.

Notice that $\mathbb{E}_{x' \sim \pi}[f(x,x')]$ is $F[\pi]$ where $F$ is the integral operator with a uniform reference measure. Therefore, the right-hand side of the closure Equation \eqref{eqn: closure equation} is contained in the range of the integral operator $F$ restricted to all densities in $\mathcal{R}(\Theta)$. Thus, a parametric family of densities $\mathcal{R}(\Theta)$ is invariant under the replicator dynamic if and only if:
\begin{equation} \label{eqn: span contains range}
    F[\pi(\cdot;\theta)] \in \text{span}(\nabla_{\theta} \log(\pi(\cdot;\theta))) \text{ for all } \theta \in \Theta
\end{equation}
and admits a unique parametric solution if the set of functions $\{\partial_{\theta_j} u(\cdot;\theta)\}_{j=1}^r$ are linearly independent for all $\theta \in \Theta$. A simpler sufficient condition is $ \text{range}(F) \subseteq \text{span}(\nabla_{\theta} \log(\pi(\cdot;\theta))) \text{ for all } \theta \in \Theta$ 
which can easily be ensured if $u(\cdot;\theta)$ is parametrized using a linear combination of basis functions that span the $\text{range}(F)$. 

These observations suggest the following parametric form:
\begin{equation} \label{eqn: affine parametrization}
    u(\cdot;\theta) = u_0(\cdot) + \sum_{j=1}^{r} \theta_j b_j(\cdot), \quad u_0(x) = \log(\pi(x,0))
\end{equation}
where $\mathcal{B} = \{b_j\}_{j=1}^{r}$ is a collection of linearly independent basis functions $b_j:\Omega \rightarrow \mathbb{R}$ whose span contains the range of the operator $F$. If $F$ is not degenerate, choose a complete basis $\mathcal{B}$ so that the sequence of partial sums of the projection of any function in the range of $F$ onto a truncation of $\mathcal{B}$ converges to that function as the dimension of the basis diverges. Equation \eqref{eqn: affine parametrization} defines an exponential family \cite{efron1978geometry,efron2022exponential} with sufficient statistic $b(\cdot)$,  natural parameters $\theta$, and base measure $\pi(x,0)$. Like any exponential family, Equation \eqref{eqn: affine parametrization} expresses a collection of distributions that can be formed by an exponential tilting of some base measure.

After adopting \eqref{eqn: affine parametrization}, the closure equations reduce to:
\begin{equation} \label{eqn: closure for affine parameterization}
    \sum_{j=1}^r b_j(x) \frac{\mathrm d}{\mathrm d t} \theta_j(t) = F[\pi(\cdot;\theta(t))](x) \text{ for all } x \in S.
\end{equation}

Any parameterization of $u$ of the form \eqref{eqn: affine parametrization} is an affine parameterization. Since the offset $u_0$ is determined by the initial distribution, the parameterization itself is determined by the choice of basis $\mathcal{B}$. We will say that an affine parameterization with basis $B$ is minimal for $S$ if it uses as few dynamic degrees of freedom as possible, that is, if the span of $\mathcal{B}$ could not be made smaller while allowing exact closure (Equation \eqref{eqn: closure for affine parameterization}) for all initial distributions with support $S$.

\begin{snugshade}
\noindent \textbf{Theorem 4: [Invariant Families]} Let $(\Omega,f)$ denote a functional form game, $S \subseteq \Omega$ a support set, and $\pi(x,t)$ is a density over $\Omega$ with support $S$ that evolves according to the continuous replicator equation, \eqref{eqn: replicator equation}. Then an affine parameterization of the log-density $u(\cdot;\theta) = \log(\pi(\cdot;\theta))$ defines a minimal invariant family of densities for initial distributions with support $S$ if and only if $\text{range}(F) = \text{span}(\mathcal{B})$ where $\mathcal{B}$ is the basis used in the affine parameterization and $F$ is the integral operator defined by the kernel $f$ and support set $S$. The parameters of the affine parameterization provide an exact closure of the replicator dynamic. The corresponding parameter dynamics are uniqely defined if and only if the basis functions are linearly independent.  
\end{snugshade}

\noindent \textbf{Proof Outline:} Sufficiency is established by \eqref{eqn: span contains range}. Necessity is established by varying the initial distribution.  By choosing a fixed reference distribution with support $S$, then perturbing about the reference, it is easy to show that the range of $F$ over all distributions with a fixed support has dimension equal to the rank of $F$ on that support. Thus, if we seek an affine parameterization that admits an exact closure for all initial distributions with shared support, then the affine closure must have dimension at least equal to the rank of $F$. Minimality is established by showing that, if the range of $F$ is a proper subset of the span of $\mathcal{B}$, then there exists a choice of basis functions such that at least one parameter remains constant over time. Then, since the affine subspace is shifted by the initial distribution, that parameter will remain zero for all time. Thus, the associated direction can be dropped from the parameterization. Uniqueness is established by linear independence of the basis functions defining the linear closure equation, \eqref{eqn: closure for affine parameterization}. $\square$

\vspace{0.05 in}
Theorem 4 provides clear instructions for selecting an affine parameterization given the range of $F$. For example, if $f(x,x')$ is a polynomial function of $x$ and $x'$ with maximal degree $m$ in $x$, then the range of $F$ consists of polynomial functions of $x$ with maximal degree $m$ and those polynomials provide a minimal, sufficient, affine parameterization of the log-density dynamics. Then, $\pi(x,t)$ is constrained to the form $\pi(x,0)\exp\left(\sum_{j=0}^m \sum_{|\alpha| < j} \theta_j(t) x^{\alpha} \right)$ where $\alpha$ is a multi-index. If $f(x,x')$ is linear, $m = 1$, and the time-varying exponential term must be linear in $x$. In particular, using $f(x,x') = x - x'$ recovers the transitive solution presented in Section \ref{sec: special cases}. Similarly, if $f$ is quadratic, then $m = 2$, so 
the log-density is quadratic in $x$ at all times $t$. The only family of densities whose support is unbounded, and whose logarithms are quadratic functions of their argument are Gaussian distributions. Thus, as stated in Section \ref{sec: special cases}, when $f$ is quadratic, the family of Gaussian distributions is invariant under the replicator dynamic. 


\begin{snugshade}
\noindent \textbf{Corollary 4.1: [Rank and Dynamic Dimension]} If $(\Omega,f)$ is a functional form game, and $\pi(x,t)$ is a density over $\Omega$ that evolves according to the continuous replicator equation, \eqref{eqn: replicator equation}, with support $S$, then the replicator equation admits an exact closure using $r \geq \text{rank}(F)$ degrees of freedom, where the rank is evaluated over all distributions supported on $S$. 
\end{snugshade}

Corollary 4.1 establishes the minimal number of degrees of freedom needed to express all solutions to the replicator dynamic with a fixed support. In particular, given a support set $S$, exact closure is possible using as many degrees of freedom as the rank of the integral operator $F$ over $S$. If $F$ is degenerate on support $S$, then this rank is finite. If $F$ is not degenerate on $S$, then the rank is infinite, so exact finite dimensional closure is impossible. In that case, any attempt at closure will require an approximation.

\begin{snugshade}
\noindent \textbf{Corollary 4.2: [Finite Dimensional Closure]} If $(\Omega,f)$ is a functional form game, and $\pi(x,t)$ is a density over $\Omega$ that evolves according to the continuous replicator Equation \eqref{eqn: replicator equation}, with support $S$, then the replicator equation admits an exact finite-dimensional closure if and only if the integral operator $F[\cdot]$ has finite rank on the domain $S$. 
\end{snugshade}



Theorem 4, and Corollaries 4.1 and 4.2 suggest a canonical choice of basis for a minimal affine parameterization of the continuous replicator equation. Recall that, if $\text{rank}(F)$ is finite given support $S$, then the performance function $f$ admits an expansion onto a finite linear combination of disc games and each disc game expands into a combination of separable functions of the form $\text{disc}(y^{(k)}(x),y^{(k)}(x'))$ where $\{y^{(k)}\}_{k=1}^{\text{rank}(F)/2}$ are the disc-game embedding functions. The embedding functions are orthogonal with respect to the reference measure used for embedding, so are uncorrelated when applied to draws from the reference measure. Therefore, the disc-game embedding functions are linearly independent. Moreover, since $f$ is a linear combination of separable functions formed as an outer product of the disc-game embedding functions, the range of $F[\cdot]$ must be contained in the span of the embedding functions. 
%
%

Therefore, the embedding functions form a linearly independent basis for the range of $F$ and the affine parameterization:
\begin{equation} \label{eqn: disc-game parameterization}
    \pi(x;\theta) = e^{u(x;\theta)}, \quad u(\cdot;\theta) = u_0(x) + \sum_{k=1}^{\text{rank}(F)/2} \theta_{1}^{(k)} y_1^{(k)}(x) + \theta_{2}^{(k)} y_2^{(k)}(x), \quad u_0(x) = \log(\pi(x,0))
\end{equation}
is minimal, allows exact closure with respect to the parameters $\theta$, and admits uniquely defined parameter dynamics. 
We will show that the geometric interpretation of the disc-game will provide useful insight into the parameter dynamics implied by Equation \eqref{eqn: disc-game parameterization}.

If, given the support set $S$, $F$ is not degenerate, then no such finite-dimensional closure exists. Nevertheless, the affine parameterization proposed in Equation \eqref{eqn: disc-game parameterization} still provides optimal approximation to the replicator dynamic for any finite number of parameters $r$. 
%
%
Truncate the expansion at some even $r < \infty$. This parameterization would provide exact closure for the truncated disc-game approximation to $f$, $f^{(r)} = \sum_{k=1}^r \text{disc}(y^{(k)},y^{(k)})$. Recall that any truncated disc-game approximation provides an optimal approximation to $f$ among all rank $r$ functions (see Lemma 3). 
The error in the approximation to $f$ is exactly equal to the sum of squares of the dropped eigenvalues, so is finite, and arbitrarily small for sufficiently large $r$. Therefore, while exact, finite-dimensional closure is impossible for all $f$, exact finite-dimensional closure is possible for an optimal sequence of approximations to $f$ which converge to $f$ in the limit as $r$ diverges.

So, from now on, we will assume that $f$ is either finite rank, or, has been replaced by a sufficiently accurate, finite rank, approximation before applying the replicator dynamic. 

\subsubsection{Parameterization} \label{sec: representation}

Equation \eqref{eqn: affine parametrization} selects a parameterization that is both sufficient and minimal. This parameterization required two choices. First, the disc game embeddings depend on an arbitrarily selected reference measure. Second, the parameterization is centered at the initial distribution. Both choices require some discussion since the first is arbitrary, and the second is confounding. The chosen parameterization also maps from a set of bounded functions to an unbounded domain. It is important to distinguish how boundaries of the original set of densities are encoded via diverging sequences of parameters. We will address each point in turn. 

\begin{figure}[t]
    \centering
    \includegraphics[trim = 50 120 40 150, clip, width = \textwidth]{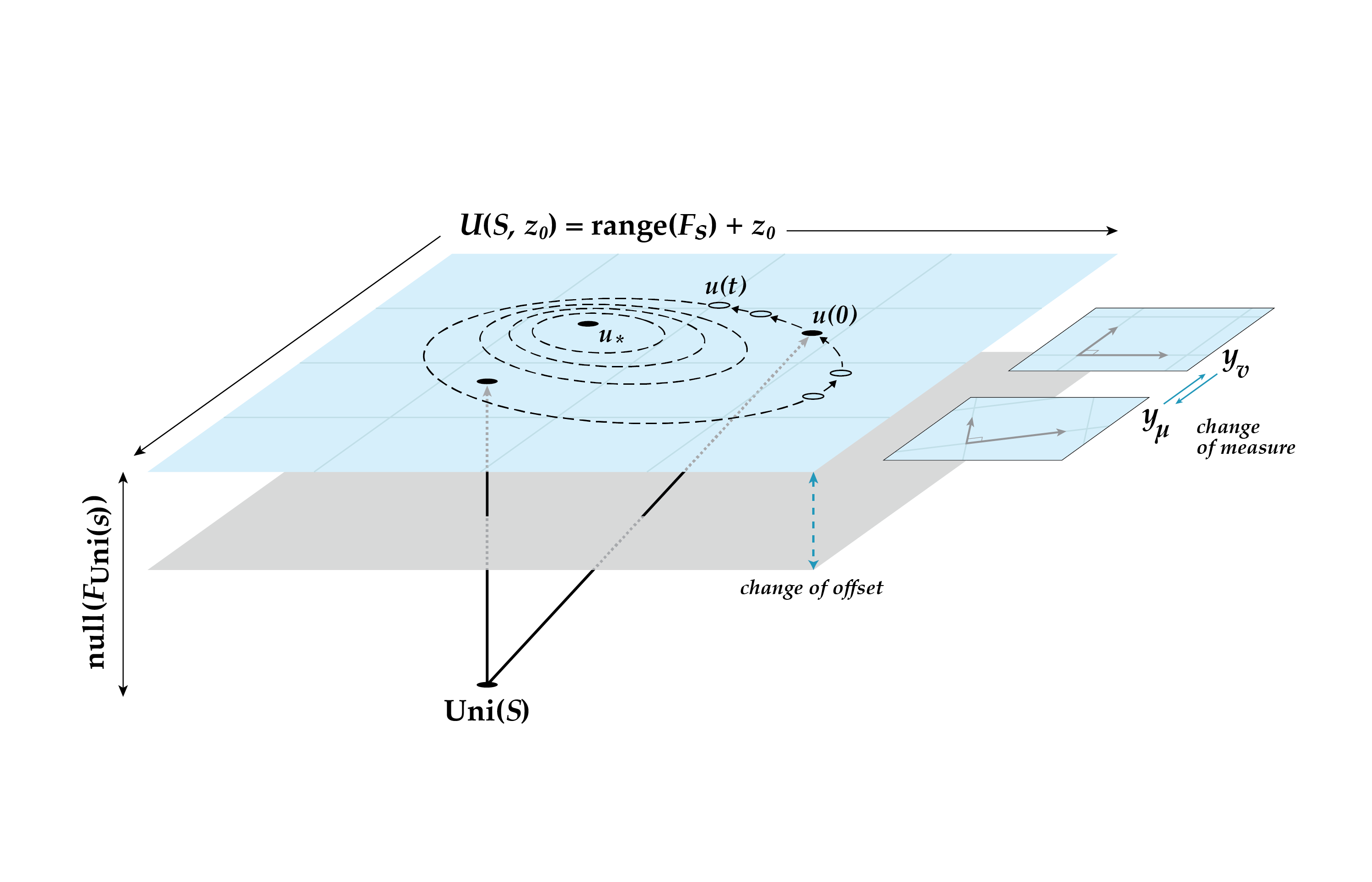}
    \caption{The invariant subspace $\mathcal{U}(\nu,u_0)$. The blue shaded parallelogram represents the subspace. The range of $F_S$ is represented by the pair of vectors spanning the subspace. The null space of $F$ given the uniform measure on $S$ is represented by the vertical direction perpendicular to the subspace. The oval marked $\text{Uni}(S)$ represents the constant function on $S$. The vector labeled $z_0$ is the null vector fixing the offset of $\mathcal{U}$. Changing the offset shifts the subspace along a direction contained in the null space. The vector labeled $u_0$ points to an initial log-density, $u(0)$. The concentric ovals represent sample orbits of the log density over time, $u(t)$. These remain within the invariant subspace. They orbit an equilibrium $u_*$ corresponding to a Nash equilibrium that is fully mixed over $S$. The offset fixing the position of the subspace can be specified by $u_0$, $x_0$, or $u_*$ (when it exists).  Distributions with support contained in, but not equal to $S$ correspond to infinite limits where $u$ diverges. The dashed grey lines represent the coordinates imposed by the choice of reference measure $\nu$, and the ensuing embedding functions $y_{\nu}$. Changing reference measure amounts to changing this basis via an invertible linear transformation, as illustrated in the two small panels to the right of the main panel.   }
    \label{fig: Invariant Subspace Schematic}
\end{figure}

To start, let:
\begin{equation}
    \mathcal{U}(\nu,u_0) = u_0 + \text{range}(F_{\nu})
\end{equation}
denote the affine subspace of log-densities associated with an initial distribution $\exp(u_0)$ and reference measure $\nu$. Figure \ref{fig: Invariant Subspace Schematic} illustrates the subspace, and its dependence on $u_0$ and $\nu$. The choice of $u_0$ fixes the offset of the affine subspace from the zero function. It centers the parameterization, since, when $\theta = 0$, $u(\cdot;0) = u_0(\cdot)$.  
We will show that $\mathcal{U}$ is invariant to changes in the reference measure that preserve its support and to changes in the offset parallel to the range of $F_{\nu}$.  The component of the offset perpendicular to the range of $F_{\nu}$ encodes the conditional distribution over each outcome equivalent class. 

\begin{snugshade}
    \noindent \textbf{Lemma 9: [Change of Reference Measure]} $\mathcal{U}(\nu,u_0)$ is invariant to changes in $\nu$ that preserve the support of $\nu$ when $\nu$ is continuous. Moreover, two affine parameterizations \eqref{eqn: affine parametrization} defined with respect to continuous measures $\nu$ and $\mu$ that share the same support are equivalent up to an invertible linear transformation.
\end{snugshade}

\noindent \textbf{Proof:} By Lemma 5, $\text{range}(F_{\nu}) = \text{range}(F_{\mu})$ if $\nu$ and $\mu$ are both continuous and share the same support set. Therefore, $\mathcal{U}(\nu,u_0) = \mathcal{U}(\mu,u_0)$. Let $y_{\nu}$ and $y_{\mu}$ denote the embedding functions for the reference measures $\nu$ and $\mu$. Since both span the same subspace, any linear combination of $y_{\nu}$ can be expressed with a linear combination of $y_{\mu}$. Since both sets are linearly independent the two affine parameterizations are equivalent up to an invertible linear transformation. $\square$
\vspace{0.05 in}


Since changing measure amounts to an invertible linear transformation of the parameter space, parameter dynamics associated with two measures that share the same support are topologically equivalent. We will use this fact to adapt the choice of reference measure to best suit the dynamical question of interest.

In contrast, changing $u_0$ changes the distribution associated with the origin of the parameter space. Changing the origin preserves the affine subspace if the log ratio of the corresponding distributions lies in the range of $F$.
\begin{snugshade}
    \noindent \textbf{Lemma 10: [Change of Center]} $\mathcal{U}(\nu,u_0) = \mathcal{U}(\nu,v_0)$ if and only if $u_0 - v_0 \in \text{range}(F_{\nu})$. 
\end{snugshade}

\noindent \textbf{Proof:} The subspaces $\mathcal{U}(\nu,u_0)$ and $\mathcal{U}(\nu,v_0)$ are parallel to $\text{range}(F)$. $\square$ 

\vspace{0.05 in}
Since the affine subspace $\mathcal{U}(\nu,u_0)$ is invariant to shifts in $u_0$ parallel to the range of $F_{\nu}$, the particular affine subspace can be represented by the choice of an element in the nullspace of the adjoint of $F_{\nu}$. Under the inner product $\langle \cdot,\cdot \rangle_{\nu}$ the adjoint of $F_{\nu}$ is $-F_{\nu}$, so the nullspace of the adjoint is the nullspace of $F_{\nu}$. Thus, it is sufficient to fix $u_0\setminus\text{range}(F_{\nu}) = \text{Proj}_{\text{null}(F_{\nu})}(u_0)$. In other words, $\mathcal{U}(\nu,u_0) = \mathcal{U}(\nu,v_0)$ if $\text{Proj}_{\text{null}(F_{\nu})}(u_0) = \text{Proj}_{\text{null}(F_{\nu})}(v_0)$.

Since $\mathcal{U}$ is invariant to changes in measure with fixed support $S$, and changes in $u_0$ parallel to the range of $F_{\nu}$, we can instead write $\mathcal{U}(S,z_0)$ where $S$ is the support of some continuous measure $\nu$, and $z_0$ is a null-vector for  $F_{\nu}$. Since the particular choice of $\nu$ does not matter, we could, for example, adopt the uniform measure over $S$, $\text{Uni}(S)$. Adopting the uniform measure is suggestive. All steady-state distributions with support $S$ must be contained in $\text{null}(F_{\text{Uni}(S)})$. 
We will show that, each $\mathcal{U}(S,z_0)$ either admits a unique steady-state distribution with support $S$, or does not. Thus, we could index invariant families by the steady-states they admit and the choice of offset amounts to a choice of steady state. 

The offset used to fix a particular invariant family is also closely related to the information discarded when reducing by outcome equivalency. Recall that $x$ is outcome equivalent to $x'$ if $f(x,x'') = f(x',x'')$ for almost all $x'' \in \Omega$. We can extend the same definition over any set $S$; $x \sim_{S} x'$ if $x$ and $x'$ return the same payout against almost all $x'' \in S$. 
\begin{snugshade}
    \noindent \textbf{Lemma 11: [Outcome Equivalent Types are Embedded Together]} Let $y(\cdot)$ denote a disc-game embedding defined with a reference measure supported on $S$. Then $y(x) = y(x')$ if and only if $x \sim_S x'$
\end{snugshade}

\noindent \textbf{Proof: } Lemma 11 is a special case of Lemma 6 with $\delta = 0$. If $x \sim_S x'$ then $f(x,x'') = f(x',x'')$, so $g(x) = g(x')$ for all functions $g \in \text{range}(F)$. Since the embedding functions span the range of $F$, each embedding function is contained in the range of $F$. This establishes the forward directon. For the reverse, note that, $f(x,x') = \text{disc}(y(x),y(x'))$. So, if $y(x) = y(x')$, then $x \sim_S x'$. $\square$

Lemma 11 shows that disc game embedding automatically reduces by outcome equivalency. All outcome equivalent types are mapped to the same disc game coordinate and each disc game coordinate specifies a unique equivalency class. The disc game coordinates index the equivalency classes. Therefore, given a distribution $\pi_X$ over traits $x$, the corresponding distribution $\pi_Y$ over the embedded traits, $y(X)$ automatically marginalizes over each equivalence class. 

We will see that the parameter dynamics depend only on the marginal distribution over each equivalency class, i,e.~the distribution of the disc game coordinates $y(X)$. Therefore, the parameter dynamics are invariant to changes in $\pi_X$ that preserve the marginal distribution over the classes, $\pi_Y$. Thus, the parameter dynamics are both independent of the conditional distribution within each equivalency class, and preserve those conditional distributions as the parameters evolve. It follows that, all of the information specifying the conditional distributions over each equivalency class is absorbed in the static component of the affine parameterization, i.e. the offset $z_0$. 

To make this notion precise, notice that, any two distributions with logarithms in the same affine subspace must have a log-ratio in the range of $F$. Since the range of $F$ is constant within each equivalency class, they must also share the same conditional distributions over each equivalency class.

\begin{snugshade}
\noindent \textbf{Lemma 12: [Invariant Affine Subspaces]} Two initial distributions can only have logarithms in the same affine subspace if they share the same support, and conditional distributions in each outcome equivalency class.
\end{snugshade}

So, each invariant family of affine log-densities corresponds to the choice of a support $S$, and an offset $z_0$ in the nullspace of $F$ for a reference measure supported on $S$. The choice of the offset fixes a unique steady state distribution within the class, when it exists. It also fixes the conditional distribution over each outcome equivalency class. These conditionals do not need a dynamical degree of freedom since they have no influence on, and are preserved by, the dynamic.

Finally, consider the nonlinear relationship between the original space of possible densities, and the set of possible parameters. The space of densities with supports contained in $S$ is an infinite-dimensional simplex. The set of densities with support equal to $S$ lies on the interior of that simplex. 
If the parameters are finite, then the corresponding density is nonzero everywhere in $S$. Therefore, the set of densities with finite parameters, $\{\pi(x;\theta) \mid \|\theta\| < \infty\}$, is contained in the interior of the set of densities. Accordingly, we will call all finite parameter vectors interior. The boundary of the parameter space corresponds to limiting densities produced by a sequence of diverging parameters. These include densities with smaller support, and densities that diverge on sets of measure zero. This language extends the standard terminology on the simplex (c.f.~\cite{akin1984evolutionary}). 

\begin{snugshade}
    \noindent \textbf{Definition: [Interior]} A parameter vector $\theta$ is \textit{interior} if $\|\theta\| < \infty$. 
\end{snugshade}

\subsubsection{Hamiltonian Parameter Dynamics}

Suppose that $(\Omega, f)$ is a functional form game, that $\pi(x,0)$ is a density over $\Omega$ with support $S$ and that $f$ is finite rank on $S$. Finally, suppose that $\pi(x,t)$ is subject to the continuous replicator Equation \eqref{eqn: replicator standardized}. Parameterize $\pi(x,t)$ according to the affine parameterization \eqref{eqn: disc-game parameterization} with basis functions set to the disc-game embedding functions given a reference measure $\nu$ supported on $S$. Then, the parameter dynamics must solve the closure equation \eqref{eqn: closure for affine parameterization}. Solving the closure equation for $\frac{\mathrm d}{\mathrm d t} \theta(t)$ produces a system of ODE's that can be elegantly expressed in terms of either the best response direction to the population centroid in disc-game space, or, the total population size as a function of the parameters.

\vspace{0.6 in}

\begin{snugshade}
    \noindent \textbf{Theorem 5: [Replicator Parameter Dynamics]} Given $(\Omega, f)$, a support set $S$, a reference measure $\nu$ supported on $S$, and an initial density $\pi(\cdot,0)$ supported on $S$, define:
    \vspace{-0.05 in}
    \begin{enumerate}
        \item The rank $r = \text{rank}(F) < \infty$ given the operator $F$ over support $S$. 
        \vspace{-0.05 in}
        \item The embedding functions $y(x) = [y_1(x),y_2(x), \hdots, y_{r}(x)]$ where $y(\cdot)$ is a column-vector valued function, with pairwise block-indexing $y^{(k)}(x) = [y^{(k)}_1(x),y^{(k)}_2(x)] = [y_{2k-1}(x),y_{k}(x)]$, and their corresponding parameters $\theta = [\theta_1,\theta_2,\hdots,\theta_r]$.
        \vspace{-0.05 in}
        \item The exponential family $\pi(x;\theta) = \exp(u(x;\theta))$, $u(x;\theta) = \log(\pi(x,0)) + \theta \cdot y(x)$.
    \end{enumerate}

    \noindent Then, the replicator equation admits an exact $r$-dimensional closure. The corresponding parameter dynamics can be expressed using:
    \vspace{-0.01 in}
    \begin{enumerate}
        \item The rotation matrices $R$, the $2 \times 2$, $90^{\circ}$ rotation matrix as defined in \eqref{eqn: real Schur form}, and $U$, the $r \times r$, block diagonal with diagonal blocks equal to $R$ as defined in \eqref{eqn: U matrix}.
        \vspace{-0.05 in}
        \item The population centroid in disc-game space:
        \vspace{-0.1 in}
        \begin{equation}
            \bar{y}[\pi] = \mathbb{E}_{X \sim \pi}[y(X)]
        \end{equation}

        \vspace{-0.1 in}
        \item The self-play vector field which specifies the optimal training direction in response to the opponent $y$:
        \begin{equation}
            v(y) = U y
        \end{equation}

        \vspace{-0.1 in}
        \item The total population size as a function of the parameters:
        \begin{equation} \label{eqn: total pop function}
            P(\theta) = \int_{x \in \Omega} \pi(x;\theta) dx.
        \end{equation}
    \end{enumerate}

    Then, the parameters obey the system of ODE's:
    \begin{equation} \label{eqn: replicator parameter dynamics}
        \frac{\mathrm d}{\mathrm d t} \theta = v(\bar{y}[\pi(\cdot;\theta)]) = U \nabla_{\theta} P(\theta) \xrightarrow{\text{blockwise, } k \leq r/2} \begin{cases} & \frac{\mathrm d}{\mathrm d t} \theta^{(k)}_1 = \hspace{0.11 in} \bar{y}^{(k)}_2[\pi(\cdot;\theta)] = \hspace{0.11 in} 
 \partial_{\theta^{(k)}_2} P(\theta) \\
            & \frac{\mathrm d}{\mathrm d t} \theta^{(k)}_2 = -\bar{y}^{(k)}_1[\pi(\cdot;\theta)] = -\partial_{\theta^{(k)}_1} P(\theta) \end{cases} 
    \end{equation}

\end{snugshade}

\noindent \textbf{Proof Outline:} The first equality in Equation \eqref{eqn: replicator parameter dynamics} is established by substituting the affine parameterization of $u$ into the replicator equation, and simultaneously expanding $f$ into a linear combination of outer products of the embedding functions. Since each outer product is separable, and expectations are linear, the expectation $\mathbb{E}_{X \sim \pi}[f(x,X)]$ decomposes into the sum $y(x)^{\intercal} U \bar{y}[\pi]$. Then, since the embedding functions are independent, the closure Equation \eqref{eqn: closure for affine parameterization} is solved by matching the coefficients of the expansion of each side on the basis formed by the embedding functions.

The second form involving the gradient of the total population size with respect to the parameters follows from the observation that the total population size, as a function of the parameters, is the Laplace transform of the initial population distribution in the disc-game space The Laplace transform of a distribution is its moment-generating function:
\begin{equation}
    P(\theta) = \int_{x \in \Omega} \pi(x;\theta) dx = \int_{x \in \Omega} \pi(x,0) e^{\theta \cdot y(x)} dx = \mathbb{E}_{X \sim \pi_X(\cdot,0)}[e^{\theta \cdot y(X)}] = \mathbb{E}_{Y \sim \pi_Y(\cdot,0)}[e^{\theta \cdot Y}].
\end{equation}

Derivatives of moment-generating functions recover moments, i.e.~ expectations, against the original density.   Straightforward calculus establishes that:
\begin{equation} \label{eqn: centroid from Hamiltonian}
    \bar{y}[\pi(\cdot;\theta)] = \mathbb{E}_{X \sim \pi(\cdot;\theta)}[y(X)] = \nabla_{\theta} P(\theta)
\end{equation}
thus establishing the equivalence of the two forms in Equation \eqref{eqn: replicator parameter dynamics}. Appendix Section \ref{app: proof of parameter dynamics} provides a detailed proof. $\square$

\vspace{0.05 in}

\begin{figure}[t]
    \centering
    \includegraphics[trim = 150 50 150 50, clip, width = \textwidth]{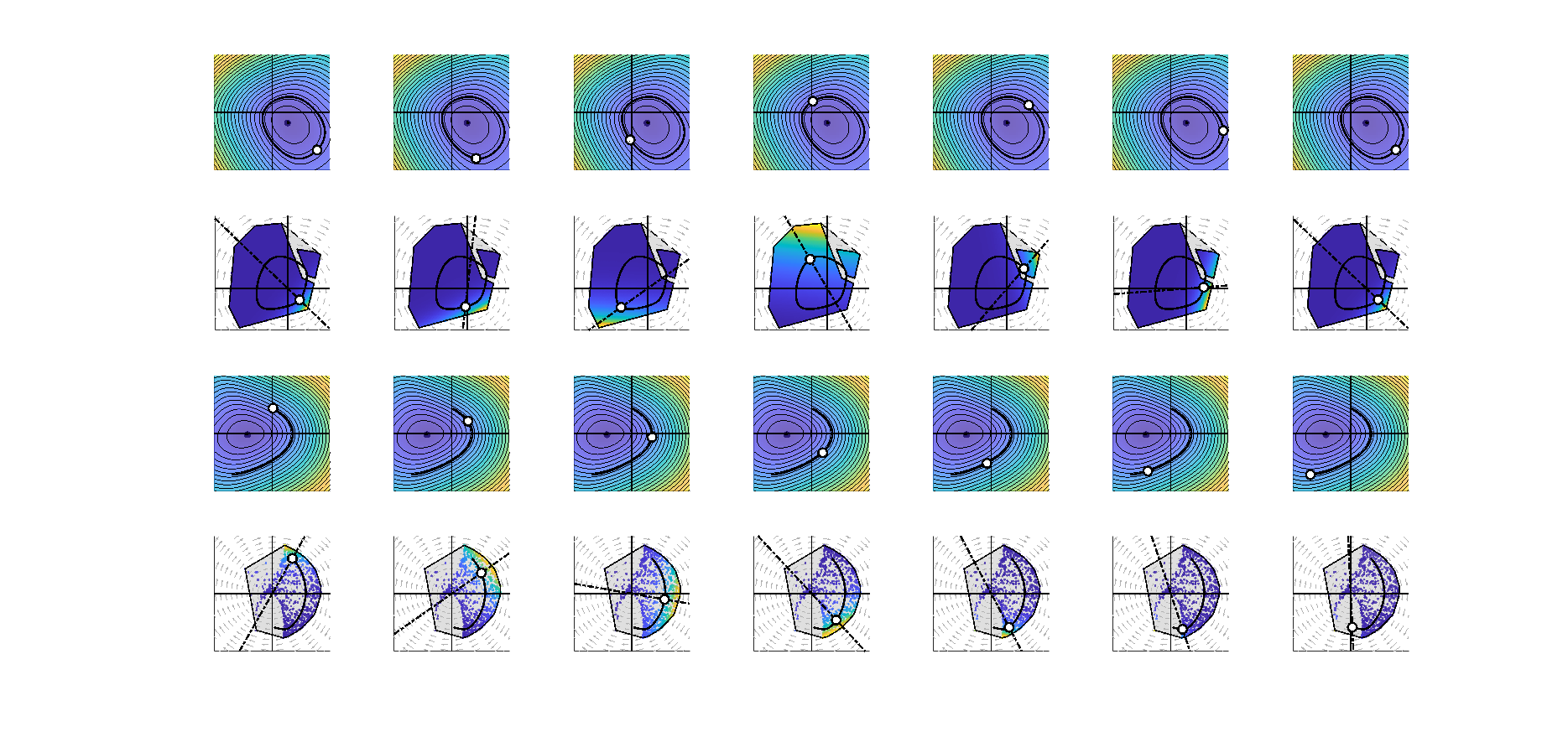}
    \caption{Autonomous centroid and Hamiltonian parameter dynamics for two example games. The top two rows correspond to the example game illustrated in Figures \ref{fig: Embedding Visualization}, \ref{fig: Exponential Tilting}, and \ref{fig: Hamiltonian and Adaptive}. The bottom two rows correspond to the IPD example illustrated in Figure \ref{fig: IPD Disc Game Embedding}. The first and third rows show orbits in the parameter space. The second and fourth rows show the corresponding sequence of population distributions in the embedding space, formed by exponentially tilting the base measure on $\Psi$. Each column represents the population at a different, equally spaced time points. The heatmap in the first and third rows show the total population function $P(\theta)$. Solutions to the replicator equation orbit on level sets of the total population. A sample orbit is shown. The white circle represents the parameters at different time points. In the second and fourth rows, the axes correspond to disc game coordinates, the colored region is $\Psi$, and the grey dashed region is its convex hull. The grey vector field is the optimal local training vector field $v$. Color represents the density of $y(X)$ given $X \sim \pi(\cdot,t)$.  The white circle represents the population centroid. The dashed line passing through the centroid and the origin separates types whose frequency is increasing (on the downstream side of the dashed line), and whose frequency is decreasing (on the upstream side of the dashed line).}
    \label{fig: Exponential Tilting}
\end{figure}

Equation \eqref{eqn: replicator parameter dynamics} is disarmingly simple. No matter the specific game or trait space, the parameters $\theta$ simply evolve in the best response direction to the centroid $\bar{y}$. Equivalently, the parameter dynamics are entirely defined by their constant of motion, the scalar-valued function $P(\theta)$ which returns the total population size. In other words, Equation \eqref{eqn: replicator parameter dynamics} is a normal form dynamic. It reduces the space of possible dynamics by grouping together functional form games, and invariant subspaces, that admit equivalent total population functions $P$.

The simplicity of \eqref{eqn: replicator parameter dynamics} highlights characteristics of the replicator dynamic that are shared by all choices of $f$ and $\Omega$, so allows highly generic analysis. It also obscures behaviors that depend on the specific choices of $f$ and $\Omega$. While Equation \eqref{eqn: replicator parameter dynamics} may appear entirely independent and invariant to game specific details, it is not independent of $f$ or $\Omega$. Rather, it expresses the game specific dependences in the structure of the function $P(\cdot)$, or, in the parametric relation between $\theta$ and $\bar{y}$. This approach is useful as it provides alternative interpretations of the dynamic that highlight characteristic behaviors. For example, we will use Equation \eqref{eqn: replicator parameter dynamics} to show that $\bar{y}(t)$ always obeys an adaptive dynamics equation, no matter the choice of $f$ and $\Omega$, and without any approximation.

To start, we offer some intuition.
The first form of the parameter dynamics specified by \eqref{eqn: replicator parameter dynamics} shows that the parameters, $\theta$ always move in the best response direction to the embedded population centroid. In a sense, this equation imitates simultaneous gradient ascent in the disc-game space (see Appendix Section \ref{app: explicit solutions}). 
In simultaneous gradient ascent, individual agents shift in the best response direction to the centroid \cite{balduzzi2018mechanics}. In the replicator dynamic, the \textit{parameters} shift in the best response direction to the centroid.

Like simultaneous gradient ascent, Equation \eqref{eqn: replicator parameter dynamics} encodes feedbacks that produce oscillation. Increasing $\theta^{(k)}_1$ multiplies the population distribution in disc-game space by the exponential function $\exp(\theta^{(k)}_1 y^{(k)}_1)$, so biases the distribution to move more mass to larger $y^{(k)}_1$. Types with large $y^{(k)}_1$ are best countered by types with large, negative, $y^{(k)}_2$. So, the best response direction to a type with large $y^{(k)}_1$ points in a direction that rapidly decreases $y^{(k)}_2$.
Then, following Equation \eqref{eqn: replicator parameter dynamics}, increasing $\theta^{(k)}_1$ leads to decreasing $\theta^{(k)}_2$. Repeating the same argument shows that decreasing $\theta^{(k)}_2$ decreases $\theta^{(k)}_1$. This establishes a self-regulating feedback loop. Increasing $\theta^{(k)}_1$ decreases $\theta^{(k)}_2$, which decreases $\theta^{(k)}_1$, which increases $\theta^{(k)}_2$, and so on. The result is a nonlinear oscillator equation, with restoring feedbacks mediated by the relationship between the parameters $\theta$, ensuing distribution over disc-game space, $\pi_Y(\cdot;\theta)$, and resulting centroid $\bar{y}[\pi(\cdot;\theta)] = \mathbb{E}_{Y \sim \pi_Y(\cdot;\theta)}[Y]$.

A word on coupling. The parameter dynamics nearly decouple by block, since the update to $\theta^{(k)}$ only depends on the $k^{th}$ pair of coordinates $\bar{y}$. The blocks don't fully decouple since $\bar{y}^{(k)}$ is a function of the distribution $\pi(x;\theta)$, which depends on all $\theta$ for each $x$. When the reference distribution used in the disc-game embedding is chosen appropriately, then linearizing Equation \eqref{eqn: replicator parameter dynamics} about the reference distribution decouples the blocks. In this case, the cross-block coupling arises in third and higher-order terms of the Taylor expansion of the exponential function used to convert from the log-density $u$ back to the density $\pi$. 

Oscillators are frequently Hamiltonian. Simultaneous gradient ascent in response to a zero-sum normal form game on finite strategy spaces is always a Hamiltonian dynamic \cite{balduzzi2018mechanics}, as is the replicator equation for generic bimatrix games on finite strategy spaces \cite{akin1984evolutionary,hofbauer1996hamiltonian}. The second equality in Equation \eqref{eqn: replicator parameter dynamics} establishes that the replicator parameter dynamics are also Hamiltonian.  

Hamiltonian dynamics were introduced to study mechanical systems, where the state of a system can be specified by a list of position and momenta coordinates, $x,p$, whose dynamics conserve a Hamiltonian function $H(x,p)$ (c.f.~ \cite{hofbauer1996hamiltonian,perelomov1990integrable}). 
At its simplest, a Hamiltonian dynamic sets $\frac{\mathrm d}{\mathrm d t} x = \partial_p H(x,p)$ and $\frac{\mathrm d}{\mathrm d t} p = - \partial_x H(x,p)$ \cite{hamilton1833general,mackay2020survey}. Equation \eqref{eqn: replicator parameter dynamics} assumes this form with Hamiltonian equal to the total population size, $P$, ``position" variables, $\theta^{(k)}_1$, and ``momentum" variables, $\theta^{(k)}_2$. In particular, Equation \eqref{eqn: replicator parameter dynamics} is block Hamiltonian, with separate blocks corresponding to separate disc-games. 

\begin{figure}[t]
    \centering
    \includegraphics[trim = 100 0 100 0, clip, width = \textwidth]{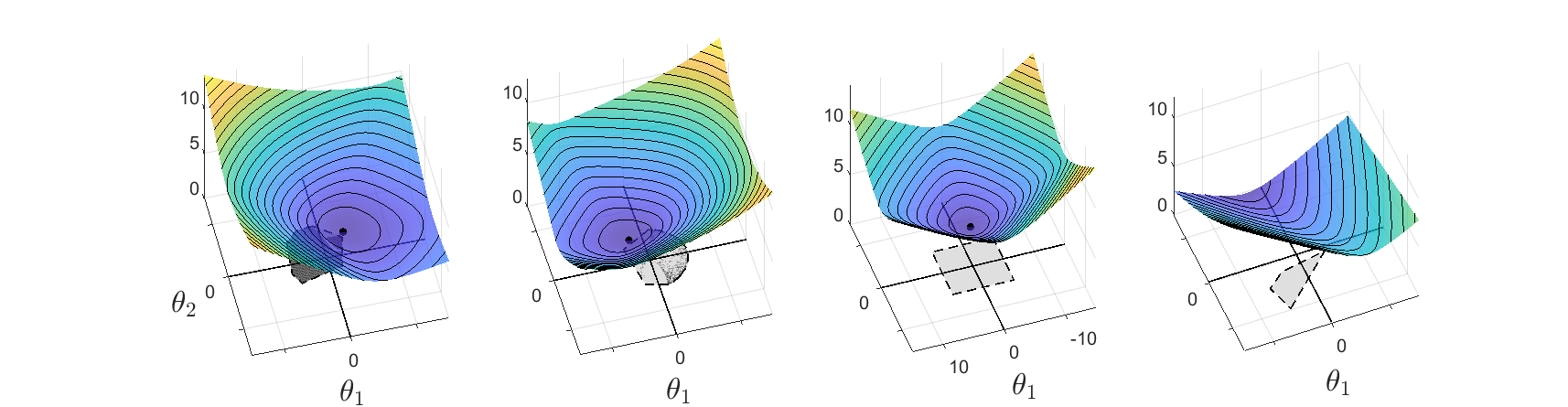}
    \caption{Four sample Hamiltonians, $P(\theta)$. The convex hull of the range of the embedding map, $\Psi$, is shown in light grey with dashed boundaries. The base measure $\nu$ is represented by dark grey scatter points in the first two panels. The last two use a uniform measure. The vertical axis represents the Hamiltonian on a log base 10 scale. Notice that the Hamiltonians are all convex functions. The Hamiltonians grow slowest in the direction along which the boundary of $\Psi$ is closest to the origin, and grow exponentially (linearly in the log scale) away from their minima. The first example matches the examples provided in Figures \ref{fig: Embedding Visualization}, \ref{fig: Exponential Tilting}, \ref{fig: Hamiltonian and Adaptive}, and \ref{fig: Adaptive}. The second matches the IPD embedding illustrated in Figure \ref{fig: IPD Disc Game Embedding}.  The third example shows that the Hamiltonian develops approximately polygonal level sets away from its minimizer when $\Psi$ is polygonal. The fourth example uses a $\Psi$ whose convex hull does not enclose the origin. In this case, the Hamiltonian does not admit an interior minimizer, so sample trajectories may escape to infinity (see Figure \ref{fig: Hamiltonian and Adaptive}). In this case there exist dominated types that will be driven to extinction.}
    \label{fig: Hamiltonians}
\end{figure}

The fact that the parameter dynamics use the total population, $P(\cdot)$, as their Hamiltonian distinguishes our analysis from standard analyses. For example, Akin, Losert, and Hofbauer demonstrated that the replicator dynamic is Hamiltonian given a zero-sum normal form game with finitely many strategies, with Hamiltonian equal to the Kullback-Liebler divergence from an interior equilibrium \cite{akin1984evolutionary,hofbauer1996hamiltonian}. See \cite{boone2019darwin} for an extension to polymatrix games and \cite{bailey2019multi} for an extension to generic follow-the-regularized-leader dynamics. While more complex constants of motion are a powerful analytic tools (c.f.~\cite{mertikopoulos2018cycles}), the simplicity of $P(\cdot)$ is appealing. The total population size has immediate biological relevance, especially when payout is interpreted as per-capita growth rate, and is the most obvious quantity conserved by the replicator dynamic. It is surprising that, the total population is, as a function of the parameters, not only conserved, but also encodes all the game information needed to define the parameter dynamics. The total population size is also analytically convenient since, like the log partition function of any exponential family, it carries rich information about the distribution in its derivatives. Figure \ref{fig: Hamiltonians} presents four sample Hamiltonians.
  

\subsubsection{Dynamical Characterization} \label{sec: analysis}


Here we show that the Hamiltonian form for the parameter dynamics admits strong global analysis. 
Each fact can be easily established using the disc game representation. For all proofs, see Appendix Section \ref{app: parameter dynamics proofs}.

\begin{snugshade}
    \noindent \textbf{Lemma 13: [Properties of the Hamiltonian]} The Hamiltonian function $P(\theta)$ defined in Equation \eqref{eqn: total pop function} satisfies the following properties:
    \begin{enumerate}
        \item The mixed partial derivatives of the Hamiltonian correspond to raw moments of the embedded population distribution. Let $p(\cdot;\theta) = \pi(\cdot;\theta)/P(\theta)$. Let $\alpha = [\alpha_1,\alpha_2,\hdots \alpha_r] \in \mathbb{Z}^r$ be a multi-index. Let $|\alpha| = \sum_j \alpha_j$ and $x^\alpha = \prod_{j} x_j^{\alpha_j}$. Then:
    \begin{equation} \label{eqn: derivatives and moments}
      \partial_{\theta}^{\alpha} P(\theta) = P(\theta) \mathbb{E}_{X \sim p(\cdot;\theta)}[y(X)^{\alpha}] = P(\theta)  \mathbb{E}_{Y \sim p_Y(\cdot;\theta)}[Y^{\alpha}].
    \end{equation}
    In particular, the Hessian of $P(\theta)$ equals both the Gram matrix for the embedding functions $y(\cdot)$ with respect to the inner product $\langle \cdot,\cdot \rangle_{\pi(\cdot;\theta)}$, and, equivalently, the covariance in the embedding of a randomly selected agent:
    \begin{equation} \label{eqn: Hessian is Gram is Covariance}
        [\partial_{\theta}^2 P(\theta)]_{ij} = \langle y_i(\cdot),y_j(\cdot) \rangle_{\pi_Y(\cdot,;\theta)} = P(\theta) \left(\text{Cov}_{Y \sim p_Y(\cdot;\theta)}[Y_i,Y_j] + \mathbb{E}_{Y \sim p_Y(\cdot;\theta)}(\theta)[Y_i] \mathbb{E}_{Y \sim p_Y(\cdot;\theta)}[Y_j]\right).
    \end{equation} 
    If we adopt a constant interaction rate rather than a linear interaction rate model, then the Hamiltonian is $\log(P(\theta))$ and its partials are the centered moments of $Y \sim p_Y$ without any extra scaling against the population size.
    \item The Hamiltonian is strictly convex. 
    \item The Hamiltonian either admits a unique global minimizer at some finite $\theta_*$, or no such minimizer exists. If such a minimizer exists, then $P(\theta)$ is radially unbounded.

    \end{enumerate}
\end{snugshade}

\noindent \textbf{Proof Outline: } The relation between partial derivatives of the Hamiltonian and the moments follows by directly differentiating Equation \eqref{eqn: total pop function} with respect to $\theta$. The Hamiltonian is convex since its Hessian at $\theta$ is a covariance matrix. All covariance matrices are positive semi-definite, so the Hessian is positive definite everywhere. Strict convexity follows from the orthogonality of the embedding functions under an appropriate change of reference measure $\nu$. $\square$

\vspace{0.05 in}
\noindent \textbf{Remark: } The convexity of the $P(\theta)$ is useful since Hamiltonian dynamics mix on level sets of their Hamiltonian. Since the level sets of a convex function are boundaries of convex regions, solutions to the parameter dynamics mix on the boundaries of convex regions. If a level set is bounded, then the parameters remain bounded for all time. If a level set is unbounded, then the parameters may escape to infinity (as observed in the transitive and quadratic special cases). 

When $P(\theta)$ admits a unique global minimizer, it is radially unbounded, so its level sets are bounded. Then, the parameters mix on the boundaries of compact sets. These observations will be used to show that all interior equilibria for the parameter dynamics are isolated, Lyapunov stable centers, so the parameters behave as coupled oscillators. More precisely, Equation \eqref{eqn: derivatives and moments} will be used to show that, up to linearization, parameters near an interior equilibria satisfy a harmonic oscillator equation while higher-order terms couple the oscillators.  $\square$

\pagebreak

\begin{snugshade}
    \noindent \textbf{Corollary 5.1: [Dynamics]} The parameters $\theta(t)$ satisfy the following dynamical (non-steady state) properties:
    \begin{enumerate}
        \item The total population size $P(\theta(t))$ is conserved, $P(\theta(t)) = P(\theta(0)) = P$.
        
        \item If the interaction rate is time-autonomous, $\rho(P,t) = \rho(P)$, then changes of interaction rate can be encoded via a transformation of the Hamiltonian. Let $R(P) = \int \rho(P)/P dP$. Then, for any time-autonomous interaction rate:
        \begin{equation}
            \frac{\mathrm d}{\mathrm d t} \theta(t) = U \nabla_{\theta} R(P(\theta)).
        \end{equation}
        In particular, if the interaction rate is constant, then $\frac{\mathrm d}{\mathrm d t} \theta(t) = \rho U \nabla_{\theta} \log(P(\theta)).$

        \item The parameter dynamics are incompressible, i.e.~volume-preserving. Let $\Theta(0)$ denote a closed set of initial parameters $\theta$, and let $V(0)$ equal the volume enclosed by $\Theta(0)$. Let $\Theta(t)$ denote the set of parameters at time $t$ after applying the replicator dynamic. Let $V(t)$ represent volume enclosed by $\Theta(t)$. Then, $V(t) = V(0)$. 

        \item If $P(\cdot)$ admits a unique global minimizer, and $\theta(0)$ is finite, then $\theta(t)$ remains finite for all times $t \geq 0$. That is, the interior is invariant under the replicator dynamic.

        \item If $P(\cdot)$ admits a unique global minimizer, then the parameter dynamics are Poincar\'e recurrent, and, if $r \leq 2$, then the dynamics are periodic.

        \end{enumerate}
\end{snugshade}

\noindent \textbf{Proof Outline:} All Hamiltonian dynamics preserve their Hamiltonian. In this case, since the Hamiltonian is the total population size, the total population size is conserved. The change of interaction rate formula is established via the chain rule. Volume preservation is established by showing that the divergence of the right-hand-side of Equation \eqref{eqn: Hamiltonian parameter dynamics} is zero at all $\theta$. The interior is time-invariant under the replicator dynamic when $P(\theta)$ admits a global minimizer since, when $P(\theta)$ admits a global minimizer it is radially unbounded, so has bounded level sets. If the level sets are bounded, then trajectories are bounded and volume preserving, so Poincar\'e recurrent \cite{barreira2006poincare,poincare1889probleme}. If $r = 2$, then, the dynamics are periodic since the level sets of the Hamiltonian are closed, nested, orbits that fill a plane.  $\square$

\vspace{0.05 in}

\noindent \textbf{Remark: } 
Fact 3 is a statement of Liouville's theorem for Hamiltonian systems \cite{liouville1838note}. A similar result is established in \cite{flokas2020no} for finite strategy sets (discrete trait space). The converse - the replicator is volume preserving if and only if the game is zero-sum - is established in \cite{boone2019darwin}. For related discussions on constants of motion and volume preservation in evolutionary game theory see \cite{hofbauer1998evolutionary,legacci2024geometric,weibull1997evolutionary}.

Volume preservation is interesting since it bans the existence of any bounded attractor \cite{boone2019darwin}. A bounded attractor is a set of parameters that can be enclosed in a bounded basin of attraction. A basin of attraction is an open set containing the attractor such that any trajectory initialized in the basin converges onto the attractor. Since the dynamic is volume preserving, no such basin can exist. Otherwise, any set that covers the attractor that is contained inside the basin would need to preserve volume while contracting.

It follows that there are no isolated, attracting equilibria of the parameter dynamics with finite parameter values. Running the same argument in reverse time implies that there are also no isolated, interior, unstable equilibria. In other words, there are no isolated sources or sinks with finite parameter values.

Thus, all isolated attractors in the space of densities, in both forward and reverse time, must exist as limiting densities that can be approached via a diverging sequence of parameters. The closure of the set of densities with finite parameters includes both densities with smaller support than the initial support $S$, and singular densities such as delta distributions (monomorphic populations). Trajectories that diverge along an unbounded level set can converge to attracting singular densities. For example, if $f$ is transitive with a rating function $r$ that admits a unique global maximizer, then the delta distribution at that maximizer is globally attracting.  For the stability analysis of monomorphic distributions, see \cite{cressman2004dynamic,cressman2006stability,oechssler2001evolutionary,oechssler2002dynamic}. More generally, diverging parameter trajectories can represent a process of iterated dominance, in which dominated types are sequentially eliminated from the support \cite{fudenberg1998learning}. $\square$

\vspace{0.05 in}
\noindent \textbf{Remark: } Facts 3 and 4 also establish Fact 5; when the Hamiltonian admits a global minimizer, the corresponding parameter dynamics are Poincar\'e recurrent. A dynamical system is Poincar\'e recurrent if solution trajectories return arbitrarily close to almost all initial condition infinitely often. Equivalently, a solution trajectory initialized in any open set will return to that set (recur) in finite time \cite{poincare1889probleme,caratheodory1919wiederkehrsatz}. Thus, Poincar\'e recurrent systems exhibit irregular oscillations. If $r = 2$ then those oscillations are periodic \cite{boone2019darwin}. If $r > 2$, then the recurrent cycles may mix chaotically  \cite{mertikopoulos2018cycles,sato2002chaos}.  $\square$

\vspace{0.05 in}
 Let's return, for a moment, to paradigm comparison. At the outset, we argued that the disc-game would provide a more accurate intuition for learning in zero-sum games than either optimization, or equilibria. Fact 5 offers strong support for this argument. 

First, it ensures that, unless initialized at a steady state, almost all initial conditions lead to solutions that wander forever without converging. Second, it ensures that motion along solution trajectories fail to optimize any continuous function, indeed, fail basic notions of directed adaptation. 

Often learning or selective dynamics are attributed direction by identifying some property of the population that changes consistently over time. For example, one might argue that adapting populations will exhibit increasing specialization, complexity, or rational behavior (c.f. \cite{mailath1998people}). To formalize this idea, let $h$ represent a continuous functional that accepts distributions and returns a real number. If $h[\pi(\cdot,t)]$ is either monotonically increasing or decreasing in time, then the dynamics are plainly directed by $h$. More weakly, $h$ can lend a sense of directed adaptation if $\inf_{s > t}\{h[\pi(\cdot,s)]\}$ is increasing, or $\sup_{s > t}\{h[\pi(\cdot,s)]\}$ is decreasing, along some sequence of times $t$. In the first case, $h(t) = h[\pi(\cdot,t)]$ does not increase monotonically, but does increase in the sense that, there exists a sequence of times, $\{t_j\}_{j=1}^{\hdots}$ chosen so that $h[\pi(\cdot,t_j)] \leq h[\pi(\cdot,s)]$ for all $s > t_j$, and such that $h[\pi(\cdot,t_j)]$ is monotonically increasing. In the second case the same statements hold, only for a sequence of decreasing upper bounds. 

When the dynamics are Poincar\'e recurrent, no such continuous function can exist, since the distribution will return arbitrarily close to its initial condition in finite time. Thus, $h[\pi(\cdot,t)]$ will return arbitrarily close to $h[\pi(\cdot,0)]$ infinitely often. It follows that, when an interior equilibrium exists, the replicator dynamics do not optimize \textit{any} continuous function of the population distribution. 

\begin{snugshade}
\noindent \textbf{Lemma 14: [No Objective]} If $\pi(\cdot,t)$ is Poincar\'e recurrent, then there is no continuous, non-constant functional $h$ such that:
\begin{enumerate}
\item $h(t) = h[\pi(\cdot,t)]$ is monotonically increasing or decreasing,
\vspace{-0.02 in}
\item $\inf_{s > t}\{h(s)\}$ is monotonically increasing, or $\sup_{s > t}\{h(s)\}$ is monotonically decreasing.
\end{enumerate}
\end{snugshade}

\noindent \textbf{Remark:} The restriction to continuous functionals in Lemma 14 is necessary if the dynamics are recurrent but aperiodic. For example, if we do not require continuity, then we could define a monotonically increasing functional by choosing a fixed initial distribution for every solution path, then setting $h[\pi]$ equal to the time it would require for a distribution initialized at the corresponding $\pi_0$ to reach $\pi$. Nevertheless, the restriction to continuous functions does not discard much, since any function that is monotonic along recurrent trajectories must span its full range of possible values on every open set containing any point on the trajectory. It is hard to imagine such a function offering a substantive notion of direction. To adopt such a function would  assign meaning to arbitrarily small perturbations of the distribution that preserve its support. Such sensitivity should disqualify any associated interpretation. $\square$
\vspace{0.05 in}

These conclusions are, perhaps, surprising, given that the replicator dynamic can be derived from a multiplicative weight updating scheme, so corresponds to a continuous implementation of the canonical no-regret learning process \cite{legacci2024geometric}. However, the fact that the replicator equation fails to converge to all interior equilibria is actually a generic characteristic of no-regret learning dynamics shared by all follow-the-regularized-leader dynamics \cite{flokas2020no}. For similar proofs of recurrence in zero-sum games with fully mixed equilibria, and a more general proof for incompressible games, see \cite{boone2019darwin,mertikopoulos2018cycles,piliouras2014optimization}.

The fact that no-regret dynamics circulate about fully mixed equilibria in zero-sum games emphasizes a subtler distinction between the optimization paradigm and learning in multi-agent settings. In optimization problems, step-wise guarantees ensure that the optimizer moves towards its goal. Sufficiently strong step-wise guarantees ensure pointwise guarantees for the limit of the path. These relations are more than ancillary features; they are often the reason optimization problems can be solved iteratively.

The replicator dynamic satisfies both a step-wise and a path-wise rationality guarantee. It is ``step-wise" rational since the distribution changes in the self-play direction subject to a Shashahani, or Fisher information, metric \cite{mertikopoulos2018riemannian}. In this sense, it always adapts optimally in response to the current population. The replicator dynamic is also path-wise rational since it is a no-regret dynamic \cite{flokas2020no,legacci2024geometric}. No-regret is a path-wise guarantee. It ensures that a trajectory is, in the long run, rational, since the learner does not, in hindsight, regret following it. If the trajectory converges to a fixed distribution, then that distribution must satisfy a static, pointwise, rationality criteria. Namely, there cannot exist an alternate distribution with positive expected payout against the fixed distribution. Otherwise, the learner would regret settling there. Thus, convergence converts a path-wise guarantee into a pointwise guarantee at the end of the path. In contrast, when a no-regret learner fails to converge, then the guarantee remains path-wise. There is no alternate fixed distribution that would have a positive expected payout against a distribution selected at random from the \textit{entire} trajectory. 

In this case, step-wise, and path-wise, rationality, offer no guarantee that any distribution reached by the process is itself rational in a static, pointwise sense. Nor is any distribution reached more rational than those visited before. Rather, when there exists an interior equilibria, no type is dominated, so every type can be justified as an exploiter of some other type \cite{akin1980domination}. More strongly, every distribution can be justified by the path it lies on. It is an optimal response to some other distribution (its immediate antecedent) in a set of distributions that is collectively not exploitable. Yet, it is not, individually, better justified than any other distribution on the path. Learning does not order specific distributions as more or less rational since the trajectory recurs. The dynamic will return arbitrarily close to every distribution it visits infinitely often, so every distribution visited can be viewed as both a predecessor, and a successor, to every other distribution visited. There exist rational learning trajectories moving in either direction between any pair of distributions sharing a path. So, rather than progressing towards an increasingly optimal distribution, the dynamic orbits ambivalently among a set of distributions that are collectively unexploitable. The process could be said to be constantly learning but to never learn, since the sequence of distributions visited do not acquire any property that distinguishes progress along the trajectory. The resulting solution concept, then, is the entire orbit \cite{bailey2019multi}. Since almost all initial conditions produce a recurrent orbit, the resulting solution concept should consist of the foliation of the space, that is, the collection of all orbits, not a collection of equilibria. This is categorically \textit{not} the intuition conveyed by optimization. It is the intuition suggested by the disc-game, whose essence suggests circulation about a center. 

These observations also impact the classification and interpretation of equilibria. They are not characteristic solutions since they are neither attracting nor optimal. Equilibria are special only in that they are balance points where the dynamic rests. That said, the equilibria still provide a useful window into the shape of orbits, since each equilibria offers the simplest possible orbit - a point - and, linearization about the equilibria characterizes orbits near that point. Using Lemmas 12 and 13, we will show that every interior equilibria corresponds to a center, all the other solution trajectories orbit that center on compact, concentric shells, and, near the equilibria, the replicator dynamic behaves like a system of independent simple harmonic oscillators.


\begin{snugshade}
\noindent \textbf{Corollary 5.2: [Steady-States]} Let $\pi_*$ denote a steady-state distribution of the continuous replicator dynamic and let $\theta_*$ denote an equilibrium of the corresponding parameter dynamics. A distribution $\pi(\cdot;\theta)$ with finite $\theta$ is a steady-state $\pi_*$ if and only if $\theta$ is an equilibrium $\theta_*$.  The equilibria obey the following properties:
\begin{enumerate}
    \item All interior equilibria $\theta_*$ are critical points of the Hamiltonian, $\nabla_{\theta} P(\theta)|_{\theta = \theta_*} = 0$. 
    \item The parameter dynamics either admit no interior equilibrium, or admits a unique, isolated, interior equilibrium corresponding to the global minimizer of the Hamiltonian.
    \item If an interior equilibrium exists, then it is a globally Lyapunov stable center for the parameter dynamics.
    \item Up to linearization, the parameter dynamics about any interior equilibrium are topologically equivalent to a collection of $r/2$ independent harmonic oscillators with frequencies $\{P \omega_k\}_{k=1}^{r/2}$ where $\omega_k$ equals the magnitude of the $2k-1^{st}$ eigenvalue of $F_{p_*}$, for $p_* = \pi(\cdot;\theta_*)/P(\theta_*)$. 
\end{enumerate}
\end{snugshade}

\noindent \textbf{Proof Outline: }  The embedding functions are linearly independent so $\frac{\mathrm d}{\mathrm d t} \pi(x,t) = 0$ if and only if $\frac{\mathrm d}{\mathrm d t} \theta(t) = 0$, and $\frac{\mathrm d}{\mathrm d t} \theta(t) = 0$ if and only if $\nabla_{\theta} P(\theta) = 0$. Since the Hamiltonian is strictly convex, all critical points correspond to minimizers, and the Hamiltonian either admits a unique global minimizer or no critical point exists. If an interior equilibrium exists, then $P(\theta)$ admits a finite global minimizer so is radially unbounded. So, any perturbation off the equilibrium produces a trajectory that mixes on a the boundary of a compact neighborhood containing the equilibrium. Thus, the equilibrium is neutrally stable, i.e. is a center for the dynamic.  Linearization is discussed in a subsequent remark. $\square$

\vspace{0.05 in}

\noindent \textbf{Remark: } When $r = 2$, minima of $P(\theta)$ correspond to centers, since the level sets of $P(\theta)$ are nested circuits in $\mathbb{R}^2$ enclosing the equilibrium. Rotating the gradient vector field about the minimizer by ninety-degrees produces a circulating vector field that circulates in the direction of the applied rotation around these loops. $\square$

\vspace{0.05 in}

\noindent \textbf{Remark: } For general $r$, we linearize to investigate the parameter dynamics near an equilibrium. Suppose that $\theta_*$ is a interior equilibrium for the parameter dynamics. The Jacobian of the right-hand side of Equation \eqref{eqn: Hamiltonian parameter dynamics} at $\theta_*$ equals $P(\theta_*)$ times the row-rotated Hessian of the total population function, or, equivalently, the row-rotated covariance in $y(X)$ when $X \sim p_* = p(\cdot;\theta_*)$:
    \begin{equation} \label{eqn: jacobian is rotated covariance}
        J(\theta)  = P(\theta_*) U \text{Cov}_{Y \sim p_Y(\cdot;\theta)}[Y].
    \end{equation}
    Therefore, the linearized dynamic is:
    \begin{equation} \label{eqn: linearized dynamic}
        \frac{\mathrm d}{\mathrm d t} (\theta(t) - \theta_*) =  J(\theta_*) (\theta(t) - \theta_*) + \mathcal{O}(\| \theta(t) - \theta_*\|^2) \simeq P(\theta_*)  U \text{Cov}_{Y \sim p_Y(\cdot;\theta_*)}[Y] (\theta(t) - \theta_*).
    \end{equation}

    By a standard result, the matrix $J$ is similar to a skew-symmetric matrix, so has purely imaginary eigenvalues \cite{arnold2012geometrical,hofbauer1996hamiltonian}. Here, we can adopt a more immediately contextual argument. The linearized dynamic \eqref{eqn: linearized dynamic} is equivalent, under a change of parameterization, to a decoupled system of simple harmonic oscillator equations with oscillation frequency determined by the eigenvalues of the operator $F_{p_*}[\cdot]$.

    Recall that the set of embedding functions used as a function basis for the range of $F$ depends on a choice of reference measure. As long as two reference measures share the same support, then the range of $F$ is fixed, so changing the reference measure changes the particular basis used to span the range of $F$, but does not change the range. Accordingly, any pair of affine parameterizations arising from different choices of reference measure correspond to a different choice of basis. Then, the parameters associated with two reference measures that share the same support, are related by an invertible linear mapping (see Section \ref{sec: representation}). It follows that, parameter dynamics associated with two different choices of reference measure are topologically equivalent provided the reference measures have the same support.

    If $\pi_*$ is an interior steady-state, then it has support $S$, as does the reference measure $\nu$. Since parameter dynamics whose reference measures share their support are topologically equivalent we are free to change the reference measure. Set $\nu = \pi_*$. Then, applying Equation \eqref{eqn: Hessian is Gram is Covariance}, and recalling that the eigenfunctions $y$ are orthogonal with respect to the reference measure yields:
    %
    %
    \begin{equation}
        J(\theta_*) = U D_{\omega} = P(\theta_*) W
    \end{equation}
    where $W$ is the matrix used in the real Schur form of $F_{p_*}$ (see Equation \eqref{eqn: real Schur form}), and $\omega_{k} = \|\lambda_{2k}(F_{p_*})\|$ equal the magnitudes of the eigenvalues of $F_{p_*}$.

    Therefore, when $\nu = \pi_*$, the linearized parameter dynamics decouple blockwise, and each block satisfies a harmonic oscillator equation with frequency $\omega$ . Let $\Delta \theta = \theta - \theta_*$. Rescaling time by $P/P(\theta_*)$ to recover the correct interaction rate for initial population $P$ yields:
    \begin{equation} \label{eqn: linearization is harmonic oscillator}
        \frac{\mathrm d}{\mathrm d t} \Delta \theta^{(k)}_1(t) \simeq P \omega_k \Delta \theta^{(k)}_2(t), \quad \frac{\mathrm d}{\mathrm d t} \theta^{(k)}_2(t) \simeq -P \omega_k \Delta 
 \theta^{(k)}_1(t) 
    \end{equation}

    Equation \eqref{eqn: linearization is harmonic oscillator} makes the analogy between the replicator dynamic and mechanical oscillators concrete. Up to linearization, the parameter dynamics near any interior equilibrium behave as decoupled harmonic oscillators, with one oscillator per pair of embedding coordinates. Collectively, solutions to the linearized equations orbit on $r/2$-dimensional tori. The solutions are periodic if the eigenvalues $\omega_k$ are all integer multiples of the smallest eigenvalue. Otherwise, generic solutions are quasi-periodic. $\square$ 
    
    \vspace{0.02 in}
    \noindent \textbf{Remark:} Coupling between the oscillators enters via higher order derivatives.  These correspond to higher order moments of the embedded population distribution (see Equation \eqref{eqn: derivatives and moments}). 
    %
    Therefore, the oscillators are coupled by the cross-block third and higher order moments of the distribution $\pi_Y(\cdot;\theta_*)$. When these moments vanish, the oscillators decouple to the corresponding order. For example, if at $\pi_*$, $Y^{(k)}$ and $Y^{(k')}$ are independent for all $k \neq k'$, then the blocks evolve independently. The same holds in reverse since, if the coordinates are not dependent, then there exists a nonzero moment coupling the parameter dynamics. So, given an interior equilibrium $\theta_*$, $\theta^{(k)}(t)$ decouple if and only if, when setting $\nu = \pi_*$, the embedding coordinates corresponding to distinct blocks are independent when sampling from $\pi_*$.  Otherwise, nonzero coupling produces rich dynamics, including chaos \cite{sato2002chaos}. 


All of the eigenvalues of $J(\theta_*)$ are purely imaginary, so $\theta_*$ is not a hyperbolic equilibrium. Then, the Hartman-Grobman theorem \cite{hartman1960lemma,hartman1960local} does not apply, so solutions to the exact dynamic need not be topologically equivalent to the linearized dynamic near interior equilibria. In principle, even small higher-order coupling terms could change the dynamics arbitrarily close to an interior equilibrium. The Lyapunov-center theorem \cite{lyapunov1992general} ensures that the linearization is characteristic. Namely, there exist $r/2$, two-dimensional manifolds, each composed of periodic orbits, intersecting the equilibrium. The $k^{th}$ manifold is tangent to $2k-1^{st}$ and $2k^{th}$ unit directions, and contains orbits with period approximately equal to $2 \pi/\omega_k$ \cite{hofbauer1996hamiltonian}. More strongly still, the Kolmogorov-Arnold-Moser theorem ensures that most of the toroidal solutions of the linearized dynamic survive nonlinear coupling up to diffeomorphism, provided the linearized trajectory is sufficiently far from periodic, and the system is initialized close to the equilibrium \cite{arnold2013mathematical,hofbauer1996hamiltonian,moser2001stable}. $\square$
\vspace{0.02 in}

\begin{figure}[t]
    \centering
    \includegraphics[trim = 130 180 100 140, clip, scale = 0.43]{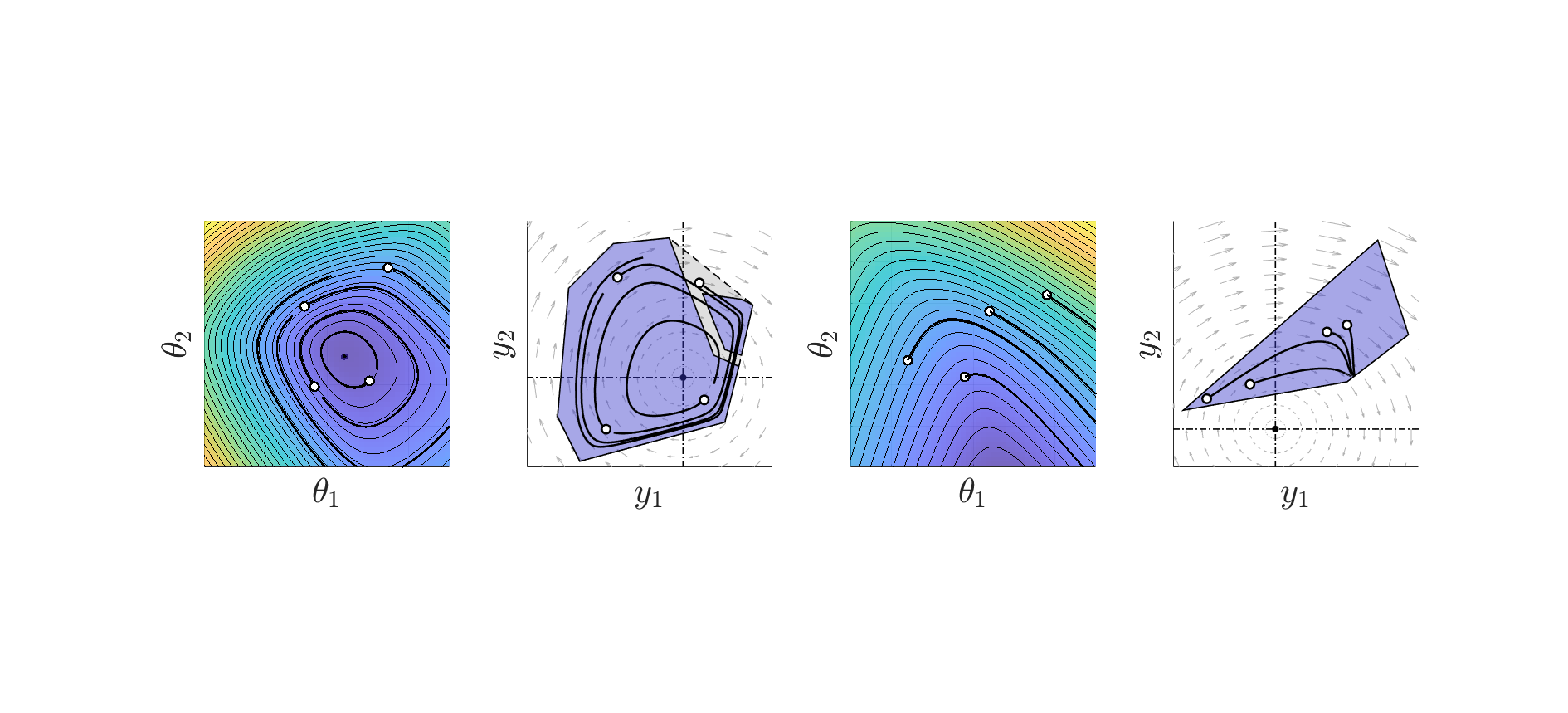}
    \caption{Hamiltonian parameter dynamics, and the associated centroid dynamics, for two example games. The examples match the first and last Hamiltonians shown in Figure \ref{fig: Hamiltonians}. The odd panels illustrate the parameter dynamics. The coordinates, $\theta$, correspond to the parameters responsible for fixing the distribution over $\Psi$. The base measure is uniform on $\psi$. The four trajectories shown are each initialized at the white scatter points lying at the start of each curve. The even panels illustrate the centroid dynamics. The coordinates, $y$, represent a pair of disc game coordinates. The blue shaded region $\Psi$ represents the the range of the embedding. The grey shaded region enclosing it is its convex hull. The grey circulating vector field represents the optimal local training vector field $v$. In the first example, the parameters orbit an equilibrium, $\theta_*$, along the level sets of the Hamiltonian function $P(\theta)$. This equilibrium corresponds to a fully mixed Nash equilibrium, and it exists since the convex hull of $\psi$ includes the origin (see Corollary 5.3). In the second example the parameters follow level sets of the Hamiltonian to infinity. As the parameters diverge, the centroid approaches the boundary of $\Psi$. In the first case, the convex hull of $\psi$ contains the origin. In the second it does not. }
    \label{fig: Hamiltonian and Adaptive}
\end{figure}
\begin{figure}[t]
    \centering
    \includegraphics[trim = 180 50 120 50, clip, width = \textwidth]{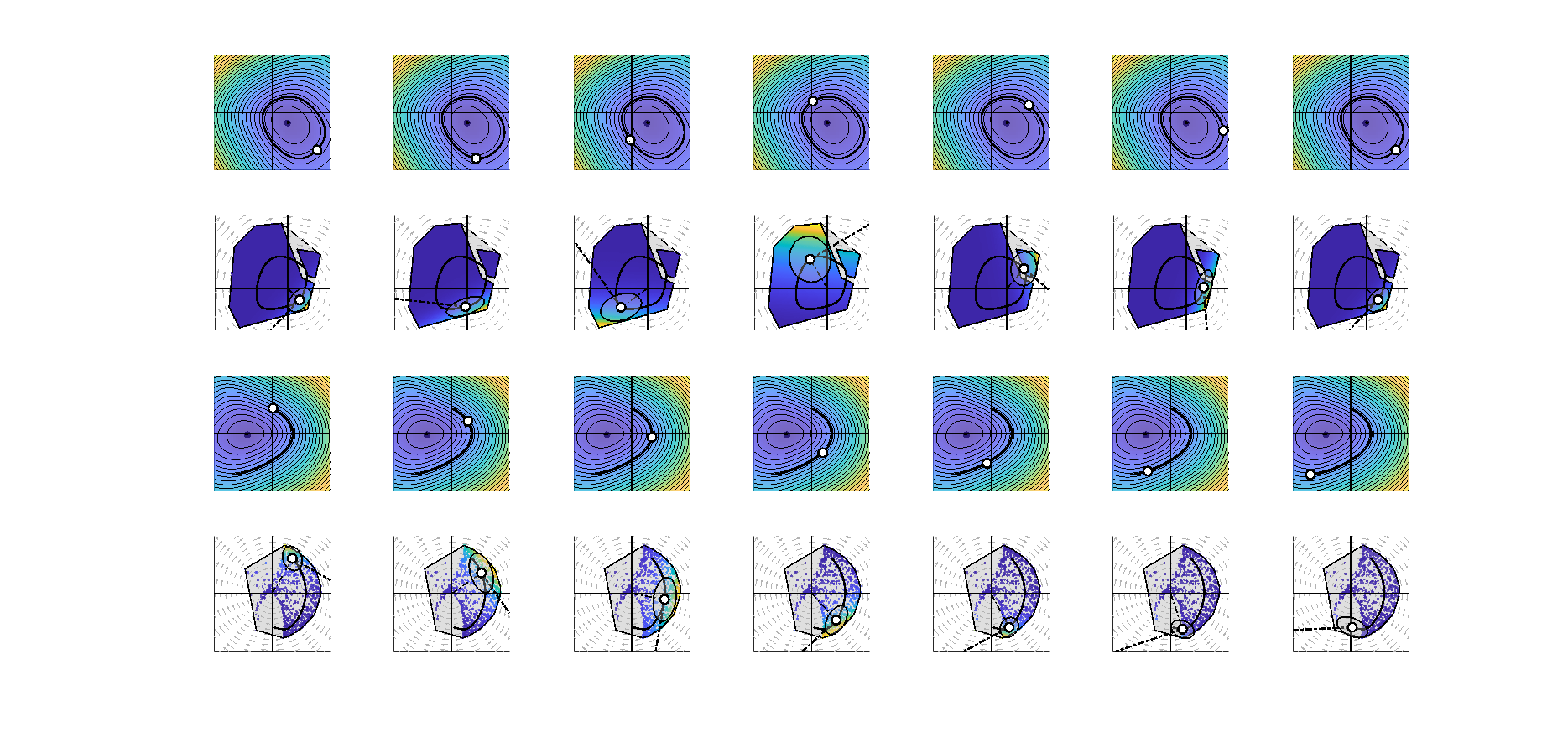}
    \caption{Autonomous centroid and Hamiltonian parameter dynamics for two example games. The top two rows correspond to the example game illustrated in Figures \ref{fig: Embedding Visualization}, \ref{fig: Exponential Tilting}, and \ref{fig: Hamiltonian and Adaptive}. The bottom two rows correspond to the IPD example illustrated in Figure \ref{fig: IPD Disc Game Embedding}. The first and third rows show orbits in the parameter space. The second and fourth rows show the corresponding sequence of population distributions in the embedding space. The white circles mark the population centroids and parameters at a sequence of evenly spaced times (separate columns). The shaded oval centered at each centroid represents the population covariance. The bold dashed line leaving the centroid marks the optimal local training vector field, $v$, evaluated at the centroid. The solid black line represents the realized orbit formed by multiplying the covariance against $v$ (see Equation \eqref{eqn: centroid autonomous adaptive dynamics}). Notice that, as the centroid approaches the boundary of $\text{conv}(\Psi)$, the covariance compresses so that its major axes are tangent to the boundary, while its minor axes are perpendicular to the boundary. Then, applying the covariance to $v$ compresses the components of $v$ pointing out of the boundary, and elongates those parallel to the boundary. This bends the self-play orbit so that the centroid remains within the convex hull of $\Psi$. For examples, see the second and fifth columns in the second row. It also slows the orbit near corners in the convex hull of $\Psi$ (see the last three columns in the bottom row). When the centroid is well removed from the boundary, the population distribution is approximately equivariant, so the centroid dynamics align to the self-play orbit (align to $v$). }
    \label{fig: Adaptive}
\end{figure}

We have established that the parameter dynamics obey a clean dichotomy. Given $f$ and support $S$, the Hamiltonian either admits a unique global minimizer or does not. If such a minimizer exists, it corresponds to an isolated interior equilibrium and a steady-state supported on all of $S$. The equilibrium is a globally Lyapunov stable center about which all perturbations mix on compact shells and is locally approximated by a system of decoupled simple harmonic oscillators. Higher-order terms couple the oscillators without eliminating the oscillating character of solutions, which return arbitrarily close to all initial conditions infinitely often. Figures \ref{fig: Exponential Tilting}, \ref{fig: Hamiltonian and Adaptive}, and \ref{fig: Adaptive} illustrate a recurrent examples. If there is no interior minimizer then all steady-state distributions correspond to parameter sequences that diverge to a boundary \cite{boone2019darwin,mertikopoulos2018cycles}. Given a finite trait space, these are the only generic options. Either, there exists a fully mixed equilbria, which the replicator dynamics orbit but fail to approach \cite{boone2019darwin}, or there is no fully mixed equilibria, in which case there exists at least some type that is dominated by all other types and that is driven to extinction under the replicator dynamic \cite{akin1980domination,akin1984evolutionary}. 

There is a straightforward geometric test that distinguishes these two cases after embedding. Recall that the origin plays a special role in a disc-game since it corresponds to a neutral type that has no advantage or disadvantage over any other type (see Box 1). When included in the convex hull of $\Psi$, it corresponds to a fully mixed evolutionary stable strategy (ESS), since any ESS in a symmetric two-player game must receive the same payout against all opponents, including itself \cite{bishop1978generalized, smith1982evolution}, against which it muxt receive a payout of zero. Recall also that any population embedded in a single disc game competes transitively if the convex hull of the embedded population does not contain the origin. Here, we show that the same test can be extended to arbitrary $r$, and distinguishes the form of the parameter dynamics.

\begin{snugshade}
\noindent \textbf{Corollary 5.3: [Interior Equilibria, Invariant Interior, and the Range of the Embedding]} Suppose that $(\Omega,f)$ is a functional form game with skew-symmetric $f$, $S$ is a support set, and $\pi(x,t)$ is a population distribution supported on $S$ evolving according to the replicator equation, \eqref{eqn: replicator standardized}. Let $\nu$ denote an arbitrary reference measure that has support equal to $S$. Let $\Psi = \{y_{\nu}(x)|x \in S\} \subseteq \mathbb{R}^r$ denote the image of the set $S$ under the disc-game embedding. Suppose that $\Psi$ is bounded.
\vspace{0.1 in}

\noindent Then, the replicator parameter dynamics admit an interior equilibrium if and only if the convex hull of $\Psi$ contains the origin in its interior.
\end{snugshade}

\noindent \textbf{Proof Outline:} First, notice that $\theta_*$ is an equilibrium if and only if $v(\bar{y}[\pi(\cdot;\theta_*)]) = U \bar{y}[\pi(\cdot;\theta_*)] = 0$. The rotation matrix $U$ is invertible, so $\theta_*$ is an equilibrium if and only if the corresponding centroid, $\bar{y}_* = \bar{y}[\pi(\cdot;\theta_*)] = 0$. 

Necessity is straightforward. If $\Psi$ does not contain the origin inside its convex hull, then, there is no distribution on $\Psi$ with centroid at the origin. To show sufficiency, we demonstrate that $P(\theta)$ is radially unbounded when the origin is in the interior. For details, see Appendix Section \ref{app: cor 5.3}. $\square$

Figure \ref{fig: Hamiltonian and Adaptive} illustrates the geometric test. The geometric test formalizes the intuition that, a population evolving in response to any combination of disc games, will orbit recurrently unless it is restricted to a subset of disc game space that excludes competitive cycles.

Importantly, the test is invariant to changes in the reference measure $\nu$ provided those changes preserve the support $S$. Changes of reference measure that preserve the support transform $\Psi$ by an invertible linear mapping (see Lemma 5 and 8). Any linear transformation of the origin is the origin. Open neighborhoods remain open under invertible linear transformations. Therefore, the origin is either inside, or outside, the interior of the convex hull of $\Psi_{\nu}$ for all $\nu$ with support $S$.


To conclude our analysis, we show that the centroid $\bar{y}$ in disc-game space obeys an adaptive dynamics equation:
\begin{snugshade}
    \noindent \textbf{Corollary 5.4: [Adaptive Dynamics]} Suppose that $(\Omega,f)$ is a functional form game with skew-symmetric $f$, $S$ is a support set, and $\pi(x,t)$ is a population distribution supported on $S$ evolving according to the replicator equation, \eqref{eqn: replicator standardized}. Then, for any reference measure $\nu$ supported on $S$, the trait centroid of the embedded population, $\bar{y}(t) = \mathbb{E}_{X \sim \pi(\cdot,t)}[y(X)]$, obeys the adaptive dynamics equation:
    \begin{equation} \label{eqn: centroid adaptive dynamics}
        \frac{\mathrm d}{\mathrm d t} \bar{y}(t) = \text{Cov}_{Y \sim \pi_Y(\cdot,t)}[Y] v(\bar{y}(t)) 
    \end{equation}
    where $v(y) = W y$ is the self-play vector field for the disc-game. 
\end{snugshade}

\noindent \textbf{Proof:} Proceed directly. 
$$
\begin{aligned}
\frac{\mathrm d}{\mathrm d t} \bar{y}(t) & = \frac{\mathrm d}{\mathrm d t} \int_{x \in \Omega} y(x) \exp(\theta(t) \cdot y(x)) \pi(x,0) dx = \int_{x \in \Omega} y(x) \left(y(x) \cdot \frac{\mathrm d}{\mathrm d t} \theta(t)) \right) \exp(\theta(t) \cdot y(x)) \pi(x,0) dx \\ & = \left[\int_{x \in \Omega} y(x) y^{\intercal}(x) \pi(x;\theta(t)) dx \right] \frac{\mathrm d}{\mathrm d t} \theta(t) 
= \text{Cov}_{Y \sim \pi_Y(\cdot,t)}[Y]  v(\bar{y}(t))
\end{aligned}
$$
where the last equality follows from the parameter dynamics and the fact that $v(\bar{y})$ is orthogonal to $\bar{y}$. $\square$

Equation \eqref{eqn: centroid adaptive dynamics} establishes that the centroid of the embedded population obeys a non-autonomous adaptive dynamics equation. Notice that this adaptive dynamics equation does not require any approximation, time-scale separation, does not restrict $f$ or $\pi$ to limited parametric forms, and does not require a monomorphic limit. In contrast \cite{nowak2004evolutionary}, it is both exact and general.

The equation appears non-autonomous since the positive definite matrix $\text{Cov}_{Y \sim \pi_{Y}(\cdot,t)}[Y]$ depends on the embedded distribution, which is changing in time, as specified by the parameters $\theta(t)$. Thus, as written, it is not clear whether Equation \eqref{eqn: centroid adaptive dynamics} can be solved without first solving for the parameter dynamics. 
Equation \eqref{eqn: centroid adaptive dynamics} is, in fact, autonomous, since the embedded centroid $\bar{y}$ uniquely determines the parameters $\theta$ \cite{efron2022exponential}. 

\begin{snugshade}
\noindent \textbf{Lemma 15: [Parameters From Centroid]} Given an affine parameterization \eqref{eqn: affine parametrization} the map from $\theta$ to $\bar{y}$ is invertible.
\end{snugshade}

\noindent \textbf{Proof: } By definition, $\bar{y} = \mathbb{E}_{X \sim \pi_X(\cdot;\theta)}[Y(X)] = \mathbb{E}_{Y \sim \pi_Y(\cdot;\theta)}[Y]$. Therefore, the parameters $\theta$ uniquely determine a centroid $\bar{y}$. To show that the mapping, $\bar{y}(\theta)$ is invertible, recall that $\bar{y}(\theta) = \nabla_{\theta} P(\theta)$ (see Equation \eqref{eqn: centroid from Hamiltonian} in Theorem 5). Lemma 13 establishes that $P(\theta)$ is strictly convex. For any strictly convex function, the mapping from argument to gradient is invertible, so, $\bar{y}$ uniquely determines the gradient $\nabla_{\theta} P(\theta)$, and the gradient uniquely determines the parameters $\theta$. $\square$

\vspace{0.05 in}
\noindent \textbf{Remark:} Therefore, for a fixed parameterization, we can write $\theta(\bar{y})$, and:
\begin{equation} \label{eqn: centroid autonomous adaptive dynamics}
     \frac{\mathrm d}{\mathrm d t} \bar{y}(t) = \text{Cov}_{Y \sim \pi_Y(\cdot;\theta(\bar{y}))}[Y]  v(\bar{y}(t)) 
\end{equation}
which is an \textit{autonomous} adaptive dynamics equation.

Since the density $\pi_X(\cdot;\theta)$ is uniquely determined by the parameters, the parameters are uniquely determined by the embedded centroid $\bar{y}$, and the embedded centroid obeys an autonomous system of equations, the centroid dynamics could, in principle, be solved independently, then inverted to recover the parameter and distributional dynamics. Solving for the inverse map is equivalent to solving a maximum likelihood estimation problem for all possible observed sufficient statistics \cite{efron2022exponential}. In practice, this is infeasible since the inverse map $\theta(\bar{y})$ is intractable. However, the fact that such a map exists implies that the centroid dynamics are a sufficient and minimal representation of the distributional dynamics. Therefore, any continuous replicator dynamic with separable $f$ is fully specified by an autonomous adaptive dynamics equation in the disc game space. In this sense, disc-game embedding radically simplifies the replicator dynamic.  In the disc-game space, the continuous replicator dynamic \textit{is} adaptive dynamics! $\square$

\vspace{0.05 in}
\noindent \textbf{Remark:} The adaptive dynamics equation, closely mirrors a self-play dynamic in the disc-game space since the initial direction of motion proposed on the right-hand side is $v(\bar{y})$, the best response direction when training in response to $\bar{y}$. In fact, Equation \eqref{eqn: centroid autonomous adaptive dynamics} \textit{is} a self-play dynamic, with respect to a Riemannian metric whose metric tensor is the covariance matrix appearing in \eqref{eqn: centroid autonomous adaptive dynamics} \cite{mertikopoulos2018riemannian}. 

The covariance in Equation \eqref{eqn: centroid autonomous adaptive dynamics} is responsible for restricting solutions to the adaptive dynamics equation to the interior of the convex hull of $\Psi$. Consider a sequence of $\bar{y}$ approaching the boundary of the convex hull of $\Psi$. Recall that $\pi_Y(y;\theta) = \pi_Y(y;0) e^{\theta \cdot y}$. To move the centroid arbitrarily close to a point on the boundary, $y_*$, the parameters $\theta$ must be chosen to concentrate the mass of $\pi_Y$ along the boundary of the convex hull of $\Psi$ passing through the $y_*$. Thus, as $\bar{y}$ approaches $y_*$, the covariance converges to a singular matrix, whose nullspace contains all directions normal to the boundary of the convex hull of $\Psi$ at $y_*$. As a result, as $\bar{y}$ approaches the boundary of the convex hull of $\Psi$, multiplication by the covariance increasingly suppresses components of $v(\bar{y})$ pointing out of the convex hull. Instead, the product skews $v(\bar{y})$ towards directions tangent to the boundary of the convex hull of $\Psi$ (see Figure \ref{fig: Adaptive}). $\square$

\vspace{0.05 in}
\noindent \textbf{Remark:} Like the parameter dynamics, the centroid dynamics match the intuition implied by the disc-game model. Namely, that solutions to the replicator equation should circulate unless barred by the boundaries of $\Psi$. 

If the convex hull of $\Psi$ includes the origin, then $\bar{y}$ can approach it. A distribution whose centroid is near the origin is near to an interior equilibrium, $\pi_*$. If we set the reference measure to $\pi_*$, then the disc-game coordinates are uncorrelated at the equilibrium, so $\text{Cov}_{Y \sim \pi_Y(\cdot;\theta(\bar{y}))}[Y] = D_{\omega}$, the diagonal matrix with diagonal entries $\omega_1, \omega_1, \omega_2, \omega_2, \hdots, \omega_{r/2}. \omega_{r/2}$. Then, linearizing equation \eqref{eqn: centroid autonomous adaptive dynamics} about $\bar{y} = 0$ gives the familiar system of decoupled oscillator equations corresponding to self-play in a disc game: $\frac{\mathrm d}{\mathrm d t} \bar{y}(t) \simeq D_{\omega} v(\bar{y}(t)) = D_{\omega} U \bar{y}(t) = W \bar{y}(t)$. Solutions to the self-play equation circulate independently in each pair of disc game coordinates at rates specified by $\omega$ and $P$. If the convex hull of $\Psi$ does not include the origin, then the solutions to the adaptive dynamics equation are non-recurrent since the parameter dynamics are non-recurrent. Indeed, solutions to the adaptive dynamics equation satisfy exactly the same dichotomy that solutions to the parameter dynamics obey since the mapping between $\theta$ and $\bar{y}$ is differentiable and invertible. Thus, the collection of solution trajectories to equations \eqref{eqn: Hamiltonian parameter dynamics} and \eqref{eqn: centroid autonomous adaptive dynamics} are diffeomorphic (see Figure \ref{fig: Hamiltonian and Adaptive}). $\square$
\vspace{0.02 in}

Why does the population centroid obey an autonomous adaptive dynamics equation in the disc game space? 

We present two arguments. The first shows that the desiderata chosen to specify the disc game embedding are precisely the constraints needed to ensure that the centroid in the transformed space obeys an autonomous adaptive dynamics equation. The second shows that the equivalence between the replicator dynamic and adaptive dynamics is an immediate consequence of a more familiar equivalence between the replicator dynamic and self-play in population games.

First, since the conversion into the disc game space preserves outcome distinct agents, the continuous replicator equation in the original trait space induces a continuous replicator equation over the population distribution in the disc game space:
$$
\begin{aligned}
    \partial_t \pi_Y(y,t) & = \int_{x \in T^{-1}(y)} \partial_t  \pi(x,t) dx = \int_{x \in T^{-1}(y)} \mathbb{E}_{X' \sim \pi_x}[f(x,X')] \pi(x,t) dx \\ & =  \mathbb{E}_{X' \sim \pi_x}[f(x(y),X')] \int_{x \in T^{-1}(y)} \pi(x,t) dx =\mathbb{E}_{Y' \sim \pi_Y}[g(y,Y')] \pi_Y(y,t)
    \end{aligned}
$$
where $x(y)$ is any member of the equivalence class $T^{-1}(y)$.

Under a disc game embedding, $g(\cdot, \cdot)$ is a bilinear function, so:
\begin{equation} \label{eqn: change in population from centroid}
\partial_t \pi_Y(y,t) = g(y,\bar{y}(t)) \pi_Y(y,t)
\end{equation}

It follows that, if $\pi_Y(y,0)$ is known, then the population density $\pi(y,t)$ can be recovered at any time using $\bar{y}(t)$ alone:
\begin{equation}
    \pi_Y(y,t) = \exp\left( \int_{s = 0}^t g(y,\bar{y}(t)) \right) \pi_Y(0,t).
\end{equation}

That is, given an initial distribution $\pi_Y$ in the disc game space, the trajectory of the centroid fully specifies the distribution at all future times. Since the current centroid fixes the rate of change in the distribution, and the rate of change in the distribution fixes the centroid, $\tfrac{\mathrm d}{\mathrm d t}  \bar{y}(t) = \int \partial_t \pi(y,t) y dy$, the centroid in the disc game space must obey an autonomous equation. Notice that this property is shared by any coordinate transformation that preserves outcome distinct classes, and that satisfies desideratum 1.

Next, recall that, if $g$ is bilinear then $g(y,y')$ can always be expressed $g(y,y') = y \cdot v(y')$ where $v(y')$ is the optimal training vector field evaluated at $y'$ (see Lemma 2). Then, under any coordinate transformation that satisfies desideratum 1 and preserves the distinctions between outcome distinct classes:
\begin{equation}
    \partial_t \pi_Y(y,t) = \left( y \cdot v(\bar{y}) \right) \pi_Y(y,t).
\end{equation}

Integrating to recover the rate of change in the centroid gives the autonomous adaptive dynamics equation \eqref{eqn: centroid autonomous adaptive dynamics}:
$$
    \frac{\mathrm d}{\mathrm d t}  \bar{y}(t) = \int_{y \in \Psi} y y^{\intercal} v(\bar{y}(t)) \pi_Y(y,t) dy = \left( \int_{y \in \Psi} y y^{\intercal} \pi_Y(y,t) dy \right) v(\bar{y}(t)) = \text{Cov}_{Y \sim \pi_Y(\cdot;\theta(\bar{y}))}[Y]  v(\bar{y}(t)).
$$

\begin{snugshade}
\noindent \textbf{Lemma 16: [Bilinearity and Autonomous Adaptive Dynamics]} If $T$ is a mapping such that outcome distinct classes remain distinct after applying $T$, and that satisfies desideratum 1, then, given a population evolving according to a replicator dynamic, the centroid in the transformed coordinates must obey an autonomous adaptive dynamics equation.
\end{snugshade}

Thus, the desiderata used to reach the disc game embedding coincide with the desiderata needed to ensure that the centroid, after the transformation, obeys an autonomous adaptive dynamics equation. This result realizes the promise made at the end of Section \ref{sec: desiderata} and helps justify our choice of desiderata.

Next, consider the bilinear population game with strategy space equal to all distributions supported on $\Omega$, and payout $f[\pi,\pi'] = \mathbb{E}_{X \sim \pi, X' \sim \pi'}[f(X,X')] = \iint f(x,x') \pi(x) \pi'(x) dx.$ Consider a self-play dynamic defined by setting $\partial_t \pi(x,t)$ proportional to $\nabla_{\nu} f[\nu,\pi(\cdot,t)]|_{\nu = \pi(\cdot,t)}$. The functional gradient is defined by maximizing the increase in payout constrained to a neighborhood of $\pi(\cdot,t)$ \cite{mertikopoulos2018riemannian}. In the Shashahani metric, self-play recovers the replicator dynamic \cite{akin2013geometry,mertikopoulos2018riemannian,shahshahani1979new}.  Since the exponential family formed by the embedding functions is a sufficient and minimal parameterization, the parameters of the exponential family, $\theta$, must also move along the self-play gradient of the population payout in the metric formed by applying the Shashahani metric to distributions parameterized by Equation \ref{eqn: disc-game parameterization}. 

Since the mapping from parameters to centroid is diffeomorphic for exponential families \cite{efron2022exponential}, the population centroid in the disc game space also provides a sufficient and minimal parameterization, so must also obey some form of self-play dynamic. To recover its form, note that the population payout can also be expressed directly using the population centroids in the disc game space. By bilinearity, $f[\pi, \pi'] = \mathbb{E}_{X \sim \pi, X' \sim \pi'}[f(X,X')] = \mathbb{E}_{Y \sim \pi_Y, Y' \sim \pi'_Y}[Y^{\intercal} U Y] = \bar{y} \cdot U \bar{y}' = \bar{y} \cdot v(\bar{y}')$. It follows that, no matter the base measure, the self-play gradient with respect to the centroid is $\nabla_{\bar{y}} f[\pi_y, \pi'_y]|_{\pi = \pi'} = v(\bar{y})$. This gradient was computed using the Euclidean metric in the disc game space. The centroid dynamics must be a self-play dynamic for the population game, but in the metric implied by mapping from centroid back to distribution, then applying the Shashahani metric. Changing the metric used to compute a gradient transforms it by a positive definite linear transformation, so, under the replicator dynamic, the centroid dynamics in the disc game space must take the form:
\begin{equation} \label{eqn: adaptive last time}
    \frac{\mathrm d}{\mathrm d t}  \bar{y}(t) =  C(\bar{y}(t)) v(\bar{y}(t))
\end{equation}
where $C(\cdot)$ is a matrix valued function that returns positive definite matrices for $\bar{y}$ in the interior of $\text{conv}(\Psi)$. Equation \eqref{eqn: adaptive last time} takes the generic adaptive dynamics form. Thus, the equivalence between the replicator and adaptive dynamics represents the generic equivalence between replicator dynamics and self-play dynamics in the population game.

\subsubsection{Solutions for Simplifying Geometries}

The disc game approach converts strategic problems into geometric problems by changing the game representation. In this section, we will consider example cases with embedded geometries $\Psi$ that simplify the continuous replicator dynamic. 

To initialize a continuous replicator dynamic in the embedding space we need to fix the number of disc games and the embedding of the initial population distribution, $\pi_Y(\cdot,0)$. Since we are now working exclusively in the disc game space, we can drop the extra subscript $Y$, and will use $\pi_0$ to represent the initial population distribution. Since the continuous replicator dynamic is support preserving, we will assume, without loss of generality, that $\Psi$ equals the support of $\pi_0$. Then, the choice of $\Psi$ may be expressed implicitly by the choice of $\pi_0$. 

Like any multivariate dynamical system, the parameter dynamics associated with the continuous replicator equation are easiest to solve when the parameters decouple. When they decouple, the system may be broken into a series of independent, lower-dimensional systems. These may be solved independently. 

Let $y^{(k)} = [y_{2k-1},y_{2k}]$ denote the pair of disc game coordinates assigned to the $k^{th}$ game. Similarly, let $\theta^{(k)} = [\theta_{2k-1},\theta_{2k}]$ denote the parameters associated with the $k^{th}$ game. When do the parameter dynamics, $\theta(t)$ decouple?

\begin{snugshade}
\noindent \textbf{Lemma 17: [Decoupled Parameter Dynamics]} Consider $Y \sim \pi_0$. If the set of component disc games $D = \{1,2,\hdots, r/2\}$ admits a partition $\mathcal{K} = \{K_j\}$ into disjoint subsets such that $Y^{(k)}$ is independent of $Y^{(k')}$ when $k$ and $k'$ are members of different sets in the partition, then each set of parameters, $\Theta_j = \{\theta^{(k)}, k \in K_j\}$ evolves independently of each other set of parameters, and, if $Y \sim \pi(\cdot,t)$, then $\{Y^{(k)}, k \in K_j\}$ remain independent of $\{Y^{(k)}, k \in K_{j'}\}$ for all $t$.  
\end{snugshade}

\noindent \textbf{Proof Outline:} Under the independence assumptions the base measure factors into a product of marginal measures. The exponential tilt also factors into marginal terms, so the integral defining the Hamiltonian factors. As a result, the gradient of the log Hamiltonian separates, decoupling the parameter dynamics. For details see Appendix section \ref{app: decoupled dynamics}. $\square$

\begin{snugshade}
\noindent \textbf{Corollary 17.1: [Fully Decoupled Parameter Dynamics]} Consider $Y \sim \pi_0$. If $Y^{(k)}$ is independent of $Y^{(k')}$ for all $k \neq k'$, then each pair of parameters $\theta^{(k)}$ evolves independently of each other pair of parameters, and, if $Y \sim \pi(t)$, then $Y^{(k)}$ remains independent of $Y^{(k')}$ for all $t$. 
\end{snugshade}

Corollary 17.1 relates problems with many disc game components to replicator dynamics on a single disc game. If we construct $\pi_0$ as a product of marginal measures assigned to each disc game, then the parameters in each disc game evolve independently, so may be solved without studying the other games. For the remainder of this section, we will focus on replicator dynamics in a single disc game. These may be lifted to quasi-periodic dynamics in more complex games by Corollary 17.1.

At simplest, $\pi_0(y)$ is an exponential tilt of a rotationally symmetric base measure, $\pi_0(y) = \exp(\theta(0) \cdot y) \nu(y)$, where $\nu(y)$ is only a function of $\|y\|_2$. For example, consider $\nu$ uniform on a ball centered at zero, or $\nu$ set equal to a centered normal with covariance proportional to an identity. Allowing exponential tilts includes all $\pi_0$ that are non-centered normals with covariance proportional to an identity. 

\begin{snugshade}
\noindent \textbf{Lemma 18: [Rotational Symmetry]} Let $\Psi \in \mathbb{R}^2$ be the domain for a single disc game, and suppose that $\pi_0$ is an initial population distribution that is an exponential tilt of a base measure $\nu$ that is symmetric under rotations. If the population evolves according to a continuous replicator dynamic, then its parameters, using base measure $\nu$, orbit on concentric circles at a constant rate with period equal to $2 \pi \|\theta(0)\|_2/\|\nabla_{\theta} P(\theta(0))\|_2$. In the parametrization with base measure $\pi_0$, the parameters orbit on concentric ellipses with period $2 \pi \|\theta(0)\|_2/\|\nabla_{\theta} P(\theta(0))\|_2$.
\end{snugshade}

\noindent \textbf{Proof:} If $\pi_0$ is an exponential tilt of $\nu$ where $\nu$ is rotationally symmetric then, with $\nu$ as the base measure, $P(\theta)$ must be invariant to rotations of $\theta$, so is also only a function of $\|\theta\|_2$. Then, since $P(\theta)$ is the Hamiltonian, and Hamiltonian dynamics follow level sets of their Hamiltonian function, the parameters must evolve along concentric circles. For remaining details, see Appendix section \ref{app: rotational symmetry}. $\square$

The continuous replicator dynamics simplify further in the case when all components of $Y \sim \pi_0$ are independent. Then, $\pi_0$ fully factors into a product of marginals so each separate pair of parameters evolves independently. The parameters associated with the same disc game remain coupled through the dynamics, however, the coupling simplifies.

Consider a single disc game, with a base measure that factors into a product of marginal measures, $\pi_0(y) = \pi_0^1(y_1) \pi_0^2(y_2)$. Then, the Hamiltonian fully factors, $P(\theta_1,\theta_2) = \prod_{j=1}^2 P_j(\theta_j)$ where  $P_j(s) = \int_{y_j} e^{s y_j} d \pi_0^j(y_j)$ is the moment generating function  (Laplace transform) of the marginal measure $\pi_0^j$. Then, the Hamiltonian parameter dynamics reduce to:
\begin{equation} \label{eqn: hamiltonian for independent ys}
    \frac{d}{dt} \theta_1(t) = P g_2(\theta_2), \quad \frac{d}{dt} \theta_2(t) =  - P g_1(\theta_1) \quad \text{ where } \quad
    g_j(s) = \frac{d}{ds} \log(P_j(s)))
\end{equation}
is the slope of the cumulant generating function and $P = P(\theta(0))$ is the initial population size. For a detailed derivation see Appendix section \ref{app: separable base measures}.

Equation \eqref{eqn: hamiltonian for independent ys} is especially simple when the marginals for $Y_1$ and $Y_2$ are identical. Then we can suppress the subscripts, and, scaling time by the total population size, reduce to a normal form nonlinear oscillator equation:
\begin{equation} \label{eqn: parameter dynamics normal form nonlinear oscillator}
    \frac{d}{dt} \theta_1(t) \propto g(\theta_2(t)), \quad \frac{d}{dt} \theta_2(t) \propto - g(\theta_1(t)).
\end{equation}

Suppose, for example, that $\pi_0$ is a Laplace distribution, so has marginals of the form $\frac{1}{2} \exp(- |y_j|)$. Then:
\begin{equation} \label{eqn: normal form Laplace oscillator}
    \frac{d}{dt} \theta_1(t) \propto \left(\frac{2}{ (1 - \theta_2(t)^2)}\right) \theta_2(t), \quad \frac{d}{dt} \theta_2(t) \propto -\left(\frac{2}{ (1 - \theta_1(t)^2)} \right) \theta_1(t).
\end{equation}

Equation \eqref{eqn: normal form Laplace oscillator} defines a separable elliptic oscillator, that approaches a simple harmonic oscillator for small $\theta$. It is well defined for all $\theta$ in the unit square $[-1,1]^2$.  The solution trajectories are the level sets of the Hamiltonian function $((1 - \theta_1^2)(1 - \theta_2^2))^{-1}$. The dynamical system admits an analytic solution for all initial conditions. 

\begin{figure}[t]
    \centering
    \includegraphics[trim = 90 20 100 40, clip, scale = 0.35]{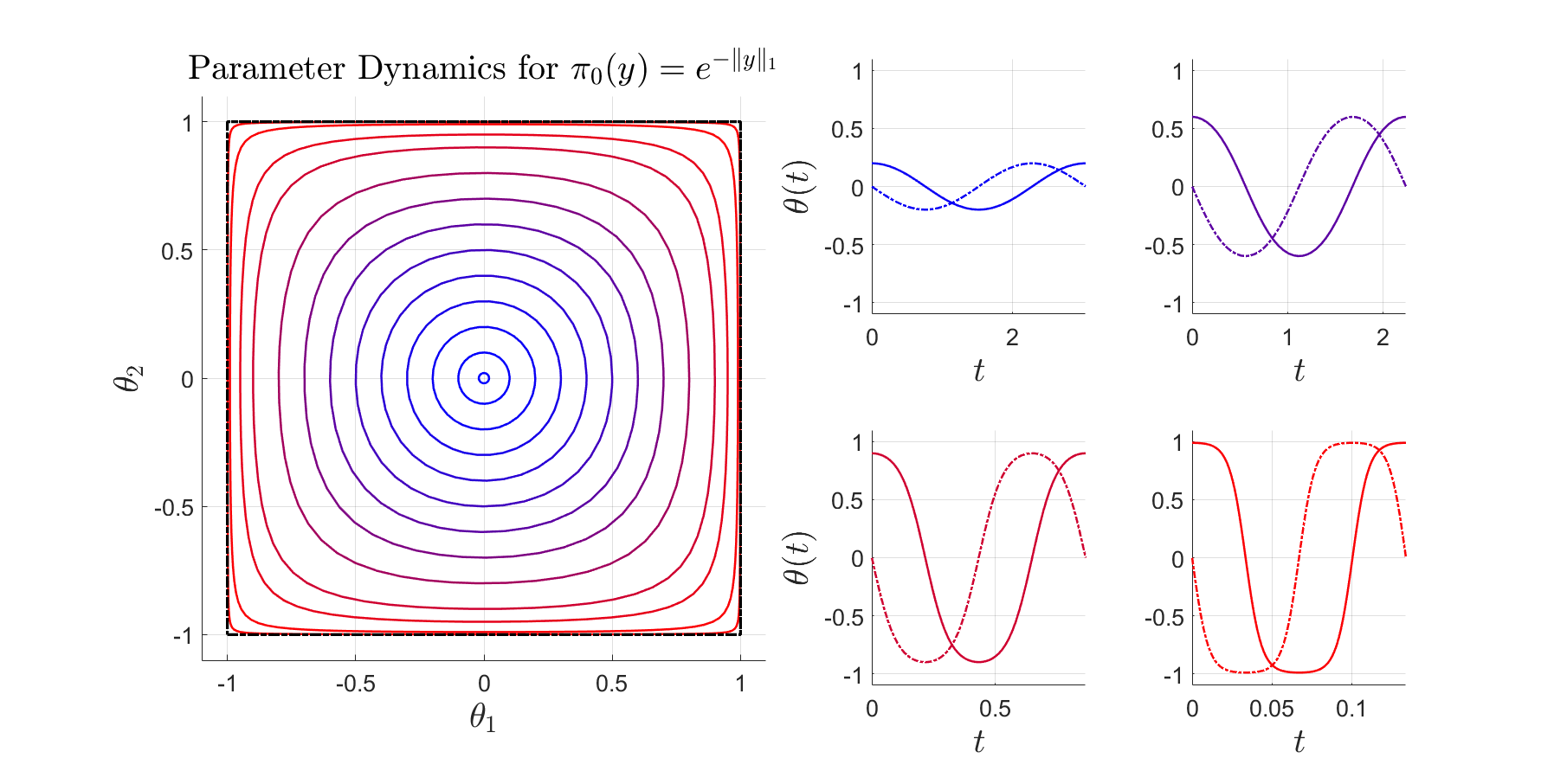}
    \caption{Parameter dynamics for a continuous replicator equation defined in a single disc game, with a Laplace initial distribution. The time traces correspond to initial amplitudes $a = [0.2, 0.6, 0.9, 0.99]$. Solid lines are $\theta_1$ and dashed are $\theta_2$. Each time trace covers one full period. Notice that, for small amplitudes, the parameters orbit on approximately circular level sets, but for large amplitudes the level sets approach the square defining the boundary of valid parameters. Also notice that, for small amplitude, the parameters move at a constant angular rate while, for large amplitude, each orbit dwells near the corners of the square, before transitioning rapidly to a new corner when near an edge of the square. In all cases the parameters orbit clockwise, i.e.~in the direction of advantage in a disc game.  }
    \label{fig: Laplace oscillator}
\end{figure}

Since the solutions are all periodic orbits in the square $[-1,1]^2$ we can adopt convenient initial conditions. Set $\theta_1(0) = 0$ and $\theta_2(0) = a > 0$, where $a$ represents the amplitude of the oscillator when it crosses a coordinate axis. Then, every solution to equation \eqref{eqn: separable form for elliptic oscillator} lies on an orbit of the form:
\begin{equation}
    \theta_1(t) = a \text{sn}(2 P t| a^2), \quad \theta_2(t) = a \frac{\text{cn}(2 P t|a^2)}{\text{dn}(2 P t|a^2)} 
\end{equation}
where $\text{sn}$, $\text{cn}$, and $\text{dn}$ are the elliptic sine, elliptic cosine, and delta amplitude functions. Each is an example of a Jacobi elliptic function \cite{dixon1894elementary,kovacic2016jacobi}. For the detailed derivation, calculation of the period, and solution for arbitrary initial conditions, see Appendix section \ref{app: separable base measures}.

The corresponding orbits are shown in Figure \ref{fig: Laplace oscillator} for amplitudes ranging from $0.1$ to $0.999$. Notice that, small amplitudes produce approximately circular orbits, while large amplitudes produce approximately square orbits. Also notice that, when the amplitude is small, the parameters orbit at an approximately constant velocity while, when the amplitude is large, the parameters dwell near the corners of the square. 

This behavior is generic when the convex hull of $\Psi$ is polygonal. For example, if the base measure separates into a product of semi-circle measures, then $\Psi$ is square, $g(s) = I_2(s)/I_1(s)$ where $I_j(s)$ is the modified Bessel function of the first kind, and the oscillator equations produce polygonal parameter orbits similar to heteroclinic motion about the boundary of a square. The Hamiltonians plotted in Figure \ref{fig: Hamiltonians} show that the level sets of the Hamiltonian approach polygons when far from the minimizer of the Hamiltonian for a variety of initial domains. Since the parameter orbits, and centroid orbits, are diffeomorphic, and large exponential tilts move the centroid near to the boundary of $\Psi$, these limits correspond to the limiting orbits of the adaptive dynamics equation, in the limit where the centroid approaches the boundary of the convex hull of $\Psi$ (c.f.~Figure \ref{fig: Hamiltonian and Adaptive}). We will show that, when the convex hull of $\Psi$ is polygonal, then the parameters approach polygonal orbits at large Hamiltonian values, and, the corners of the convex hull of the embedding space, $\Psi$, are key geometric features that generate the limiting polygonal orbit. Moreover, we will show that, as the centroid approaches the boundary, then the centroid dynamics slow down near corners of $\Psi$. Therefore, the centroid behaves like a heteroclinic oscillator near the boundaries of $\Psi$.

To study these orbits, we focus on a single disc game, assume that $\Psi$ is radially bounded so that its convex hull is compact, that the convex hull contains the origin, and that the convex hull is a polygon. Polygonal $\Psi$ occur in any normal form game with a finite strategy space since, in a normal form game, the payout is bilinear, so the embedding maps are linear functions, and $\Psi$ is a linear function of the space of mixed strategies, which is itself a convex polytope \cite{strang2022principal}. In this case, the vertices of the embedding correspond to pure strategies. To simplify the geometry of the parameter orbits we will also assume that the base measure is uniform over $\Psi$. 

The orbits of interest correspond to solutions of the parameter dynamics in a limit where the value of the Hamiltonian on the orbit diverges, or, equivalently, where $\text{inf}_t(\|\theta(t)\|_2)$ diverges. We will also work, without loss of generality, after scaling time by the initial population size, or, equivalently, with a constant interaction rate model. Then, $\frac{d}{dt} \theta(t) = U \bar{y}(\theta)$ where $\bar{y}(\theta)$ is the population centroid in the disc game space associated with the normalized distribution $p(y,\theta) = \pi(y,\theta)/P(\theta)$.

\begin{figure}[t]
    \centering

    \centering

    \begin{subfigure}[b]{0.41\textwidth}
    \centering
    \includegraphics[trim = 90 160 50 110, clip, width = \textwidth]{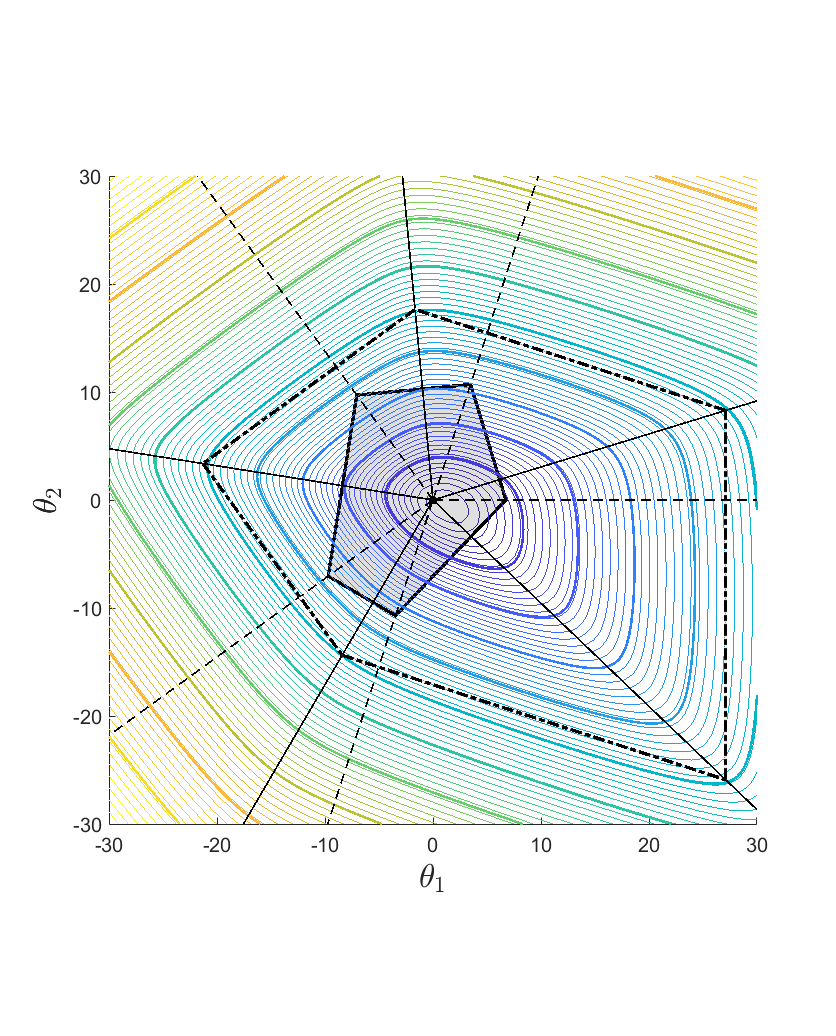}
    \end{subfigure}
    \hfill
    \begin{subfigure}[b]{0.2065\textwidth}
    \centering

    \begin{subfigure}[b]{\textwidth}
        \centering
        \includegraphics[trim = 150 68 175 40, clip, , width = \textwidth]{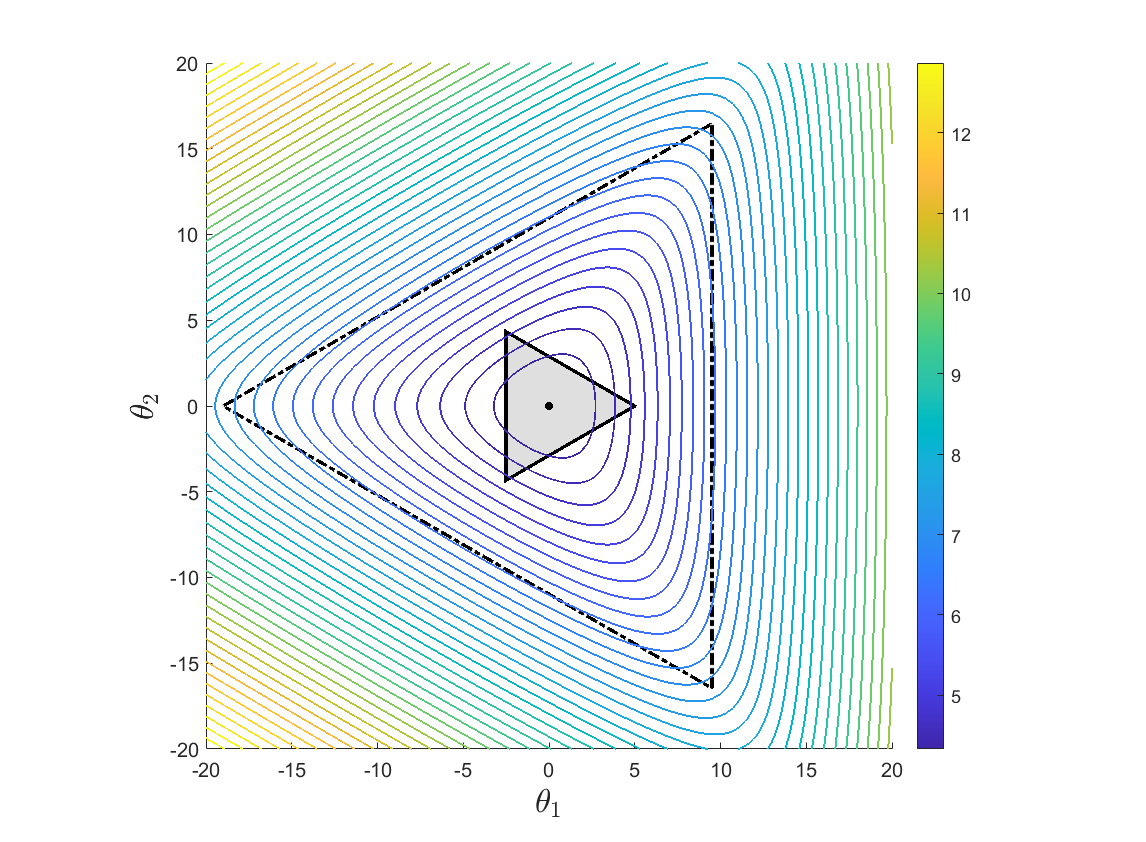}
    \end{subfigure}

    \vspace{0.25em}

    \begin{subfigure}[b]{\textwidth}
        \centering
        \includegraphics[trim = 150 68 175 40, clip, , width = \textwidth]{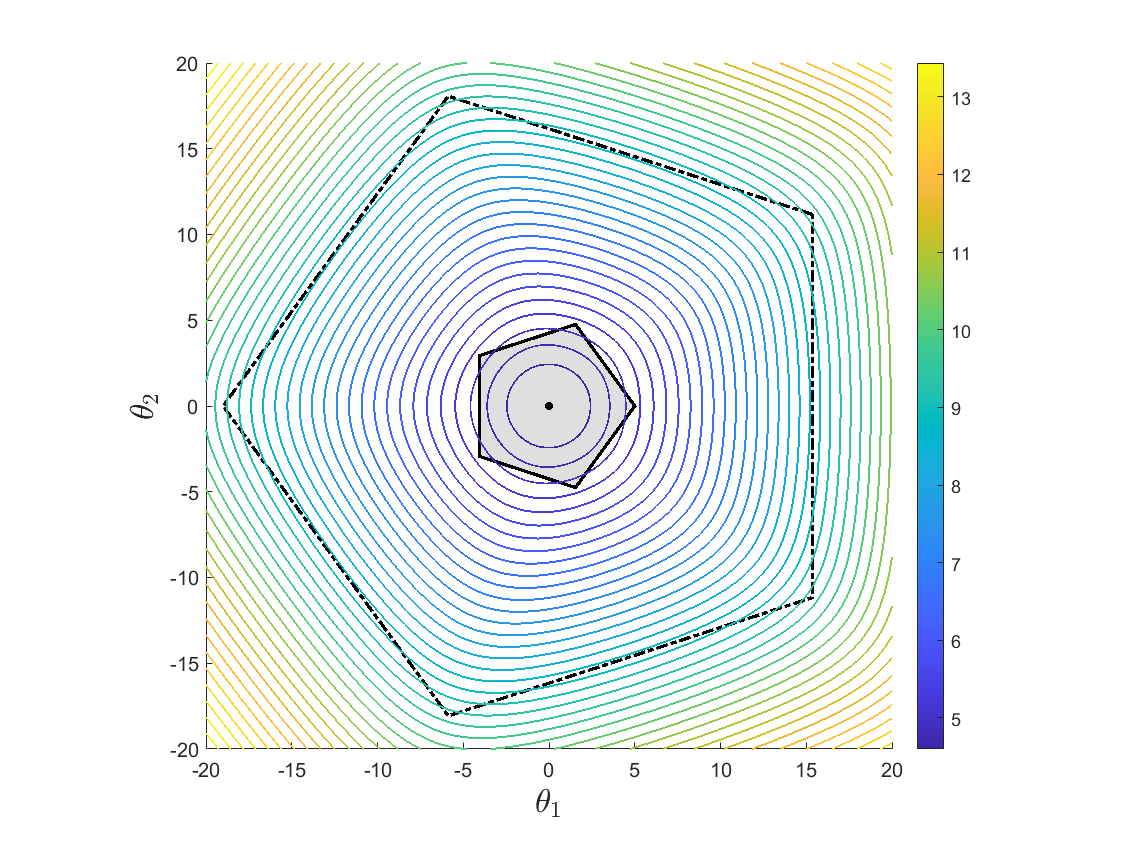}
    \end{subfigure}
    
    \end{subfigure}
    \hfill
    \begin{subfigure}[b]{0.35\textwidth}
    \centering
    \centering
    \includegraphics[trim = 100 0 100 40, clip, width = \textwidth]{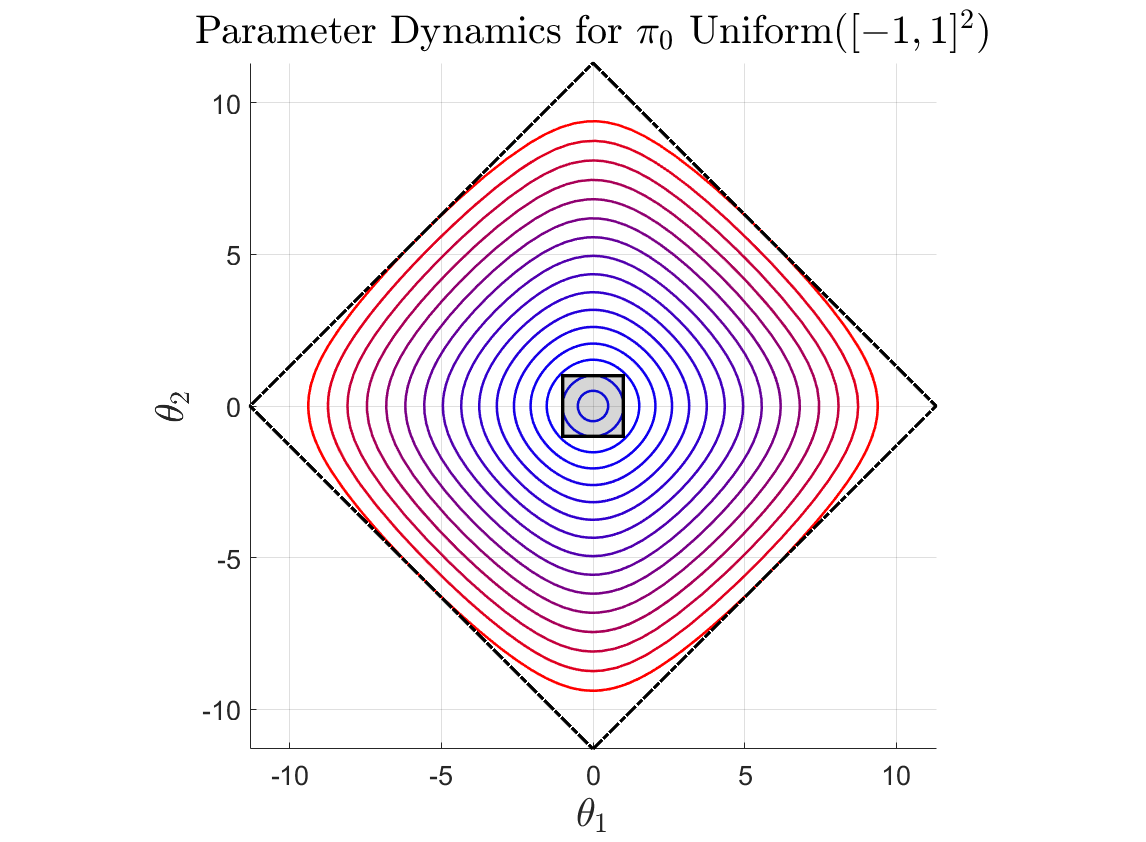}
    \vspace{0.5em}
    \end{subfigure}
    
    \caption{\textbf{Left:} Solution trajectories for the parameter dynamics of a population evolving under the continuous replicator dynamic (colored curves, colored by value of the Hamiltonian) with initial measure equal to a uniform distribution over a polygon $\Psi$ (shaded grey region). The dot-dashed outer polygon is dual to $\Psi$. Its vertices lie on rays normal to the boundaries of $\Psi$. Its boundaries are perpendicular to rays passing through the vertices of $\Psi$. Notice that trajectories with large amplitudes (large Hamiltonian) converge to forms similar to the dual polygon. \textbf{Middle:} Solutions for two regular polygonal regions (triangles and pentagons). \textbf{Right:} Solution trajectories for $\pi_0$ uniform on the square $[-1,1]^2$.}
    \label{fig: Dual Polygons}
\end{figure}

Let $\text{conv}(\Psi)$ denote the convex hull of $\Psi$, and $C = \{c_1,c_2,\hdots, c_m\}$ denote the corners of the convex hull. Consider $\frac{d}{dt} \theta(t)$ at $t = 0$ if $\theta(0) = a \hat{\theta}_0$ in the limit as the amplitude, $a$, diverges. Then $p(y;a \hat{\theta}_0)$ approaches a measure that places all of its mass on the set of $y \in \Psi$ that maximize $\hat{\theta}_0 \cdot  y$. When $\Psi$ is polygonal, the maxima are solutions to a linear programming problem. These are restricted to the vertices of the domain for generic $\hat{\theta}_0$, i.e.~the corners $C$. More generally, since $\Psi$ has a polygonal convex hull, $\Psi$ is contained inside its convex hull, and all vertices of the convex hull are elements of $\Psi$, the set of $y$ that maximize $\hat{\theta}_0 \cdot  y$ is restricted to the corners $C$ of $\text{conv}(\Psi)$ for almost all $\hat{\theta}_0$. 

If $\Psi$ includes a linear boundary segment connecting two corners of its convex hull, then all $y$ on the boundary maximize $\hat{\theta}_0 \cdot y$ for $\hat{\theta_0}$ normal to the boundary. Since the normal to the boundary is uniquely defined, and $\Psi$ has, at most $m$ linear boundary segments that match a boundary of its convex hull, there are, at most, $m $ values of $\hat{\theta}_0$ for which the argument maximizer of $\hat{\theta}_0 \cdot y$ is non-unique. Let $\hat{N} = \{\hat{n}_1,\hat{n}_2,\hdots, \hat{n}_{m}\}$ denote the collection of normals to the boundaries of $\text{conv}(\Psi)$. 

List the boundaries of the convex hull sequentially. Then, the sequence of normal directions $\{\hat{n}_1,\hat{n}_2,\hdots, \hat{n}_{m}\}$ partition $\mathbb{R}^2$ into $m$ angular regions, bounded by the rays $\{\{s \hat{n}_j\}_{s \geq 0}\}_{j=1}^{m}$. If the ray $\{s \hat{\theta}_0\}_{s > 0}$ points into the region bounded by $\{s \hat{n}_j \}_{s \geq 0}$ and $\{s \hat{n}_{j+1}\}_{s \geq 0}$, where indices are counted mod $m$, then $\hat{\theta}_0 \cdot y$ is maximized at the vertex contained between edges $j$ and $j+1$. 

Then, for each corner $c_j$, we can identify an angular region, bounded by the rays parallel to $\hat{n}_{j-1}$ and $\hat{n}_j$, such that, if $\hat{\theta}_0$ lies inside the region, then $p(y,a \hat{\theta}_0)$ approaches a delta distribution at the corner $c_j$. In this case, $\lim_{a \rightarrow \infty} \bar{y}(a \hat{\theta}_0) = c_j$. Since this limit holds for all rays pointing into the $j^{th}$ region, the level sets of the Hamiltonian approach straight line segments parallel to $U c_j$ inside the angular region. Since the angular regions generated by the corners partition all angles about the origin, and since all level sets of the Hamiltonian form a closed orbit, in the limit of large initial $\theta$, the parameter orbits must approach a series of similar polygons. 

\begin{snugshade}
\noindent \textbf{Lemma 19: [Dual Polygons]} Suppose that $\text{conv}(\Psi)$ is a bounded polygon, $P$, and $\pi_0$ is uniform on $\Psi$. Then, the parameters orbits approach, in the limit of large amplitude, a dual polygon, $P'$, with an edge for every vertex of $P$, and a vertex for every edge of $P$. The dual polygon is defined by the orthogonality constraints:
\begin{enumerate}
    \item The vector from the origin to the $j^{th}$ vertex of $P'$ is perpendicular to the $j^{th}$ edge of $P$.
    \item The vector from the origin to the $j^{th}$ vertex of $P$ is perpendicular to the $j^{th}$ edge of $P'$.
\end{enumerate}
\end{snugshade}

Such a dual exists for all $P$. It equals the polar reciprocal of $P$ with respect to the unit circle \cite{barvinok2025course,brondsted2012introduction}. Figure \ref{fig: Dual Polygons} shows two polygons related by polar reciprocation, and an example set of parameter orbits converging to the dual of an irregular polygonal domain $\Psi$. 

If $\Psi$ is a regular polygon, then the dual polygon associated with limiting orbits is a regular polygon of the same degree, rotated so that its edges are bisected by the vertices of $\Psi$. For example, the square boundary observed for a Laplace base measure is the dual to the diamond-shaped level sets of the Laplace density. The middle column of Figure \ref{fig: Dual Polygons} shows example trajectories for a uniform measure on a regular triangle and pentagon.


The time it takes to traverse each edge of the limiting polygon diverges as the initial parameter amplitude, $a$, diverges since the length of the segment diverges linearly in the initial amplitude $a$, but, the rate of travel approaches the distance from the associated corner of $\Psi$ to the origin, which is bounded. As a consequence, the centroid dynamics slow down near the corners of $\text{conv}(\Psi)$ with the time spent near a corner diverging linearly in the parameter amplitude $a$. In the limit, each corner acts as an equilibrium, since, if the normalized population is initialized with centroid at the corner, it must be monomorphic, so, by support conservation, cannot change over time. 

For example, consider a uniform base measure on the square $[-1,1]^2$. Then, the Hamiltonian factors, $P(\theta) = (\sinh(\theta_1)/\theta_1) (\sinh(\theta_2)/\theta_2)$ so the parameter dynamics simplify (See Appendix section \ref{app: heteroclinic orbits and dual polygons}). For large $\theta_1$ and $\theta_2$, the dynamics are approximated by:
\begin{equation}
    \frac{d}{dt} \theta_1(t) = \text{sign}(\theta_2(t)) + \mathcal{O}(|\theta_2(t)|^{-1}), \quad \frac{d}{dt} \theta_2(t) = -\text{sign}(\theta_1(t)) + \mathcal{O}(|\theta_1(t)|^{-1}).
\end{equation}

Then, ignoring the small corrections, $\theta_1(t)$ and $\theta_2(t)$, follow diamond-shaped orbits at a constant speed equal to $\sqrt{2}$ (see the right-most column of Figure \ref{fig: Dual Polygons}). Each side of the diamond corresponds to a corner of the original square domain. If the parameters are initialized with amplitude $a$ (e.g.~they pass through the point $[a,0]$), then each side of the diamond has length $\sqrt{2} a$ and is traversed at rate $\sqrt{2}$. Then, the time spent on any side is asymptotic to its amplitude, $a$. Therefore, the time the centroid $\bar{y}$ spends near each corner diverges as the centroid orbit approaches the boundary.


These observations suggest that the boundary of $\text{conv}(\Psi)$ should play an organizing role in explaining continuous replicator dynamics. When the initial distribution is concentrated near a boundary of $\Psi$, or places its centroid near a boundary of $\text{conv}(\Psi)$, then the boundary of the embedding space determines the shape of the parameter dynamics, and, the corners of the embedding space act like the semistable equilibria of a heteroclinic orbit. 

Extending the intuition developed with polygons, we should expect regions where the boundary is sharply curved to act like corners whose incident edges form an acute angle. These produce long edges in the dual polygon, so require a long time to transit. Therefore, the centroid will slow down when it approaches a point on the boundary of the convex hull of $\psi$ with large curvature. In contrast, regions with low curvature will act like edges in the polygonal approximation, so will not slow the centroid.

\subsection{Generalization}

A successful paradigm should provide accurate intuition for a family of related problems. Here, we show that the conclusions drawn in the context of the continuous replicator dynamic extend to a family of dynamics that: (a) are equivalent to self-play in the population game using a different metric over the space of unnormalized densities, (b) allow additional frequency dependence in the per-capita growth rates of types, and (c) extend to metapopulation models that mix inhomogeneously. 

\subsubsection{Additional Frequency Dependence}

Following \cite{mertikopoulos2018riemannian}, we will derive a general class of evolutionary dynamics by introducing a Lagrangian function, then selecting the dynamic to instantaneously maximize the Lagrangian. We will construct the Lagrangian as a mixture of two terms. The first is a fitness flux \cite{mustonen2010fitness}. Consider a time evolving population $\pi(\cdot,t)$, changing at rate $\dot{\pi}(\cdot,t) = \frac{d}{dt} \pi(\cdot,t)$. Then the fitness flux is defined by the bilinear form:
\begin{equation}
    \phi[\pi,\dot{\pi}; f] = \lim_{\Delta t \rightarrow 0} \frac{1}{\Delta t} f[\pi + \Delta t \dot{\pi},\pi] = \langle \dot{\pi}(\cdot), F[\pi](\cdot)\rangle = \int_{x,x' \in \Omega \times \Omega} \dot{\pi}(x) f(x,x') \pi(x) dx' dx.
\end{equation}

The second is a cost-of-motion that penalizes rapid changes in the distribution. It evaluates the rate of change, squared, of the distribution given an implicitly defined Riemannian metric over the space of distributions. We will consider costs of motion of the form:
\begin{equation}
    K[\pi,\dot{\pi};g] = \frac{1}{2} \langle \dot{\pi}(\cdot), G_{\pi}[\pi](\cdot) \rangle = \frac{1}{2} \int_{x,x' \in \Omega \times \Omega} \dot{\pi}(x) \frac{\delta(x - x')}{g(\pi(x))} \dot{\pi}(x') dx dx' = \frac{1}{2} \int_{x \in \Omega} \frac{\dot{\pi}(x)^2}{g(\pi(x))} dx
\end{equation}
for $g: \mathbb{R}^{+} \rightarrow \mathbb{R}^+$, and where $g(\pi) = 0$ if and only if $\pi = 0$. We will restrict our attention to $g$ continuous, monotonically increasing, $\mathcal{O}(\pi)$ as $\pi \rightarrow 0$ and as $\pi \rightarrow \infty$, possibly with different bounding constants. These ensure that the ensuing dynamic is support preserving in finite time for any game with bounded payouts.

Changing $g$ varies the metric structure imposed on the space of distributions. For example, setting $g(\pi) = \pi$ returns the Shahshahani, or Fisher information, metric \cite{mertikopoulos2018cycles}. Then $K[\pi,\dot{\pi}]$ is the squared rate of change in the symmetrized Kullbach-Liebler divergence (Jensen-Shannon distance) between $\pi + \dot{\pi} \Delta t$ and $\pi$. 

Next, construct the Lagrangian:
\begin{equation}
    L[\pi,\dot{\pi};f,g,\lambda] = \phi[\pi,\dot{\pi};f] - \lambda K[\pi,\dot{\pi};g].
\end{equation}

Then, since the cost of motion is strictly convex in $\dot{\pi}(x)$ over the support of $\pi(x)$, maximizing the Lagrangian with respect to $\dot{\pi}$ given $\pi$ is equivalent to computing the self-play gradient of the population payout in the metric induced by the cost of motion \cite{mertikopoulos2018cycles}. To maximize the Lagrangian, set its variational gradient with respect to $\dot{\pi}$ to zero:
\begin{equation}
    \nabla_{\dot{\pi}} L[\pi,\dot{\pi}_*(\pi);f,g,\lambda](x) = F[\pi](x) - \lambda \frac{1}{g(\pi(x))} \dot{\pi}_*(\pi) = 0.
\end{equation}

The ensuing generalized replicator dynamic takes the form:
\begin{equation} \label{eqn: freq dep replicator}
    \partial_t \pi(x,t) \propto \mathbb{E}_{X' \sim \pi(\cdot,t)}[f(x,X')] g(\pi(x,t)) = \mathbb{E}_{X' \sim \pi(\cdot,t)}[f(x,X')] \frac{g(\pi(x,t))}{\pi(x,t)} \pi(x,t).
\end{equation}

Varying the metric over the space of distributions by varying $g$ amounts to varying the formula for the per capita growth rate of type $x$ as a function of its expected payout and frequency. The function $g$ controls the growth rate. Setting $g(\pi) = \pi$ recovers the standard replicator dynamic. Otherwise, the per-capita rates scale in the frequency of each type according to the ratio $g(\pi(x,t))/\pi(x,t)$. A dynamic of the form \eqref{eqn: freq dep replicator} applies whenever the ratio of the growth rate of a type to its expected payout is a nonnegative function of the frequency of the type. This is a useful generalization since saturating per capita rates are a common feature in ecological models.

Choosing $g(\pi) = \mathcal{O}(\pi)$ ensures that these per capita rates remain finite for all $\pi$, although they may limit to different constants as $\pi$ approaches zero or approaches infinity. In particular, if $g(\pi) = \mathcal{O}(\pi)$ and $f$ is bounded then $F[\pi](x)$ is bounded, so $\pi(x,t)$ can be bounded above and below by an envelope that grows exponentially in $t$. Since no exponential function can reach zero in finite time, or diverge in finite time, the dynamic will preserve the support of $\pi$, and will not introduce singularities that place finite mass at a single type in finite time. 

Allowing nonlinear $g(\cdot)$ allows additional modeling freedom. For example, if $g$ is monotonically increasing and concave down, then less occupied types will change more rapidly than highly occupied types in response to game outcomes. This could model a setting where low density types can change abundance more rapidly than high density types, when rates of change are measured per capita. 

Given $g$, define $h^{-1}$ and $h$:
\begin{equation}
    \frac{d}{du} h^{-1}(u) = g(h^{-1}(u)), \quad   h(\pi) = \int_{1}^{\pi} \frac{1}{g(s)} ds.
\end{equation}

Since $g(\pi) = \mathcal{O}(\pi)$, $h(\pi) \rightarrow -\infty$ as $\pi \rightarrow 0$ and $h(\pi) \rightarrow \infty$ as $\pi \rightarrow \infty$. Since $g$ is nonnegative, and positive for all positive entries $h$ and $h^{-1}$ are monotonically increasing. Since $g$ is finitely valued and continuous, $h$ and $h^{-1}$ are continously differentiable. It follows that $h$ defines an diffeomorphic map from the positive half-line to the real line, and $h^{-1}$ maps from the whole real line to the positive half line. If, in addition, $g$ is monotonically increasing, then $h$ is concave and $h^{-1}$ is convex.

Let $u(x,t) = h(\pi(x,t))$. In the standard replicator equation $g$ is the identity function, so $1/g(\pi) = 1/\pi$ and $h(\cdot) = \log(\cdot)$. Alternately, the saturating function $g(\pi) = \pi/(1 + \pi)$ yields $h(\pi) = \int_{1}^{\pi} 1 + 1/s ds = (\pi - 1) + \log(\pi)$ so $h^{-1}(u) = W(e^{u + 1})$ where $W$ is the Lambert $W$, or product-log function. The function $g(\pi) = \pi^2/(1 + \pi)$ yields $h^{-1}(u) = 1/W(e^{1 - u})$. This models a situation where per capita growth rates are asymptotically proportional to expected payout, but are subject to an Allee effect that suppresses growth at low densities \cite{boukal2002single,mccarthy1997allee,stephens1999allee}. Classic examples include increased difficulty finding mates, decreased chance of pollination, loss of genetic diversity, or the collapse of social groups \cite{courchamp1999inverse,stephens1999allee}. In the following theory, $h$ will replace the log, and $h^{-1}$ the exponential, in the exponential family parameterization used to study the replicator equation.

Then:
\begin{equation}
    \partial_t u(x,t) = \mathbb{E}_{X'\sim h^{-1}(u(\cdot,t))}[f(x,X')] = F[h^{-1}(u(\cdot,t))] \in \text{range}(F).
\end{equation}

It follows that we can build an invariant family of densities by expressing $u(x,t)$ as a linear combination of the embedding functions:
\begin{equation} \label{eqn: freq dep parameterization}
    u(x,t) = \theta(t) \cdot y(x) + u_0(x), \quad \pi(x,t) = h^{-1}(u(x,t)))
\end{equation}
where $u_0(x)$ is the projection of $u(x,0) = h(\pi(x,0))$ onto the orthogonal complement of the range of $F$ with respect to the $\langle \cdot, \cdot \rangle_{\nu}$ inner product: 
\begin{equation}
    u_0(x) = h(\pi(x,0)) - \text{Proj}_{\text{range}(F)}[h(\pi(x,0))].
\end{equation}

\begin{snugshade}
\noindent \textbf{Theorem 6: } Suppose that $\pi$ obeys the generalized replicator dynamic specified in equation \eqref{eqn: freq dep replicator}. Let $h^{-1}$ be the inverse of $h = \int^\pi 1/g(s) ds$. Then the parameters $\theta$ of the parameterization \eqref{eqn: freq dep parameterization} obey a Hamiltonian dynamic of the form:
\begin{equation}
    \frac{d}{dt} \theta(t) = U \mathbb{E}_{X \sim \pi(\cdot;\theta(t))}[y(X)] = U \nabla_{\theta} H(\theta)
\end{equation}
with Hamiltonian:
\begin{equation} \label{eqn: generalized Hamiltonian}
    H(\theta) = \int_{x \in \Omega} \int_{0}^{\theta \cdot y(x) + u_0(x)} h^{-1}(u) du
\end{equation}
\end{snugshade}

\noindent \textbf{Proof Outline:} The proof follows immediately from the construction of the Hamiltonian, using the same techniques applied to prove Theorems 4 and 5 for the standard case when $h^{-1}(u(x;\theta,u_0))$ defined an exponential family. See Appendix Section \ref{app: theorem 6}. $\square$

\begin{figure}[t]
    \centering
    \includegraphics[trim = 80 0 80 40, clip, scale = 0.47]{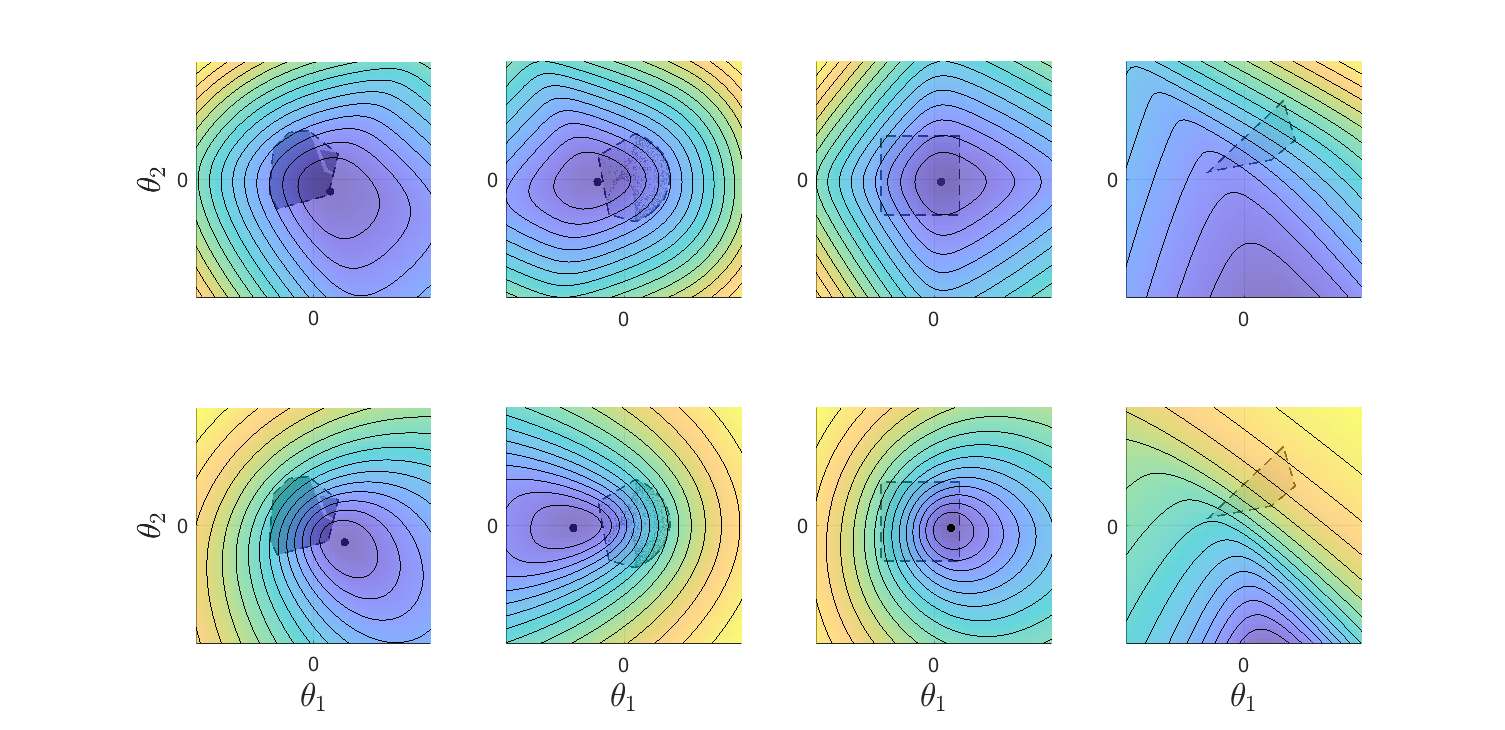}
    \caption{The Hamiltonian functions for four different games (columns), and for $g(p) = p$ (top row), versus $g(p) = p/(1 + p)$ (bottom row). The games are represented by the image of $\Omega$ under the embedding map, $\Psi$, whose boundary is shown as the dashed region in each plot. The second column corresponds to the IPD example (see Figure \ref{fig: IPD Disc Game Embedding}). The Hamiltonians are similar near their minimizers, since $p \simeq p/(1+p)$ for small $p$, but differ in shape for large $\theta$. The level sets of the Hamiltonians generated by $g(p) = p$ are more polygonal, with sharper corners, while the level sets of the Hamiltonians generated by $g(p) = p/(1+p)$ are smoother, with fewer pronounced corners. This follows since $g(p) = p/(1+p)$ saturates for large $p$, so $h^{-1}(u)$ is nearly linear in $u$ for large $u$.  }
    \label{fig: Hamiltonian Comparison}
\end{figure}

Equation \eqref{eqn: generalized Hamiltonian} can be used to compute the Hamiltonian for different functions $g$. Setting $g$ to the identity recovers the total population size as illustrated before: $1/g(s) = 1/s$ so $h(\pi) = \log(\pi)$ so $h^{-1}(u) = \exp(u)$ so $\int^{p} h^{-1}(u) = \exp(u)$. Importantly, $H(\theta)$ need not equal $P(\theta)$ for generic $g$, so the relationship between the total population size and the Hamiltonian is a special feature of the standard replicator dynamic. Indeed, the generalized dynamic need not conserve the total population size. Figure \ref{fig: Hamiltonian Comparison} illustrated the four example Hamiltonians for $g(\pi) = \pi$ and $g(\pi) = \pi/(1 + \pi)$ for the same cases illustrated in Figure \ref{fig: Hamiltonians}.

We can reuse the same sequence of arguments established in the standard case to prove essentially the same sequence of results for the general case. 

First, the gradient and Hessian of the Hamiltonian are:
\begin{equation}
\begin{aligned}
    & \nabla_{\theta} H(\theta) = \mathbb{E}_{X \sim \pi(\cdot;\theta)}[y(X)], \quad \textbf{H}_{ij}(\theta) = [\partial_{\theta}^{2} H(\theta)]_{i,j} = \int_{x \in \Omega} g(\pi(x;\theta)) y_i(x) y_j(x) dx.
\end{aligned}
\end{equation}

The Hamiltonian is evidently convex since $v^{\intercal} \textbf{H}(\theta) v = \int_{x \in \Omega} g(\pi(x;\theta)) (v \cdot y(x))^2 dx$, and the integral of any nonnegative function is greater than or equal to zero. It is also strictly convex given some constraints on $g$.

\begin{snugshade} 
\noindent \textbf{Lemma 20: [Strict Convexity]} If $f$ is bounded, $0 \leq g(\pi)$ with equality if and only if $\pi = 0$, and $g(\pi) = \mathcal{O}(\pi)$ as $\pi \rightarrow 0$ and $\pi \rightarrow \infty$, then the Hamiltonian function $H(\cdot)$ is a strictly convex, second-differentiable function of $\theta$ for all interior $\theta$.
\end{snugshade}

\noindent \textbf{Proof Outline:} Convexity is established by nonnegativity of the integrand. Existence is guaranteed by bounding $f$, as, if $f$ is bounded, then the embedding functions are bounded. If $g(\pi) = \mathcal{O}(\pi)$, then the dynamic is support-preserving, so $\pi$ is supported on $\Omega$ for all $\theta(t)$ reachable in finite time from a density with initial support $\Omega$. Then, $g(\pi(x;\theta))$ is nonzero for all $x$ and any $\theta(t)$ reachable in finite time. If $g(\pi) = \mathcal{O}(\pi)$ then $g(\pi) \leq C \pi$ for some $C$, so the integral $\int_{x \in \Omega} g(\pi(x;\theta))$ is finite. Let $\pi^g(x;\theta) \propto g(\pi(x;\theta))$ denote a normalized density supported on all of $\Omega$. Then, the Hessian is proportional to the second moment tensor for $y(X)$ when $X \sim \pi^g$. This tensor is positive definite by construction of the embedding map. See Appendix Section \ref{app: lemma 18} for details and proof with a more general theorem statement. $\square$

 If the Hamiltonian is strictly convex, then it is radially unbounded if it admits an isolated interior equilibrium, so the parameter dynamics satisfy the same dichotomy between recurrent and boundary-seeking dynamics observed in the standard case when $g$ is the identity function.

\begin{snugshade}
\noindent \textbf{Corollary 6.1: [Recurrence]} Suppose that the conditions of Lemma 20 hold. Then, the parameter dynamics either admit an isolated, interior equilibrium that corresponds to the choice of $\theta$ such that $\mathbb{E}_{X \sim \pi(\cdot;\theta)}[y(X)] = 0$, so correspond to a fully mixed NE, or no such equilibrium exists. The former case requires $0 \in \text{conv}(\Psi)$ and is impossible if the convex hull of $\Psi$ does not contain the origin. If an isolated interior equilibrium exists then it is a center and the parameter dynamics are Poincar\'e recurrent.
\end{snugshade}

\noindent \textbf{Proof Outline:} Under the conditions of Lemma 20, the Hamiltonian is strictly convex so either admits a unique, isolated, interior minima, which is an isolated equilibrium for the parameter dynamics, or it does not, and all parameter orbits escape to infinity. The remaining statements follow by the same arguments used for the standard replicator dynamic. For details, see Appendix Section \ref{app: corollaries 6} $\square$

%





Suppose that such an equilibrium exists and denote it $\theta_*$. The Jacobian at $\theta_*$ is $U \textbf{H}(\theta_*)$. Then, up to linearization, the parameter dynamics obey a simple Harmonic oscillator equation for a sequence of $\lfloor r/2 \rfloor$ independent oscillators. To find the frequency/period of each oscillator, recall that changing the reference measure without changing its support transforms the parameter dynamics by an invertible linear transformation, so preserves the topology of the orbits and their period. It follows that, up to linearization, the parameters obey the approximate dynamic:
\begin{equation} 
        \frac{\mathrm d}{\mathrm d t} \Delta \theta^{(k)}_1(t) \simeq  P(\theta_*) \omega_k \Delta \theta^{(k)}_2(t), \quad \frac{\mathrm d}{\mathrm d t} \theta^{(k)}_2(t) \simeq - P(\theta_*) \omega_k \Delta 
 \theta^{(k)}_1(t) 
\end{equation}
where $\Delta \theta = \theta - \theta_*$, and where $\omega_k$ is the magnitude of the $2k-1^{st}$ eigenvalue of $F_{\nu}$ for $\nu = \frac{1}{P(\theta_*)}\pi(x;\theta_*)$. 

Since the generalized dynamic is a self-play dynamic for the population game, it can be reduced to adaptive dynamics in the disc game coordinates. Corollary 6.2 establishes this fact for the generalized replicator dynamic using the same arguments developed for the standard case. Any population evolving under a generalized replicator dynamic, also obeys an adaptive dynamics equation in the disc game space, and, the centroid dynamics in the disc game coordinates are an equivalent specification of the full distributional dynamic given the initial distribution. 

\begin{snugshade}
\noindent \textbf{Corollary 6.2: [Adaptive Dynamics]} Let let $\bar{y}(\theta) = \mathbb{E}_{X \sim \pi(\cdot;\theta)}[y(X)]$.

\begin{equation}
    \frac{d}{dt} \bar{y}(t) = \textbf{H}(\theta(\bar{y}(t);u_0)) v(\bar{y}(t))
\end{equation}
where $\theta(\bar{y};u_0)$ is the unique solution to the equation $\mathbb{E}_{X \sim \pi(\cdot;\theta,u_0)}[y(X)] = \bar{y}$ given $u_0(x) = h(\pi(x,0))$.
\end{snugshade}

\noindent \textbf{Proof:} Given $\bar{y}(\theta) = \mathbb{E}_{X \sim \pi(\cdot;\theta)}[y(X)]$,
$$
\frac{d}{dt} \bar{y}(\theta(t)) = \frac{d}{dt} \mathbb{E}_{X \sim \pi(\cdot;\theta)}[y(X)] = \frac{d}{dt} \nabla_{\theta} H(\theta(t)) = \textbf{H}(\theta(t)) \frac{d}{dt} \theta(t) = \textbf{H}(\theta(t)) v(\bar{y}(t))
$$
where $\textbf{H}(\theta)$ is the Hessian of the Hamiltonian and where $v(y) = U y$ is the optimal self-play vector field. Since the Hessian is strictly convex, the matrix $\textbf{H}(\theta(t))$ is positive definite, so the centroid $\bar{y}(t)$ obeys an adaptive dynamics equation. Since $H(\theta)$ is strictly convex, the map between $\theta$ and $\nabla_{\theta} H(\theta) = \bar{y}(\theta)$ is a bijection. Therefore, the parameters obey an autonomous adaptive dynamics equation. $\square$

\subsubsection{Metapopulations and Inhomogeneous Mixing}

Real populations may not mix homogeneously. Consider a metapopulation model involving $l$ patches. Patches need not represent spatial clusters of agents. Rather, we will assume that patches represent any partitioning of the total population that satisfies two conditions. First, payout is conditionally independent of patch membership given agent attributes. This can be accomplished trivially by adding all attributes used to assign patch membership (e.g. all covariates that influence interactions rates) into the list of attributes $\Omega$, however this approach is semantic, since agents in different patches must be drawn from distributions with disjoint support. More practically, it means that the covariates responsible for determining interaction rate have no impact on the outcome of an interaction, conditional on a set of attributes $\Omega$ that \textit{can be shared by agents in different patches}. Second, the interaction probabilities between agents is constant in agent type conditional on their patch membership. 

Let $\pi_i(x,t)$ denote the density of type $x$ at time $t$ in patch $i$, and $P_i(t)$ the total population of the $i^{th}$ patch. The population within a patch may change over time.

Suppose that individuals are paired at random, where the frequency of interactions between individuals in patch $i$ with individuals in patch $j$, per unit time, is given by $m_{ij} P_i(t) P_j(t)$ for some $m_{ij} > 0$. Since the rate cannot depend on the ordering of the indices, the mixing matrix $M$ must be symmetric. The diagonal entries are intra-patch mixing rates, and the off-diagonal entries are inter-patch mixing rates.  

Suppose that the per capita growth rate of type $x$ in patch $i$ is proportional to the average total payout received by an individual of type $x$, in patch $i$, averaging over the interaction rate with each possible opponent, times a frequency dependent term $g(\pi)/\pi$. Then, the metapopulation will obey a block-structured, generalized replicator dynamic of the form: 
\begin{equation} \label{eqn: metapopulation replicator}
    \partial_t\pi_i(x,t) = \left[\sum_{j=1}^l m_{ij} \mathbb{E}_{X'\sim \pi_j(\cdot,t)}[f(x,X')] \right] g(\pi_i(x,t))
\end{equation}

This model collapses back to the homogeneous case if $m_{ij} = m_{lk}$ for all $i,j,k,l$, with interaction rate $\rho(P) = P$. If $M$ is an identity, then it reduces to $l$ independent populations. 

Since the rates are all defined per-capita, or with growth rate function $g(\pi) = \mathcal{O}(\pi)$, the dynamic is support preserving within each patch. Then, dividing across by the density yields the equivalent dynamic:
\begin{align}
   \partial_t h(\pi_i(x,t))) = \sum_{j=1}^l m_{ij} \mathbb{E}_{X'\sim \pi_j(\cdot,t)}[f(x,X')]
\end{align}
where $h= \log$ when $g(\pi) = \pi$. 

To make use of common embeddings, assume that for all $i\leq l$, $\pi_i(x,0)$ share support $S$, and, without loss of generality, assume that $S = \Omega$. Then, parameterize the distribution in each patch separately:
\begin{equation} \label{eqn: metapopulation parameterization}
    \pi_i(x,t) = h^{-1}(u_0^{(i)}(x) + \theta^{(i)}(t) \cdot y(x)).
\end{equation}

\begin{snugshade}
\noindent \textbf{Theorem 7:} Suppose that $\{\pi_i(\cdot,t)\}_{i=1}^l$ obey a metapopulation, generalized replicator dynamic of the form specified in equation \eqref{eqn: metapopulation replicator}, and each population is initialized with the same support. Then, the parameters $\theta = [\theta^{(1)},\theta^{(2)},\hdots,\theta^{(l)}]$ of the parameterization \eqref{eqn: metapopulation parameterization}, obey a Hamiltonian dynamic of the form:
\begin{equation} \label{eqn: metapopulation Hamiltonian}
    \frac{d}{dt} \theta(t) = (M \otimes U) \nabla_{\theta} H(\theta(t))
\end{equation}
where $\otimes$ is the Kronecker product, and $H$ is the Hamiltonian function:
\begin{equation}
H(\theta) = \sum_{i=1}^l H^{(i)}(\theta^{(i)},u_0^{(i)}), \quad H^{(i)}(\theta^{(i)}) = \int_{x \in \Omega} \int_{0}^{\theta^{(i)} \cdot y(x) + u^{(i)}_0(x)} h^{-1}(u).
\end{equation}
\end{snugshade}

\textbf{Proof:} The reduction to equation \eqref{eqn: metapopulation Hamiltonian} follows the same steps used before. The Kronecker product is a bookkeeping artifact. It remains to show that the specified dynamic is Hamiltonian. 

Any dynamic of the form $\frac{d}{dt} \theta = A \nabla_{\theta} H(\theta)$ is Hamiltonian if $A$ is a square, real-valued, skew-symmetric matrix. Indeed, expanding $A$ into its Schur form, then changing coordinates via the orthonormal matrices $Q$, recovers a block Hamiltonian dynamic of the kind we derived for the homogeneous population case. In this setting, block-diagonalizing the matrix $M \otimes U$ is unhelpful, since it introduces a linear transformation that mixes parameters assigned to separate patches. 

So, to show that the dynamic is Hamiltonian it suffices to show that $M \otimes U$ is skew symmetric. This follows immediately since $M$ is symmetric, and $U$ is skew symmetric:
\begin{equation}
    (M \otimes U)^{\intercal} = M^{\intercal} \otimes U^{\intercal} = M \otimes (- U) = - (M \otimes U).
\end{equation}

It follows that the parameter dynamics are Hamiltonian. $\square$

Since the metapopulation Hamiltonian is a sum of Hamiltonian functions that all share the same characterization when there exists, or does not exist, a fully mixed NE (when the origin is contained, or is not contained, inside the convex hull of $\Psi$), every qualitative result established before holds for the metapopulation extension. In particular,

\begin{snugshade}
\noindent \textbf{Corollary 7.1: [Recurrence]} The parameter dynamics either admit an isolated, interior equilibrium that corresponds to a fully mixed NE, or no such equilibrium exists. If an isolated interior equilibrium exists, then it is a center and the parameter dynamics are Poincar\'e recurrent. An isolated interior equilibrium exists if and only if the origin is contained inside the convex hull of $\Psi$. 
\end{snugshade}

\noindent \textbf{Proof Outline:} All arguments follow exactly as before. If such an equilibrium exists, then the Hamiltonian is a sum of radially unbounded, strictly convex functions, so is radially unbounded and strictly convex. In that case, the dynamics are volume preserving and bounded, so are recurrent. For details, see Appendix Section \ref{app: corollaries 7} $\square$

As in the homogeneous case, the metapopulation parameter dynamics can also be reduced by studying the collection of population centroids in the embedding space. 

\begin{snugshade}
\noindent \textbf{Corollary 7.2: [Mixed Adaptive Dynamics]} Let $\bar{y}_i(t) = \mathbb{E}_{X \sim \pi_i(\cdot,t)}[y(X)]$. Let $\bar{y}(t) = [\bar{y}_1(t),\bar{y}_2(t),\hdots,\bar{y}_l(t)]$ denote the concatenation of all the centroids, $\theta(\bar{y})$ the map from the centroid back to the parameters, and $\textbf{H}$ the Hessian of the Hamiltonian. Let $v(y) = U y$ denote the optimal training vector field that directs self-play. Let $m_i = \sum_{j = 1}^k m_j$ denote the total per capita interaction rate for patch $i$. Let $\hat{M} = \text{diag}(m_1^{-1},m_2^{-1},\hdots,m_l^{-1}) M$ denote the row-stochastic version of $M$. Then, the centroids obey an autonomous dynamic of the form:
\begin{equation} \label{eqn: mixed adaptive}
    \frac{d}{dt} \bar{y}_i(t)  = m_i  \textbf{H}(\theta(\bar{y}(t))) v\left( [\hat{M} \bar{y}(t) ]_i \right). 
\end{equation}
\end{snugshade}

\noindent \textbf{Proof Outline:} See Appendix Section \ref{app: corollaries 7}. $\square$

\vspace{0.02 in}
\noindent \textbf{Remark:} In the standard case when $g(\pi) = \pi$, $\textbf{H}$ can be replaced with the raw second moment tensor $\mathbb{E}_{Y \sim \pi_y(\cdot;\theta(y))}[Y Y^{\intercal}]$. In this case, the positive definite matrix entering the mixed adaptive dynamics equation is closely related to the population covariance in after embedding. $\square$

\vspace{0.02 in}
\noindent \textbf{Remark:} Equation \eqref{eqn: mixed adaptive} is to simultaneous gradient ascent as adaptive dynamics is to self-play. The right most term is the self-play gradient, or optimal training vector, evaluated at a weighted average of the centroids of each embedded population. The weights in the average differ by the focal patch $i$, and are specified by the terms in the mixing matrix. Since the optimal training vector field is linear, this is equal to a weighted average of the optimal training vectors associated with each patch. This averaged training vector is then multiplied by the covariance of the embedded population in the $i^{th}$ patch to recover the rate of change of the $i^{th}$ centroid. $\square$

The notation in equation \eqref{eqn: mixed adaptive} hides two nuances. First, the centroids are really the expected value of $y(X)$ when $X$ is drawn from $\pi_i$ times the population $P_i$. This is kept implicit since we adopted the definition that $\mathbb{E}_{X \sim \pi}[f(X)] = \int_{x}\pi(x) f(x) dx$ regardless the normalization of $\pi$. Second, each $\pi_i(\cdot;\theta)$ function is different depending on $\pi_i(x,0)$. It follows that the inverse $\theta^{(i)}(\bar{y})$ is also a different function for each patch $i$.

Since the dynamic is autonomous, is uniquely defined up to the initial distributions in each patch, and since we can recover the parameters in every patch from their centroids, the metapopulation version of the replicator equation also admits an exact representation as simultaneous (mixed) adaptive dynamics equation in the embedded centroids, for any payout function $f$. 

\subsubsection{Discussion}

Theorems 6 and 7, and their corollaries, show that the essential qualitative description of evolutionary population dynamics in symmetric, zero-sum games, as systems of coupled oscillators, holds beyond the special cases of self-play, adaptive dynamics, simultaneous gradient ascent, or the standard replicator equation. This characterization does not require homogeneous mixing, nor does it require that per capita growth rates are strictly proportional to expected payouts. Indeed, generic no-regret learning dynamics are Hamiltonian, and Poincar\'e recurrent, in zero sum, symmetric games with fully mixed equilibria (c.f. \cite{bailey2019multi,boone2019darwin,flokas2020no,legacci2024geometric,mertikopoulos2018cycles}). Our results are consistent with this more general collection of theories. They differ in the use of the disc game embedding to recover a useful parameterization of densities, to provide a geometric condition distinguishing recurrent dynamics from iterated dominance dynamics, and a generic, yet explicit conversion to adaptive dynamics in a coordinate system intrinsic to the game.

The generality of these conclusions in the setting of symmetric, two-player, zero-sum games suggests that the key phenomena, iterated domination followed by recurrent dynamics, are essential properties of the class of games, not the specification of the learning dynamic. All of these dynamics are Hamiltonian because, after an appropriate transformation, the right hand side of the dynamic is a skew-symmetric linear transformation of an input that depends on the current population. In all cases, the skew symmetry of the transformation was a direct consequence of the skew-symmetry of the payout, and would not hold if the payout was not skew-symmetric, that is, given a variable-sum game.

A paradigm should provide incisive intuition within a clearly defined class of problems. A paradigm may be misused if it is unclear which problems it does not address. Barring adversarially constructed games with infinite rank, and infinite variance under all continuously distributed populations, a disc game embedding exists if and only if the underlying game is skew symmetric. This symmetry is the only substantive restriction on its existence. The results illustrated here suggest that the embedding provides qualitatively accurate intuition for a broad class of problems that share the same sharply defined scope: learning dynamics in symmetric, two-player, \textit{constant-sum }games. The embedding does not exist for variable-sum games, where its misapplication as a mental guide may not provide accurate intuition. 

\subsection{Efficient Simulation Leveraging the Latent Space} \label{sec: numerics}

The latent space form, \eqref{eqn: replicator parameter dynamics},  offers promising new avenues for simulating the continuous replicator equation, \eqref{eqn: replicator equation}.

Direct simulation of the continuous replicator equation is expensive since, to evaluate the rate of change in the population density at each supported $x$ requires performing an integral over the full support of the population. Let $t$ denote the number of traits. Then the cost of evaluating the rate of change of the density at any $x$ will grow exponentially in $t$. This exponential cost is incurred at every quadrature node where the density is computed. So, for general $f$, the cost for direct implementation of the continuous replicator equation in the original trait space is exponential in $2 t$. Consequently, even relatively low dimensional settings exceed reasonable simulation budgets.

If the rank, $r$, of $f$ is less than $t$, then simulation in the latent space may be cheaper. What is the cost of simulating the parameter dynamics, $\tfrac{\mathrm d}{\mathrm d t}  \theta = U \nabla_{\theta} P(\theta)$?

First, at each time step, we only need to update $r$ parameters to update the density. In contrast, without a functional form, we need to update the density at each quadrature node explicitly. Second, to update each parameter, we need to evaluate the partial derivative of $P(\theta)$ with respect to one parameter. In total, the cost per time step is equivalent to computing $\nabla_{\theta} P(\theta)$, then permuting and negating the entries of the gradient.

Recall that $\nabla_{\theta} P(\theta) = \mathbb{E}_{Y \sim \pi_Y(\cdot;\theta)}[Y]$ where $\pi_Y(y;\theta) = \exp(\theta \cdot y) \pi_Y(y,0)$ and $\pi_Y(y,0)$ is the distribution of $y(X)$ for $X \sim \pi(x,0)$. The cost of evaluating the expectation of each coordinate is equivalent to the cost of performing quadrature in the parameter space. Quadrature incurs an exponential cost in $r$. Therefore, instead of paying an exponential cost in $2 r$, we pay an exponential cost in $r$, times $r$. The required integrals can be all conveniently collected into a single quadrature call using an auto-differentiation routine. 

All that remains is evaluation of the base measure $\pi_Y(y;0)$ given by embedding the initial distribution $\pi(x;0)$, or some other base measure in $x$. The population distribution in the embedding space is a simple exponential tilting of the base measure pushed forward by the embedding. Since the base measure remains fixed, evaluating the base measure is a one-time cost. The additional cost to push the base measure forward is distributed across each time step. 

So, disc game embedding promises a simple method for fast simulation of the continuous replicator equation whenever the rank $r$ is sufficiently small. Push the chosen base measure forward. Then, evolve equation \eqref{eqn: replicator parameter dynamics} using:
\begin{enumerate}
    \item A symplectic integrator for each time step,
    \item An auto-differentiator to evaluate $\nabla_{\theta} P(\theta)$ from a single evaluation of $P(\theta)$, and
    \item A chosen quadrature scheme for evaluating $P(\theta)$.
\end{enumerate}
Notice that only the push forward requires integrals over the original $t$ dimensional space. Thus, after pushing forward, the simulation cost is essentially independent of $t$. As a back of the envelope comparison, an $n$-step simulation in the original space has cost $\mathcal{O}(\exp(2 t))$ per time step, while simulation in the latent space has cost $\mathcal{O}(\exp(t + r))/n + \mathcal{O}(r \exp(r))$. So, in the limit of large $n$, simulation in the latent space can be performed at a cost that is entirely independent of the dimension of the original space. We leave detailed numerical development and testing to separate work.

In practice, it is more stable to apply the same algorithm using $\log(P(\theta))$ as the Hamiltonian, then scaling time by the initial population size to recover the desired dynamic.

\section{Conclusions}

This paper offers three contributions. First, we have established an axiomatic justification for disc-game embedding that motivate its use for studying learning dynamics. Second, we have shown that most skew-symmetric games admit a disc-game embedding and that the embedding allows accurate approximation for essentially all games. All previous work leveraging disc game embeddings either assumed a bilinear payout function or only applied the embedding pointwise to sampled data \cite{balduzzi2019open,balduzzi2018mechanics,balduzzi2018re,strang2022principal}. The generality and interpretability of the embedding recommend its use as a mental model for evolution subject to a zero-sum game. 

Third, we have shown that the intuition provided by disc-game embedding accurately characterizes an important class of learning dynamics in zero-sum games. Specifically, the disc-game embedding functions offer a useful function basis when studying generalized, continuous replicator dynamics in populations that could mix inhomogeneously. This function basis offers analytical insight by identifying a normal form dynamic and speeds computation. Using the disc-game embedding basis, we derived the necessary conditions for exact finite-dimensional closure, showed that the number of dynamical degrees of freedom needed corresponds to the rank of the payout, and derived two intuitive forms for the associated parameter dynamics. These results completely generalize the restricted parameteric solutions presented in \cite{cebra2024almost,cressman2004coevolution,Cressman_b}. We showed that the parameter dynamics are highly structured, so admit strong global analysis. In particular, the replicator dynamic reduces to a Hamiltonian system of coupled oscillators, driven by adaptive dynamics in the embedding space. These results directly recover parallel results for finite strategy sets, such as Poincar\'e recurrence in the presence of a fully mixed equilibria \cite{boone2019darwin,mertikopoulos2018cycles,piliouras2014optimization}. In addition, they establish an exact correspondence between replicator/multiplicative weight dynamics and adaptive dynamics in the disc game space. 


For the reasons demonstrated in this paper, we recommend disc-game embedding as a canonical latent space representation for populations evolving in response to a symmetric, zero-sum, game. 


\section*{Acknowledgements} The authors thank Jade Sheng for her assistance creating schematic figures, as well as Stefano Allesina and Mary Silber for their insightful support. The authors also thank Yucong Liu for his contributions applying the Hilbert-Schmidt theorem to games.

\bibliographystyle{siam}
\bibliography{References.bib}

\pagebreak

\pagebreak
\section{Appendix}

\subsection{An Example Game} \label{app: example game}

Consider the following population extension of the iterated prisoner's dilemma game, chosen for illustration:

\begin{enumerate}
    \item \textbf{Pairwise, single-round game: } The game is subdivided into individual rounds between individual pairs of agents. Each individual round, between each individual pair, is an instance of the canonical prisoner's dilemma game (PD). Each agent must choose to either cooperate or defect. If both agents cooperate, then they both receive 0 units of utility, if both defect, they receive the -1 units of utility, if one defects and the other cooperates, the defecting agent receives 1 unit of utility, while the cooperating agent receives -2 unit of utility. The agents make their choices simultaneously and cannot communicate during each individual round.
    \item \textbf{Iterated game:} Instead of playing a single round, each pair of agents play multiple rounds. We adopt a variant of the iterated prisoner's dilemma (IPD) in which the total number of rounds is random (and drawn from a geometric distribution). We set the expected number of rounds to 50. The final payout to a pair after playing IPD is the sum of the utilieis they received in each round. IPD admits rich strategies since agents who can effectively cooperate, and punish defection, may outcompete dishonest opponents. However, dishonest opponents may exploit cooperative opponents if they can mislead the cooperative opponent. Thus, IPD provides an interesting example of decision problems involving cooperation and deception \cite{axelrod1981evolution,chong2007iterated}. 
    \item \textbf{Sufficient Parameterization:} If agents can remember all actions played over $m$ rounds, then, since each round has $2^2 = 4$ possible action combinations, the number of total strings an agent may remember is of size $4^m$. A complete policy specification for agents with perfect memory would consist of a function mapping from each of the $4^m$ possible histories to a probability of cooperating. This would represent a sufficient parameterization, with attribute space $\Omega = [0,1]^{4^m}$. While this attribute space is sufficiently rich to express all perfect memory agents, it is evidently opaque. Agents with memories of length 10 would have policies defined by $10^6$ attributes.
    \item \textbf{Auto-regressive Policies:} For interpretability, we adopt a simpler policy parameterization. Agents policies are parameterized by three numbers: an innate preference for cooperation over defection, $p_*$, a memory or learning rate parameter, $\alpha$, which controls how quickly agents adapt their policies and the time-scale with which the influence of old information decays, and a reactive parameter, $\gamma$, which controls how, and with what severity, agents react to their opponents actions. At each round, an agent possesses a probability of cooperating, $p(j)$. On the first round, the agent uses their innate preference $p(1) = p_*$. On subsequent rounds, their policies update autoregressively: 
    $$
    p(k+1) = \alpha p(k) + (1 - \alpha) \left((1 - |\gamma|) p_* + |\gamma| \tfrac{1}{2}(1 + \text{sign}(\gamma) \Delta p(a_-(k)) \right)
    $$
    where $a_-(k)$ is the opponents action on round $k$, and $\Delta p (a)$ returns $+1$ for cooperative actions and $-1$ for defections.

    \item{\textbf{Parameters:}} The attribute space $\Omega$ is $[0,1] \times [0,1] \times [-1,1]$. 

    Agents with a large innate preference $p_*$ are inherently cooperative or trusting. Agents with a small $p_*$ are inherently uncooperative or distrusting. 

    Agents with large $\alpha$ react slowly to new information. Agents with $\gamma = 0$ do not react to their opponents actions. Agents with large positive $\gamma$ react by imitating their opponent. The larger $\gamma$, the more they imitate, and the less strongly they revert towards their innate preference. $\gamma = 1$ returns ``tit-for-tat" like strategies \cite{axelrod1981evolution}, where an agent imitates their opponent in order to reward sustained cooperation, but defects immediately following any defection to punish their opponent for defection. Agents of this kind attempt to enforce a cooperative social contract with their opponent, so are ``law-enforcing". 

    Agents with negative $\gamma$ react to their opponent by adopting the opposing action. If their opponent cooperates, they see an opportunity, and defect. If their opponent defects, they attempt to re-establish trust, and cooperate. This strategy aims, in essence, to trick the opponent. This strategy has advantages. For example, when an inherently trusting law-enforcing agent ($p_* = 1, \gamma = 1)$ is paired with an inherently distrusting and deceptive agent $(p_* = 1, \gamma = -1)$, the sequence of plays will proceed deterministically according to repetitions fo the sequence (C,D), (D,D), (D,C), (C,C). On the whole, this balances the payouts received to both agents. However, the deceptive agent is always ahead or tied, while the law-enforcing agent is always behind or tied. The cumulative payout sequences are: (-2,1), (-3,0), (-2,-2), (-1,-1). This means that, if the game is stopped randomly, the deceptive agent is expected to receive a higher payout than their law-enforcing opponent. 
    
    \item \textbf{Payout:} All of the methods developed in this paper treat zero-sum, or constant-sum games. The cumulative IPD payout is not constant-sum, as illustrated in the example discussed above. Indeed, skew-symmetrizing the IPD payout directly by setting payout equal to the difference in cumulative utilities eliminates the component of the game that contributes its strategic interest. Namely, the shared reward for cooperating, and the shared cost of defection.  

    IPD was introduced to study the evolution of cooperative behavior. So, we consider an evolutionary instance of the game. When pairing types $x$ and $x'$, we initialize a well-mixed population, half of whose agents are of type $x$, and half of type $x'$. The population updates through a Moran process \cite{moran1958random}. At each time step of the Moran process, the members of the population are broken into randomly assigned pairs. They play the IPD game, and are assigned a fitness equal to their cumulative payouts. On the next round, we draw a new population of the same size, where each child is randomly assigned a parent from the previous population. The probability distribution over parents is set to a soft-max of their cumulative fitness. This has the effect of rewarding cooperation since agents who effectively cooperate with their opponent both receive large cumulative payouts, so produce more children in the subsequent round. The process terminates when the one type goes extinct, and the dominating type is declared the winner. The probability that the dominating type is $x$, when paired with $x'$ is the fixation probability, $p_{\text{fix}}(x,x';N,\lambda)$ where $N$ and $\lambda$ are the total initial population size and the softmax parameter (20, and 1 in the experiments reported).

    Then, the payout $f(x,x')$ is set equal to $2 \times (p_{\text{fix}}(x,x';N,\lambda) - 1/2).$ This payout is skew-symmetric by construction, but encodes the evolutionary reward for cooperative behavior in addition to the evolutionary reward for successful defection. 
    
\end{enumerate}

We adopted this example for demonstration since it illustrates a number of the key points of the paper. First, it does not reduce to a simple hierarchy. Second, while it uses a low-dimensional attribute space, with interpretable parameters, the exact mapping between attributes and outcome is opaque. Direct calculation of $f$ is nontrivial, since it requires accounting for all possible play trajectories, and all possible trajectories of the Moran process. Direct simulation is easy to construct, but is similarly limited since it requires repeatedly sampling outcomes between populations of types $x$ and $x'$ to estimate the fixation probability, each outcome requires running an evolutionary process, and each time step of the evolutionary process encodes many repetitions of the IPD process, one for every pair of agents. Even the IPD process itself is stochastic under the autoregressive policies. As such, while it seems plausible that, given the relatively simple individual components, a clear strategic interpretation should exist, it is non-obvious how to recover or represent it. 

Figure 1 shows the resulting payout matrices for a sample population of 800 agents, with population distribution $\pi$ given by independently sampling $p_* \sim \text{Beta}(0.7,0.7)$, $\alpha \sim \text{Beta}(1,1)$ and $\gamma = 2\times(\eta - 1/2)$ where $\eta \sim \text{Beta}(3,3)$. 

Figure 4 illustrates the leading disc game for the resulting embedding. This disc game accounts for 90\% of the overall variance in performance. The next two disc games account for 6\% and 2\% of the variance respectively. 

\subsection{Proofs of Embedding Results}

\subsubsection{Proof of Lemma 1} \label{app: lemma 1}

\begin{snugshade}
     \noindent \textbf{Lemma 1: [Opponent Independent Learning Implies Transitivity]} If there exists a transformation $T$ such that $g$ satisfies desiderata (1) and $\Psi$ is connected\footnote{Can we generalize to drop this constraint?}, then $f$ is perfectly transitive.
 \end{snugshade}

 \noindent \textbf{Proof: } Suppose that desiderata (1) applies, and $\Psi$ is connected. Then $v(y,y') = v(y)$, and, by the fundamental theorem of calculus,
 \begin{equation}
 \begin{aligned}
     & g(y,y') - g(y_0,y') = \int_{w = y_0}^{y} v(w) dw \\
     & g(y,y') - g(y,y_0') = -\int_{w = y_0'}^{y'} v(w) dw. 
\end{aligned}
 \end{equation}
 The latter equality holds since $f$ is skew-symmetric, thus $g$ must also be skew-symmetric. If $g$ is skew-symmetric, then reversing the order of the arguments converts the first expression into the second.
 
 Also, by skew-symmetry $g(y_0,y_0') = 0$ if $y_0 = y_0'$ for any $y_0$. Therefore, adding zero:
 \begin{equation}
     g(y,y') = g(y,y') - g(y_0,y_0) = (g(y,y') - g(y_0,y')) + (g(y_0,y') - g(y_0,y_0)) = \int_{w = y_0}^{y} v(w) dw - \int_{w = y_0}^{y'} v(w) dw.
 \end{equation}

Let $r(y) = \int_{w=y_0}^{y} v(w) dw$. Then $g(y,y') = r(y) - r(y')$ for all pairs $(y,y') \in \Psi \times \Psi$, so $g$ is perfectly transitive. If $g$ is perfectly transitive, then $f$ is perfectly transitive. $\square$

\subsubsection{Proof of Lemma 2} \label{app: lemma 2}

\begin{snugshade}
    \noindent \textbf{Lemma 2: [Student Independent Learning Implies Bilinearity]} If $v(y,y')$ does not depend on $y$, then $g(y,y')$ is a bilinear function of $y$ and $y'$ of the form:
    \begin{equation} \label{appeqn: bilinear form}
        g(y,y') = \left[y^{\intercal},1 \right] G \left[ \begin{array}{c} y' \\ 1 \end{array} \right]
    \end{equation}
    where $G$ is skew-symmetric.
\end{snugshade}

\noindent \textbf{Proof:} Suppose that the optimal training vector is student independent. Then, $v(y,y')$ can be rewritten, $v(y')$ , since $v$ does not depend on $y$. By definition, $\nabla_w g(w,y')|_{w = y} = v(y,y') = v(y')$. That is, the gradient of $g(y,y')$ with respect to $y$ depends only on $y'$. So, for each fixed $y'$, $g(y,y')$ is an affine function of $y$. Since $g$ is skew symmetric, it must also be affine in $y'$ for any fixed $y$. It follows that $g$ must take the form:
\begin{equation}
    g(y,y') = y^{\intercal} G y' + a^{\intercal} y + b^{\intercal} y' + c
\end{equation}
for some choice of the matrix $G$, vectors $a$ and $b$, and vector $c$. Since $g$ is skew-symmetric, $g(0,0) = 0$. It follows that $c = 0$. 

Then, by skew symmetry:
\begin{equation}
    y^{\intercal} G y' + a^{\intercal} y + b^{\intercal} y' = g(y,y') = - g(y',y) = - y'^{\intercal} G y - a^{\intercal} y' - b^{\intercal} y.
\end{equation}

Subtracting the left-hand side from the right-hand side leaves:
\begin{equation}
    y^{\intercal} (G + G^{\intercal}) y' + (a + b)^{\intercal} (y + y') = 0 \text{ for all } y,y'
\end{equation}

Picking $y = y' = t z$ such that $(a + b)^{\intercal} z \neq 0$, then varying $t$ produces a quadratic function of $t$. A quadratic function of $t$ is only zero for all $t$ if its coefficients equal zero. This requires that the projection of $a + b$ onto the range of possible $y$ is $0$, so the projection of $a$ onto the range of possible $y$ equals the projection of $-b$ onto the range of possible $y$. Any remaining component vanishes in the inner-product, thus may be dropped. Therefore, $a = -b$. 

Then, matching the bilinear terms requires:
\begin{equation}
    y^{\intercal}(G + G^{\intercal}) y' = 0
\end{equation}
for all pairs $y$ and $y'$. This requires $G = -G^{\intercal}$, so the matrix $G$ must be skew-symmetric. Then:
\begin{equation}
    g(y,y') = y^{\intercal} G y' + a^{\intercal} (y - y').
\end{equation}

We can extend this to a bilinear form when $a \neq 0$ by expanding $G$. Set:
\begin{equation}
    G \leftarrow \left[\begin{array}{cc} G & -a \\ a^{\intercal} & 0 \end{array} \right]
\end{equation}

Then, $G$ is skew-symmetric, and $g(y,y')$ takes the form \eqref{appeqn: bilinear form}. $\square$

Notice that the linear term $a^{\intercal}(y - y') = a^{\intercal} y - a^{\intercal} y'$ is transitive with $r(y) = a^{\intercal} y$. Thus, if the bilinear term is zero, then the game is perfectly transitive. When the bilinear term is not zero, then the game is not perfectly transitive, thus the second desiderata is more general than the first, which requires perfect transitivity. 


\subsubsection{Proof of Theorem 3} \label{app: theorem 3}

\begin{snugshade}
    \textbf{Theorem 3: [Disc Game Embeddings Are Sufficient]} Any coordinate transformation satisfying desiderata (2) using finitely many coordinates is related to a disc game embedding by a linear transformation.
\end{snugshade}

\textbf{Proof:} Suppose that there exists a transform $T$ satisfying desiderata (2). Then $g(y,y') = f(T^{-1}(y),T^{-1}(y'))$ is bilinear. All bilinear functions, skew-symmetric functions admit a disc-game embedding as, all finite-dimensional, skew-symmetric, bilinear functions may be expressed as a linear combination of finitely many separable functions, thus are degenerate. All degenerate functions are disc-game embeddable.

To find a particular embedding, suppose that:
\begin{equation}
    g(y,y') = \left[y^{\intercal},1\right] G \left[\begin{array}{c} y' \\ 1 \end{array}\right].
\end{equation}
for some skew-symmetric $G$. Then, Schur-decompose $G$, and regroup terms:
\begin{equation}
\begin{aligned}
    f(x,x') & = g(T(x),T(x')) = \left[T(x)^{\intercal},1\right] G \left[\begin{array}{c} T(x') \\ 1 \end{array}\right] = \left[T(x)^{\intercal},1\right] Q D_{\omega} U D_{\omega}^{1/2} Q^{\intercal}  \left[\begin{array}{c} T(x') \\ 1 \end{array}\right] \\
    & = \left(D_{\omega}^{1/2} Q^{\intercal}  \left[\begin{array}{c} T(x) \\ 1 \end{array}\right] \right)^{\intercal} U \left(D_{\omega}^{1/2} Q^{\intercal}  \left[\begin{array}{c} T(x') \\ 1 \end{array}\right] \right)
\end{aligned}
\end{equation}

If we let:
\begin{equation}
    z(x) = D_{\omega}^{1/2} Q^{\intercal} \left[\begin{array}{c} T(x') \\ 1 \end{array}\right]
\end{equation}
then:
\begin{equation}
    f(x,x') = \sum_{k} \text{disc}(z^{(k)}(x),z^{(k)}(x'))
\end{equation}
where $z(x)$ is a linear transformation of $y(x)$. When this transform is invertible, then $y$ may be recovered from $z$ by a linear transformation. If not, then $G$ is not full rank, so has a nontrivial nullspace, and any restriction of $y$ to a subspace perpendicular to the nullspace of $G$, produces a valid coordinate system satisfying desiderata (2), that may be recovered from $z$ via a linear transformation. $\square$
\vspace{0.1 in}

\subsubsection{Proof of Lemma 6} \label{app: lemma 6}

\begin{snugshade}
    \textbf{Lemma 6: [Outcome Continuity]} If $|f(x,x'') - f(x',x'')| \leq \delta$ for almost all $x''$ with respect to $\nu$, or $\|f(x,\cdot) - f(x',\cdot)\|_{\nu} = \mathbb{E}_{X'' \sim \nu}[(f(x,x'') - f(x',x''))^2] \leq \delta^2$ then $\|y^{(k)}(x) - y^{(k)}(x')\|_2 \leq |\lambda_{2k-1}|^{-1/2} \delta$.
\end{snugshade}

\textbf{Proof:} First, if $|f(x,x'') - f(x',x'')| \leq \delta$ for almost all $x''$ with respect to $\nu$, then $\mathbb{E}[(f(x,X'') - f(x',X'')^2] \leq \delta^2$ when $X'' \sim \nu$. Therefore, if the lemma statement holds for the two-norm $\|f(x,\cdot) - f(x',\cdot)\|_{\nu}$ it will also hold for the infinity norm. Both cases can be shown by writing $y^{(k)}$ in terms of the eigenfunctions of $F_{\nu}$.

Recall that $y^{(k)}(x)$ is the vector in $\mathbb{R}^2$ whose coordinates correspond to $\sqrt{|\lambda_{2k-1}|} \phi_k(x)$ in the complex plane. Therefore, $\|y^{(k)}(x) - y^{(k)}(x')\|_2 = \sqrt{|\lambda_{2k-1}|}\|\phi_{2k-1}(x) - \phi_{2k-1}(x')\|_2$. We state our bounds blockwise, but not coordinate-wise, since the embedding maps are only uniquely defined up to rotation within each block.

Now, since $\phi_j = \lambda_j F_{\nu}[\phi_j]$:
\begin{equation}
\begin{aligned}
    \|\phi_j(x) - \phi_j(x')\|^2 & = \frac{1}{|\lambda_{j}|^2} \|\mathbb{E}_{X'' \sim \nu}[(f(x,X'') - f(x',X'')) \phi_j(X'')]\|^2 \\
    & \leq \frac{1}{|\lambda_{j}|^2} \mathbb{E}_{X'' \sim \nu}[|f(x,X'') - f(x',X'')|^2] \mathbb{E}_{X'' \sim \nu}[\|\phi(X'')\|^2] \leq \frac{\delta^2}{|\lambda_j|^2}
\end{aligned}
\end{equation}
where the second step follows by the Cauchy-Schwarz inequality for the inner product $\langle \cdot, \cdot \rangle_{\nu}$, and the last step follows from the facts that the eigenfunctions are normalized with respect to $\| \cdot\|_{\nu}^2$. 

Then:
\begin{equation}
    \|y^{(k)}(x) - y^{(k)}(x')\|_2 = \sqrt{|\lambda_{2k-1}|}\|\phi_{2k-1}(x) - \phi_{2k-1}(x')\| \leq \frac{\delta}{\sqrt{|\lambda_{2k-1}|}.} \quad \blacksquare
\end{equation}

\subsection{Alternate Derivation of the Disc Game Embedding} \label{app: vector field}

We derived the disc-game embedding by seeking a coordinate transformation that satisfied desiderata (2). Since we will propose disc-game embedding as a canonical representation for learning dynamics, it is important to highlight other possible rationales for disc-game embedding. 

For example, consider a single agent involved in a training process. Let $\{x(j)\}_{j=1}^{n}$ represent the attributes of the agent at some sequence of times. With each step, the agent measures their relative improvement in performance by evaluating $ f(x(j+1),x(j))$. At the end of the training process, they evaluate their actual improvement via $f(x(n),x(1))$. It is natural to expect that, their accumulated incremental advantage, $\sum_{j=1}^{n-1} f(x(j+1),x(j))$ might approximate their ultimate advantage, $f(x(n),x(1))$. However, there is no guarantee that this approximation is accurate. Moreover, there is no obvious way to predict the size of the error given the sequence of agent types.

The error in the approximation, $\sum_{j=1}^{n-1} f(x(j+1),x(j)) - f(x(n),x(1))$ is an example of a curl. By convention, set $x(n+1) = x(1)$. Then, the sequence $\{x(j)\}_{j=1}^n$ form a cycle, $\mathcal{C}$ of agent types. By the skew-symmetry of $f$, the difference between accumulated incremental advantage, and ultimate advantage is $\text{curl}_{\mathcal{C}}[f] = \sum_{j=1}^{n} f(x(j+1),x(j))$. 

Let $T$ denote an invertible coordinate change. Then $\text{curl}_{\mathcal{C}}[f] =  \text{curl}_{T(\mathcal{C})}[g]$ where $y = T(x)$, and $g(y,y') = f\cdot(T^{-1} \times T^{-1})$. Let's choose $T$ so that the curl is easy to predict from the sequence of embedded coordinates $\{y(j)\}_{j=1}^n$. For example, we might seek a coordinate change such that, the curl is zero, for all sets of colinear agents. Then, when $\{y(j)\}_{j=1}^n$ all lie on some line in the embedding space, the cumulative incremental advantage received by stepping along the line will correctly predict the ultimate advantage an agent at one end possesses over the other. 

\begin{snugshade}
    \noindent \textbf{Desiderata:} Identify a transform $T$ such that:
    \begin{enumerate}
        \item[4.] $\text{curl}_{\mathcal{C}}[g] = 0$ for all cycles $\mathcal{C} = \{y(j)\}_{j=1}^{n+1}, y(n+1) = y(1)$ such that $\{y(j)\}_{j=1}^{n+1}$ are colinear .
    \end{enumerate}
\end{snugshade}

Like desiderata (2), desiderata (4) restricts the class of possible $g$. 
\begin{snugshade}
    \noindent \textbf{Lemma A.1: } A symmetric, zero-sum, differentiable functional form game satisfies desiderata (4) if and only if it is a vector field game. A symmetric, zero-sum, differentiable functional form game is a vector field game if and only if its payout function $g$ can be satisfies:
    \begin{equation}
        g(y,y') = \int_{z=y'}^{y} v_{\text{self}}(z) \cdot dz
    \end{equation}
    where the path integral is evaluated over the line segment frim $y'$ to $y$. 
\end{snugshade}

In a vector-field game, the payout $y$ receives against $y'$ can be expanded as a path-integral against the optimal self-play gradient along the line segment linking $y'$ to $y$. Note that this is, in essence, a fundemental theorem of calculus. The ultimate advantage gained by training along a line equals the accumulated advantage gained by each training step. Indeed, when $y$ is close to $y'$ and $g$ is differentiable, then the path-integral provides an asymptotically accurate approximation to $g(y,y')$ by linearization. However, although the path integral is a fundamental theorem in essence, it is not the fundamental theorem, and does not apply in general. It fails in general since $g(y,y')$ is simply a function evaluation in the product space $\Psi \times \Psi$, while the path integral is evaluated over a sequence of different local neighborhoods of $(z,z)$ in the product space, where $z$ ranges along the line from $y'$ to $y$. When $y'$ is far from $y$, the set of local neighborhoods along the line segment from $y'$ to $y$ may be far removed from the point $(y,y')$. 

\begin{snugshade}
    \noindent \textbf{Theorem A.2: [Vector Field Games]} The set of vector-field games is:
    \begin{enumerate}
        \item a vector space of functions on the product space,
        \item non-empty and isomorphic the set of continuous vector fields on $\Psi$,
        \item a proper subset of the set of all symmetric, zero-sum games on $\Psi \times \Psi$ (there exist non-vector-field games).
    \end{enumerate}
    
    Moreover, an analytic symmetric, zero-sum, functional form game is not a vector-field game if, for any $n \neq m$, both greater than zero, there exists a directional derivative $\partial_t^n \partial_s^m g(z+tv, z+sv) \neq 0$  at some $z \in \Psi \subseteq \mathbb{R}^d$, for all directions $v \in \mathbb{R}^d$.
\end{snugshade}

It follows that, all quadratic polynomial payouts define vector-field games. Generic third-order polynomials do not define vector-field games. The disc game is bilinear, so is a vector-field game. The disc game is a particularly convenient vector-field game since by Stoke's theorem, the curl over any cycle of competitors equals twice the sum over each pair of consecutive coordinates, of the signed-area traced out by the cycle in that pair of coordinates. Thus, loops with large areas correspond to sequences of competitors where the accumulated incremental advantage is a poor predictor of ultimate advantage. 

A functional form game $(\Omega, f)$ is vector-field representable if there exists an invertible, differentiable coordinate transformation $T$ such that $(\Psi, g)$ is a vector-field game. Then, accumulated incremental advantage equals true advantage accrued over paths corresponding to straight lines in $\Psi$. These paths are geodesics with respect to Euclidean distance in the embedding coordinates. Then, $f(x,x')$ equals the path integral over the geodesic linking $x'$ to $x$ against the self-play vector field $v_{\text{self}}$ in $\Omega$. Note that there may not be a geodesic that remains in $\Omega$ connecting all pairs $x,x'$ if $\Psi$ is not connected and convex. 

A priori, it is unclear which games are vector-field representable. This class is at least as large as the set of all continuous vector fields on $\Omega$, and should be much larger since we did not enforce many constraints on $T$. 

Here, the observation that essentially all games of interest are disc-game embeddable offers a clear answer. If a game is disc-game embeddable, then, since the disc-game is a vector-filed game, the original game is vector-field representable, and the disc-game embedding provides a sufficient transform $T$. Therefore, the disc-game embedding accomplishes desiderata (4) for essentially all games of interest. Note that desiderata (4) is less restrictive than desiderata (2), so there may be other, transforms that satisfy deisderata (4) that are not affine transformations of a disc game embedding. Nevertheless, the disc-game embedding offers a sufficient construction for almost all games of interest, and, as shown in the main text, offers more than a vector-field representation alone.

\subsection{Additional Learning Dynamics} \label{app: dynamics}

Evolutionary game theory studies the dynamics of populations evolving subject to an underlying game. For classical surveys, focusing primarily on economics, see \cite{friedman1991evolutionary,fudenberg1998learning,weibull1997evolutionary}.

\subsubsection{Self Play and Adaptive Dynamics}

The simplest learning dynamic is self-play. Self-play assumes a monomorphic population. In a monomorphic population, all individuals have the same type, i.e. share the same traits. Self-play may be used to study learning when the population consists of a single agent that trains against copies of itself. Thus, assuming self-play, we can replace the study of a time-evolving distribution $\pi(\cdot,t)$ over $\Omega$, with the time evolution of the traits of the single individual, $x(t)$. 

Let $v_{\text{self}}(x) = v(x,x) = \nabla_{w} f(w,x)|_{w = x}$ return the direction of fastest increase in $f$ for an agent of type $x$ paired against themselves. The vector field $v_{\text{self}}(x)$ gives the optimal training direction for an individual of type $x$ learning under self-play, when constrained to incremental adjustments. A self-play dynamic sets:
\begin{equation} \label{eqn: self play}
    \frac{d}{dt} x(t) \propto v_{\text{self}}(x(t)) = \nabla_{w} f(w,x(t))|_{w = x(t)}.
\end{equation}

Here, $\propto$ allows time-dependent proportionality, as when a designer adopts a decaying learning rate. We will generally ignore any time-dependent scaling factor since all such dynamics are equivalent under nonlinear transformations of time. 

In a training example, the gradient $\nabla_w f(w,x)|_{w = x}$ may be approximated using automatic differentiation, or by simulating many competition events between an agent of type $x$, and opponents of type $x'$ for $x' \approx x$.

Self-play is a form of adaptive dynamics. Like self-play, adaptive dynamics studies the change in a trait vector, $x(t)$, that acts as a proxy for a distribution. The form of the proxy, and specific dynamics, depend on the approximations used. In adaptive dynamics,
\begin{equation}
    \frac{d}{dt} x(t) \propto C(x,t) v_{\text{self}}(x(t))
\end{equation}
for some positive semi-definite matrix $C(x,t)$.

The choice of $C(x,t)$ depends on the method used to derive the adaptive dynamics equation. For example, if mutations are rare on the time scale of fixation, and only produce small changes in traits, then an evolving population will be near to monomorphic most of the time. Then, $x(t)$ may be chosen to represent the traits of the majority of the population or the centroid of the distribution of traits. Then $x(t)$ changes due to a combination of random mutation and the fixation of mutant types. By adopting a long time scale relative to both fixation and mutation, an analyst can predict the expected dynamics of $x(t)$ by averaging over many possible mutations and fixation events. Then  $C(x,t)$ depends on the distribution of mutations. 

Alternately, adaptive dynamics equations can be derived by assuming a highly concentrated population. That is, a population consisting of mostly similar individuals. In this case, the dynamics of $x(t)$ can be approximated using closure approximations. For example, if the original population obeys a replicator dynamic, is Gaussian distributed, and is driven by a quadratic payout function $f$, then the time-evolution of the centroid of the population obeys an adaptive dynamics equation. In this setting, $C(x,t)$ is the covariance in the traits of the population. 

\subsubsection{Fictitious Self-Play}

Fictitious self-play adapts self-play by recording the history of an evolving agent with type $x(t)$. Instead of adapting $x(t)$ along the direction of best response to the current type, fictitious self-play adapts along the average best response direction against the full past of $x(t)$. By incorporating the history of $x(t)$, fictitious self-play can avoid training in cycles when presented with rock-paper-scissors-type games that involve advantage cycles. 

Specifically, fictitious self-play sets:
\begin{equation} \label{eqn: fictitious self-play}
    \frac{d}{dt} x(t) \propto \int_{s = 0}^t \kappa(t,s) v(x(t),x(s)) ds 
\end{equation}
where $\kappa(t,s)$ is a kernel used to weight the contribution of time $s$ to the update of the agent at time $t$. 

Different choices of $\kappa$ produce different dynamics. For example, if $\kappa = \frac{1}{t}$ weights all past agent types uniformly when choosing the update at time $t$, while $\kappa(t,s) \propto k(t - s)$ for some monotonically decreasing function $k$ weights the average to attend more to recent types than past types. 

\subsubsection{Simultaneous Gradient Ascent}

Fictitious self-play updates an agent by adapting in the average best-response direction with respect to all the past variants of the agent. Given a population containing multiple types, the average over past types can be replaced with an average against all types contained in the current population.

Consider a discrete population containing $n$ individuals of types $\{x(j,t)\}_{j=1}^n$. Then, the population obeys simultaneous gradient-ascent if every individual adapts in their average best-response direction against opponents drawn from the current population:
\begin{equation} \label{eqn: simultaneous grad ascent discrete}
    \frac{d}{dt} x(j,t) \propto \frac{1}{n}\sum_{i \neq j} v(x(j,t),x(i,t))
\end{equation}

Alternately, consider a continuous population represented by a density function $\pi(\cdot,t)$. Then, if the population obeys simultaneous gradient ascent:
\begin{equation} \label{eqn: simultaneous grad ascent continuous}
    \partial_t \pi(x,t) \propto - \nabla_x \cdot \left( \mathbb{E}_{Y \sim \pi(\cdot,t)}[v(x,Y)] \pi(x,t) \right) = -\mathbb{E}_{Y \sim \pi(\cdot,t)}\left[ \nabla_{x} \cdot (v(x,Y) \pi(x,t)) \right].
\end{equation}

Like fictitious self-play, simultaneous gradient ascent can be extended to incorporate past agents by extending the average to run over the full population and over all past times $s \leq t$.

\subsection{Explicit Solutions} \label{app: explicit solutions}

\subsubsection{Advective Dynamics}

Here, we present explicit solutions for self-play, fictitious self-play, and simultaneous gradient ascent on discrete populations. Our solutions are restricted to the interior of the set of admissable agents $\Psi$. 

First, a note of caution. The learning dynamics that advect agents along average optimal training vectors (self-play, fictitious self-play, simultaneous gradient ascent), cannot, barring very restrictive assumptions, be reduced to transformations of a set of canonical solutions in an intrinsic space since the transformed dynamics retain a nontrivial dependence on distances in the original (extrinsic) space. Thus, even if essentially all payouts can be replaced with the disc game, different choices of the original coordinate system lead to different dynamics in the latent space. 

Let $J(x)$ denote the Jacobian of the transformation from $x$ to $y$. Then, by the chain rule
\begin{equation}
\begin{aligned}
    & \frac{d}{dt} y(t) = \frac{d}{dt} T(x(t)) = J(x(t)) \frac{d}{dt} x(t) \\
    & v(x,x') = \nabla_w f(w,x')|_{w = x} = \nabla_w T(w)^{\intercal} U T(x')|_{w = x} = J(x)^{\intercal} U T(x') = J(x)^{\intercal} U y' = J(x)^{\intercal} v_{\text{self}}(y').
\end{aligned}
\end{equation}

Therefore, any learning dynamic that equates $\frac{d}{dt} x(t)$ to a weighted average of $v(x,x')$ over some set of $x'$ of the form:
\begin{equation} \label{eqn: average training response}
    \frac{d}{dt} x(t) \propto \mathbb{E}_{x'}\left[ v(x,x') \right]
\end{equation}
sets:
\begin{equation} \label{eqn: generalized adaptive dyanmics}
 \frac{d}{dt} y(t) = \left[ J(x(t)) J^{\intercal}(x(t)) \right] \mathbb{E}_{y'} \left[ v_{\text{self}}(y') \right] = C(x(t)) \bar{v}_{\text{self}}(t) = C(x(t)) v_{\text{self}}(\bar{y}(t)) 
\end{equation}
where $C(x) = J(x) J(x)^{\intercal}$ is the metric tensor relating differential distances in the embedded coordinate system to differential distances in the original coordinate system, $x(t) = T^{-1}(y(t))$, and $\bar{y}(t) = \mathbb{E}_{y'}[y'] = \mathbb{E}_{x'}[T(x')]$ is the centroid in the embedded coordinates of the distribution of agents used in the average \eqref{eqn: average training response}. The metric tensor $C(x)$ is positive definite when $T$ is invertible, so Equation \eqref{eqn: generalized adaptive dyanmics} generalizes the adaptive dynamics equation by averaging the optimal self-play training field over a population of possible opponents. 

Since \eqref{eqn: generalized adaptive dyanmics} depends on the metric tensor $C$, the solutions to \eqref{eqn: generalized adaptive dyanmics} depend on the specific mapping $T$ from the original coordinates $x$ into the embedded coordinates $y$. Therefore, even though the coordinate change $T$ replaces arbitrary $f$ with a fixed payout function (the disc game), the solutions to \eqref{eqn: generalized adaptive dyanmics} still vary with the choice of the original coordinate system. The dependence on the metric tensor $C$ vanishes if $J(x)$ has orthogonal columns, and consecutive pairs of columns of $J(x)$ share a constant norm for all $x$. In that case, $C$ is diagonal, with constant diagonal entries grouped in consecutive pairs, and, since $v_{\text{self}}(y) = U y$ where $U$ is block-diagonal, equation \eqref{eqn: generalized adaptive dyanmics} can be broken into a set of decoupled two-dimensional learning dynamics, each of the same form, with time-constant determined by the norm of the columns of $J$. Unfortunately, the set of transformations $T$ whose Jacobians are orthogonal is very small. For example, Louiville's theorem states that, the set of smooth conformal maps (transformations whose Jacobians are orthogonal matrices) in $d > 2$ is restricted to Mobius transformations (affine transformations, and transformations between circles and lines).\footnote{Is there a reasonable symmetry of $f$ that implies the embedding is conformal?} 

Therefore, learning dynamics of the form \eqref{eqn: average training response} do not admit a canonical solution up to a coordinate change for a reasonable generic set of payout functions. Therefore, when seeking explicit solutions, we will focus on the case when learning dynamics of the form \eqref{eqn: average training response} are applied in the disc game space directly. This significantly restricts the generality of the following explicit solutions. In particular, they limit the use of these results to artificial training settings where a rational designer can choose the learning dynamic.

\begin{enumerate}
\item{\textbf{Self-Play:}} Suppose that $y(t) = T(x(t))$ obeys a self-play ODE in the embedded coordinate space:
\begin{equation} \label{eqn: self play disc}
    \frac{d}{dt} y(t) = \lambda(t) v_{\text{self}}(y(t))
\end{equation}
for some $\lambda(t) > 0$.

Without loss of generality, we assume that $\lambda(t) = 1$. All solutions to \eqref{eqn: self play disc} are equivalent to the solution for $\lambda(t) = 1$ up to a monotonic change in the time variable. Since $v_{\text{self}}(y) = U y$ when $f(y,y') = \text{disc}(y,y')$, equation \eqref{eqn: self play disc} decouples blockwise:
\begin{equation} \label{eqn: self-play oscillator}
    \begin{aligned}
        & \frac{d}{dt} y_{2k-1}(t) = y_{2k}(t), \quad \frac{d}{dt} y_{2k}(t) = - y_{2k-1}(t).
    \end{aligned}
\end{equation}

Equation \eqref{eqn: self-play oscillator} is the phase space equation for a simple harmonic oscillator. As long as $y_{2k-1}$ and $y_{2k}$ remain on the interior of $\Psi$, it is solved by:
\begin{snugshade}
\begin{equation}
    y^{(k)}(t) = [y_{2k-1}(t), y_{2k}(t)] = \|y^{(k)}(0)\| [\cos(t - s) , -\sin(t - s)]
\end{equation}
\end{snugshade}
for the phase $s$ satisfying the initial conditions. In other words, each pair of consecutive disc-game coordinates evolve independently, and follow circular path segments at a unit rate as long as the circular path segment does not intersect the boundary of $\Psi$. 

\item{\textbf{Fictitious Self-Play:}}  Suppose that $y(t) = T(x(t))$ obeys a fictitious self-play dynamic in the embedded coordinate space. For example, consider the dynamic:
\begin{equation} \label{eqn: fictitious self play disc}
    \frac{d}{dt} y(t) = \frac{1}{t} \int_{s=0}^t v_{\text{self}}(y(s)).
\end{equation}

Since $v_{\text{self}}(y) = U y$ is linear:
\begin{equation} \label{eqn: fictitious lag}
    \frac{d}{dt} y(t) = U \bar{y}(t), \quad \bar{y}(t) = \frac{1}{t} \int_{s=0}^t y(s) ds
\end{equation}
where $\bar{y}(t)$ is the centroid of the full past trajectory $\{y(s)\}_{s=0}^t$. In, other words, if $y(t)$ evolves according to fictitious self-play in the latent space, then $y(t)$ always moves along the optimal training response direction to the average past agent. 

The lag equation \ref{eqn: fictitious lag} can be written as a first-order system of coupled ODE's by differentiating $\bar{y}(t)$:
\begin{snugshade}
\begin{equation} \label{eqn: fictitious coupled}
    \frac{d}{dt} y(t) = U \bar{y}(t), \quad \frac{d}{dt} \bar{y}(t) = \frac{1}{t} ( y(t) - \bar{y}(t) ), \quad \bar{y}(0) = y(0).
\end{equation}
\end{snugshade}
%
%
%
The factor of $1/t$ in Equation \eqref{eqn: fictitious coupled} accounts for the fact that, as $t$ increases, the average agent $\bar{y}(t)$ over the full past history becomes less sensitive to the new agent added at time $t$. In other words, as $t$ increases, $\bar{y}(t)$ moves more slowly. 

%
%

Differentiating \eqref{eqn: fictitious coupled}  with respect to $t$ a second-time produces a single second-order amplified oscillator equation:
\begin{equation}
    \frac{d^2}{dt^2} y(t) = U \frac{d}{dt} \bar{y}(t) = \frac{1}{t} U (y(t) - \bar{y}(t)) = \frac{1}{t} \left( U y(t) - \frac{d}{dt} y(t) \right). 
\end{equation}

Rearranging, and separating blockwise yields the system inhomogeneous of second-order equations:
\begin{equation} \label{eqn: fictitious second order}
\begin{aligned}
    t \frac{d^2}{dt^2} y_{2k-1}(t) = y_{2k}(t) - \frac{d}{dt} y_{2k-1}(t) \\ 
    t \frac{d^2}{dt^2} y_{2k}(t) = - y_{2k-1}(t) - \frac{d}{dt} y_{2k}(t) \\ 
\end{aligned}
\end{equation}

Equation \ref{eqn: fictitious second order} is almost solved by a logarithmic spiral with phase increasing proportional to $\sqrt{2 t}$.  That is $\theta_k(t) = \tan^{-1}(y_{2k-1}(t)/y_{2k}(t)) \simeq \sqrt{2 t} + \theta_k(0)$, and $r_k(t) = \|y^{(k)}(t)\| \simeq c_3 \exp( \theta_k(t))$ for some choice of constants. This solution is inexact, since, for small $t$, $\bar{y}(t) \approx y(t)$. Thus, for small $t$, the fictitious play solution imitates the self-play solution. Indeed, at $t = 0$, the two solutions must align, which is impossible if the radius is increasing at $t = 0$. The radius only starts to grow exponentially in the phase once $y(t)$ has separated from $\bar{y}(t)$. For example, if initialized from $y_k(0) = \bar{y}_k(0) = [1,0]$, then numerically solving the dynamic produces $\theta_k(t) \approx \sqrt{t/2}$ and $\log(r_k(t)) \approx (\theta_k(t)^2 - \theta_k(t)/\sqrt{2} - \sqrt{2} )/(\theta_k(t) + 2)$ to small errors. Both approximations are essentially exact for $t > 10$. Notice that $\log(r_k(t))$ is very nearly proportional to $\theta_k(t)$ for large $t$, but, for small $t$, is approximately constant in $t$. The solution initialized at $[1,0]$ is generic since the initial average $\bar{y}(0)$ will always match the initial agent type $y(0)$, the disc game is rotationally symmetric, so changing the initial phase only rotates the solution, and, the dynamic is linear, so scaling by some scalar $r_0$ only scales the corresponding solution by $r_0$.

The solution spirals outwards since the average $\bar{y}(t)$ lags behind $y(t)$, and, as $y$ increases, responds more slowly to $y(t)$. As $\bar{y}(t)$ slows in response, $y(t)$ travels along longer path segments that are approximately linear, with direction fixed by $\bar{y}(t)$. 

Like our solution for self-play, this solution is only valid until $y(t)$ intersects the boundary of $\Psi$. Notice that, fictitious self-play leads the sequence of agents to explore outwards, so is likely to encounter the boundary if $\Psi$ is bounded, as is usually the case when $f$ is bounded.

Equation \ref{eqn: fictitious second order} can be solved exactly using power series expansion. Without loss of generality, set $y^{(k)}(0) = [1,0]$ and $\bar{y}^{(2k)}(0) = y^{(k)}(0)$. Then, $\frac{d}{dt} y^{(k)}(0) = [0,-1]$. Let $z(t) = y_{2k-1}(t)$ and $w(t) = y_{2k}(t)$. Then, if we expand:
\begin{equation}
    w(t) = \sum_{n=0}^{\infty} \frac{w^{(n)}_{0}}{n!} t^n, \quad z(t) = \sum_{n=0}^{\infty} \frac{z^{(n)}_{0}}{n!} t^n
\end{equation}
substituting into equation \eqref{eqn: fictitious lag}, and matching coefficients yields the recursion:
\begin{equation}
    w_0^{(n+1)} = \frac{1}{n+1} z_0^{(n)}, \quad z_0^{(n+1)} = -\frac{1}{n+1} w_0^{(n)}.
\end{equation}

Substituting $w_0^{(0)} = 1$, $z_0^{(0)} = 0$ gives the correct first order derivatives at $t = 0$, $w_0^{(1)} = 0$, $z_0^{(0)} = -1$. Then, closing the recursion yields the power series expansion:
\begin{equation}
    \begin{aligned}
        & w(t) = 1 + 2 \sum_{k=1}^{\infty} \frac{(-1)^k}{(2k+1) (2k)!^2} t^{2k} \\ 
        & z(t) = 2 \sum_{k=1}^{\infty} \frac{(-1)^k}{(2k) (2k-1)!^2} t^{2k-1} \\
    \end{aligned}
\end{equation}

These power series close, producing the explicit solution for $y^{(k)}(0) = [1,0]$:
\begin{snugshade}
\begin{equation} \label{eqn: fictitious self play solution}
\begin{aligned}
    & y_{2k-1}(t) =  \sqrt{\frac{1}{2t}} \left( \text{ber}_1(2 \sqrt{t}) -  \text{bei}_1(2 \sqrt{t})\right) \\
    & y_{2k}(t) = \sqrt{\frac{1}{2t}} \left( \text{ber}_1(2 \sqrt{t}) +  \text{bei}_1(2 \sqrt{t})\right) \\
\end{aligned}
\end{equation}
\end{snugshade}
where $\text{bei}_1(x)$ and $\text{ber}_1(x)$ are the Kelvin bei and ber functions. These are, respectively, the real and imaginary parts of the Bessel function of the first kind, $J_1$. Specifically $\text{bei}_1(2 \sqrt{t}) = \text{Imag}(J_1((-1 + i) \sqrt{2 t}))$ and $\text{ber}_1(2 \sqrt{t}) = \text{Real}(J_1((-1 + i) \sqrt{2 t}))$. All other solutions can be found by first scaling the solution to match the initial radius, then rotating to match the initial phase.

\item{\textbf{Simultaneous Gradient Ascent:}} Suppose that a population of $n$ agents obeys simultaneous gradient ascent in the latent space. Let $y(j,t)$ denote the coordinates of the $j^{th}$ agent at time $t$. Then, for all $j$:
\begin{equation}
    \frac{d}{dt} y(j,t) = \lambda(t) \mathbb{E}_{I \neq j}\left[v(y(I,t)) \right] = \frac{\lambda(t)}{n-1} \sum_{i \neq j} v(y(i,t)) = \frac{\lambda(t)}{n-1} U \sum_{i \neq j} y(i,t).
\end{equation}

Without loss of generality, let $\lambda(t) = \frac{n-1}{n}$. Then:
\begin{equation}
    \frac{d}{dt} y(j,t) = U \left( \bar{y}(t) - \frac{1}{n} y(j,t) \right)
\end{equation}
where $\bar{y}(t) = \frac{1}{n} \sum_{i=1}^{n} y(i,t)$. This equation can be solved efficiently by noting that the motion of the average $\bar{y}$, and the difference between each agent and the average, $w(j,t) = y(j,t) - \bar{y}$, decouple, and can be solved independently.

Solving for the average:
\begin{equation}
    \frac{d}{dt} \bar{y}(t) = \frac{1}{n} \sum_{j=1}^n \frac{d}{dt} y(j,t)) = \frac{1}{n} U \sum_{j=1}^n \left(\bar{y}(t) - \frac{1}{n} y(j,t) \right) = \left(1 - \frac{1}{n} \right) U \bar{y}(t). 
\end{equation}

Then the centroid obeys the self-play equation with time constant $1 - \tfrac{1}{n}$. Therefore, $\bar{y}(t)$ traces a circular path at rate $1 - \tfrac{1}{n}$. 

Let $w(j,t) = y(j,t) - \bar{y}(t)$. Then:
\begin{equation}
    \frac{d}{dt} w(j,t) = \frac{d}{dt} y(j,t) - \frac{d}{dt} \bar{y}(t) = U \left( \bar{y}(t) - \frac{1}{n} y(j,t) \right) - \left(1 - \frac{1}{n} \right)U \bar{y}(t) = - \frac{1}{n} U (y(j,t) - \bar{y}(t)) = -\frac{1}{n} U w(j,t).
\end{equation}
Therefore, $w(j,t)$ obeys a time-reversed self-play equation with time constant $1/n$. 

Combining these results produces an epicyclic solutions where the centroid of the population moves clockwise at rate $1 - \tfrac{1}{n}$, and where each individual orbits counterclockwise about the centroid at rate $\tfrac{1}{n}$:
\begin{snugshade}
    \begin{equation} \label{eqn: epicycles}
        y^{(k)}(j,t) = \left[ \begin{array}{c} y_{2k-1}(j,t) \\ y_{2k}(j,t) \end{array} \right] = \| \bar{y}^{(k)}(0)\| \left[ \begin{array}{c} \cos( \tfrac{n-1}{n}(t - \bar{s}^{(k)})) \\ -\sin( \tfrac{n-1}{n}(t - \bar{s}^{(k)})) \end{array} \right] + \| y^{(k)}(j,0) - \bar{y}^{(k)}(0)\| \left[ \begin{array}{c} \cos( \tfrac{1}{n}(t - s^{(k)}(j))) \vspace{0.03in} \\ \sin( \tfrac{1}{n}(t - s^{(k)}(j))) \end{array} \right]
    \end{equation}
\end{snugshade}
for appropriate phase offsets $\bar{s}$ and $s$.

The continuous population equation can be solved by taking the limit as $n$ approaches infinity. As $n$ approaches infinity the motion of the centroid approaches the self-play solution, while the rate of backward rotation about the centroid converges to zero. Thus, in the limit as $n$ approaches infinity, every agent in the population obeys a self-play solution centered at their initial offset from the population centroid. Equivalently, the entire distribution translates in a circular orbit at a unit rate. So, while averaging against the past, as in fictitious self play, leads to rapid outward pressure towards the boundary of $\Psi$, averaging over a current population leads to a bulk dynamic that is, in essence, equivalent to self-play.

Like the self-play, and fictitious self-play solutions, the epicyclic solution \eqref{eqn: epicycles} only applies when all the individuals are on the interior of $\Psi$. 

\end{enumerate}

The three solutions developed above are each special cases of more general dynamics. For example, changing the kernel used for time-averaging would change the fictitious self-play solution, as would any interaction with the boundary of $\Psi$. Nevertheless, these results provide simple intuition, and demonstrate some of the analytic advantages of the disc game payout. In each case, we used bilinearity to replace an average over optimal response directions with the optimal response direction to an average agent, used the block structure of the disc game to isolate separate pairs of coordinates that evolve independently, then used the linearity of the optimal response vector field, to solve the associated dynamical system. More realistic dynamics may not admit such clean analysis, however, the advantages leveraged here remain useful, even when the dynamic cannot be explicitly solved, or decoupled between pairs of coordinates. We use a detailed case study of the replicator equation to illustrate the efficacy of the disc-game embedding for dynamics that do not admit such easy solutions. 

Our choice to focus on interior solutions is limiting since the particulars of different games are encoded in the boundaries of $\Psi$, and, if the dynamic was defined in a different initial coordinate system, in the metric tensor associated with the coordinate change. For example, if the underlying game is perfectly transitive, then $\Psi$ is a line segment, in which case $\Psi$ has no proper interior. Developing explicit solutions for different boundary geometries is an interesting avenue for future work.

\subsubsection{Replicator Dynamics: The Transitive Case} \label{app: transitive solution}

Suppose that $f(x,x') = r(x) - r(x')$ for some rating function $r:\Omega \rightarrow \mathbb{R}$. Then, $f$ is a perfectly transitive function \cite{strang2022network}. When $f$ is perfectly transitive, the replicator equation can be solved exactly for any initial distribution. In particular (see Appendix section \ref{app: transitive solution}),
\begin{equation}
    \pi(x,t) \propto e^{t P r(x)}\pi(x,0), \text{ where } P = P(0).
\end{equation}

To confirm the solution, differentiate with respect to time. Let $Z(t) = \int_{x \in \Omega} e^{t P r(x)} \pi(x,0) dx$ be the proportionality factor needed to hold the population fixed. Then:
$$
\partial_t \pi(x,t) = \partial_t \frac{e^{t P r(x) }}{Z(t)} \pi(x,0) = \left(r(x) - \partial_t \log(Z(t)) \right) \frac{e^{t P r(x)}}{Z(t)} \pi(x,0) = \left(r(x) - \partial_t \log(Z(t)) \right) \pi(x,t).
$$

Expanding the logarithmic derivative:
$$
\partial_t \log(Z(t)) = \frac{\partial_t Z(t)}{Z(t)} = \int_{x \in \Omega} \frac{\partial_t e^{t P r(x)}}{Z(t)} \pi(x,0) = \int_{x \in \Omega} r(x) \frac{ e^{t P r(x)}}{Z(t)} \pi(x,0) = \mathbb{E}_{X' \sim \pi(\cdot,t)}[r(X')].
$$

Then:
$$
\partial_t \pi(x,t) = \left(r(x) - \partial_t \log(Z(t)) \right) \pi(x,t) = \mathbb{E}_{X' \sim \pi(\cdot,t)}[r(x) - r(X')] \pi(x,t) = \mathbb{E}_{X' \sim \pi(\cdot,t)}[f(x,X')] \pi(x,t)
$$
so the proposed solution satisfies the replicator equation. $\square$

\subsection{Generic Parameter Dynamics Proofs} 

\subsubsection{Proof of Result 1} \label{app: proof of parameter dynamics}

Suppose that $(\Omega, f)$ is a functional form game, that $\pi(x,0)$ is a density over $\Omega$ with support $S$. Suppose that $f$ is of finite rank on $S$. Finally, suppose that $\pi(x,t)$ is subject to the continuous replicator equation. Parameterise $\pi(x,t)$ according to the exponential family parameterisation with basis functions set to the disc game embedding functions for $f$ and some reference measure $\nu$ supported on $S$. Then, the parameter dynamics must solve the closure equation. Simplifying the closure equation gives:
\begin{equation}
    \sum_{k=1}^{\text{rank}(F)} y^{(k)}_1(x) \frac{d}{dt} \theta_1^{(k)}(t) + y^{(k)}_2(x) \frac{d}{dt} \theta_2^{(k)}(t) = \sum_{k=1}^{\text{rank(F)}/2} y^{(k)}_1(x) \mathbb{E}_{X \sim \pi(\cdot;\theta(t))}[y^{(k)}_2(X)] - y^{(k)}_2(x) \mathbb{E}_{X \sim \pi(\cdot;\theta(t))}[y^{(k)}_1(X)] \text{ for all } x \in S. 
\end{equation}

Notice that both the left and right hand sides are a linear combination of the disc game embedding functions. Since the disc game embedding functions are linearly independent, the left and right hand sides match for all $x \in S$ if and only if the coefficients of the combination on the left and right-hand sides match. 

Therefore, the parameters obey the closed system of $r = \text{rank}(F)$ ODE's:
\begin{equation} \label{eqn: parameters blockwise}
    \text{ for all } k \in [1,2,\hdots,\text{rank}(F)/2] \quad \begin{cases} & \frac{d}{dt} \theta^{(k)}_1(t) = \mathbb{E}_{X \sim \pi(\cdot;\theta(t))}[y^{(k)}_2(X)] \\
    & \frac{d}{dt} \theta^{(k)}_2(t) = -\mathbb{E}_{X \sim \pi(\cdot;\theta(t))}[y^{(k)}_1(X)]
    \end{cases}
\end{equation}

Expressed vector-wise:
\begin{equation}
     \frac{d}{dt}\left[ \begin{array}{c} \theta^{(k)}_1(t) \\  \theta^{(k)}_2(t)\end{array} \right] = R \mathbb{E}_{X \sim \pi(\cdot;\theta(t))}\left[\begin{array}{c} y^{(k)}_1(X) \\  y^{(k)}_2(X)  \end{array} \right] \text{ for all } k \in [1,2,\hdots,\text{rank}(F)/2].
\end{equation}
where $R$ is the two-by-two ninety degree rotation matrix. 

Let $\theta^{(k)} = [\theta_1^{(k)},\theta_2^{(k)}]$ and $y^{(k)} = [y_1^{(k)},y_2^{(k)}]$. Then, recall that $v(y) = R y$ is the optimal self-play vector field for a disc game. Then, the parameter dynamics can be expressed blockwise:
\begin{equation} \label{eqn: parameter dynamics disc game vector field}
    \frac{d}{dt}\theta^{(k)}(t) = \mathbb{E}_{X \sim \pi(\cdot;\theta(t))}[v(y^{(k)}(X))] = v(\bar{y}^{(k)}(t)) \text{ where } \bar{y}^{(k)}(t) = \mathbb{E}_{X \sim \pi(\cdot;\theta(t))}[y^{(k)}(X)]
\end{equation}
for all  $k \in [1,2,\hdots,\text{rank}(F)/2]$.

Equation \eqref{eqn: parameter dynamics disc game vector field} shows that the parameters, $\theta$ always move in the best response direction to the centroid of the population in the disc game coordinates. In a sense, this equation imitates simultaneous gradient ascent in the disc-game space, whose right-hand side is also equal to the optimal self-play vector field of the population centroid in the disc game space (see Appendix section \ref{app: explicit solutions}). We will show that, like simultaneous gradient ascent for normal form games on finite strategy spaces, equation \eqref{eqn: parameter dynamics disc game vector field} defines a Hamiltonian dynamic, so may be interpreted as a system of coupled nonlinear oscillator equations.

To show that the parameter dynamics are Hamiltonian, define the total population function:
\begin{equation}
    P(\theta) = \int_{x \in \Omega} \pi(x;\theta) dx = \int_{x \in \Omega} \pi(x,0) \exp(\sum_{k=1}^r \theta^{(k)} \cdot y^{(k)}(x)) = \mathbb{E}_{y^{(k)}(X) \mid X \sim \pi(\cdot,0)}[\exp(\sum_{k=1}^r \theta^{(k)} \cdot y^{(k)}(x))] = M_{\pi_{y}(\cdot,0)}[\theta].
\end{equation}

The total population function is the Laplace transform of the initial population distribution in the disc game coordinates. The Laplace transform of a distribution is its moment generating function, $M$. Derivatives of a moment generating function recover moments of the corresponding distribution.

In this case, the centroid of the population distribution in the disc-game space can be expressed as the gradient of the total population function with respect to the parameters $\theta$:
$$
\begin{aligned}
    & \bar{y}_j(t) = \mathbb{E}_{X \sim \pi(\cdot,t)}[y_j(X)] = \int_{x \in \Omega} y_j(x) \pi(x,t) dx = \int_{x \in \Omega} y_j(x) e^{\sum_i \theta_i(t) y_i(x) } \pi(x,0) dx \\ 
    & = \int_{x \in \Omega} \partial_{\theta_j} e^{\sum_i \theta_i(t) y_i(x)} \pi(x,0) dx = \int_{x \in \Omega} \partial_{\theta_j} \pi(x;\theta(t)) dx = \partial_{\theta_j} P(\theta).
\end{aligned}
$$

Therefore:
\begin{equation}
    \bar{y}(t) = \mathbb{E}_{X \sim \pi(\cdot;\theta(t))} = \nabla_{\theta} P(\theta(t)).
\end{equation}

Then, equation \eqref{eqn: parameter dynamics disc game vector field} admits the block Hamiltonian form:
\begin{equation}\label{eqn: block Hamiltonian parameter dynamics}
\begin{aligned}
    & \frac{d}{dt}\theta_1^{(k)}(t) = \partial_{\theta^{(k)}_2} P(\theta(t)), \quad
    & \frac{d}{dt}\theta_2^{(k)}(t) = -\partial_{\theta^{(k)}_1} P(\theta(t)) \\
\end{aligned}    
\end{equation}
for all  $k \in [1,2,\hdots,\text{rank}(F)/2]$. 

In vector form:
\begin{equation} \label{eqn: Hamiltonian parameter dynamics}
    \frac{d}{dt} \theta(t) = U \nabla_{\theta} P(\theta(t))
\end{equation}
where $U$ is the block diagonal matrix with two-by-two diagonal blocks equal to the ninety-degree rotation matrix $R$. $\square$

\subsubsection{Properties of the Hamiltonian Dynamic} \label{app: parameter dynamics proofs}

\begin{enumerate}
    \item \textbf{Population Conservation:} Multiplication by $U$ rotates by 90 degrees. So, $\frac{d}{dt} \theta = U \nabla_{\theta} P(\theta)$ is orthogonal to $\nabla_{\theta} P(\theta)$. Therefore, the parameters always move tangent to the level sets of $P$, thus conserve $P$.

    \item \textbf{Interaction Rate and Hamiltonian:} Let $R(P) = \int \rho(P)/P dP$. Then, for any time-autonomous interaction rate:
    \begin{equation}
        U \nabla_{\theta} R(P(\theta)) = U \frac{\rho(P)}{P} \nabla_{\theta} P(\theta) =  \frac{\rho(P)}{P} U\nabla_{\theta} P(\theta) = \frac{d}{dt} \theta(t).
    \end{equation}

    \item \textbf{Volume Preservation:} The Hamiltonian dynamics are also volume preserving. Let $\Theta(0)$ denote a closed set of parameters $\theta$, and let $V(0)$ equal the volume of the region enclosed by $\Theta(0)$. Let $\Theta(t)$ denote the set of parameters formed by applying the parameter dynamic to every initial parameter vector in $\Theta(0)$. Let $V(t)$ represent volume of the region enclosed by $\Theta(t)$. Then, $V(t) = V(0)$.

    The proof follows via the divergence theorem. The rate of change in the volume at any given time is the integral over the interior of $\Theta(t)$ of the divergence of the vector field defining the rate of change of the parameters. Hamiltonian dynamics are divergence free. In particular,
    $$
    \nabla_{\theta} \cdot \frac{d}{dt} \theta(\theta) = \nabla_{\theta} \cdot U \nabla_{\theta} P(\theta) = \sum_{k=1}^{r/2} \partial^2_{\theta^{(k)}_1,\theta^{(k)}_2} P(\theta) - \partial^2_{\theta^{(k)}_2,\theta^{(k)}_1} P(\theta)  = 0.
    $$

    Notice that, the divergence vanishes since the divergence restricted to each pair of coordinates vanishes. It follows that, the parameter dynamics are also volume preserving for any set of parameters $\Theta(0)$ that only vary over subsets of the parameter pairs. Suppose, for instance, that the $\theta \in \Theta(0)$ only differ in a subset of the coordinates $J \subseteq [1,2,\hdots r]$. Let $|J|$ denote the number of coordinate  along which the parameters vary, and $V(t)$ denote $2|J|$ dimensional volume. Then, $V(t)$ is also conserved, $V(t) = V(0)$, provided $2k \in J$ if $2k - 1 \in J$. That is, if $J$ is a set of consecutive coordinate pairs. In particular, if $\Theta(0)$ is constant in all parameters except for one parameter pair, then $V(t)$ is an area. Then, the parameter dynamics are area preserving in each pair of parameters corresponding to paired disc-game embedding coordinates.

    \item \textbf{No Bounded Attractors:} A bounded attractor is a set of finite parameters that can be enclosed in an open neighborhood of finite parameters such that any parameter trajectory initialized in the neighborhood converges onto the attractor. This neighborhood is the basin of attraction of the attractor. Since the dynamic is volume preserving, no such attractor can exist. 
    
    The proof proceeds by contradiction. Suppose such an attractor existed. Initialize a set of parameters inside the basin of attraction of the attractor, $\Theta(0)$, such that $\Theta(0)$ encloses the entire attractor. Then, by definition of the basin of attraction, $\Theta(t)$ converges to the attractor as time progresses. However, since the attractor is contained inside $\Theta(0)$, $\Theta(t)$ cannot converge onto the attractor without losing volume.

    It follows that no isolated, finite attractor exists. If such an attractor existed it would admit a Lyapunov function, and level sets of the Lyapunov function would enclose a volume that would have to both decrease over time, and remain constant.

    \item \textbf{Interior Steady States are Critical Points:} The rotation matrix $U$ is unitary, so its kernel is trivial. Therefore, $U \nabla_{\theta} P(\theta) = 0$ if and only if $\nabla_{\theta} P(\theta) = 0$. So, all interior equilibria correspond to critical points of the total population function. 

    All interior steady state distributions correspond to interior equilibria since the set of basis functions are linearly independent. Therefore, if $\frac{d}{dt} \theta(t) \neq 0$, then $\partial_t \pi(x,t) \neq 0$ for some $x$.

    By manipulating the Jacobian of $U \nabla_{\theta} P(\theta)$ about a critical point, we can show that maxima and minima of $P(\theta)$ correspond to centers in the parameter dynamics, while saddle points remain saddles. Minima correspond to centers whose direction of rotation aligns with the self-play vector field. Maxima correspond to centers whose direction of rotation runs backwards against the self-play vector field.

    \item \textbf{Convexity:} The total population function $P(\theta)$ is convex. 
    
    \noindent \textbf{Proof:} The entries of the Hessian are,
    $$
    \begin{aligned}
     \partial^2_{\theta_i, \theta_j} P(\theta) = \int_{x \in \Omega} \partial^2_{\theta_i,\theta_j} e^{\theta \cdot y(x)} \pi(x,0) dx = \int_{x \in \Omega} y_i(x) y_j(x) e^{\theta \cdot y(x)} \pi(x,0) dx = \int_{x \in \Omega} y_i(x) y_j(x) \pi(x;\theta) dx.
    \end{aligned}
    $$

    Therefore, the Hessian of the total population function at $\theta$ equals both the Gram matrix for the embedding functions $y(\cdot)$ with respect to the inner product $\langle \cdot,\cdot \rangle_{\pi(\cdot;\theta)}$, and, equivalently, the covariance in the embedding coordinates of an agent drawn at random from the population specified by the parameters $\theta$:
    \begin{equation}
        [\partial_{\theta}^2 P(\theta)]_{ij} = \langle y_i(\cdot),y_j(\cdot) \rangle_{\pi_Y(\cdot,;\theta)} = \text{Cov}_{Y \sim \pi_Y(\cdot;\theta)}[Y_i,Y_j].
    \end{equation}

    Both Gram matrices and covariance matrices are positive semi-definite. It follows that the Hessian of the total population function is positive semi-definite at all $\theta$. Any function whose Hessian is positive semi-definite at all inputs is convex. Therefore, $P(\theta)$ is a convex function. $\square$

    \item \textbf{Strict Convexity:} More strongly, the total population function is strictly convex. 
    
    \textbf{Proof:} Recall, a function is strictly convex if its Hessian is positive definite everywhere. Since $P$ is convex, it is sufficient to show that the Hessian is invertible everywhere. 

    The Hessian at $\theta$ equals the Gram matrix $\langle y(\cdot), y(\cdot) \rangle_{\pi_Y(\cdot;\theta)}$. Recall that, for any finite $\theta$, $\pi(\cdot;\theta)$ is supported everywhere on $S$, the support of $\pi_0$. Recall also that the embedding functions $y$ depend on the reference measure $\nu$, but changing the reference measure amounts to a linear change of basis when the support of the measure is fixed. Since $\nu$ was chosen with support $S$, we can change measure from $\nu$ to $\pi(\cdot;\theta)$. 
    
    Let $\phi$ denote the coordinates of the new parameterization. Then, $\theta(\phi) = (Y^{(\nu)}_{\pi(\cdot;\theta)})^{\intercal} \phi$. Since the matrix $Y^{\nu}_{\mu}$ responsible for the change of basis is invertible, the Hessian of $P(\theta)$ is invertible at $\theta(\phi)$ if and only if the Hessian of $P(\phi)$ is invertible.

    The Hessian of $P(\phi)$ is the Gram matrix for the embedding functions  defined with the reference measure $\pi(\cdot;\theta)$. The embedding functions are always orthogonal under the inner product defined by their reference measure. So, the Hessian for $P(\phi)$ is diagonal entries, with diagonal entries equal to the norms of the embedding functions. Since none of the embedding functions are identically zero, the Hessian of $P(\phi)$ is diagonal with nonzero entries, so is invertible. It follows that the Hessian of $P(\theta)$ is invertible for all finite $\theta$, so $P$ is strictly convex. $\square$

    \item \textbf{All Interior Equilibria are Isolated:} The total population function is strictly convex. Any interior critical point of a strictly convex function is a unique, isolated, global minimizer of the function.

    \item \textbf{Radially Unbounded:} A function is radially unbounded if it diverges whenever the norm of the argument diverges. So, to show that $P(\cdot)$ is radially unbounded, it is enough to show that $P(\theta)$ diverges if $\|\theta\|$ diverges. We will show that, if an interior equilibrium exists, then $P(\cdot)$ is radially unbounded.

    \textbf{Proof:} If an interior equilibrium exists, then $P(\cdot)$ is strictly convex and admits a global minimizer. A strictly convex function is strictly convex along any line segment. Consider a ray in parameter space, of the form $\theta(s) = \theta_* + \phi s$ for $s \geq 0$ and any $\phi \in \mathbb{R}^r$. Since $\theta_*$ is an isolated equilibrium, it is the unique global minimizer of $P(\cdot)$. It follows that $P(\theta(s))$ must be monotonically increasing in $s$. Any monotonically increasing, strictly convex function of $s$ must diverge as $s$ approaches infinity. Therefore, $P(\cdot)$ diverges along every ray leaving the equilibrium. Since the equilibrium is on the interior, and thus $\|\theta_*\| < \infty$, $P(\cdot)$ also diverges along every ray leaving the origin. Thus, $P(\cdot)$ is radially unbounded. $\square$

    \item \textbf{The Interior is Invariant:} If $P(\cdot)$ is radially unbounded, then its level sets are all bounded, as, for any $p \in [P(\theta_*), \infty)$, there exists a minimal radius $R(p)$ such that $P(\theta) \leq p$ for any $\theta$ such that $\|\theta\| \leq R(p)$. Bounded convex sets are compact, so, when an interior equilibrium exists, the level sets of $P$ are boundaries of nested compact regions. It follows that, if an interior equilibrium exists, then the parameters mix on compact sets.

    \item \textbf{All Interior Equilibria are Centers:} The total population function $P(\theta)$ is strictly convex so does not admit any maxima or saddles. Moreover, any interior local minimizer of $P(\theta)$ is an isolated global minimizer. Minima of $P(\theta)$ correspond to centers, since rotating the gradient vector field about a minimizer by ninety-degrees produces a circulating vector field that circulates in the direction of the applied rotation. More strongly, if an interior equilibrium exist, then perturbations off the equilibrium mix on the boundaries of nested compact sets containing the equilibrium, so neither converge to or diverge from the equilibrium. This remains true for perturbations of any finite size. In other words, all interior equilibria are globally Lyapunov stable centers.

    \item \textbf{Poincar\'e Recurrence:}  If $P(\theta)$ admits a global minimizer then, as it is strictly convex, it is radially unbounded, so its level sets are bounded. Any volume-preserving dynamic whose trajectories are restricted to bounded sets is Poincar\'e recurrent \cite{caratheodory1919wiederkehrsatz,poincare1889probleme}. $\square$

\end{enumerate}

\subsubsection{Proof of Corollary 5.3} \label{app: cor 5.3}

\begin{snugshade}
\noindent \textbf{Corollary 5.3: [Interior Equilibria, Invariant Interior, and the Range of the Embedding]} Suppose that $(\Omega,f)$ is a functional form game with skew-symmetric $f$, $S$ is a support set, and $\pi(x,t)$ is a population distribution supported on $S$ evolving according to the replicator equation. Let $\nu$ denote an arbitrary reference measure that has support equal to $S$. Let $\Psi = \{y_{\nu}(x)|x \in S\} \subseteq \mathbb{R}^r$ denote the image of the set $S$ under the disc-game embedding. 
\vspace{0.1 in}

\noindent Suppose that $\Psi$ is bounded. Then, the replicator parameter dynamics admit an interior equilibrium, and the parameters mix on compact sets, if and only if the convex hull of $\Psi$ contains the origin in its interior.
\end{snugshade}

\noindent \textbf{Proof:} Note that $\theta_*$ is an equilibrium if and only if $v(\bar{y}[\pi(\cdot;\theta_*)]) = U \bar{y}[\pi(\cdot;\theta_*)] = 0$. The rotation matrix $U$ is invertible, so $\theta_*$ is an equilibrium if and only if the corresponding centroid, $\bar{y}_* = \bar{y}[\pi(\cdot;\theta_*)] = 0$. In other words, $\theta_*$ is an equilibrium if and only if the corresponding distribution of traits is centered after embedding in the disc-game space. Recall that the origin plays a special role in disc-games since it corresponds to a neutral type that has no advantage or disadvantage over any other type. 

Let $\Psi = \{y(x) \mid x \in S\}$ denote the image of the support set $S$ after embedding. Let $\text{Conv}(\Psi)$ denote the convex hull of the image of $S$ after embedding. If $\text{Conv}(\Psi)$ does not contain the origin, then no distribution over $\Psi$ is centered. More strongly, since, for finite $\theta$, $\pi_Y(\cdot;\theta)$ is supported on all of $S$, $\bar{y}(\theta)$ is contained in the interior of the convex hull of $\Psi$ for all finite $\theta$. Therefore, if the interior of the convex hull of $\Psi$ does not contain the origin, then the parameter dynamics do not admit interior equilibrium.

To show the converse, we will show that $P(\cdot)$ is radially unbounded if the origin is contained inside the interior of the convex hull of $\Psi$. If $P(\cdot)$ is radially unbounded, then it cannot admit an infimum at infinity, must have compact level sets, and must admit an interior global minimizer.

First, $P(\cdot)$ is radially unbounded if and only if $\log(P(\cdot))$ is radially unbounded since $P(\cdot)$ is positive valued, and $\log(P)$ is monotonically increasing and unbounded as $P$ diverges. 

To show that $\log(P)$ is radially unbounded, note that $\nabla_{\theta} \log(P(\theta)) = \frac{1}{P(\theta)} \mathbb{E}_{Y \sim \pi(\cdot;\theta)}[Y]$ is the centroid of the normalized trait distribution in the embedding space. Let $\{\theta(s) = s \phi\}_{s > 0}$ denote a ray leaving the origin. Then, let $p(\cdot;s) = \frac{1}{P(\theta(s))} \pi_Y(\cdot;\theta(s))$. Let $\bar{y}(s) = \mathbb{E}_{Y \sim p(\cdot;s)}[Y]$. Then, since $\pi_Y(y;\theta) =  \pi_Y(y,0) \exp(\theta \cdot y)$, $\lim_{s \rightarrow \infty} \bar{y}(s) = y_*(\Psi,\phi)$ where $y_*(\Psi,\phi)$ is the intersection of the boundary of the convex hull of $\Psi$ with the ray $\{y(s) = \phi s\}_{s > 0}$. Then, if $\phi$ is normalized, $y_*(\Psi,\phi) = d(\Psi,\phi) \phi$ where $d(\Psi,\phi)$ is the distance from the origin to the boundary of $\Psi$ along the direction $\phi$. By assumption, $d(\Psi,\phi) > 0$ for all $\phi$. 

Choose $\phi$ normalized. Then, for any ray leaving the origin of parameter space along the direction $\phi$, $\lim_{s \rightarrow \infty} \partial_s \log(P(\theta(s))) = \lim_{s \rightarrow \infty} \bar{y}(s) \cdot \phi = d(\Psi,\phi) > 0$. Then, along every ray leaving the origin, there exists a sufficiently large distance $s_*(\Psi,\phi)$ such that, $\partial_s \log(P(\theta(s))) > 0$ for all $s > s_*(\Psi,\phi)$. Therefore, $\log(P(\theta)$ diverges along all rays leaving the origin, so is radially unbounded. It follows that $P$ is radially unbounded. $\square$

\subsubsection{Proof of Theorem 6} \label{app: theorem 6}

\begin{snugshade}
\noindent \textbf{Theorem 6: } Suppose that $\pi$ obeys a generalized replicator dynamic with frequency dependent growth rates $g(\pi)$. Let $h^{-1}$ be the inverse of $h = \int^\pi 1/g(s) ds$.  Then, $h^{-1}(u) = \sum_{n=0}^{\infty} h_n u^n$, the parameters $\theta$ obey a Hamiltonian dynamic of the form:
\begin{equation}
    \frac{d}{dt} \theta(t) = U \mathbb{E}_{X \sim \pi(\cdot;\theta(t))}[y(X)] = U \nabla_{\theta} H(\theta)
\end{equation}
with Hamiltonian:
\begin{equation}
    H(\theta) = \int_{x \in \Omega} \int_{0}^{\theta \cdot y(x) + u_0(x)} h^{-1}(u) du
\end{equation}
\end{snugshade}

\noindent \textbf{Proof:} Expanding $f$ using the disc game decomposition gives:
$$
y(x) \cdot \frac{d}{dt} \theta(t) = y(x) \cdot U \int_{x' \in \Omega} y(x') h^{-1}(\theta(t) \cdot y(x') + u_0(x)) dx'
$$

Since the disc game embedding functions are linearly independent, the coefficients defining the combination of embedding functions on each side must match. Therefore:
$$
\frac{d}{dt} \theta(t) = U \int_{x'} y(x') h^{-1}(\theta(t) \cdot y(x') + u_0(x)) dx' = U \int_{x'} y(x') \pi(x';\theta(t)) dt = U \mathbb{E}_{X' \sim \pi(\cdot;\theta(t))}[y(X')] = v(\bar{y}(\theta(t))).
$$
So, just as before, the rate of change in the parameters equal the optimal response vector evaluated at the centroid of the embedded population. This result will always hold when $\mathbb{E}_{X' \sim \pi}[f(x,X')]$ is expanded onto the embedding functions.

It remains to show that the expectation can be expressed as the gradient of a scalar-valued function of $\theta$. Define the function:
\begin{equation}
    H(\theta) = \int_{x \in \Omega} \int_{0}^{\theta \cdot y(x) + u_0(x)} h^{-1}(u) du 
\end{equation}

Then:
$$
\partial_{\theta_j} H(\theta) = \int_{x \in \Omega} \partial_{p} \left[\int_{0}^p h^{-1}(u) du \right]_{p = \theta \cdot y(x) + u_0(x)} \partial_{\theta_j} \theta \cdot y(x) dx = \int_{x \in \Omega} h^{-1}(\theta \cdot y(x) + u_0(x)) y_j(x) dx = \mathbb{E}_{X \sim \pi(\cdot;\theta(t))}[y(X)]. 
$$

It follows that the generalized replicator dynamic is Hamiltonian in its parameters with Hamiltonian $H$. $\square$

\subsubsection{Proof of Lemma 20} \label{app: lemma 18}

\begin{snugshade}
\noindent \textbf{Lemma 20: [Strict Convexity]} If $f$ is bounded, $0 \leq g(\pi)$ with equality if and only if $\pi = 0$, and $g(\pi) = \mathcal{O}(\pi)$ as $\pi \rightarrow 0$ and $\pi \rightarrow \infty$, then the Hamiltonian function $H(\cdot)$ is a strictly convex, second-differentiable function of $\theta$ for all interior $\theta$.
\end{snugshade}

\noindent \textbf{Proof:} The Hamiltonian has Hessian entries are:
$$
\begin{aligned}
    \partial^2_{\theta_i \theta_j} H(\theta) & = \partial_{\theta_i} \int_{x \in \Omega} h^{-1}(\theta \cdot y(x) + u_0(x)) y_j(x) dx \\ & = \int_{x \in \Omega} \frac{d}{du} h^{-1}(u)|_{u = \theta \cdot y(x) + u_0(x)} y_i(x) y_j(x) dx = \int_{x \in \Omega} \left[ \frac{d}{dp} h(p) \right]_{p = h^{-1}(\theta \cdot y(x)) + u_0(x)}^{-1} y_i(x) y_j(x) \\
    & = \int_{x \in \Omega} \left[1/g(p) \right]^{-1}_{p = h^{-1}(\theta \cdot y(x) + u_0(x))} y_i(x) y_j(x) dx = \int_{x \in \Omega} g(h^{-1}(\theta \cdot y(x) + u_0(x)) y_i(x) y_j(x) \\ & = \int_{x \in \Omega} g(\pi(x;\theta)) y_i(x) y_j(x).
\end{aligned}
$$

It follows that, the Hessian of the Hamiltonian is:
\begin{equation}
\textbf{H}_{ij}(\theta) = [\partial_{\theta}^{2} H(\theta)]_{i,j} = \int_{x \in \Omega} g(\pi(x;\theta)) y_i(x) y_j(x) dx.
\end{equation}
if the integral converges. 

If $f$ is bounded, then the embedding functions are bounded so $y(X)$ has finite moments of all order, for any distribution defined on $\Omega$. It remains to show that the integral defines a finite, positive-definite matrix for all interior $\theta$ reachable by the dynamic in dinite time. Showing that the integral is finite establishes second-differentiability. Showing that it is positive definite establishes strict convexity. 

Suppose that $g(\pi)$ is chosen to be $\mathcal{O}(\pi)$. Then, the support of $\pi$ is preserved under the dynamic, and $\int_{x \in \Omega} g(p(x)) dx < \infty$ whenever $\int_{x \in \omega} p(x) dx < \infty$ for all density functions $p$ since $\int_{x \in \Omega} g(p(x)) dx < C \int_{x \in \omega} p(x) dx$ for some $C < \infty$. Define:
$$
P^g(\theta) = \int_{x \in \Omega} g(\pi(x;\theta)) dx, \quad \pi^g(x;\theta) = \frac{1}{P^g(\theta)} g(\pi(x;\theta)).
$$
Then, since $g$ is nonnegative valued, $\pi^g(x;\theta)$ is a normalized density function. Then:
$$
\textbf{H}(\theta) = P^g(\theta) \mathbb{E}_{X \sim \pi^g(\cdot;\theta)}[y(X) y(X)^{\intercal}] \succeq P^g(\theta) \text{Cov}_{X \sim \pi^g(\cdot;\theta)}[y(X)] = P^g(\theta) \text{Cov}_{Y \sim \pi^g_y(\cdot;\theta)}[Y].
$$

It follows that $\textbf{H}$ is positive definite whenever the covariance of $Y \sim \pi^g_y(\cdot;\theta)$ is full rank (has rank $r$). The covariance is full rank whenever the convex hull of the support of $\pi^g_y$ is a set of nonzero measure in $\mathbb{R}^r$. Since $g(p) = 0$ if and only if $p = 0$, the support of $\pi^g_y$ is the support of $\pi_y$. Since $g$ is chosen so that the dynamic is support preserving, the support of $\pi_Y$ is $\Psi$. Suppose that the convex hull of $\Psi$ has nonzero measure in $\mathbb{R}^r$. Then $Y \sim \pi^g_y(\cdot;\theta)$ must as well, so $\textbf{H}$ must be positive definite for all $\theta$ reachable under the dynamic. It follows that $H$ is strictly convex on all $\theta$ reachable under the dynamic from an initial condition supported on $\Omega$. 

The last constraint on $\Psi$ fails for transitive functions, or if the collection of embedding functions that include the constant function, as when one of the disc game components is perfectly transitive. Since the embedding functions are linearly independent, at most one is constant, so the convex hull of $\Psi$ is, generically contained within a linear subspace of codimension at most one. The latter condition, however, is sufficient, not necessary. 

The Hessian is:
$$
\textbf{H}(\theta) = P^g(\theta)\left(\bar{y}^g(\theta) \bar{y}^g(\theta) + \text{Cov}_{Y \sim \pi^g_y(\cdot;\theta)}[Y]\right) \text{ where } \bar{y}^g(\theta) = \mathbb{E}_{Y}
$$
so $v^{\intercal} \textbf{H}(\theta) v = 0$ requires $v \perp \bar{y}^g(\theta)$. Then, $v^{\intercal} H(\theta) v = v^{\intercal} \text{Cov}_{Y \sim \pi^g_y(\cdot;\theta)}[Y] v = \text{Var}_{Y \sim \pi^g_y(\cdot;\theta)}[v \cdot Y] = 0$ if and only if the convex hull of the support of $\pi^g_y(\cdot;\theta)$ is contained in an affine subspace of codimension one that is \textit{perpendicular} to $\bar{y}^g(\theta)$. Since $\bar{y}^g(\theta)$ is the expected value of $Y$ under $\pi^g_y(\cdot;\theta)$, it is contained inside the convex hull of the support of $\pi^g_y(\cdot;\theta)$. The only linear subspace of codimension one, that is perpendicular to a member of the subspace, is a subspace that passes through zero. However, $\Psi$ cannot be contained in a linear subspace passing through the origin, since such $\Psi$ could be rotated to include a constant function that equals zero for all $x$. The function $\phi_k(x) = 0$ is never a valid eigenfunction, nor is $y_k(x) = 0$ a valid basis function. It follows that $\Psi$ cannot be contained inside a linear subspace passing through the origin, so $\textbf{H}(\theta)$ is positive definite, so the Hamiltonian is strictly convex for all $\theta$ reachable under the dynamic from an initial condition supported on $\Omega$.  $\square$

\subsubsection{Proof of Corollary 6.1} \label{app: corollaries 6}

\begin{snugshade}
\noindent \textbf{Corollary 6.1: [Recurrence]} Suppose that the conditions of Lemma 20 hold. Then, the parameter dynamics either admit an isolated, interior equilibrium that corresponds to the choice of $\theta$ such that $\mathbb{E}_{X \sim \pi(\cdot;\theta)}[y(X)] = 0$, so correspond to a fully mixed ESS, or no such equilibrium exists. The former case requires $0 \in \text{conv}(\Psi)$ and is impossible if the convex hull of $\Psi$ does not contain the origin. If an isolated interior equilibrium exists then it is a center and the parameter dynamics are Poincar\'e recurrent.
\end{snugshade}

\noindent \textbf{Proof:} Assume that either condition of Lemma 20 hold, or, more generally, that $H$ is strictly convex and second-differentiable. Then, since the Hessian is strictly convex it either admits a unique, isolated, interior minima, which is an isolated equilibrium for the parameter dynamics, or it does not, and all parameter orbits escape to infinity. 

Suppose that there exists an isolated interior equilibrium $\theta_*$. This requires that:
\begin{equation}
    \mathbb{E}_{X \sim \pi(\cdot;\theta)}[y(X)] = \nabla_{\theta} H(\theta) = 0.
\end{equation}

Under the conditions of Lemma 20, the total population size $P(\theta) = \int_{x \in \Omega} h^{-1}(\theta \cdot y(x) + u_0(x)) dx$ is nonzero for finite $\theta$ (it only vanishes if $u(x,\theta)$ diverges in finite time, which is impossible when $f$ is bounded, as, if $f$ is bounded, then the rate of change in $\theta$ is bounded). It follows that, the unnormalized expectation can only equal zero if the normalized expectation equals zero. The normalized expectation lies inside the convex hull of $\Psi$, so, if the Hamiltonian admits an interior minima, then the convex hull of $\Psi$ must contain the origin.

Then, since the Hamiltonian is strictly convex and second-differentiable, $H(\theta)$ is bounded from below by its quadratic Taylor approximation about $\theta_*$:
$$
H(\theta) \geq H(\theta_*) + \frac{1}{2} (\theta - \theta_*)^{\intercal} \textbf{H}(\theta_*) (\theta - \theta_*) = H^{(2)}(\theta;\theta_*)
$$ 

Since $\textbf{H}(\theta)$ is positive definite, the quadratic approximation $H^{(2)}(\theta;\theta_*)$ is radially unbounded, so the Hamiltonian is radially unbounded.

It follows that the level sets of the Hamiltonian are the boundaries of a sequence of nested convex sets, and each level set is the boundary of a compact set. Then, the dynamics are both volume-preserving and bounded ($\theta(t)$ remains finite for all $t$ if $\theta(0)$ is finite). In that case, the dynamics are Poincar\'e recurrent. $\square$

\subsubsection{Proof of Corollaries 7.1 - 7.2} \label{app: corollaries 7}

\begin{snugshade}
\noindent \textbf{Corollary 7.1: [Recurrence]} The parameter dynamics either admit an isolated, interior equilibrium that corresponds to a fully mixed ESS, or no such equilibrium exists. If an isolated interior equilibrium exists, then it is a center and the parameter dynamics are Poincar\'e recurrent. An isolated interior equilibrium exists if and only if the origin is contained inside the convex hull of $\Psi$. 
\end{snugshade}

\noindent \textbf{Proof:} First, Hamiltonian dynamics are volume-preserving, so the parameter dynamics do not admit any bounded attractors or sources. This rules out any isolated interior sinks or sources. So, as for homogeneous mixing, the  dynamics may only admit attractors at the boundary, and all isolated interior equilibria are centers.

More strongly, since the Hamiltonian $H(\theta)$ is a sum of finitely many functions which are all strictly convex and second-differentiable, the combined Hamiltonian is strictly convex and second-differentiable. It follows that the system admits at most one isolated interior equilibrium. This equilibrium exists under the same geometric conditions on $\Psi$ established for the homogeneous population case. If the convex hull of $\Psi$ contains the origin, then the component Hamiltonians all admit a unique, isolated global minimzer, so their sum does as well. If the origin lies outside the convex hull of $\Psi$, then none of the component Hamiltonians admit an interior minimizer, so their sum does not either. 

If such an equilibrium exists, then all of the component Hamiltonians are radially unbounded. It follows that the composite Hamiltonian is radially unbounded, so the parameters orbit on the boundaries of compact sets. Then, since the dynamic is volume-preserving, it must also be Poincar\'e recurrent. $\square$

\begin{snugshade}
\noindent \textbf{Corollary 7.2: [Mixed Adaptive Dynamics]} Let $\bar{y}_i(t) = \mathbb{E}_{X \sim \pi_i(\cdot,t)}[y(X)]$. Let $\bar{y}(t) = [\bar{y}_1(t),\bar{y}_2(t),\hdots,\bar{y}_K(t)]$ denote the concatenation of all the centroids, $\theta(\bar{y})$ the map from the centroid back to the parameters, and $\textbf{H}$ the Hessian of the Hamiltonian. Let $v(y) = U y$ denote the optimal training vector field that directs self-play. Let $m_i = \sum_{j = 1}^k m_j$ denote the total per capita interaction rate for patch $i$. Let $\hat{M} = \text{diag}(m_1^{-1},m_2^{-1},\hdots,m_K^{-1}) M$ denote the row-stochastic version of $M$.

\noindent Then, the centroids obey an autonomous dynamic of the form:
\begin{equation}
    \frac{d}{dt} \bar{y}_i(t)  = m_i  \textbf{H}(\theta(\bar{y}(t))) v\left( [\hat{M} \bar{y}(t) ]_i \right). 
\end{equation}
\end{snugshade}

\noindent \textbf{Proof:} Let $\bar{y}_i(t) = \mathbb{E}_{X \sim \pi_i(\cdot,t)}[y(X)]$. Then:
\begin{equation}
\begin{aligned}
    \frac{d}{dt} \bar{y}_i(t) & = \frac{d}{dt} \nabla_{\theta_i} H^{(i)}(\theta^{(i)}(t),u_0^{(i)}) = \textbf{H}^{(i)}\left(\theta^{(i)};u_0^{(i)}\right) \frac{d}{dt} \theta^{(i)}(t) = \textbf{H}^{(i)}\left(\theta^{(i)};u_0^{(i)}\right) \sum_{j=1}^K m_{ij} U \bar{y}_j(t) \\ & = \textbf{H}^{(i)}\left(\theta^{(i)};u_0^{(i)}\right) \sum_{j=1}^K m_{ij} v(\bar{y}_j(t)) = m_i  \textbf{H}^{(i)}\left(\theta^{(i)};u_0^{(i)}\right) v\left(\frac{1}{m_i} \sum_{j=1}^K m_{ij} \bar{y}(t) \right)
\end{aligned}
\end{equation}
where $m_i = \sum_{j=1}^K m_{ij}$. 

As before, this equation appear non-autonomous, since the covariance is expressed as a function of the parameters. However, since the component Hamiltonians are all strictly convex, the mapping between the parameters in each patch, and the embedding centroid of each patch, is invertible. $\square$

\subsection{Solutions for Special Geometries} \label{app: special geometries}

\subsubsection{Decoupled Parameter Dynamics} \label{app: decoupled dynamics}



Let $y^{(k)} = [y_{2k-1},y_{2k}]$ denote the pair of disc game coordinates assigned to the $k^{th}$ game. Similarly, let $\theta^{(k)} = [\theta_{2k-1},\theta_{2k}]$ denote the parameters associated with the $k^{th}$ game. When do the parameter dynamics, $\theta(t)$ decouple blockwise so that $\theta^{(k)}$ evolves independently of $\theta^{(k')}$ for all $k \neq k'$? 

\begin{snugshade}
\noindent \textbf{Lemma 17: [Decoupled Parameter Dynamics]} Consider $Y \sim \pi_0$. If the set of component disc games $D = \{1,2,\hdots, r/2\}$ admits a partition $K = \{K_j\}$ where $K_j \subseteq D$, $\cup_{j} K_j = D$, and $K_j \cap K_{j'} = 0$ if $j \neq j'$ such that $Y^{(k)}$ is independent of $Y^{(k')}$ when $k$ and $k'$ are members of different sets in the partition, then each set of parameters, $\Theta_j = \{\theta^{(k)}, k \in K_j\}$ evolves independently of each other set of parameters, and, if $Y \sim \pi(\cdot,t)$, then $\{Y^{(k)}, k \in K_j\}$ remain independent of $\{Y^{(k)}, k \in K_{j'}\}$ for all $t$.  
\end{snugshade}

\noindent \textbf{Proof:} Suppose that the independence assumption holds.  Then, $\pi_0$ factors, $\pi_0(y) = \prod_{j} \pi_0^{(K_j)}(y^{(K_j)})$ and, by necessity, that $\Psi$ can be expressed as a Cartesian product, $\Psi = \Psi^{(K_1)} \times \Psi^{(K_2)} \times \hdots = \prod_j \Psi^{(K_j)}$. Then, the Hamiltonian $P(\theta)$ also factors:
\begin{equation}
    P(\theta) = \int_{y \in \Psi} e^{\theta \cdot y} d \pi_0(y) = \prod_{j} \int_{y^{(K_j)} \in \Psi^{(K)}} e^{\theta^{(K_j)} \cdot y^{(K_j)}} d\pi_0^{(K_j)}(y^{(K_j)}) = \prod_j P^{(K_j)}(\theta^{(K_j)}). 
\end{equation}

Since the Hamiltonian is conserved under the dynamic, scaling time by the population size scales the dynamical system by $P(\theta)$, which replaces the Hamiltonian with $\log(P(\theta))$. If we can show that the time scaled dynamics decouple, then the original dynamics must also decouple.

Consider the time-scaled dynamics:
\begin{equation}
    \frac{d}{dt} \theta(t) \propto U \nabla_{\theta} \log(P(\theta)) = U \nabla_{\theta} \sum_j \log(P^{(K_j)}(\theta^{(K_j)})) = U \sum_j \nabla_{\theta} \log(P^{(K_j)}(\theta^{(K_j)}))
\end{equation}

Since $\log(P^{(K_j)}(\theta^{(K_j)}))$ depends on only the parameter blocks $k \in K_j$, its gradient is zero for all blocks $k' \notin K_j$. Since $U$ only relates pairs of rows belonging to the same block, multiplication by $U$ preserves this property. It follows that $\frac{d}{dt} \theta^{(K_j)}(t)$ depends only on the parameters $\theta_{(K_j)}$, and does not depend on any parameters belonging to a different component of the partition. 

If $Y \sim \pi(\cdot,t)$, then separate components of the partition remain independent at all $t$ since $\pi(y,t) = e^{\theta(t) \cdot y} \pi_0(y) = \prod_{j} e^{\theta^{(K_j)}(t) \cdot y^{(K_j)}} \pi_0^{(K_j)}(y^{(K_j)}$. $\square$

\begin{snugshade}
\noindent \textbf{Corollary 17.1: [Fully Decoupled Parameter Dynamics]} Consider $Y \sim \pi_0$. If $Y^{(k)}$ is independent of $Y^{(k')}$ for all $k \neq k'$, then each pair of parameters $\theta^{(k)}$ evolves independently of each other pair of parameters, and, if $Y \sim \pi(t)$, then $Y^{(k)}$ remains independent of $Y^{(k')}$ for all $t$. 
\end{snugshade}

\noindent \textbf{Proof:} The proof is a direct application of Lemma 17. $\square$


\begin{snugshade}
\noindent \textbf{Corollary 17.2: [Fully Decoupled Parameter Dynamics are Quasi-Periodic]} Consider $Y \sim \pi_0$. If $Y^{(k)}$ is independent of $Y^{(k')}$ for all $k \neq k'$, and the origin is contained in the convex hull of the support of $\pi_0$, then each pair of parameters $\theta^{(k)}(t)$ evolve periodically.
\end{snugshade}

\noindent \textbf{Proof:} The proof is immediate from previous results. If the assumptions of 18.1 hold, then each pair of parameters evolve independently. If the origin is contained in the convex hull of $\pi_0$, then there exists a fully mixed Nashed equilibria with the same support as the initial distribution, so, by Corollary 5.3, the dynamics are recurrent. All two-dimensional, recurrent dynamics are periodic. $\square$ 

\subsubsection{Rotationally Symmetric Base Measures} \label{app: rotational symmetry}


\begin{snugshade}
\noindent \textbf{Lemma 18: [Rotational Symmetry]} Let $\Psi \in \mathbb{R}^2$ be the domain for a single disc game, and suppose that $\pi_0$ is an initial population distribution that is an exponential tilting of a base measure $\nu$ that is symmetric under rotations. If the population evolves according to a continuous replicator dynamic, then its parameters, using base measure $\nu$, orbit on concentric circles at a constant rate with period equal to $2 \pi \|\theta(0)\|_2/\|\nabla_{\theta} P(\theta(0))\|_2$. In the parametrization with base measure $\pi_0$, the parameters orbit on concentric ellipses with period $2 \pi \|\theta(0)\|_2/\|\nabla_{\theta} P(\theta(0))\|_2$.
\end{snugshade}

\noindent \textbf{Proof:} If $\pi_0$ is an exponential tilt of $\nu$ where $\nu$ is rotationally symmetric then, with $\nu$ as the base measure, $P(\theta)$ must be invariant to rotations of $\theta$, so is also only a function of $\|\theta\|_2$. Then, since $P(\theta)$ is the Hamiltonian, and Hamiltonian dynamics follow level sets of their Hamiltonian function, the parameters must evolve along concentric circles. The period of each orbit is equal to the circumference of the circle divided by the rate of travel about the circle, which must also be constant, and equals $\|\nabla_{\theta} P(\theta)\|_2$ at $\theta_0$. Changing base measure corresponds to a linear transformation of the parameter space, so, using any other base measure that is absolutely continuous with respect to $\nu$ produces concentric elliptical orbits. Periods of orbits are invariant under invertible changes of coordinates. $\square$



\subsubsection{Separable Base Measures} \label{app: separable base measures}


The continuous replicator dynamics simplify further in the case when every component of $Y \sim \pi_0$ are independent. Then, $\pi_0$ fully factors into a product of marginals:
\begin{equation}
    \pi_0(y) = \prod_{j = 1}^r \pi_0^r(y_r).
\end{equation}

In this case, each separate pair of parameters evolves independently, but the parameters associated with the same disc game remain coupled through the dynamics. 

Consider a single disc game, with a base measure that factors into products of measures on its marginals (under some rotation of the coordinates). For example, we could use $\pi_0$ equal to a uniform distribution on a rectangular region, or a Laplace distribution, or any other product of marginal measures. Then:
\begin{equation}
    \pi_0(y) = \pi_0^1(y_1) \pi_0^2(y_2).
\end{equation}

In this case, the Hamiltonian fully factors:
\begin{equation}
    P(\theta_1,\theta_2) = \prod_{j=1}^2 P_j(\theta_j) \text{ where } P_j(s) = \int_{y_j} e^{s y_j} d \pi_0^j(y_j).
\end{equation}
Notice, $P_j(s)$ is the moment generating function, or Laplace transform of, the marginal measure $\pi_0^j$ for the $j^{th}$ embedding coordinate. 

Now we can write:
\begin{equation}
\partial_{\theta_1} P(\theta) = (\partial_{\theta_1} P_1(\theta_1)) P_2(\theta_2) = \frac{\partial_{\theta_1} P_1(\theta_1)}{P(\theta_1)} P_1(\theta_1) P_2(\theta_2) = (\partial_{\theta_1} \log(P_1(\theta_1)) P(\theta) \text{ and } \partial_{\theta_2} P(\theta) = (\partial_{\theta_2} \log(P_2(\theta_2)) P(\theta).
\end{equation}

This form is convenient since, $\log(P_j(s)) = K_j(s)$ is the cumulant generating function for the $j^{th}$ marginal of the base measure. Moreover, $P(\theta)$ is conserved by the dynamic, so $P(\theta(t)) = P(\theta(0)) = P$, the total population size. Then, the Hamiltonian parameter dynamics reduce to:
\begin{equation} \label{appeqn: hamiltonian for independent ys}
\begin{aligned}
    & \frac{d}{dt} \theta_1(t) = P g_2(\theta_2), \quad \frac{d}{dt} \theta_2(t) =  - P g_1(\theta_1)
\end{aligned}
\end{equation}
where:
\begin{equation}
    g_j(s) = \frac{d}{ds} K_j(s) = \frac{d}{ds} \log(P_j(s)))
\end{equation}
is the slope of the cumulant generating function. 

Equation \eqref{appeqn: hamiltonian for independent ys} is especially simple when the marginals for $Y_1$ and $Y_2$ are identical. Then we can suppress the subscripts, and, scaling time by the total population size, reduce to a normal form, nonlinear oscillator equation:
\begin{equation}
    \frac{d}{dt} \theta_1(t) \propto g(\theta_2(t)), \quad \frac{d}{dt} \theta_2(t) \propto - g(\theta_1(t)).
\end{equation}
%

In the main text, we considered an example where $\pi_0$ is a Laplace distribution, so has marginals $\pi_0^j(y_j) = \frac{1}{2 \lambda} \exp(- \lambda |y_j|)$. In this section, we will work out the solution presented in the main text.

The cumulant generating function for a Laplace distribution is:
\begin{equation}
    K(s) = -\log(1 - (s/\lambda)^2) \text{ for } |s| < \lambda
\end{equation}
so:
\begin{equation}
    g(s) = \frac{2 s/\lambda^2}{1 - (s/\lambda)^2} = 2 \frac{s}{\lambda^2 - s^2} \text{ for } |s| < \lambda. 
\end{equation}

Then, the system of coupled equations is:
\begin{equation}
    \frac{d}{dt} \theta_1(t) \propto \frac{2}{\lambda^2 - \theta_2(t)^2} \theta_2(t), \quad \frac{d}{dt} \theta_2(t) \propto -\frac{2}{\lambda^2 - \theta_1(t)^2} \theta_1(t).
\end{equation}

We can simplify further by changing variables via $\theta = \lambda \phi$. Then:
\begin{equation} \label{appeqn: normal form Laplace oscillator}
    \frac{d}{dt} \phi_1(t) \propto \frac{2}{ (1 - \phi_2(t)^2)} \phi_2(t), \quad \frac{d}{dt} \phi_2(t) \propto -\frac{2}{ (1 - \phi_1(t)^2)} \phi_1(t).
\end{equation}
where the constant of proportionality (time-rescaling) needed to recover the original dynamic is $P/\lambda^2$. We will from now on, assume without loss of generality, that $\lambda = 1$, and that we are working after rescaling time by $P$ (i.e.~using normalized distributions when computing expectations). 

Equation \eqref{appeqn: normal form Laplace oscillator} defines a separable elliptic oscillator, that, as expected, approaches a simple harmonic oscillator for small $\phi$. It is well defined for all $\phi$ in the unit square $[-1,1]^2$.  The solution trajectories are the level sets of the Hamiltonian function $((1 - \phi_1^2)(1 - \phi_2^2))^{-1}$, so, by conservation of the Hamiltonian, we can solve directly for $\phi_2$ as a function of $\phi_1$ and visa versa. In particular, 
\begin{equation}
\phi_2(\phi_1,P) = \pm \sqrt{1 - (P(1 - \phi_1^2))^{-1}}
\end{equation}
where $P$ is the value of the Hamiltonian along the level set. The level sets of the Hamiltonian converge to circles near the origin, and to squares as the parameters $\phi$ approach the boundary of the $[-1,1]^2$ square. 

The dynamical system admits an analytic solution for all initial conditions. To solve the system, rearrange the conservation constraint $(1 - \phi_2^2)^{-1} = P (1 - \phi_1^2)$. Then:
\begin{equation} \label{eqn: separable form for elliptic oscillator}
    \frac{d}{dt} \phi_1(t) = 2 \frac{\phi_2(t)}{1 - \phi_2(t)^2} = 2 P \phi_2(t) (1 - \phi_1(t)^2) = \pm 2 \sqrt{P(1 - \phi_1^2)(P(1 - \phi_1^2) - 1)}
\end{equation}
with sign determined by the initial conditions. In particular, $\text{sign}(\frac{d}{dt} \phi_1(t)) = \text{sign}(\phi_2(t))$. 

Equation \eqref{eqn: separable form for elliptic oscillator} is separable, so $\phi_1(t)$ must satisfy the integral equation:
\begin{equation}
    \int^{\phi_1(t)} \frac{du}{2 \sqrt{P(1 - u^2)(P(1 - u^2) - 1)}} = t 
\end{equation}
where $P = 1/((1 - \phi_1(0)^2)(1 - \phi_2(0)^2))$ is the initial population size. The integral on the right hand side is an elliptic integral that can be closed using Jacobi elliptic functions. To close the integral, substitute $u = \sin(\psi)$ so that $1 - u^2 = \cos(\psi)^2$ and $du = \cos(\psi) d\psi$. Then $P(1 - u^2) = P \cos^2(\psi)$ and $P(1 - u^2) - 1 = P \cos^2(\psi) - 1$. Then, the integral is:
\begin{equation}
\begin{aligned}
    \int^{\psi_1(t)}\frac{\cos(\psi)d\psi}{2 \sqrt{P \cos(\psi)^2(P \cos(\psi)^2 - 1)}} & = \int^{\psi_1(t)}\frac{d\psi}{2 \sqrt{P (P \cos(\psi)^2 - 1)}}  = \int^{\psi_1(t)}\frac{d\psi}{2 \sqrt{P (P - 1) - P^2 \sin(\psi)^2}} \\
    & = \frac{1}{2 \sqrt{P(P - 1)}} \int^{\psi_1(t)}\frac{d\psi}{\sqrt{1 - \frac{P}{P-1} \sin(\psi)^2}} \\
    &  = \frac{k}{2 \sqrt{P(P - 1)}} \int^{\psi_1(t)}\frac{d\psi}{\sqrt{k^2 - \sin(\psi)^2}} = \frac{1}{2P} \int^{\psi_1(t)} \frac{d\psi}{\sqrt{k^2 - \sin(\psi)^2}}
    \end{aligned}
\end{equation}
where $k^2 = (P - 1)/P$ and $\psi_1(t) = \arcsin(\phi_1(t))$. 

Finally, to bring the integral into the standard elliptic form, let $\sin(\psi) = k \sin(\zeta)$ so that $\psi = \arcsin(k \sin(\zeta))$, $d \psi = (1 - k^2 \sin(\zeta)^2)^{-1/2} k \cos(\zeta) d \zeta$ and $(k^2 - \sin(\psi)^2)^{-1/2} = (k^2(1 - \sin(\zeta)^2))^{-1/2} = k^{-1} \cos(\zeta)^{-1}$. Then, Equation \eqref{eqn: separable form for elliptic oscillator} produces the integral equation:
\begin{equation}
    \int^{\zeta_1(t)} \frac{d\zeta}{\sqrt{1 - k^2 \sin(\zeta)^2}} = F(\zeta_1(t)|k^2) = 2 P t + C \text{ where } \zeta_1(t) = \arcsin\left( \frac{\phi_1(t)}{k} \right)
\end{equation}
and where $C$ is a constant of integration to be determined by the initial conditions and $F(\cdot|\cdot)$ is the incomplete elliptical integral of the first kind. 

Inverting, and enforcing the conservation constraint, returns the solution:
\begin{equation}
    \phi_1(t) = k \text{sn}(2Pt + C|k^2), \quad \phi_2(t) = \text{sign}(\phi_2(0)) k \frac{\text{cn}(2P t + C|k^2)}{\text{dn}(2 P t + C|k^2)} \text{ for } \begin{cases} & P = ((1 - \phi_1(0)^2)(1 - \phi_2(0)^2))^{-1} \\
    & k = \sqrt{(P - 1)/P} \\
    & C = F(\arcsin(\phi_1(0)/k|k^2))\end{cases}
\end{equation}
and where $\text{sn}$, $\text{cn}$, and $\text{dn}$ are the elliptic sine, elliptic cosine, and delta amplitude functions. Each is an example of a Jacobi elliptic function \cite{dixon1894elementary,kovacic2016jacobi}. 


Since the solutions are all periodic orbits in the square $[-1,1]^2$ we can adopt convenient initial conditions. Set $\phi_1(0) = 0$ and $\phi_2(0) = a > 0$, where $a$ represents the amplitude of the oscillator where it crosses a coordinate axis. Then $C = 0$, $\text{sign}(\phi_2(0)) = 1$, $P = (1 - \phi_2(0)^2)^{-1}$, and $k = \sqrt{1 - 1/P} = \sqrt{1 - (1 - \phi_2(0)^2)} = \phi_2(0) = a$. Then, every solution to equation \eqref{eqn: separable form for elliptic oscillator} lies on an orbit of the form:
\begin{equation}
    \phi_1(t) = a \text{sn}(2 P t| a^2), \quad \phi_2(t) = a \frac{\text{cn}(2 P t|a^2)}{\text{dn}(2 P t|a^2)}. 
\end{equation}

The corresponding orbits are shown in Figure \ref{fig: Laplace oscillator} for amplitudes ranging from $0.1$ to $0.999$. Notice that, small amplitudes produce approximately circular orbits, while large amplitudes produce approximately square orbits. Also notice that, when the amplitude is small, the parameters orbit at an approximately constant velocity while, when the amplitude is large, the parameters dwell near the corners of the square, and move rapidly when near its edges. We explore this heteroclinic behavior in the main text, and in more detail in Appendix section \ref{app: heteroclinic orbits and dual polygons}.

The period of the orbits can also be computed in closed form since the Jacobi function $\text{sn}(\cdot|k^2)$ has period $4 K(k)$ where $K(k) = F(\pi/2|k^2) = \int_{0}^{\pi/2} (1 - k^2 \sin(\zeta)^2)^{-1/2} d\zeta$ is the complete elliptic integral of the first kind.\footnote{Not the cumulant generating function.} Therefore, the parameters orbit with amplitude dependent period:
\begin{equation}
    T(a) = 2 \frac{K(a)}{P} = 2(1 - a^2) K(a) = 2 (1 - a^2) \int_{0}^{\pi/2} \frac{d \zeta}{\sqrt{1 - a^2 \sin(\zeta)}}.
\end{equation}

For small initial amplitudes $T(a) \simeq \pi$. This period is consistent with the period of the harmonic oscillator $\frac{d}{dt} \phi_1 = 2 \phi_2, \frac{d}{dt} \phi_2 = -2 \phi_1$ generated by linearizing the nonlinear oscillator about the origin. As $a$ approaches 1 from below (large initial amplitudes), the period approaches zero, with $T(a) \sim \frac{1}{2}(1 - a^2) |\ln(1 - a^2)|$. In this example the period of the oscillator converges to zero as the initial amplitude $a$ approaches 1, since, when $a$ approaches 1, $\theta$ approaches $\lambda$, so the expected embedding coordinate $\bar{y}$, whose magnitude sets the rate of motion of the parameters, diverges.

\subsubsection{Heteroclinic Dynamics Near Boundary} \label{app: heteroclinic orbits and dual polygons}

When initialized near the minimizer of the Hamiltonian, the parameters behave like a simple harmonic oscillator. Their orbits are elliptical, and, for the appropriate choice of base measure, circular. However, when initialized far from the minimizer, the parameter orbits are typically nonelliptical. The solutions shown in Figure \ref{fig: Laplace oscillator} provide an example. The Hamiltonians plotted in the main text, show that, when the domain $\Psi$ is polygonal, then the level sets of the Hamiltonian approach polygons, so the parameters approach polygonal orbits far from the minimizer. Since the parameter orbits, and centroid orbits, are diffeomorphic, and large exponential tilts move the centroid near to the boundary of $\Psi$, these limits correspond to the limiting orbits of the adaptive dynamics equation, in the limit where the centroid approaches the boundary of the convex hull of $\Psi$. We will show that, when this boundary is polygonal, then the corresponding parameter dynamics also produce polygonal orbits, and, the corners of the convex hull of the embedding space, $\Psi$, are key geometric features that generate the limiting polygonal orbit. Moreover, we will show that, as the centroid approaches the boundary, then the centroid dynamics slow down near corners of $\Psi$, so the centroid dynamics behave like a heteroclinic oscillator near the boundaries of $\Psi$.

To study these orbits, we focus on a single disc game, assume that $\Psi$ is radially bounded so that its convex hull is compact, the convex hull contains the origin and is a polygon. To highlight the geometric relationship between the level sets of the Hamiltonian for large Hamiltonian values, and the polygonal convex hull, we will also assume that the base measure is uniform over $\Psi$. 

\begin{figure}[t]
    \centering
    \includegraphics[trim = 40 2 80 20, clip, width = 0.6\textwidth]{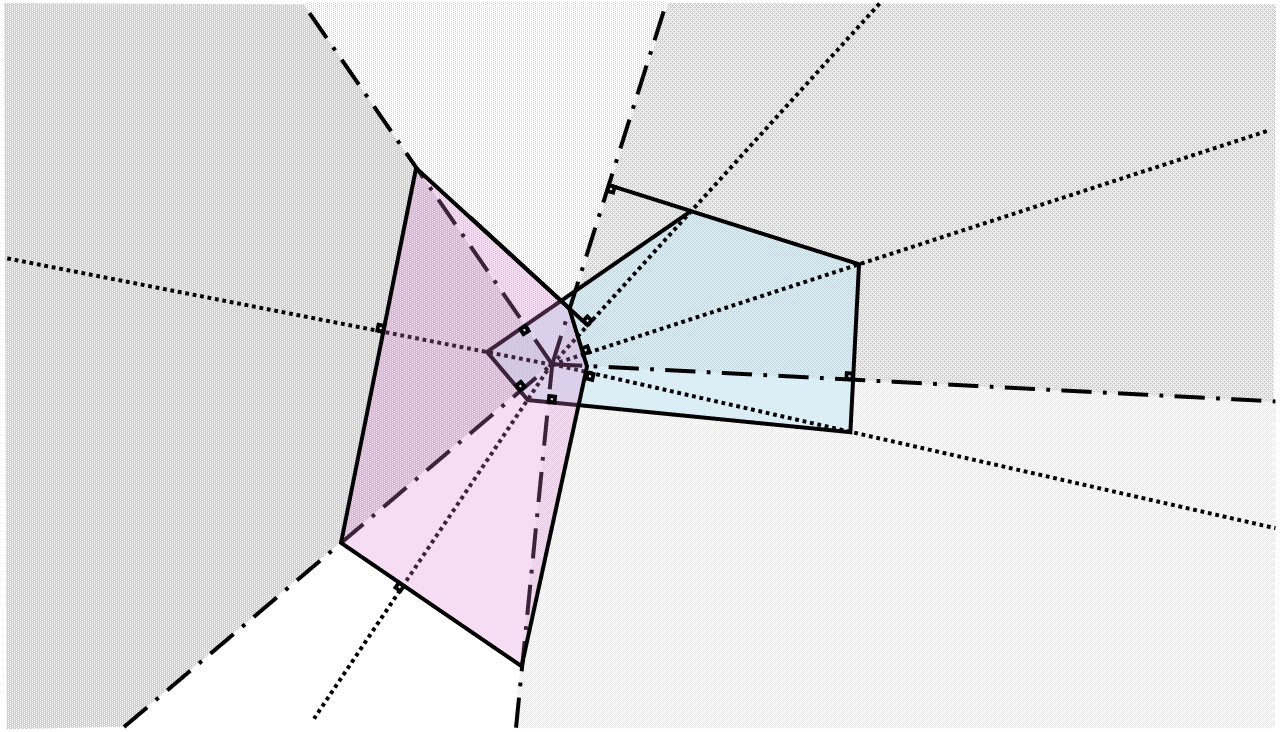}
    \caption{\textbf{Left:} Dual polygons (blue and purple) related by polar reciprocation. The vertices of each polygon lie on rays perpendicular to a side of the other polygon. \textbf{Right:} Solution trajectories for the parameter dynamics of a population evolving under the continuous replicator dynamic (colored curves, colored by value of the Hamiltonian) with initial measure equal to a uniform distribution over a polygon $\Psi$ (shaded grey region). The Dot-dashed outer polygon is dual to $\Psi$. Notice that trajectories with large amplitudes (large Hamiltonian) converge to forms similar to the dual polygon.}
    \label{appfig: Duals}
\end{figure}

The orbits of interest correspond to solutions of the parameter dynamics in a limit where the value of the Hamiltonian on the orbit diverges, or, equivalently, where $\text{inf}_t(\|\theta(t)\|_2)$ diverges. These two conditions are equivalent when $\psi$ is bounded, and its convex hull contains the origin since in this situation the Hamiltonian is defined for all $\theta$, is radially unbounded, convex, and has a unique global minimizer for interior $\theta$. 

Recall that the parameters obey the Hamiltonian dynamic:
\begin{equation}
    \frac{d}{dt} \theta(t) = U \nabla_{\theta} P(\theta) \propto U \nabla_{\theta} \log(P(\theta)) = U \bar{y}(\theta)
\end{equation}
where $\bar{y}(\theta) = \mathbb{E}_{Y \sim p(\theta)}[Y]$ where $p(\cdot,\theta)$ is the normalized measure $\pi(\cdot,\theta)/P(\theta)$. The constant of proportionality relating both sides is the population size $P$. The two sides produce identical trajectories, so we will adopt the convention that all expectations are computed against the normalized population. This corresponds to an agent-based mixing model with constant per capita interaction rates. This convention simplifies the limiting analysis since $P$ is convex, so $\|\frac{d}{dt} \theta(t)\| = \|\nabla_{\theta} P(\theta(t))\|_2$ may diverge as $\|\theta(t)\|_2$ diverges. In contrast, if we use constant per capita interaction rates, then, if $\Psi$ is bounded, $\|\bar{y}(\theta)\|_2$ is bounded, so the parameter dynamics may produce orbits with finite rates as $\|\theta(0)\|_2$ diverges.

Then, following the main text argument, each corner $c_j$ generates an angular region, bounded by the rays parallel to $\hat{n}_{j-1}$ and $\hat{n}_j$, where $\hat{n}_j$ is the normal to the $j^{th}$ boundary of $\text{conv}(\psi)$, such that, if $\hat{\theta}_0$ lies inside the region, then $p(y,a \hat{\theta}_0)$ approaches a delta distribution at the corner $c_j$. In this case, $\lim_{a \rightarrow \infty} \nabla_{\theta} \log(P(\theta))|_{\theta = a \hat{\theta}_0} = \lim_{a \rightarrow \infty} \bar{y}(a \hat{\theta}_0) = c_j$. Since this limit holds for all rays pointing into the $j^{th}$ region, the level sets of the Hamiltonian, and the parameter trajectories, approach straight line segments parallel to $U c_j$, bounded by their intersections with the rays $\{s \hat{n}_{j-1}\}_{s \geq 0}, \{s \hat{n}_{j}\}_{s \geq 0}$. Since the solutions are periodic, the collection of parameter trajectories form a series of concentric orbits. Then, since the angular regions generated by the corners partition all angles about the origin, in the limit of large initial $\theta$, the parameter orbits must approach a series of similar polygons. 

The associated polygon is dual to the polygon $\text{conv}(\Psi)$. Figure \ref{appfig: Duals} shows two examples. The dual has one edge for every vertex in $\text{conv}(\Psi)$, and, as a consequence, one corner for every edge of $\Psi$. These corners correspond to the finite set of initial directions $\hat{\theta}_0$ for which $p(\cdot,a \hat{\theta}_0)$ does not approach a delta distribution at a corner of $\Psi$. 

The limiting dual polygon is similar to the dual of $\text{conv}(\Psi)$ constructed by:
\begin{enumerate}
    \item Partition $\mathbb{R}^2$ into angular segments bounded by rays parallel to the normal vectors of each side of $\text{conv}(\Psi)$.
    \item Pick an initial point away from the origin, within the first angular segment. 
    \item Extend that point into a line segment, perpendicular to the vector pointing to the corner of $\text{conv}(\Psi)$ linking the two boundaries defining the angular region, extended until it intersects with the boundaries of the angular region.
    \item Then, from each point where the line segment enters a new region, extend it by adding a new line segment, perpendicular to the corner of $\text{conv}(\Psi)$ assigned to the associated angular region, and extended until the line segment intersects both boundaries of the angular region. 
    \item Iterate until the process produces a closed orbit. 
\end{enumerate}

Let $P = \text{conv}(\Psi)$ denote the original polygon. Then, this procedure produces a dual polygon, $P'$, where:
\begin{enumerate}
    \item The vector from the origin to the $j^{th}$ vertex of $P'$ is perpendicular to the $j^{th}$ edge of $P$.
    \item The vector from the origin to the $j^{th}$ vertex of $P$ is perpendicular to the $j^{th}$ edge of $P'$.
\end{enumerate}

\begin{figure}[t]
    \centering
    \includegraphics[trim = 150 68 175 40, clip, scale = 0.3]{Figures/Triangle_Oscillator.png}
    \includegraphics[trim = 150 68 175 40, clip, scale = 0.3]{Figures/Pentagon_Oscillator.png}
    \includegraphics[trim = 150 68 175 40, clip, scale = 0.3]{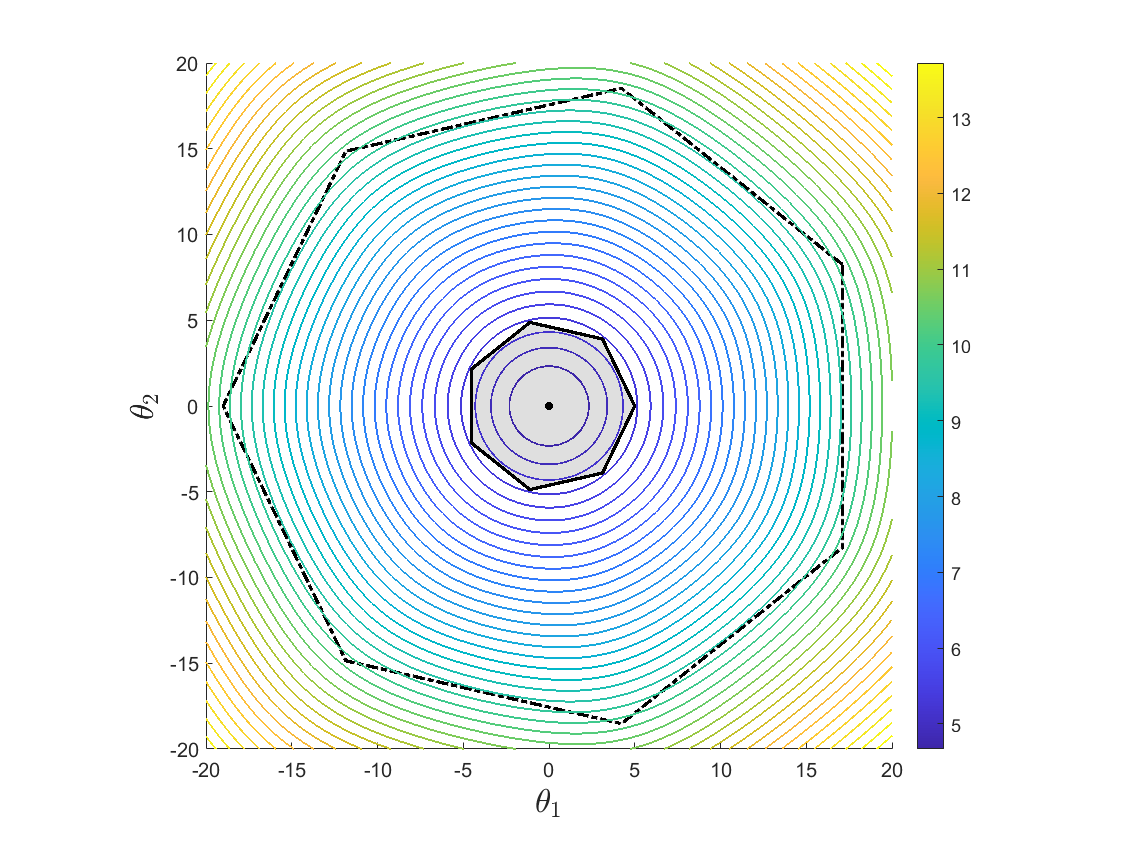}
    \caption{Parameter dynamics for a continuous replicator equation defined in a single disc game, with a base measure equal to a uniform distribution on regular polygons (the shaded grey triangle, pentagon, and septagon). For an irregular example, see Figure \ref{appfig: Duals}. The orbits are colored by Hamiltonian value. Orbits at large values of the Hamiltonian (large amplitude) approach a polygonal form. The outer polygon (dot-dashed outer polygon) is the dual to the domain $\Psi$ represented in grey. Notice that, as the number of vertices increases, the polygon approaches a circle, so the base measure develops an approximate rotational symmetry. As a result, the parameter trajectories converge to concentric circles as the degree of a regular polygon increases.}
    \label{fig: Regular polygons}
\end{figure}

If $\Psi$ is a regular polygon, then the dual polygon associated with limiting orbits is a regular polygon of the same degree, rotated so that its edges are bisected by the vertices of $\Psi$, and so that its vertices bisect the edges of $\Psi$. For example, the square boundary observed for a Laplace base measure is the dual to the diamond shaped level sets of the Laplace density (see Figure \ref{appfig: Duals}). Figure \ref{fig: Regular polygons} shows example trajectories for a uniform measure on a regular triangle, pentagon, and septagon.

For example, consider a uniform base measure on the square $[-1,1]^2$. Then, the Hamiltonian factors, $P(\theta) = (\sinh(\theta_1)/\theta_1) (\sinh(\theta_2)/\theta_2)$ so the parameter dynamics simplify as outlined in Section \ref{app: separable base measures}. Then:
\begin{equation}
    \frac{d}{dt} \theta_1(t) \propto \coth(\theta_2(t)) - \frac{1}{\theta_2(t)}, \quad \frac{d}{dt} \theta_2(t) \propto -\coth(\theta_1(t)) + \frac{1}{\theta_1(t)}.
\end{equation}
with constant of proportionality equal to the initial population size. As usual, converting to a constant per capita interaction rate sets the constant of proportionality to one. If $\theta_1$ and $\theta_2$ are both large, then $\coth(\theta_j) \simeq \text{sign}(\theta_j) + \mathcal{O}(e^{-2 \theta_j})$ and $1/\theta_j$ is small. So, for large $\theta_1$ and $\theta_2$, the dynamics are approximated by:
\begin{equation}
    \frac{d}{dt} \theta_1(t) = \text{sign}(\theta_2(t)) + \mathcal{O}(|\theta_2(t)|^{-1}), \quad \frac{d}{dt} \theta_2(t) = -\text{sign}(\theta_1(t)) + \mathcal{O}(|\theta_1(t)|^{-1}).
\end{equation}

\end{document}